\documentclass[final]{unmeethesis}
\usepackage{BQB}
\usepackage[square,sort,comma,numbers]{natbib}

\begin{document}

\frontmatter


	\title{Open systems dynamics for propagating quantum fields}
	\author{Ben Quinn Baragiola}
	\degreesubject{Ph.D., Physics}
	\degree{Doctor of Philosophy \\ Physics}
	\documenttype{Dissertation}
	\previousdegrees{B.A., University of New Mexico, 2004\\B.S., University of New Mexico, 2005}
	\date{July 2014}

	\maketitle
	\makecopyright

\begin{dedication}

for Frank Kelly, \\you who knew me first.
\end{dedication}
  
\begin{acknowledgments}

Whether or not one is a physicist, we and our natural world are subject to \emph{all} the laws of physics.  Particles dance here and there as waves interfere; every bit of everything interacting with every other bit.  This notion of all-pervasive interconnection does not come naturally, as we rely primarily on our purely local senses and the intuitions that follow.  The first stage of training as a physicist involves abandoning our instinct in favor of universal natural laws.  
But yet, how does this not complicate our description of the world to the point where computation and prediction are impossible?  The second stage is learning to forget what we have just learned and to treat problems using only the essential complications.  This task is delicate, indeed.  For me it spanned the majority of my years of graduate school during which I encountered so many influential and unforgettable people.  These are but a few.

I was initially culled from an exploratory undergraduate quantum mechanics course by JM Geremia, who must be thanked for leading me into theoretical physics before making his mysterious exit.  From my first day as a greenhorn in his lab and then beyond, no one walked beside me more patiently than Rob Cook. He stands surely as a friend and scientific colleague.  Several years later, I careened into the research group of my advisor, Ivan Deutsch. To him I extend my gratitude for his guidance as I learned to separate the vital from the frivolous.  It is now matter-of-fact that an equation must come accompanied by a physical understanding.  Within Ivan's group I began a close collaboration with Leigh Norris, whose innovations and meticulous calculations were the saving grace that brought us to the end of a truly challenging project.  Finally, from Josh Combes, who served as an additional, \emph{de facto} advisor, I learned not to fear the unblazed path or the complex idea castled in Byzantine formalism.  Much of this dissertation sprang directly from his inexhaustible creativity and ceaseless encouragement.  There were many other physicists whose insights, genius, friendship, and cumulative support were critical to my success.   

Beyond the physics department were my mom, dad, sister, and a huge group of friends with varied lives and interests -- architects, actors, engineers, teachers, skaters, psychologists, lovers, craftsmen, lawyers, anthropologists, doctors, musicians, motorcycle mechanics, statesmen, counselors, midnight revelers, mathematicians, mothers, fathers, doctors, builders, artists, biologists, cin\'{e}philes, coders, linguists, entrepreneurs, fellow travelers.  Thanks to you all.



%
%
%
%
%
%

\end{acknowledgments}

\maketitleabstract

\begin{abstract}

In this dissertation, I explore interactions between matter and propagating light.  The electromagnetic field is modeled as a Markovian reservoir of quantum harmonic oscillators successively streaming past a quantum system.  Each weak and fleeting interaction entangles the light and the system, and the light continues its course.  In the context of quantum tomography or metrology one attempts, using measurements of the light, to extract information about the quantum state of the system.  An inevitable consequence of these measurements is a disturbance of the system's quantum state.  These ideas focus on the system and regard the light as ancillary.  It serves its purpose as a probe or as a mechanism to generate interesting dynamics or system states but is eventually traced out, leaving the reduced quantum state of the system 
as the primary mathematical subject. 

What, then, when the state of light itself harbors intrinsic self-entanglement?  One such set of states, those where a traveling wave packet is prepared with a definite number of photons, is a focal point of this dissertation.  These $N$-photon states are ideal candidates as couriers in quantum information processing device.   In contrast to quasi-classical states, such as coherent or thermal fields, $N$-photon states possess temporal mode entanglement, and local interactions in time have nonlocal consequences.  The reduced state of a system probed by an $N$-photon state evolves in a non-Markovian way, and to describe its dynamics one is obliged to keep track of the field's evolution.  I present a method to do this for an arbitrary quantum system using a set of coupled master equations.  



Many models set aside spatial degrees of freedom as an unnecessary complicating factor.  By doing so the precision of predictions is limited. 
Consider a ensemble of cold, trapped atomic spins dispersively probed by a paraxial laser beam.  Atom-light coupling across the ensemble is spatially inhomogeneous as is the radiation pattern of scattered light.  To achieve strong entanglement between the atoms and photons, one must match the spatial mode of the collective radiation from the ensemble to the mode of the laser beam while minimizing the effects of decoherence due to optical pumping.  In this dissertation, I  present a three-dimensional model for a quantum light-matter interface for propagating quantum fields specifically equipped to address these issues.  The reduced collective atomic state is described by a stochastic master equation that includes coherent collective scattering into paraxial modes, decoherence by local inhomogeneous diffuse scattering, and measurement backaction due to continuous observation of the light.  As the light is measured, backaction transmutes atom-light entanglement into entanglement between the atoms of the ensemble.  This formalism is used to study the impact of spatial modes in the squeezing of a collective atomic spin wave via continuous measurement.  The largest squeezing occurs precisely in parameter regimes with significant spatial inhomogeneities, far from the limit in which the interface is well approximated by a one-dimensional, homogeneous model.

\end{abstract}

\tableofcontents

\listoffigures


\mainmatter

	\chapter{Introduction}

The joint power of light, matter, and their interactions for quantum information science lies in the exquisite precision with which experimentalists can control such quantum systems in the laboratory, but also in the detailed mathematical models we use to understand them.  
A quantum description of the electromagnetic field requires a countably infinite Hilbert space associated with \emph{each} mode we wish to describe.  The immensity of the Hilbert space 
has weighed heavy enough to render single-mode approximations commonplace in many cases where a quantum treatment is necessary.  However, with restricted models the predictive potential is limited.  When further accuracy in needed, ultimately we are forced to concede that nothing is a closed system and the influence of the full electromagnetic field must be included.  Accompanying the increased complexity is a richness in the models that allows for deeper understanding of the fundamental light-matter interactions.  In this dissertation, I study quantum systems interacting with propagating quantum fields, which requires such a multi-mode description.  I rely on the theory of open quantum systems, where one balances the universality of coupling to all the modes of the electromagnetic field with a subjective division into a \emph{system} and an \emph{environment}.  The power of this theory lies in placing the divider in such a way that the system can be described relatively simply and the quantum state of the environment can be ignored while while its effects on the system are retained.  The penalty is that the reduced system does not in general admit a description as a pure quantum state, rather we are obliged to use a statistical weighting over pure states -- a density matrix.  

The theoretical toolbox of open systems for quantum optics is packed with semi-adjustable gadgets and esoteric contraptions, each with its purposes.  The bulk of this dissertation is dedicated to extensions to current tools that allow a more precise description of light-matter interactions.  The first is a master equation description for a quantum system interacting with a traveling wave packet of definite photon number.  The second is a three-dimensional model for an atomic ensemble interacting with a paraxial laser field.  For dipole-trapped atomic clouds probed by a paraxial laser, such a model is necessary to fully characterize the inhomogeneous light-matter coupling.




\section{Quantum systems interacting with $N$-photon states}

Nonclassical states of light are important resources for quantum metrology \cite{GioLor11, Leroux11}, secure communication \cite{BevGrangier02}, quantum networks \cite{Kimble08, MoeMauOlm07, Agh11}, and quantum information processing \cite{Nielsen05,KLM01}. Of particular interest for these applications are traveling wave packets prepared with a definite number of photons in a continuous temporal mode.  As the generation of such states becomes technologically feasible \cite{BullColl10, Varcoe04, Yamamoto06,  Zeilinger06, Walmsley08, Silberberg10, Kuhn11, LeePatPar12, McKBocBoo04, SpecBocMuc09, KolBelDu08}, a theoretical description of the light-matter interaction becomes essential. 

A natural approach to such problems is through the input-output formalism of Gardiner and Collett \cite{Cav82,YurDen84,ColGar84,GarCol85}. A central result of input-output theory is the Heisenberg-Langevin equation of motion driven by \emph{quantum noise} that originates from the continuum of harmonic oscillator field modes \cite{GardinerBook,CleSch10}. The application of input-output theory to open quantum systems has historically been restricted to Gaussian fields \cite{GarCol85,DumGar92,GardinerBook} ---vacuum, coherent, thermal, and squeezed.  $N$-photon states are distinct from quasi-classical Gaussian states in that they feature temporal mode entanglement that manifests in temporal correlations.  This is why intensity correlation measurements are used to diagnose single-photon states.  A quantum system interacting with an $N$-photon state at time $t$ becomes entangled the entirety of the field state, including the portion that has not yet reached it.  This entanglement precludes the use of a standard Markovian master equation description of the system's reduced dynamics.  

One approach to modeling these reduced dynamics is to enlarge the Hilbert space under consideration to include a photon ``source."  For instance, a single photon of arbitrary shape can be modeled as the output of a cavity with a controllable decay rate \cite{GJNphoton}.  By feeding the output of the source into the system of interest using a cascaded systems approach \cite{Car93}, one can use a standard Markovian master equation under vacuum to describe the joint state of the source/system.  Tracing over the source then gives the reduced system state.  The cascaded approach has a straightforward physical understanding; however, it requires a source model for input states one would like to consider.  Recently, it has been shown that a host of interesting nonclassical field states can be modeled as the output of a modulated cavity, including multi-photon states and photonic cat states \cite{GouZha14}.  

An alternate approach, first developed in Ref. \cite{Zoolander} uses a system of coupled master equations.  The equation that describes the physical state couples to a set of auxiliary ``reference" states that keep track of the necessary degrees of freedom in the field.  Again, when one consider the set of states as a whole, Markovian evolution under vacuum input applies.  Similar coupled equations for Heisenberg-picture operators of a two-level atom were independently developed in Refs. \cite{DomHorRit02, WanSheSca10}.

Using a variety of methods including those discussed above, aspects of quantum systems interacting with a propagating single-photon state have been examined. In addition to master equations these include two-time correlation functions \cite{DomHorRit02}, properties of scattered light \cite{DroHavBuz00, SheFan05, DomHorRit02, Kos08, ZhoGonSun08, LonSchBus09, Roy10, ZheGauBar10, CheWubMor11, Ely12}, optimal pulse shaping for excitation \cite{StoAlbLeu07, DuBel09, WanSheSca10, StoAlbLeu10, RepSheFan10, Ely12}, and comparisons with coherent state input \cite{WanSca12}.  Such studies have rarely been applied beyond two-level systems, nor have field states where $N \gg 1$ often been considered.

In Chapter \ref{Ch::NPhotonME} of this dissertation we present a unifying method, based on the coupled master equations of Ref. \cite{Zoolander}, to describe the reduced system dynamics of an arbitrary quantum system as it interacts with a propagating $N$-photon state.  From the form of the light-matter interaction, one begins by identifying a set of field states in the Fock basis that couple to the physical state, each of which has an associated ``reference" system state.  The result is a set of intercoupled master equations that are propagated as a whole.  With this technique one can describe a wide variety of input field states including superpositions and mixtures of Fock states, spectrally correlated $N$-photon states, and multi-mode, multi-photon states.

\section{Three-dimensional quantum interface for atomic ensembles}

Atomic ensembles interacting with optical fields have proven to be powerful tools in quantum science with applications that include quantum communication \cite{DuaZol02, MatKuz04}, quantum memory \cite{FleLuk02, JulPol04, ChoKim08}, continuous variable quantum computing \cite{BraLook05}, and metrology \cite{AppMet09, VulMet}.  Measurement of the light entangled with an atomic ensemble plays a critical role in many applications, providing the necessary nonlinearity for remote entanglement \cite{DLCZ} and the backaction for quantum nondemolition (QND) spin squeezing \cite{KuzMan98, TakTak05}.  Continuous measurement of atomic ensembles has been used for the production of spin-squeezed states \cite{KuzBig00}, Faraday spectroscopy \cite{SmiJes03}, high-bandwidth magnetometry \cite{ShaRom10}, quantum state tomography \cite{RioDeu11}, and optimal phase estimation \cite{YonFur12}. 

At the heart of these protocols is the strong coupling between a quantum mode of the field and an effective collective spin of the ensemble. This coupling can generate entanglement between atoms and photons, such that measurement of the light yields strong quantum backaction on the atoms.  Photons can also enable a quantum data bus for entangling atoms with one another. Further, neutral atomic spins are a robust, controllable resource \cite{DeuJes10}.  Enhancing the atom-light interface is thus essential for improving the performance of quantum technologies and for reaching new regimes where a quantum advantage becomes manifest.  This can be achieved through confined modes such as in optical cavities~\cite{Kimble2005, VulMet, CheTho11} or waveguides in optical nanostructures~\cite{VetRau10, Waks2012, Kimble2013}. 

Strong atom-photon coupling occurs in free space when photons are indistinguishably scattered by the ensemble and interference enhances the radiation into the probe mode relative to diffuse scattering into $4\pi$ steradians~\cite{TanVul11, BieKai13}.  Early experiments demonstrated such strong coupling and entanglement in vapor cells where a one-dimensional description of plane wave modes and uniform atomic density is applicable \cite{KuzBig99, JulPol01, JulPol04}.  More recently, experiments have employed ensembles of ultracold atoms in pencil-shaped dipole traps probed by highly focused laser beams \cite{KubMit09, KosMit10, KamMul12}. When the radiation pattern of the light scattered from the average atomic ensemble is well matched with the paraxial mode of the probe, the spatial mode of the scattered photons is effectively indistinguishable from the probe.  In this case the probe mode becomes strongly entangled with a collective variable of the atomic ensemble. Such geometries have the potential to enhance the atom-photon quantum interface, but their description is more complex and requires a full treatment of scattering, diffraction, inhomogeneous coupling, and decoherence.   

Harnessing the advantages of these atomic ensembles thus demands a three-dimensional quantum theory of the underlying interaction, including both coherent coupling and quantum noise.  Significant progress has been made recently in the development of such a model. Mode matching of the scattered light to the spatial mode of the probe laser, including the effects of diffraction, has been studied using a semiclassical scattering model \cite{MulPol05}.  A rigorous field-theoretic treatment separates the mean-field classical effects from the quantum fluctuations and noise, including the spatial inhomogeneities of the atomic and light modes~\cite{SorSor08}.   Models that include spatial modes have been developed in a variety of contexts~\cite{KuzKen04, Windpassinger2008, KosMit09, SauSta10}. Applications include remote entanglement via collective Raman scattering in a DLCZ-type protocol \cite{DuaZol02, SorSor09} multi-mode quantum memories \cite{ZueGro11}.  From such studies, it is clear that not only can one-dimensional models not only fail to describe relevant coherent and incoherent effects, but they also overlook spatial degrees of freedom as a resource \cite{GroSor12, HigBuc12}.

In Chapter \ref{Ch::3DInterface} we present a theoretical model for a three-dimensional quantum interface for a cloud of multi-level alkali atoms interacting dispersively with a paraxial laser.  The model rests on a transverse spatial mode decomposition of the propagating paraxial quantum field, which allows us to identify the collective spin waves that couple to each of the field modes.  In addition to the coherent coupling that acts collectively across the ensemble, diffuse scattering of photons leads to decoherence that acts locally on the atoms at a rate proportional to the local probe intensity.  A proper accounting of the balance between coherent coupling and decoherence is especially challenging given the tensor nature of the atom-photon interaction for alkali atoms.  

 
Through the interaction, information about the quantum state of the atoms is coherently mapped onto the light as it propagates through the ensemble.  Measuring the light retrieves this information, and the atomic state can be conditioned on the measurement result.  The indistinguishability of contributions to the measured light from atoms throughout the ensemble generates entanglement.  The conditional dynamics of the collective atomic state can be formalized in a \emph{stochastic master equation}, which we derive for continuous polarimetry measurements.  The dynamics of the collective atomic state include the effects of measurement backaction, collective decoherence from unmeasured paraxial light, and local decoherence from diffuse photon scattering that gives rise to optical pumping. The model should be broadly applicable to protocols where a strong, free-space, atom-light interface is essential, and where measurement backaction may be a tool for induced atom-atom interactions.



In Chapter \ref{Ch::SpinSqueezing} we employ the three-dimensional atom-light interface to study QND squeezing of spin waves via the Faraday effect \cite{KuzBig00,KosMit10, Takano2009}. In this protocol, the key interaction is the off-resonant scattering of horizontally polarized photons into vertical polarization.  Measurement in a balanced polarimeter corresponds to a homodyne measurement of the scattered photons.  The degree of scattering into the local oscillator, defined by the paraxial laser mode, determines the measurement strength and the resulting backaction that generates spin squeezing.  However, counteracting the squeezing are the damaging effects of decoherence from diffuse scattering.  Optimal squeezing results from a geometry-dependent balance of coherent squeezing and incoherent optical pumping.  We use numerical simulations to help build physical intuition about the three-dimensional atom-light interface and to investigate how the model can be used to optimize an experimental design.  We find that the greatest squeezing occurs in parameter regimes where spatial inhomogeneities are significant, far from the limit in which the interface is well approximated by a one-dimensional, homogeneous model.


\section{Structure of the dissertation}

The remainder of this dissertation is organized as follows.  In Chapter \ref{Ch::PropagatingFields} we give a review of two quantization schemes for propagating quantum fields, both of which are used throughout this dissertation. Interactions with matter in a weak coupling regime are described with quantum stochastic differential equations (QSDEs), which are briefly reviewed\footnote{In the words of Joseph Kerchoff \cite{KerchoffThesis}, the formalism of QSDEs is presented as an ``incredibly useful...tool, not an object of study in itself."}.  In Chapter \ref{Ch::NPhotonME} we derive the master equations for systems interacting with various types of $N$-photon states, and examples are presented to aid understanding.  Chapter \ref{Ch::DispersiveInt} gives the essential details for the coupling of an off-resonant electric field to the hyperfine spin of a single alkali atom.  This description is extended to include spatial degrees of freedom for a collection of atoms in Chapter \ref{Ch::3DInterface}.  The result is a model for a three-dimensional quantum interface for atomic ensembles.  In Chapter \ref{Ch::SpinSqueezing}, this model is used to study the squeezing of spin waves in an atomic ensemble.  Numerical results point to preferable geometries for spin squeezing.  Finally, in Chapter \ref{Ch::Conclusion} we summarize the key results and provide directions for future research and enquiry.

\section{Publications and papers in preparation}
\begin{itemize}
	\item B. Q. Baragiola and J. Combes.  \emph{Quantum trajectories for systems probed with propagating Fock states}, in preparation.
	\item B. Q. Baragiola, L. Norris, E. Monta\~{n}o, P. Mickelson, P. Jessen, and I. H. Deutsch, \emph{Three-dimensional light-matter interface for spin squeezing in atomic ensembles}, Phys. Rev. A {\bfseries 89}, 033850 (2014).
	\item S. R. Sathyamoorthy, L. Tornberg, A. F. Kockum, B. Q. Baragiola, J. Combes, C. M. Wilson, T. M. Stace, and G. Johansson, \emph{Quantum nondemolition measurement of a propagating microwave photon}, Phys. Rev. Lett. {\bfseries 112}, 093601 (2014). 
	\item B. Q. Baragiola, R. L. Cook, A. M. Bra\'nczyk, and J. Combes, \emph{$N$-photon wave packets interacting with an arbitrary quantum system}, Phys. Rev. A {\bfseries 86}, 013811 (2012).
	\item B. Q. Baragiola, B. A. Chase, and JM Geremia, \emph{Collective uncertainty in partially-polarized and partially-decohered spin-1/2 systems}, Phys. Rev. A {\bfseries 81}, 032104 (2010).
	\item B. A. Chase, B. Q. Baragiola, H. L. Partner, B. T. Black, and JM Geremia, \emph{Magnetometry via a double-pass continuous quantum measurement of atomic spin}, Phys. Rev. A {\bfseries 81} 032104 (2010).
\end{itemize}

	\chapter{Propagating quantum fields} \label{Ch::PropagatingFields}

\section{Introduction}

The interaction between quantum light and matter serves as the foundation for quantum optics upon which a smorgasbord of theoretical and technological innovations rests.  When presenting such a highly developed and detailed formalism one runs the risk of falling down the rabbit hole and including far more than necessary.  Claiming to have leapt this pitfall altogether would be exceedingly dishonest, but at least by keeping this caveat in mind, I hope to have trimmed down the content to a useful yet manageable level.  The goal of this chapter is to lay out the majority of the mathematical tools that are put to specific uses in the following chapters.  For this reason a reader who finds this chapter dense and in some cases needlessly detailed may proceed to the following chapters, returning here only for reference.  
	
The description of propagating quantum fields has historically proceeded along several parallel routes.  In this chapter we will present a quantization of free-space paraxial fields from Ref. \cite{BarDeu14}, which is based on the idea of paraxial field states introduced by Deutsch and Garrison in Ref. \cite{DeuGar91}.  From this quantization scheme arises a set of creation and annihilation operators, defined with respect to a slowly varying envelope, that form the backbone of the analysis in Chapters \ref{Ch::DispersiveInt}, \ref{Ch::3DInterface}, and \ref{Ch::SpinSqueezing}.  A large body of work relies on an alternate description, that of \emph{continuous-mode quantum optics}, introduced by Blow et al. \cite{BloLou90}.  This theory has been folded into the widely-used description of light-matter interactions known as input-output theory \cite{GardinerBook}. The equivalence of the methods will be shown,  
as throughout this thesis we make use of both.
		 
For many quantum optical situations it is appropriate to make any number of simplifying approximations which can reduce the complexity of the resulting equations or transform them to more mathematically tractable forms.  As with all complex physics no model is right, what we seek is a model that is \emph{not wrong}.  We are primarily interested in electric fields which arrive, interact with a quantum system, and then propagate away, possibly towards a detector, potentially carrying with them some information acquired from the interaction.  In such a description the direction of propagation plays a special role as it becomes, in a sense which we hope to clarify, interchangeable with time.  For the situations considered in the remainder of this dissertation, we will focus on quasi-monochromatic fields coupling to quantum systems where the rotating wave approximation can be made, and the scattered fields are well described in the first Born approximation.  
	
%

\section{Classical paraxial electric fields} \label{Sec::ClassicalParaxialFields}
				
We begin with a classic description of free-space paraxial electric fields in the absence of sources or sinks. 
The classical electric field is the solution to the wave equation,
	\begin{align} \label{Eq::WaveEquation}
		\left( \nabla^2 - \frac{1}{c^2} \frac{\partial^2}{\partial t^2} \right) \mathbf{E}(\mathbf{r},t) = 0,
	\end{align}
where the electric field is represented as the real part of a complex, quasi-monochromatic vector field with carrier frequency $\omega_c$,
	\begin{align}  \label{Eq::ComplexElectricEnvelope}
		\mathbf{E}(\mathbf{r},t) = \mbox{Re}\big[ \vec{\mathcal{E}}(\mathbf{r},t) e^{i (k_c z - \omega_c t)} \big],
	\end{align}
with free-space dispersion, $\omega_c = c |k_c|$.  Within this description, the $z$-direction has already been established as a ``preferred" spatial direction.  The conditions for the slowly varying envelope approximation are
	\begin{equation} \label{Eq::ParaxialConditions}
		\bigg| \frac{\partial{\vec{\mathcal{E}}}}{\partial t} \bigg| \ll \omega_c | \vec{\mathcal{E}} |, \quad \quad \bigg| \frac{\partial{\vec{\mathcal{E}}}}{\partial z} \bigg| \ll k_c | \vec{\mathcal{E}} |.
	\end{equation}
	That is, the electric field envelope varies slowly in time compared to the carrier frequency $\omega_c$ and slowly in space compared to the wave number $k_c$.  Plugging \erf{Eq::ComplexElectricEnvelope} into \erf{Eq::WaveEquation} and neglecting terms according to \erf{Eq::ParaxialConditions}, gives the homogeneous paraxial wave equation,
	\begin{align} \label{Eq::HomParaxialWaveEquation}
		i \left( \frac{\partial}{\partial z} + \frac{1}{c}  \frac{\partial}{\partial t} \right) \vec{\mathcal{E}}(\mathbf{r}, t) = - \frac{1}{2 k_c} \nabla^2_\perp \vec{\mathcal{E}}(\mathbf{r},t),
	\end{align}
where $\nabla_\perp^2$ is the transverse Laplacian,
	\begin{align}
		\nabla_\perp^2 \equiv \frac{\partial^2}{\partial x^2} + \frac{\partial^2}{\partial y^2}.
	\end{align}

We now make an ansatz that the slowly varying envelope factors into two functions, $\vec{\mathcal{E}}(\mathbf{r},t) = \mathcal{A}(z,t) \vec{\mathcal{U}}(\mathbf{r}),$ where $\mathcal{A}(z,t)$ is the temporal pulse envelope and $\vec{\mathcal{U}}(\mathbf{r})$ is the vector-valued spatial mode function.  The paraxial wave equation \erf{Eq::HomParaxialWaveEquation} becomes separable and yields an independent differential equation for each.  

	The pulse envelope, satisfying
	\begin{align} \label{Eq::PulseEnvelope}
		\left( \frac{\partial}{\partial z} + \frac{1}{c}  \frac{\partial}{\partial t} \right) \mathcal{A}(z,t) = 0,
	\end{align}
has solutions of the form $\mathcal{A}(z,t) = f(t - z/c)$ for any function $f$ that complies with the conditions in \erf{Eq::ParaxialConditions}.  To see this one makes the substitution $\tau \equiv t - z/c$, where $\tau$ is the \emph{retarded time}.  Total differentials can be used to show
	\begin{align}
		\left( \frac{\partial}{\partial z} + \frac{1}{c}  \frac{\partial}{\partial t} \right) f(\tau) 
		& = 0.
	\end{align}
Thus, any choice of $\mathcal{A}(\tau)$ satisfies \erf{Eq::PulseEnvelope}.

We now turn to the function $ \vec{\mathcal{U}}(\mathbf{r})$ that describes the spatial dependence of the slowly varying envelope.  In homogeneous media such as free space, the transverse polarization components decouple and can be treated independently using a scalar function $\mathcal{U}(\mathbf{r}_\perp, z)$.  On occasions where the distinction is important, we explicitly separate the transverse and longitudinal coordinates within the parentheses to emphasize the fact that the longitudinal propagation coordinate $z$ plays a different role than the transverse spatial coordinates $\mathbf{r}_\perp$. The spatial function satisfies the homogeneous paraxial Helmholtz equation,
	\begin{align} \label{Eq::ParaxialHelmholtz}
		 \left( -i \frac{\partial}{\partial z} + \frac{1}{2 k_c}  \nabla_\perp^2 \right) \mathcal{U}(\mathbf{r}_\perp,z) = 0.
	\end{align}
Solutions to this equation can be decomposed in an orthonormal set of dimensionless transverse mode functions, $\{ u_{i}(\mathbf{r}_\perp,z) \}$, with mode label $i$.  Normalizing to an effective transverse area $A$, the transverse modes enjoy several properties.  First, they are orthonormal in every longitudinal plane designated by $z$,
	\begin{align} 
		\int d^2 \mathbf{r}_\perp u^*_{i} (\mathbf{r}_\perp , z) u_{j} (\mathbf{r}_\perp , z) & = A \, \delta_{i, j},\label{Eq::TransverseOrthogonality}
	\end{align}
and second, they form a complete basis in that plane	
	\begin{align}
		\sum_{i} u_{i}(\mathbf{r}_\perp , z) u^*_{i}(\mathbf{r}_\perp' , z)  &=  A \, \delta^{(2)}(\mathbf{r}_\perp-\mathbf{r}_\perp'). \label{Eq::Completeness}
	\end{align}
Between two different longitudinal planes the mode functions are interconnected by the classical propagator, which we will see in the following subsection.  The Laguerre-Gauss modes, described in Appendix \ref{Appendix::LGModes}, are one such set that we will make use of in Chapter \ref{Ch::SpinSqueezing}.

	\subsection{Classical paraxial scattering} \label{Appendix::ParaxialScattering}
	
We would like to describe the output electromagnetic field for a system probed by an input field.  The system's polarizability determines its response to the input field and the nature of the induced radiation.  In Maxwell's equations, this corresponds to an induced current, or macroscopic polarization density, that acts as a \emph{source} term in the wave equation,
	\begin{align} \label{Eq::WaveEquationWithSource}
		\left( \nabla^2 - \frac{1}{c^2} \frac{\partial^2}{\partial t^2} \right) \mathbf{E}(\mathbf{r},t) = \frac{4 \pi}{c} \frac{\partial^2}{\partial t^2} \mathbf{P}(\mathbf{r},t),
	\end{align}
Within the paraxial approximation, the slowly varying envelope is then governed by,
	\begin{align}
		& i\bigg(\frac{\partial}{\partial z}  + \frac{1}{c}\frac{\partial}{\partial t} \bigg)  \vec{\mathcal{E}}(\mathbf{r}, t) = -\frac{1}{2 k_c}\grad_\perp^2 \vec{\mathcal{E}}(\mathbf{r}, t)-2 \pi k_c \tensor{\chi}(\mathbf{r}) \cdot \vec{\mathcal{E}}(\mathbf{r}, t),
	\end{align}
where $\tensor{\chi}(\mathbf{r})$ is the spatially averaged dielectric susceptibility \cite{GarChiBook}.  
As above, a transformation to a comoving frame with the retarded time, $\tau= t-z/c$, yields a factorized solution, with the spatial function satisfying the paraxial Helmholtz equation,
	\begin{equation}
		\label{paraxEq}
		i\frac{\partial}{\partial z} \vec{\mathcal{U}}(\mathbf{r}_\perp , z)  = -\frac{1}{2 k_c}\grad_\perp^2 \vec{\mathcal{U}}(\mathbf{r}_\perp , z) - 2 \pi k_c \tensor{\chi}(\mathbf{r}_\perp, z) \cdot \vec{\mathcal{U}}(\mathbf{r}_\perp , z).
	\end{equation}

Equation (\ref{paraxEq}) is isomorphic to the time-dependent Schr\"{o}dinger equation with the propagation distance $z$ playing the role of time and the susceptibility playing the role of the potential \cite{BarDeu14}.  As such, we can define a Hilbert space of square-integrable functions in a transverse plane and use Dirac notation to express the evolution of the scalar function $\mathcal{U}(\mathbf{r}_\perp, z)$ as a function of $z$:
	\begin{align}
	 	\mathcal{U}(\mathbf{r}_\perp , z)=\langle \mathbf{r}_\perp | \mathcal{U}(z)\rangle .
		\end{align}
In representation-free operator form, the free-space propagator, $\hat{K}(z-z')$, that generates $z$-evolution, 
	\begin{align}
		\ket{\mathcal{U}(z)} = \hat{K}(z-z') \ket{\mathcal{U}(z')},
	\end{align}
for $z \geq z'$ satisfies the free-particle Schr\"{o}dinger equation in two dimensions,
	\begin{equation}
		i\frac{\partial }{\partial z} \hat{K}=\frac{\hat{\mbf{p}}_\perp^2}{2 k_c} \hat{K}.
	\end{equation}
The solution, 
	\begin{align}
		\hat{K}(z-z')= \exp \left[{-i\frac{\hat{\mbf{p}}_\perp^2}{2 k_c} (z-z')} \right],
	\end{align}
has the familiar position-space representation, using $\hat{\mbf{p}}_\perp = -i \nabla_\perp$, for the spreading of a wavepacket and Fraunhofer diffraction~\cite{Newton1982},
	\begin{align}
		K(\mathbf{r}_\perp-\mathbf{r}_\perp',z-z') &=\langle \mathbf{r}_\perp | \hat{K}(z-z') | \mathbf{r}_\perp' \rangle \nn\\
		&= \frac{-i k_c}{2\pi (z-z')}\exp\left[ \frac{ik_0 |\mathbf{r}_\perp - \mathbf{r}_\perp'|^2}{2(z-z')} \right].  \label{Eq::ParaxialPropagator}
	\end{align}
This equation for the classical \emph{paraxial propagator} can also be found by making the paraxial approximation on the three-dimensional, free-space Green's function for outgoing waves\footnote{One must be mindful of the units when performing this operation.  The free-space Green's function that solves the full Helmholtz equation has units $1/V$, where as \erf{Eq::ParaxialPropagator} has units $1/A$.  Transforming from the full wave equation, \erf{Eq::WaveEquation} to the paraxial wave equation, \erf{Eq::HomParaxialWaveEquation}, we have divided by the carrier wave number $k_c$. }.

When the spatial function is known in a transverse plane at longitudinal plane $z'$, the longitudinal evolution for a freely propagating paraxial field is found using the paraxial propagator, \erf{Eq::ParaxialPropagator}.  At longitudinal position $z$, the spatial function is given by
	\begin{align}
		\mathcal{U}(\mathbf{r}_\perp , z) &= \langle \mathbf{r}_\perp | \mathcal{U}(z)\rangle \nn \\
			&= \langle \mathbf{r}_\perp | \hat{K}(z-z') |\mathcal{U}(z')\rangle \nn \\
			& = \langle \mathbf{r}_\perp | \hat{K}(z-z') \left( \int d^2\mathbf{r}_\perp''\; \op{ \mathbf{r}''_\perp}{ \mathbf{r}''  _\perp} \right) |\mathcal{U}(z')\rangle \nn \\ 
			&= \int d^2\mathbf{r}_\perp'\; K(\mathbf{r}_\perp-\mathbf{r}_\perp',z-z') \mathcal{U}(\mathbf{r}_\perp' , z'). \label{Eq::propint}
	\end{align}
Other properties of the propagator follow from unitarity, $\hat{K}^\dag(z-z')=\hat{K}(z'-z)$, and thus
	\begin{eqnarray}
		\label{backprop}
		\mathcal{U}^* (\mathbf{r}_\perp', z') \nonumber \nonumber&=& \langle \mathbf{r}_\perp' | \hat{K}(z'-z)| \mathcal{U} (z) \rangle^* =   \langle \mathcal{U} (z)  | \hat{K}(z-z')| \mathbf{r}_\perp' \rangle \nonumber\\
& =& \int  d^2 \mathbf{r}_\perp  \mathcal{U}^* (\mathbf{r}_\perp, z) \, K(\mathbf{r}_\perp-\mathbf{r}_\perp', z-z')   .
	\end{eqnarray}
In analogy with the previous section, we define a complete basis, $\{\ket{u_{i}(z)}\}$, that can be used to express the propagator as
	\begin{equation} \label{Eq::KOperator}
		\hat{K}(z-z')= \sum_{i} \ket{u_{i}(z)} \bra{u_{i}(z')}. 
	\end{equation}
In the position representation, the dimensionless basis functions are found by projecting onto the transverse position eigenkets $\ket{\mathbf{r}_\perp}$, which have units $1/\sqrt{A},$
	\begin{align}
		u_i(z) = \sqrt{A} \ip{\mathbf{r}_\perp}{u_i(z)}
	\end{align}
are normalized to a fixed transverse area $A$, [\erf{Eq::TransverseOrthogonality}],
	\begin{equation}
		\langle u_{j}(z) |u_{i}(z) \rangle = \frac{1}{A} \int d^2\mathbf{r}_\perp u^*_{j}(z) u_{i}(z) = \delta_{i,j}.
	\end{equation}
Then, the position-space representation of the propagator, as in \erf{Eq::propint}, is
	\begin{align} \label{Eq::KPhysical}
		K(\mathbf{r}_\perp-\mathbf{r}_\perp', z-z') = \frac{1}{A}\sum_{i} u^*_{i}(\mathbf{r}_\perp', z') u_{i}(\mathbf{r}_\perp, z),
	\end{align}
with the boundary condition\footnote{In some sense, this is more of an initial condition than a boundary condition.} $K(\mathbf{r}_\perp-\mathbf{r}_\perp', 0) = \delta^{(2)} (\mathbf{r}_\perp'-\mathbf{r}_\perp) $ that follows from completeness.  

The scattering of paraxial fields thus follows in complete analogy to the scattering of nonrealistic Schr\"{o}dinger waves~\cite{Newton1982}, where the time-dependent formulation of scattering translates into $z$-dependence.  
In the first Born approximation that applies for dilute samples where multiple scattering is negligible, given an incident field (free propagating solution) $\vec{\mathcal{U}}_{\rm in} (\mathbf{r}_\perp,z)$, the total scattering solution is
	\begin{align}
		\vec{\mathcal{U}} (\mathbf{r}_\perp, z) 
  		 = \vec{\mathcal{U}}_{\rm in} (\mathbf{r}_\perp,z) + i 2 \pi k_c \int_{-\infty}^z dz' \int d^2 \mathbf{r}_\perp' K(\mathbf{r}_\perp-\mathbf{r}_\perp', z-z') \tensor{\chi}(\mathbf{r}_\perp', z')  \cdot \vec{\mathcal{U}}_{\rm in} (\mathbf{r}_\perp',z'), \label{Eq::ClassicalScattering}
	\end{align}
corresponding to the superposition of incident and reradiated fields.

%

\section{Quantization of the paraxial electric field} \label{Sec::ParaxialQuantization}
	
	Paraxial quantization follows from the slowly varying envelope approximation detailed in the previous sections \cite{DeuGar91}.  For these modes, we define the positive-frequency component of the electric field analogous to a classical beam
	\begin{equation}
		\hat{\mbf{E}}^{(+)}(\mbf{r}, t) = \sqrt{2 \pi \hbar \omega_c} \sum_\lambda \mbf{e}_\lambda \hat{\Psi}_\lambda (\mathbf{r}_\perp,z,t) e^{i(k_c z - \omega_c t)},
	\end{equation}
where $\lambda$ labels transverse polarizations, and the slowly varying envelope satisfies the equal-time commutation relations of a nonrelativistic bosonic field,
	\begin{equation}
		\big[ \hat{\Psi}_\lambda(\mathbf{r}_\perp,z,t), \hat{\Psi}^\dag_{\lambda'}(\mathbf{r}_\perp',z',t) \big] = \delta_{\lambda, \lambda'} \delta^{(2)} (\mathbf{r}_\perp - \mathbf{r}_\perp') \delta (z-z').
	\end{equation}
The appearance of space-local $\delta$-functions in the commutation relations is a reflection of the fact that the slowly varying envelope approximation smears over the nonlocal features in the exact commutation relations.  These $\delta$-functions must be understood as being coarse-grained over volumes large compared to a cubic wavelength, $\lambda_c^3$ \cite{GarChiBook}.

The free field satisfies the homogeneous paraxial wave equation,
	\begin{equation}
		i \frac{\partial}{\partial t} \hat{\Psi}_\lambda = -ic \frac{\partial}{\partial z} \hat{\Psi}_\lambda -\frac{1}{2 k_c} \grad^2_\perp \hat{\Psi}_\lambda,
	\end{equation}
which is the Heisenberg equation of motion for a forward-propagating envelope governed by the free paraxial Hamiltonian,
	\begin{equation} \label{Eq::ParaxialHamiltonian}
		\hat{H}_{\text{free}}=   \hbar \sum_\lambda \int d^3 \mbf{r} \, \hat{\Psi}^\dag_\lambda  \left(- i c \frac{\partial}{\partial z}  -\frac{1}{2 k_c} \grad^2_\perp  \right) \hat{\Psi}_\lambda .
	\end{equation}
The free field solution is thus determined by the classical propagator,
\begin{equation} \label{Eq::ParaxialFreeFieldSolution}
\hat{\Psi}_{\lambda}(\mathbf{r}_\perp, z, t) = \int d^2 \mathbf{r}_\perp' K(\mathbf{r}_\perp-\mathbf{r}_\perp',ct) \hat{\Psi}_\lambda (\mathbf{r}_\perp', z-c t,0).
\end{equation}
It then follows that the free field satisfies the general commutation relations,
	\begin{align}
		\big[ & \hat{\Psi}_\lambda(\mathbf{r}_\perp,z,t), \hat{\Psi}^\dag_{\lambda'}(\mathbf{r}_\perp',z',t')  \big] =  \delta_{\lambda, {\lambda'}} \delta \left(z-z'-c(t-t')\right)  K(\mathbf{r}_\perp - \mathbf{r}_\perp',z-z')  , 
	\end{align}
and thus equal-$z$, unequal-$t$ commutation relations,
	\begin{equation}
		\big[ \hat{\Psi}_\lambda(\mathbf{r}_\perp,z,t), \hat{\Psi}^\dag_{\lambda'}(\mathbf{r}_\perp',z,t') \big] =\frac{1}{c} \delta_{\lambda, {\lambda'}} \delta^{(2)} (\mathbf{r}_\perp - \mathbf{r}_\perp') \delta (t-t') .
	\end{equation}

	As discussed above, the paraxial field is naturally decomposed into an orthonormal set of dimensionless transverse mode functions, $\{ u_{i}(\mathbf{r}_\perp,z) \}$.  Using the completeness relation, \erf{Eq::Completeness}, we define local, slowly varying mode creation and annihilation operators for each transverse mode $i$ and polarization $\lambda$ as follows,
	\begin{align}
		\hat{\mbf{E}}^{(+)}(\mbf{r},t) = & \sqrt{2 \pi \hbar \omega_c} \sum_\lambda \mbf{e}_\lambda \hat{\Psi}_\lambda (\mathbf{r}_\perp,z,t) e^{i(k_c z - \omega_c t)} \nn \\
		= & \sqrt{2 \pi \hbar \omega_c} \sum_\lambda \mbf{e}_\lambda \int d^2 \mathbf{r}_\perp' \hat{\Psi}_\lambda (\mathbf{r}'_\perp,z,t) \delta^{(2)}(\mathbf{r} - \mathbf{r}')e^{i(k_c z - \omega_c t)} \nn \\
		= & \sqrt{2 \pi \hbar \omega_c} \sum_\lambda \mbf{e}_\lambda \sum_{i} \int \frac{d^2 \mathbf{r}_\perp'}{A} \hat{\Psi}_\lambda (\mathbf{r}'_\perp,z,t) u_i(\mathbf{r}_\perp,z) u^*_i(\mathbf{r}'_\perp,z') e^{i(k_c z - \omega_c t)} \nn \\
		= & \sum_{i,\lambda} \sqrt{\frac{2 \pi \hbar \omega_c}{c A}} \, \mbf{e}_\lambda  \,u_{i}(\mathbf{r}_\perp,z)  \, \hat{a}_{i,\lambda}(z,t) e^{i(k_c z - \omega_c t)} . \label{Eq::IvansParaxialField}
	\end{align}
The slowly varying, traveling-wave mode annihilation operator has been defined,
	\begin{equation} \label{Eq::ModeCreation}
		\hat{a}_{i,\lambda}(z,t) \equiv \int d^2 \mathbf{r}_\perp \sqrt{\frac{c}{A}} \hat{\Psi}_\lambda(\mathbf{r}_\perp,z,t) u_{i}^*(\mathbf{r}_\perp,z), 
	\end{equation}
and, along with the partner creation operator, it satisfies the free-field unequal-space, unequal-time commutation relation,
	\begin{equation} \label{Eq::ParaxialFieldCommutator}
		\big[ \hat{a}_{i,\lambda}(z,t),\hat{a}^{\dag}_{j,\lambda'}(z',t')\big] = \delta_{i, j} \delta_{\lambda, \lambda'} \, \delta(t-t'-(z-z')/c).
	\end{equation}
The mode creation operators in \erf{Eq::ModeCreation} evolve under the free-field Hamiltonian according to $\hat{a}_{i,\lambda}(z,t)=\hat{a}_{i,\lambda}(0,t-z/c)=\hat{a}_{i,\lambda}(z-ct,0)$. 

This paraxial quantization will be put to use in our model of a quantum interface for atomic ensembles in Chapter \ref{Ch::3DInterface} and its application to spin squeezing in Chapter \ref{Ch::SpinSqueezing}.  In these studies, we are specifically interested in the spatial dependence of the atom-light coupling and how it affects coherent interactions, polarimetry measurements, and decoherence.  This quantization scheme is not limited to free space and may be employed in more general situations where propagation is restricted to one dimension.  In inhomogeneous media, the boundary conditions often mix the polarization components of the electric and magnetic fields, and the mode labels do not necessarily refer to fixed polarizations.

	\subsection{Continuous-mode quantum optics} \label{Sec::CMQO}
			
In this section we briefly review the method of Blow et al. presented in the seminal paper, \emph{Continuum fields in quantum optics} \cite{BloLou90}, that takes a slightly different path to describe propagating quantum fluctuations.  This theory rests on several assumptions.  First, the field of interest is one-dimensional in the sense that it is well described by a single direction in $\mathbf{k}$-space.  The mode variables are then indexed by the magnitude of the wave vector, $k$, or equivalently by the positive angular frequencies, $\omega = c|k|$.  In such an approximation transverse effects are ignored, and a fixed transverse quantization area $A$ is assumed. Second, one considers quantization along a length $L$ large enough that the discrete quantized frequency spacing, $\Delta \omega = 2 \pi c / L$, is sufficiently small that the frequency distribution can be considered effectively continuous.  In this case the sum over wave vectors is converted to an integral,
	\begin{align} \label{Eq::KtoOmega}
		\sum_k  \rightarrow \frac{1}{\Delta \omega} \int d\omega,
	\end{align} 
and the continuous-mode creation and annihilation operators are related to the discrete-mode versions through,
	\begin{align}
		\hat{a}_k \rightarrow \sqrt{\Delta \omega} \, \hat{a}(\omega), \quad \quad \hat{a}\dg_k \rightarrow \sqrt{\Delta \omega} \, \hat{a}\dg(\omega),
	\end{align}
which yields the continuous-mode commutation relation,
	\begin{align} \label{Eq::BlowCommutation}
		[\hat{a}(\omega), \hat{a}\dg(\omega')] = \delta(\omega - \omega').
	\end{align}
The positive frequency component of the one-dimensional electric field operator is expressed via the continuous-mode creation operators as
	\begin{align}  \label{Eq::BlowEMField}
		\hat{\mathbf{E}}^{(+)}(z,t) = i \sum_\lambda \int_0^\infty d\omega \sqrt{ \frac{\hbar \omega}{c A} } \mathbf{e}_\lambda \, \hat{a}_\lambda(\omega) e^{ -i \omega (t-z/c)},
	\end{align}
where we have included a transverse polarization index $\lambda$, just as in \srf{}.  The free electromagnetic field Hamiltonian, neglecting vacuum energy terms, is
	\begin{align} \label{Eq::BlowFieldHam}
		\hat{H}_{\rm field} = \int_0^\infty d \omega \, \hbar \omega \, \hat{a}\dg(\omega) \hat{a}(\omega).
	\end{align}
	
Up to this point, nothing more has been done other than a conversion to a one-dimensional continuous theory.  We now assume the field to be sufficiently narrowband such that the spread in frequencies (bandwidth) is small compared to the carrier frequency $\omega_c$.  This brings along with it the quasi-monochromatic condition and is equivalent to the slowly varying envelope approximation.  Within this approximation the range of integration may be extended to negative frequencies without consequence, and one may define Fourier-transformed pairs of field operators,
	\begin{align} \label{Eq::FourierTransformFieldOps}
		\hat{a}(t) \equiv \frac{1}{\sqrt{2 \pi}} \int_{-\infty}^\infty d\omega \, \hat{a}(\omega) e^{-i\omega t} \quad \longleftrightarrow \quad  \hat{a}(\omega) = \frac{1}{\sqrt{2 \pi}} \int_{-\infty}^\infty dt \, \hat{a}(t) e^{i\omega t}.
	\end{align}
The Fourier-transformed field operators obey the commutation relation,
	\begin{align}
		[\hat{a}(t), \hat{a}\dg(t)] = \delta(t-t'),
	\end{align}
which follows from \erf{Eq::BlowCommutation} and \erf{Eq::FourierTransformFieldOps}.

To the extent that the field is sufficiently narrowband around a carrier frequency $\omega_c$, the electric field operator in \erf{Eq::BlowEMField} may be well approximated by making the replacement $\omega \rightarrow \omega_c$ and using the definition in \erf{Eq::FourierTransformFieldOps}\footnote{We have chosen to follow the convention in the literature and include a phase of $i$ the positive-frequency component of the electric field.  Note that in \erf{Eq::IvansParaxialField} the phase is chosen differently.}:
	\begin{align}
		\hat{\mathbf{E}}^{(+)}(z,t) = i \sum_\lambda  \sqrt{ \frac{2 \pi \hbar \omega_c}{c A} } \mathbf{e}_\lambda \, \hat{a}_\lambda(t - z/c) e^{-i\omega_c t}. 
	\end{align}
Once again we see the equivalence of the longitudinal spatial coordinate and time, which results from making the slowly varying envelope approximation along that direction.  This gives the more general unequal-space, unequal-time commutation relation for the slowly varying, free-field operators,
	\begin{align}  \label{Eq::BlowUnequalTimeCommutation}
		\big[ a_\lambda(t-z/c), a_{\lambda'}\dg(t'-z'/c) \big] = \delta_{\lambda, \lambda'} \delta(t-t' - (z-z')/c),
	\end{align}
which is identical to \erf{Eq::ParaxialFieldCommutator} in the absence of transverse spatial dependence.  The continuous-mode quantization scheme explicitly avoids writing the Hamiltonian in the time domain, but it would follow in analogy to \erf{Eq::ParaxialHamiltonian} as a one-dimensional paraxial wave equation that marries time evolution with propagation in the $z$-direction, 

 \section{Interaction with quantum systems}\label{SEC::InteractionWithQuantumSystems}

Now that a mathematical foundation for propagating quantum fields has been established from two distinct but related standpoints, we are poised to develop an understanding of how such fields interact with quantum systems.  The atom-light interaction for multi-level atoms will be treated separately and in great detail in Chapter \ref{Ch::DispersiveInt}.  For our study of $N$-photon states, we will approach this subject with the well-developed input-output formalism using quantum stochastic differential equations (QSDEs) based on the continuous-mode quantization of \srf{Sec::CMQO}, which will be reviewed here.   A foundation of rich mathematical machinery underlies the manipulation of QSDEs and their derivation from physical systems.  We only touch the surface commensurate with our purposes; an interested reader is directed to Refs.~\cite{HudPar84, GardinerBook, AccardiBook, BarchielliBook, ZolGarNotes, Gou06, WisMilBook} for a more rigorous and detailed analysis.  

	\subsection{The quantum white noise limit} \label{Sec::WhiteNoiseLimit}
	
We consider a quantum system at position $z$ interacting with a continuous-mode field described by bosonic field operators, $\hat{a}(\omega)$, satisfying the commutation relation \erf{Eq::BlowCommutation}.  In the Schr\"{o}dinger picture, where quantum states evolve and operators are stationary, the total Hamiltonian has three distinct parts,
	\begin{align} \label{Eq::TotalHam}
		\hat{H} = \hat{H}_{\rm field} + \hat{H}_{\rm sys} + \hat{H}_{\rm int}.
	\end{align}
The bare Hamiltonian of the system is left general and is designated by $\hat{H}_{\rm sys}$.  The Hamiltonian for the free field is given in \erf{Eq::BlowFieldHam},
	\begin{align} \label{Eq::ContModeFieldHam}
		\hat{H}_{\rm field} &= \int_0^\infty d\omega \, \hbar \omega \, \hat{a}\dg (\omega) \hat{a} (\omega) .
	\end{align}
The Hamiltonian describing the interaction between system and field is described by a dipole-type, linear coupling of the general form,
	\begin{align} \label{Eq::GardinerHam}
		\hat{H}_{\rm int} = - i \hbar \int_0^\infty d \omega \kappa( \omega ) \big( \hat{c} + \hat{c}\dg \big) \left( \hat{a}(\omega) - \hat{a}\dg (\omega) \right),
	\end{align}
where $\hat{c}$ is the system operator that couples to the field.  The strength of the interaction is given by $\kappa(\omega)$ which has units of $\sqrt{\mbox{frequency}}$ and is assumed to be real-valued.  For instance, if the quantum system is a two-level atom, then $\hat{c} = \op{g}{e}$ and $\kappa(\omega) = |\bra{e}\hat{d}\ket{g}| \sqrt{ \omega / \hbar c A}$.  
The electric field operators in the interaction Hamiltonian are evaluated at the position of the system, assumed to be point-like in space, chosen to be $z=0$.

We work in the interaction picture, as it gives a clear justification for making the rotating wave approximation and, for resonant interactions, greatly simplifies the form of the Hamiltonian. We now specify an interaction picture with the choice $\hat{H}_0 = \hat{H}_{\rm field} + \hat{H}_{\rm sys}$.  The field operators in the interaction picture become $\hat{a}(\omega)e^{-i\omega t}$ and system operators rotate via the bare system Hamiltonian at transition frequency, $\hat{c} e^{-i \omega_0 t}$.  Any remaining detuning between the system frequency and the carrier frequency of the field manifests in a remaining bare Hamiltonian on the system at the detuning $\Delta = \omega_c - \omega_0$. The interaction Hamiltonian in the interaction picture can then be written,
	\begin{align} \label{Eq::IntHamIntPic}
		\hat{H}_{\rm int} = - i \hbar \int_0^\infty \kappa(\omega) \left( \hat{c} e^{-i \omega_0 t} + \hat{c}\dg e^{i \omega_0 t} \right) \left( \hat{a}(\omega) e^{-i \omega t} - \hat{a}\dg (\omega) e^{i \omega t} \right).
	\end{align}
Only in the interaction picture is it clear that there are co-rotating terms whose time evolution oscillates so quickly that its effect is averaged out over system time scales.  Making the rotating wave approximation by discarding these terms, \erf{Eq::IntHamIntPic} becomes
	\begin{align} \label{Eq::IntHamRWA}
			\hat{H}_{\rm int} =& - i \hbar \int_0^\infty \kappa(\omega) \left( \hat{c}\dg \hat{a}(\omega) e^{-i (\omega-\omega_0) t} - \hat{c} \hat{a}\dg (\omega) e^{i (\omega - \omega_0) t}  \right) \\
			=& - i \hbar \kappa(\omega_0) \big(  \hat{c}\dg \tilde{b}(t) - \hat{c} \, \tilde{b}\dg(t) \big)
	\end{align} 
where the operator $\tilde{b}(t)$ has been defined,
	\begin{equation} \label{Eq::TildeB}
		\tilde{b}(t) \equiv \frac{1}{\sqrt{2\pi} \kappa(\omega_0) } \int_0^\infty d\omega \, \kappa(\omega) \hat{a}(\omega) \, e^{-i (\omega - \omega_0) t},
	\end{equation}
and has commutation relation,	
	\begin{equation} \label{Eq::BathCommutator}
		\big[\tilde{b}(t), \tilde{b}\dg(t') \big] = \int_0^\infty d\omega \left(\frac{\kappa(\omega)}{\kappa(\omega_0)} \right)^2 \frac{e^{-i(\omega-\omega_0)(t-t')}}{2\pi}.
	\end{equation}
	
It is at this point that the first Markov approximation, or \emph{quantum white noise limit}, is made \cite{GardinerBook}.  When $\kappa(\omega)$ is slowly varying around $\omega_0$, we make the approximation that the atom has a flat spectral response; mathematically this translates to making the replacement, $\kappa(\omega) \rightarrow \kappa(\omega_0)$.  The implication of the Markov approximation is that the correlation time of the field is short compared to the slowly-varying interaction time, $\tau_i  \approx 1/|\kappa(\omega_0)|^2 $.  That is, the Markov approximation amounts to coarse-graining over time scales that are long compared to the field correlation time but slow compared to system dynamics. Within this approximation, the limits of integration in \erf{Eq::TildeB} can be extended to negative frequencies, and the field can be described by the following operators,
	\begin{equation} \label{Eq::WhiteNoiseOps}
		\hat{b}(t) \equiv \frac{1}{\sqrt{2\pi} } \int_{-\infty}^\infty d\omega \,\hat{a}(\omega) \, e^{-i (\omega - \omega_0) t},
	\end{equation}
which, from \erf{Eq::BathCommutator}, obey the singular commutation relation $[\hat{b}(t),\, \hat{b}^\dag(t')] = \delta(t - t')$. For classical stochastic processes, $\delta$-correlation implies white noise, so the operators $\hat{b}(t)$ and $\hat{b}^\dagger(t)$ are dubbed \emph{quantum white noise operators}.  These operators describe the propagating quantum field that arrives at time $t$ and interacts with the system.  The modes label $t$ indexes the time at which the operator $\hat{b}(t)$ interacts with the system.  It is clear that within the white noise approximation the operators in \erf{Eq::WhiteNoiseOps} are those from continuous-mode quantization, \erf{Eq::FourierTransformFieldOps}, which in turn are analogous to those defined for free-space paraxial propagation, \erf{Eq::ParaxialFieldCommutator}.  The equivalence comes from the fact that in making the Markov approximation, we assumed a quasi-monochromatic field.

The interaction Hamiltonian in \erf{Eq::IntHamRWA} can be recast in terms of the quantum white noise operators, \erf{Eq::WhiteNoiseOps}, as
	\begin{equation} \label{Eq::WhiteNoiseDipHam}
 		\hat{H}_{\rm int}(t) =\,i \hbar \sqrt{\gamma} \big( \hat{c} \, \hat{b}^\dag(t) - \hat{c}\dg \, \hat{b}(t) \big),
	\end{equation}
where the coupling rate $\gamma$ is defined through the relation\footnote{The density of states has been included in the continuous mode quantization and gives rise to \erf{Eq::KtoOmega}.  See, for example, Ref. \cite[Ch. 1]{CarmichaelBook} or Ref. \cite[Ch. 6]{LoudenBook}.}
	\begin{align} \label{Eq::KappaGamma}
		\kappa(\omega_0) = \sqrt{\gamma/2\pi} \quad \leftrightarrow \quad  \gamma = 2 \pi |\kappa(\omega_0)|^2.
	\end{align}   
This is the fundamental interaction in input-output theory that describes the linear coupling of a quantum system to propagating fields through the operators $\hat{c}$ and $\hat{c}\dg$.  The moniker \emph{input-output theory} is explained when we consider the time evolution of a field operator $\hat{b}(t)$ via the interaction Hamiltonian.  Since $t$ labels the mode, we use a subscript ``in" to indicate the free field which arrives and interacts with the system and a subscript ``out" to indicate the field after the interaction.  The output field is generated via the Hamiltonian, \erf{Eq::WhiteNoiseDipHam}, under the assumption of weak coupling such that the first Born approximation applies,
	\begin{align} \label{Eq::InputOutput}
		\hat{b}_{\rm out}(t) = \hat{b}_{\rm in}(t) + \sqrt{\gamma} \hat{c}(t).
	\end{align}
This input-output relationship reveals how the output field becomes entangled with the system through a linear coupling.  When performing continuous measurements in time, it is these output fields that are detected.

	\subsection{Quantum stochastic differential equations} \label{SEC::QWNL}

Input-output theory has been widely used in the quantum optical community.  Recently a powerful tool, the $(S,L,H)$ formalism, has emerged that builds on input-output theory and the theory of cascaded quantum systems \cite{Car93,CarmichaelBook2} in analogy to modular circuit design in electronics.  Within the $(S,L,H)$ formalism, one identifies three operators that characterize a quantum system: the Hamiltonian $\hat{H}_{\rm sys}$, the operator $\hat{L}$ that describes linear coupling to the continuous-mode field, and the unitary scattering matrix\footnote{Within the formal theory of quantum stochastic differential equations, the general form goes back to Hudson and Parthasarathy \cite{HudPar84}.} $\hat{S}$.  Networking various quantum components through optical connections is simply a matter of combining their $(S,L,H)$ triples using a set of rules \cite{GouJam09,NurDoh09,JamGou10}.  
The underlying foundation is the theory of quantum stochastic differential equations, a mathematically rigorously formalism for the singular quantum white noise operators that arise in input-output theory.

Under the Hamiltonian in \erf{Eq::WhiteNoiseDipHam}, the system and the field undergo joint unitary evolution via the propagator (time evolution operator) $\hat{U}(t)$ which has a formal solution \cite{Bar90},
	\begin{align} 
		\hat{U}(t) & = \overleftarrow{\mathcal{T}} \left\{ \exp \left[ \frac{-i}{\hbar} \int_0^t dt' \, \hat{H}_{\rm int} (t')  \right] \right\}\\
		 	& = \overleftarrow{\mathcal{T}} \left\{ \exp \left[  \int_0^t dt' \big( \hat{L} \, \hat{b}^\dag(t') -\hat{L}\dg \hat{b}(t') \big) \right] \right\},  \label{Eq::TimeOrderedProp}
	\end{align}
where $\overleftarrow{\mathcal{T}}$ indicates time ordering and, to make a connection with the notation in the literature, we have absorbed the coupling rate into the system operator with the definition $\hat{L} \equiv \sqrt{\gamma} \hat{c}.$	

Being the quantum versions of classical white noise, the operators $\hat{b}(t)$ and $\hat{b}^\dagger(t)$ that appear in the unitary propagator, \erf{Eq::TimeOrderedProp}, bring along the difficulties of zero-mean, infinite-bandwidth noise\footnote{They are, in fact, operator densities and should never appear outside of a time integral.  This become clear in the oft-repeated example, where one attempts to calculate the expectation under vacuum $\bra{0} \hat{b}(t) \hat{b}\dg(t)\ket{0} = \delta(0)$.}.  Further, the differential $dt \, \hat{b}(t)$ is of order $\sqrt{dt}$, indicating that in the Dyson series second order terms must be kept.  For these reasons, care must be taken when defining stochastic integrals of the sort that appear in \erf{Eq::TimeOrderedProp}.  First, we define $B_t$ and $B_t\dg$ as time integrals over the quantum noises,
	\begin{align}\label{Eq::QuantNoises}
		B_t = \int_0^t dt' \, \hat{b}(t') \quad \text{and} \quad B^\dagger_t = \int_0^t dt' \, \hat{b}^\dagger(t').
	\end{align}
The subscript on the quantum noises indicates that they act only up to time $t$ and as the identity for the time interval $[t, \infty)$.  The singular nature of the quantum white noise operators can be removed by expressing \erf{Eq::TimeOrderedProp}  in terms of continuous differential increments $dB_t$ and $dB\dg_t$ of the quantum noises\footnote{Although we punctiliously label operators with hats throughout this dissertation, here we follow the literature, which does not do so for the quantum noise increments.} \cite{GarCol85, Bar86},
	\begin{align}
		dB_t \equiv B_{t+dt} - B_t \quad \text{and} \quad  dB_t\dg \equiv B\dg_{t+dt} - B\dg_t .
	\end{align} 
These are the quantum, non-commuting analogues of the classical Wiener process and are referred to generically as \emph{quantum noise increments} and are in some sense short-time averages of the quantum white noise operators \cite{DohertyNotes}.  Now equation (\ref{Eq::TimeOrderedProp}) can be recast in the form
	\begin{align} \label{Eq::TimeOrderedPropNoiseIncrements}
		\hat{U}(t) & = \overleftarrow{\mathcal{T}} \left\{ \exp \left[ \int_0^t \hat{L}  dB_{t'}\dg -\hat{L}\dg  dB_{t'} \right] \right\}. 
	\end{align}
 Giving precise mathematical meaning to \erf{Eq::TimeOrderedPropNoiseIncrements} requires a formal definition of an integral with respect to the quantum noise increments $dB_t$ and $dB\dg_t$.  Even classically, where integration is defined with respect to a classical Wiener increment $dW$, this is not a trivial task.  
	 
Two distinct but equivalent definitions of stochastic integrals exist, both in the classical and quantum domains.  The Stratonovich integral is defined,
	\begin{align}
		\mathbf{(S)} \int_{t_0}^t f({t'}) dB_{t'} = \lim_{n \rightarrow \infty} \sum_{i=0}^{n} \frac{  f(t_{i+1}) - f(t_i) }{2}\left( B_{t_{i+1}} - B_{t_i}\right).
	\end{align}
The integrand is taken as the midpoint of the function $f(t)$ in each time interval.  In the stochastic calculus associated with the quantum Stratonovich integral, differentials follow the standard rules from calculus.  For quantum (non-commuting) stochastic processes $\hat{X}$ and $\hat{Y}$:
	\begin{align}
		d(\hat{X}\hat{Y}) = (d\hat{X}) \hat{Y} + \hat{X} (d\hat{Y}).
	\end{align}
Stratonovich QSDEs arise as the natural form for the quantum white noise limit of physical processes \cite{ZolGarNotes, Gou06}.  

The It\={o} integral is defined,	
	\begin{align}
		\mathbf{(I)} \int_{t_0}^t f({t'}) dB_{t'} = \lim_{n \rightarrow \infty} \sum_{i=0}^{n} f(t_i) \left( B_{t_{i+1}} - B_{t_i} \right).
	\end{align}
The beauty of the quantum It\={o} integral stems from the fact that the integrand $f(t)$ and the operator differential $dB_t$ act on independent time intervals and therefore commute, $[f(t), dB_t] = 0$.  As a result, expectations with respect to a quantum state factorize,
	\begin{align}
		\mathbbm{E} \left[ \int_{t_0}^t f(t') dB_{t'} \right] = \int_{t_0}^t  \mathbbm{E} [f({t'})] \mathbbm{E} [dB_{t'}].
	\end{align}
In Chapter \ref{Ch::NPhotonME} we will emphatically exploit this property in the calculation of expectation values with respect to continuous-mode $N$-photon states.  In spite of the useful properties, working in It\={o} form brings the burden of its own calculus, which requires that differentials be taken to second order.  For quantum stochastic processes $\hat{X}$ and $\hat{Y}$ this means
	\begin{align}
		d (\hat{X}\hat{Y}) = (d\hat{X}) \hat{Y} + \hat{X} (d\hat{Y}) + d\hat{X} d\hat{Y}.
	\end{align}
Henceforth, we will work exclusively with QSDEs in It\={o} form, and omit further discussion of Stratonovich integrals. Now the time evolution operator in \erf{Eq::TimeOrderedPropNoiseIncrements} can be expressed as a QSDE in It\={o} form by expanding to second order
	\begin{equation}\label{Eq::UIto}
   		 d\hat{U}(t) =   \Big( \hat{L} dB^\dag_t -\hat{L}\dg dB_t  - \half \hat{L}\dg \hat{L} dt \Big) \hat{U}(t).
	\end{equation}
The first two terms describe the dipole coupling to the quantum noise increments and the third, deterministic term, known as the \emph{It\={o} correction}, is an artifact of using It\={o}-form QSDEs.

	\subsection{Including the gauge process and scattering matrix} \label{SEC::GaugeProcess}

Before moving on, we include a generalization of the Hamiltonian, \erf{Eq::WhiteNoiseDipHam}, and related unitary propagator, \erf{Eq::TimeOrderedPropNoiseIncrements}. The $(S,L,H)$ formalism includes the scattering matrix $\hat{S}$ that describes a system's response to the photon flux at time $t$.  In a single mode, as considered here, $\hat{S}$ describes a unitary coupling of a system to a two-photon process in an infinitesimal time increment, where a photon is absorbed and and immediately re-emitted.  In the interaction the system is returned to its initial state, possibly with a photon-dependent phase imprinted on it (and a state-dependent phase on the outgoing field). For example in Ref. \cite{KerMab09}, the authors consider as their basic unit a $\Lambda$-type three-level atom in a cavity.  In the limit where the cavity and excited atomic state decay quickly compared to the interaction time, they may be adiabatically eliminated\footnote{A technique for adiabatic elimination within the formalism of QSDEs can be found in Ref. \cite{BouSil08}.}, leaving effective dynamics in the two atomic ground states, $\ket{g}$ and $\ket{h}$.  In this case they find that the cavity QED system acts just as a simple, state-dependent scatterer with $\hat{S} = \op{g}{g} - \op{h}{h}.$ 

When multiple field modes are considered, as in \srf{Sec::MultipleModes}, $\hat{S}$ describes a system's response to scattering between them.  For example, a beam splitter has no internal dynamics but scatters between spatial field modes \cite{GouJam09}, while a multi-level atom in an adiabatically eliminated model  scatters between polarization modes and responds with effective ground state dynamics \cite{CookThesis}.  Other effective couplings to photon number appear in optomechanical systems \cite{Van11}.  Within the unitary time evolution operator, $\hat{S}$ couples to another fundamental quantum stochastic process,
	\begin{align}
		\Lambda_t = \int_0^t dt' \, b\dg(t') b(t'),
	\end{align}
known as the gauge process. It counts the number of photons in the field up to time $t$ and has increments
	\begin{align}\label{Eq:dLambda_yo}
		d\Lambda_t \equiv \Lambda_{t + d t} - \Lambda_t,
	\end{align}
that describe photon flux.  With this, and including a possible Hamiltonian acting on the system, $\hat{H}_{\rm sys}$, the general QSDE for the time evolution operator in one mode has the form \cite{GouJam09},
	\begin{align}   \label{Eq::QsdeUnitary}
 		d\hat{U}(t) = \bigg\{ & - \Big(\smallfrac{i}{\hbar}\hat{H}_{\rm sys} + \smallfrac{1}{2}  \hat{L}\dg \hat{L} \Big) \otimes \hat{I}_{\rm field} \, dt  - \hat{L}\dg  \hat{S} \otimes dB_t  \\
		& + \hat{L} \otimes dB\dg_t + \big(\hat{S} - \hat{I}_{\rm sys} \big) \otimes d\Lambda_t  \bigg\} \hat{U}(t). \nn
	\end{align}
Explicit tensor product notation is used here to be very clear about system and field operators within the QSDE.  

As a quick endnote, we have only briefly discussed the scattering operator $\hat{S}$ as it appears for effective couplings after adiabatic elimination and for a non-dynamic beamsplitter.  For \emph{fundamental} number coupling there is still some debate as to whether $\hat{S}$ should be found from a normally-ordered Stratonovich calculus \cite{Gou06, GouvanH07} or from a time-ordered exponential \cite{KholevoBook}, as both seem to give different results\footnote{Perplexing is that the results agree to second order in a Taylor expansion.  It may be that they are mathematically equivalent. }.

	\subsection{It\={o} Langevin equations} \label{App::ItoLangEQS}
	
The time evolution operator in \erf{Eq::QsdeUnitary} allows us to calculate the equation of motion for a system operator $\hat{X}(0) =  \hat{X} \otimes \hat{I}_{\rm field}$.  Since we work with It\={o} QSDEs, this requires taking differentials to second order,
	\begin{align} \label{Eq::HEOMArb}
		d\big(\hat{U}\dg(t)  \hat{X} \hat{U}(t) \big) = \big(d \hat{U}\dg(t) \big) \hat{X} \hat{U}(t) + \hat{U}\dg(t) \hat{X} \big(d\hat{U}(t)\big) + \big(d \hat{U}\dg(t) \big) \hat{X}\big(d \hat{U}(t)\big).
	\end{align}
When manipulating QSDEs such as \erf{Eq::HEOMArb} one encounters products of the quantum noise increments.  Under vacuum expectation the rules for these products are given by the vacuum It\={o} table \cite{GardinerBook, BarchielliBook},
	\begin{equation} \label{Eq::ItoTable}
	\begin{tabular}{cc}
		$dB_t dB_t\dg = dt$        & $dB_t d\Lambda_t = dB_t$ \\ 
		$d\Lambda_t d\Lambda_t = d\Lambda_t$              & $d\Lambda_t  dB_t\dg = dB_t\dg$,             
	\end{tabular} 
	\end{equation}	
with all other products vanishing.

With \erf{Eq::HEOMArb} and \erf{Eq::ItoTable} we can write down the It\={o} QSDE for a system operator, $\hat{X}(t_0) =  \hat{X} \otimes \hat{I}_{\rm field}$,
	\begin{align} \label{Eq::dXAppendix}
		 d \hat{X} =&\big( \smallfrac{i}{\hbar}[\hat{H}_{\rm sys}, \hat{X}]  + \mathcal{L}_L\dg[\hat{X}] \big) dt   + [\hat{L}\dg,\hat{X}]\hat{S} dB_{t}  +\hat{S}\dg[\hat{X},\hat{L}] dB\dg_{t}  + (\hat{S}\dg \hat{X}\hat{S}-\hat{X}) d\Lambda_{t},
	\end{align}
referred to as an \emph{It\={o} Langevin equation}\footnote{In the quantum filtering literature one often finds that Heisenberg-picture, time-evolved system operators are denoted as $j_t(\hat{X}) \equiv \hat{U}\dg(t) \hat{X} \hat{U}(t)$.  We proceed without this notation, as time evolution of the operators in the Heisenberg picture is assumed.}. The action of the Lindblad superoperator in the Heisenberg picture is,
	\begin{equation} \label{Eq::LinbladHeiO}
		\mathcal{L}_L\dg[\hat{X}] \equiv \hat{L}\dg \hat{X}\hat{L}- \smallfrac{1}{2} \big( \hat{L}\dg \hat{L} \hat{X} + \hat{X} \hat{L}\dg \hat{L} \big).
	\end{equation}
The first two terms in \erf{Eq::dXAppendix} describe smooth evolution from an external Hamiltonian on the system and from a Lindblad-type dissipator.  The second two terms describe the influence of quantum noise through coupling of a system operator $\hat{L}$ linearly to the field operators, e.g. dipole-type coupling. The final term arises from coupling of a system operator $\hat{S}$ to a quantity quadratic in the field operators, such as photon number. 

We can also find the Heisenberg-Langevin operators for output quantum noises, such as $B_t^{\rm out} = \hat{U}\dg(t) B_t \hat{U}(t)$.  Since \erf{Eq::QsdeUnitary} is an It\={o} form QSDE for a time-ordered exponential of the form of \erf{Eq::TimeOrderedPropNoiseIncrements}, expanding to first order gives the infinitesimal evolution operator over the interval $[t, t+dt)$ \cite{Bar90, WisMilBook},
	\begin{align}
		\hat{U}(t,t+dt) =& \hat{I}_{\rm sys} \otimes \hat{I}_{\rm field} - \Big(\smallfrac{i}{\hbar}\hat{H}_{\rm sys} + \smallfrac{1}{2}  \hat{L}\dg \hat{L} \Big) \otimes \hat{I}_{\rm field} \, dt \label{Eq::InfPropagator} \\
		& - \hat{L}\dg  \hat{S} \otimes dB_t  + \hat{L} \otimes dB\dg_t + \big(\hat{S} - \hat{I}_{\rm sys} \big) \otimes d\Lambda_t . \nn
	\end{align}
Then, the QSDE for the output quantum noise increments can be found \cite{Bar86, ZolGarNotes},
 	\begin{align}  \label{Eq::dBAppendix}
		dB_t^{\rm out} = & B^{\rm out}_{t+dt} - B^{\rm out}_t \nn \\
			= & \hat{U}\dg(t) \Big\{ \hat{U}\dg(t,t+dt) dB_{t} \hat{U}(t,t+dt) \Big\} \hat{U}(t)  \nn \\
		 	= &\hat{L} dt + \hat{S} dB_t,
	\end{align}
We perform similar calculations to find the QSDE for output photon number $\Lambda_t^{\rm out} = \hat{U}\dg(t) \Lambda_t \hat{U}(t)$,
	\begin{align}  \label{Eq::dLambdaAppendix}
		d\Lambda_t^{\rm out} &=   \hat{L}\dg \hat{L}dt +  \hat{L}\dg \hat{S} dB_t + \hat{S}\dg \hat{L} dB\dg_t  +d\Lambda_t,
	\end{align}
 where we have used the relation, $\hat{S}\dg \hat{S} = \hat{I}_{\rm sys}$.  Since \erf{Eq::dBAppendix} and  \erf{Eq::dLambdaAppendix} are in the Heisenberg picture, the system operators that appear are the time-evolved versions from \erf{Eq::dXAppendix}.  The two relations in \erf{Eq::dBAppendix} and \erf{Eq::dLambdaAppendix} are the input-output relations within the formalism of It\={o} QSDEs.

	 \subsection{Multi-mode fields} \label{Sec::MultipleModes}
	
In more general cases, we might wish to model interactions in multiple field modes, separate from the longitindual spatio-temporal continuous modes, such as polarization or transverse spatial modes.  A discussion of the underlying physical modeling is given in \srf{SEC::SingleModeInteractions}.  For multiple modes the evolution is given by the QSDE for the multi-mode time evolution operator, 
	\begin{align}  \label{a2mode_qsdeunitary}
		d\hat{U}(t) =  \bigg\{ & -  \Big(\smallfrac{i}{\hbar} \hat{H}_{\rm sys} + \smallfrac{1}{2} \sum_{i} \hat{L}_i^{\dagger} \hat{L}_i \Big) \otimes \hat{I}_{\rm field} \, dt  - \sum_{i,j} \hat{L}_{i}^\dagger \hat{S}_{ij} \otimes dB_j \\
		&+ \sum_{i} \hat{L}_i \otimes dB_i^\dagger  + \sum_{i,j} \big( \hat{S}_{ij} - \delta_{ij} \hat{I}_{\rm sys} \big) \otimes d\Lambda_{ij} \bigg\} \hat{U}(t). \nn
	\end{align}
Here, $\hat{L}_{i}$ is the linear coupling operator between the $i^{th}$ mode and the system, $\hat{H}_{\rm sys}$ is an external Hamiltonian, and the scattering matrix $\hat{S}_{ij}$ is constrained by unitarity: $\sum_k \hat{S}_{ik}\hat{S}_{jk}\dg= \delta_{ij} \hat{I}_{\rm sys}$ and $\sum_k \hat{S}_{ki}\dg \hat{S}_{kj} = \delta_{ij} \hat{I}_{\rm sys}$ (see \cite[ Appendix A]{GouGohYan08} and \cite[Sec. IV]{GouJam09} and the references therein for more details on multi-mode QSDEs). The fact that in physical situations the system couples differently to each mode is captured by the fact that the coupling operators contain the coupling rate, $\hat{L}_{i} = \sqrt{\gamma_i} \hat{c}_i$, where $\gamma_i$ is given by \erf{Eq::ModeCouplingRate}.  Note that the subscript $t$ on the quantum noises has been dropped for notational compactness in favor of the mode labels $\{i,j\} $. The multi-mode quantum noise increments,
	\begin{align}\label{a2mode_flow}
		dB_i \equiv& \int_t^{t + dt} dt' \, \hat{b}_i(t') \quad \text{and} \quad d\Lambda_{ij} \equiv \int_t^{t + d t} dt' \, \hat{b}_i^\dag(t') \hat{b}_j(t')
	\end{align}
satisfy the multi-mode vacuum It\={o} table,
	\begin{equation} \label{Eq::MultimodeItoTable}
	\begin{tabular}{cc}
		$d B_i dB_j\dg = \delta_{i,j} dt$        & $ dB_i d\Lambda_{jk} = \delta_{i,j} dB_k$ \\ 
		$d\Lambda_{ij} d\Lambda_{kl} = \delta_{j,k} d\Lambda_{il}$             & $d\Lambda_{ij} dB_k\dg = \delta_{j,k} dB_i\dg$.                  
	\end{tabular} 
	\end{equation}
Using multi-mode versions of the evolution operators, \erf{Eq::HEOMArb}, and the It\={o} table, we can write down the QSDE for a system operator \cite{BarchielliBook},
	\begin{align} \label{Eq::dXTwoMode}
		d\hat{X} = &\Big(\smallfrac{i}{\hbar} \big[\hat{H}_{\rm sys}, \hat{X} \big] + \sum_i \mathcal{L}_{L_i}[\hat{X}] \Big) dt  + \sum_{i,j} [\hat{L}_i\dg,\hat{X}]\hat{S}_{ij} dB_{j}  \\
		&+\sum_{i,j} \hat{S}_{ij}\dg[\hat{X},\hat{L}_i] dB_j\dg  + \sum_{i,j} \Big( \sum_k \hat{S}_{ki}\dg \hat{X}\hat{S}_{kj}-\delta_{ij}\hat{X} \Big) d\Lambda_{ij}. \nn 
	\end{align}
The output quantum noise increments in mode $i$ can likewise be found:
	\begin{align} \label{Eq::BOutTwoMode}
		dB_i^{\rm out} = \sum_j \hat{S}_{ij} dB_j + \hat{L}_i dt,
	\end{align}
as well as the output photon flux from mode $j$ to mode $i$,
	\begin{align} \label{Eq::LambdaOutTwoMode}
		d\Lambda_{ij}^{\rm out} =  \hat{L}_i\dg \hat{L}_j dt + \sum_k \hat{L}\dg_i \hat{S}_{jk} dB_k +  \sum_k \hat{S}\dg_{ik} \hat{L}_j dB_k +  \sum_{k,l} \hat{S}\dg_{ik} \hat{S}_{jl} d\Lambda_{kl}.
	\end{align}

In Chapter \ref{Ch::NPhotonME}, we will use the QSDE formalism laid out here to describe the interaction of a quantum system with a traveling wave packet of light prepared in a state of definite photon number.

	\chapter{Quantum systems interacting with $N$-photon states} \label{Ch::NPhotonME}


\section{Introduction}

In this chapter we present a unifying method, based on the formalism of the previous chapter, to describe the dynamics of a quantum system as it interacts with a continuous-mode $N$-photon state, as depicted in \frf{Fig::FockInteractionCartoon}.  During the interaction, these fundamentally quantum mechanical states of light become nonlocally entangled with the system.  This is a departure from the standard situation in open quantum systems where the input field interacting with the system at time $t$ is assumed to be uncorrelated both with the system and with the field at other times.  Consider the simplest situation in which the input field is prepared in a wave packet $\xi(t)$ with exactly one photon.  Classically, there are two possible paths that can have been taken by time $t$: (i) the photon has been absorbed by the system at some previous time $t' < t$, or (ii) the photon has not yet been absorbed and can be found, with certainty, in the remaining input field\footnote{The first path also bifurcates.  After the photon is absorbed, it can either remain as an excitation within the system or be reemitted into the field.}.  Quantum mechanically, these two classical options can also be in superposition.  The major obstacle to describing the reduced system's dynamics comes from keeping track of the joint system-field correlations that can arise.  The method detailed in this chapter addresses this issue and allows one to derive the master equations and output field quantities for an arbitrary quantum system interacting with any combination of continuous-mode $N$-photon states.

	\begin{figure}[!t]
    		\begin{center}
    		\includegraphics[width=1\hsize]{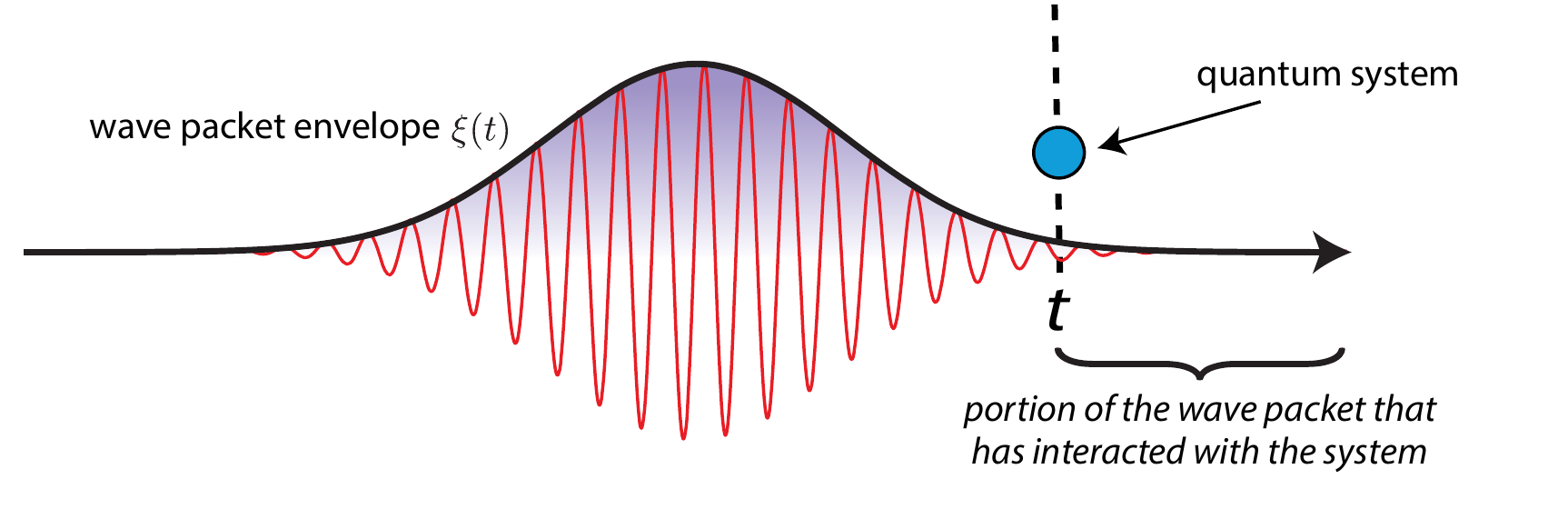}
       		 \caption[Schematic for a wave packet interacting with a quantum system]{A traveling wave packet interacting with an arbitrary quantum system.  The temporal wave packet is described by a slowly-varying envelope $\xi(t)$ which modulates fast oscillations at the carrier frequency.  We consider the case where the wave packet is prepared in a nonclassical state of definite photon number.} \label{Fig::FockInteractionCartoon}
    		\end{center}
	\end{figure}

A description of a system interacting with a traveling wave packet naturally calls for a formulation in the time domain.  The input-output theory and underlying continuous-mode quantization of the field, reviewed in Chapter \ref{Ch::PropagatingFields}, provide such a description \cite{GarCol85, YurDen84, Cav82, DumGar92, GardinerBook, Gar93, Car93}.  Often input-ouput theory is formulated for a one-dimensional electromagnetic field, although this is not a necessary restriction \cite{DumGar92}. Such effective one-dimensional models are typically applied in the context of optical cavities \cite{Aoki09} or photonic waveguides \cite{CheWubMor11, Spillane08, VetRau10, Chang07}. In this formalism the rotating wave approximation, the weak-coupling limit (the first Born approximation), and the Markov approximation are made \cite{VanHove55, vanHMab05}.  Strict enforcement of these approximations is known as the quantum white noise limit, discussed in \srf{Sec::WhiteNoiseLimit}.  The main result is a quantum stochastic differential equation (QSDE) for the unitary time evolution operator that governs the system-field dynamics. From this equation one can derive Heisenberg-picture QSDEs for system and field operators driven by white noise, also known as white-noise Langevin equations.  Taking expectations with respect to input states then gives unconditional system and field dynamics, which can be ported to quantum states in the Schr\"{o}dinger picture with a quick transformation. 

This chapter is organized as follows: we begin in \srf{SEC:mulitphoton} with the definition of a subset of $N$-photon states, known as continuous-mode Fock states, for which the input photons are uncorrelated.  To motivate the Fock-state master equations derived later in this chapter, and in particular to highlight the complications when dealing with $N$-photon states, we present an analytic solution for the particular case of a two-level atom interacting with a Fock state.  In \srf{sec_fock} we present the central result: master equations for systems interacting with continuous-mode Fock states and related equations for output field observables.  The formalism is illustrated with a variety of examples beginning with a two-level atom interacting with Fock states in wave packets of different shape.  In \srf{sec_2modefock}, we present a generalization to Fock states in multiple modes. This sets the stage for the study of many canonical problems in quantum optics. As an example, we examine the scattering of Fock states from a two-level atom.
Finally, in \srf{Sec::GeneralWavepackets} we show how these equations can be used to study general $N$-photon states, for which the spectral density function is not factorizable.  This is illustrated by a two-level atom interacting with a pulse train of two consecutive Gaussian single-photon wave packets.

\section{Continuous-mode Fock states}\label{SEC:mulitphoton}

A continuous-mode single-photon state \cite{LoudenBook, GardinerBook, BloLou90} can be interpreted as a single photon coherently superposed over many spectral modes \cite{MilburnBook,Mil08} with weightings given by the spectral density function (SDF) $\tilde{\xi}(\omega)$,
	\begin{align}  \label{Eq::SinglePhotonFreqDomain}
		\ket{1_{\xi}} & = \int d\omega \, \tilde\xi(\omega) b\dg(\omega)\ket{0}.
	\end{align}
We focus on quasi-monochromatic wave packets, where the spectral spread is much smaller than the carrier frequency, $\Delta \omega \ll \omega_c$.  This holds for optical carriers, whose bandwidths are small relative to the carrier frequency.  Then, we can define a \emph{slowly-varying envelope} rotating at the carrier frequency,
	\begin{align}
		\tilde\xi(\omega) \rightarrow \tilde \xi(\omega)e^{-i \omega_c t},
	\end{align}
where $\omega_c$ is near relevant system frequencies.  The Fourier transform of the slowly-varying envelope, $\mathcal{F}[\tilde{\xi}(\omega)] = \xi(t)$, characterizes a square-normalized temporal wave packet, $\int dt \, |\xi(t)|^2 = 1$.  In the time domain, and within the quasi-monochromatic approximation, the single-photon state in Eq. (\ref{Eq::SinglePhotonFreqDomain}) becomes \cite{LoudenBook},
	\begin{align}\label{eqBdef}
		\ket{1_{\xi}} & = \int dt' \, \xi(t') b\dg(t')\ket{0} \\
		& \equiv B\dg (\xi) \ket{0},
	\end{align} 
where we have absorbed the possible detuning from the system frequency into $\xi(t)$. The operator $B\dg (\xi)$ creates a single photon in the wave packet $\xi(t)$ and satisfies the commutation relation,
	\begin{align} \label{Eq::ModeOpCommutation}
		\big[ B(\xi), B\dg(\xi)\big] = 1.
	\end{align}
Equation (\ref{eqBdef}) can be interpreted as a superposition of instantaneous photon creation times weighted by the temporal wave packet. Since the white noise operators are defined in the interaction picture, it is clear that $\xi(t)$ is a slowly-varying temporal envelope modulating fast oscillations at the carrier frequency $\omega_c$. By focusing on quasi-monochromatic wave packets we ensure the approximations made in the quantum white noise limit are not violated.
	
By selecting a time $t$ at which to divide the integral in \erf{eqBdef}, one can clearly see that the single-photon Fock state possesses temporal mode entanglement \cite{GJNphoton},
	\begin{align} \label{Eq::TempModeDecomposition}
		\ket{1_{\xi}} & = \int_0^t dt' \, \xi(t') b\dg(t')\ket{0} +  \int_t^\infty dt' \, \xi(t') b\dg(t')\ket{0} .
	\end{align} 
That is, the photon is in a superposition of being found before time $t$ and after time $t$.  This property is not shared by vacuum, coherent, and thermal fields\footnote{Squeezed states are under investigation.}, which can all be written as a continuous-time tensor product state \cite{Par92,CookThesis}.  While not an attribute that we study in this dissertation \emph{per se}, it is this temporal entanglement that makes $N$-photon states manifestly non-classical and complicates dynamical descriptions when they interact with other systems.  

A straightforward extension leads to the definition of normalized, continuous-mode Fock states (referred to hereafter as \emph{Fock states}) in the wave packet $\xi(t)$ with $N$ photons \cite{BloLou90},
	\begin{subequations} \label{Eq::ContModeFockState}
	\begin{align}
		\ket{N_{\xi}}  &=\frac{1}{\sqrt{N!}}\left [\int dt' \, \xi(t') b\dg(t')\right]^{N} \ket{0}\\
                   &=\frac{1}{\sqrt{N!}}\left [ B\dg(\xi)\right ]^{N}\ket{0}. 
	\end{align}
	\end{subequations}
With respect to the temporal envelope $\xi(t)$, continuous-mode Fock states behave just like the single-mode Fock states with the standard normalization $1/\sqrt{N!}$.  At this point one might imagine that the description is effectively single mode, defined by the temporal envelope $\xi(t)$.  However, interactions occur at specific local times as seen in the white noise Hamiltonian in \erf{Eq::WhiteNoiseDipHam}, rather than over the entire temporal envelope.  Acting with a white-noise operator on a Fock state,
	\begin{align}
		\hat{b}(t) \ket{N_\xi} = & \frac{1}{\sqrt{N!}} \int_0^t dt_1 \dots \int_0^t dt_N \, \xi(t_1)\dots \xi(t_N)  \hat{b}(t) \hat{b}\dg(t_1) \dots \hat{b}\dg(t_N) \ket{0} \\
		= & \xi(t) \frac{N}{\sqrt{N!}} \int_0^t dt_1 \dots \int_0^t dt_{N-1} \,  \hat{b}\dg(t_1) \dots \hat{b}\dg(t_{N-1}) \ket{0} \\
		= & \sqrt{N} \xi(t) \frac{1}{\sqrt{(N-1)!}} \big[ B\dg(\xi)\big]^{N-1} \ket{0} \\
		= & \sqrt{N} \xi(t) \ket{N-1_\xi}, \label{Eq::bAction}
	\end{align}
not only lowers the photon number by one and generates a $\sqrt{N}$ factor that comes from normally ordering, but also pulls out the value of the temporal envelope at time $t$.  This can be understood by \erf{Eq::bAction} taking the inner product to find the probability density for finding a photon at time $t$,
	\begin{align} \label{Eq::FockPhotonProb}
		\mathbbm{P}(t) = \bra{N_\xi} \hat{b}\dg(t) \hat{b}(t) \ket{N_\xi} = N |\xi(t)|^2.
	\end{align}
One can similarly perform a temporal decomposition as in \erf{Eq::TempModeDecomposition} to find that the temporal correlations become even more intertwined due to the indistinguishability of the photons.  

When multiple photons are produced they are often spectrally correlated, as in spontaneous parametric down conversion, which translates to temporal correlations.  The Fock states in \erf{Eq::ContModeFockState} are a subset of these more general $N$-photon states for which the SDF is not, in general, factorizable \cite{Ou06, Roh07}.  These states are defined and treated in \srf{Sec::GeneralWavepackets}.

	\subsection{Interactions in one or many modes}	\label{SEC::SingleModeInteractions}

The quantum white noise limit of \srf{SEC::InteractionWithQuantumSystems} is derived for system-field interactions within a single spatial/polarization mode.  However, in physical situations, a system with dipole matrix element $\mathbf{d}_{eg}$ couples to each field mode with strength,
	\begin{align} \label{Eq::ModeCouplingStrength}
		\kappa_{i}(\omega) = \mathbf{d}_{eg} \cdot \vec{u}_{i}(\mathbf{r}) \sqrt{ \frac{\omega}{\hbar  A c } }, 
	\end{align}
where $\vec{u}_{i}(\mathbf{r})$ is one of a set of dimensionless, orthonormal modes, as in \srf{Sec::ClassicalParaxialFields} \cite[Ch. 2]{CarmichaelBook}.  As described in \srf{Sec::WhiteNoiseLimit}, the interaction is centered around a system frequency $\omega_0$, and the Markov approximation is made, giving a coupling rate to mode $i$, 
	\begin{align} \label{Eq::ModeCouplingRate}
		\gamma_i = 2 \pi |\kappa_{i}(\omega_0)|^2.
	\end{align}
The total decay (coupling) rate into all modes given by $\Gamma = \sum_i\gamma_i$; in free space, $\Gamma$ is the Wigner-Weisskopf spontaneous emission rate.  When the rate of coupling to a particular mode, labeled by $g$, far exceeds that to others, then the interaction is effectively single mode and $\gamma_g \approx \Gamma$.  

The single-mode approximation 
is rooted in the presumption that light in a chosen mode can be efficiently coupled to the system.  An example is the excitation of a two-level atom in free space by input light whose spatial profile matches the dipole emission pattern \cite{TeyKur08,WanSheSca10}.  To achieve proper mode matching, the input light must be prepared in a mode consisting of a symmetric combination of left- and right-propagating dipole waves \cite{WanSca12}.  Alternatively, the atom can be placed at the focus of a parabolic mirror \cite{AlbLeu13}.  This is more closely analogous to the standard physical situation for a single-mode approximation using a single-sided cavity \cite{vanHMab05, JohFio11, DilKuh12}, where there is no distinction between forward and backward propagation.  

Another widely applicable physical context where the continuous-mode formalism can be useful is that of strongly confined one-dimensional photonic or microwave waveguides\footnote{Methods for continuous-mode quantization in a dielectric waveguide are given in Refs. \cite{BloLou90, Mil95}.  Since one quantizes the photon flux, the pre-factor in the expression for the electric field, \erf{Eq::BlowEMField}, contains the group velocity in the medium.  The coupling strength, \erf{Eq::ModeCouplingStrength}, is likewise modified.} \cite{LeKHak05, VetRau10, RepSheFan10, vanLoo13, GobKim14}.  In such systems there is often significant backscattering, and one must often consider both forward- and backward-propagating guided modes.  In some cases, the coupling rates into the guided modes, $\gamma_g$, can be much larger than into all other modes, $\gamma_\perp$.  The total decay rate is modified in the presence of a dielectric and is found by summing over all modes, $\Gamma = \gamma_\perp + \sum_g \gamma_g$ ~\cite{ Chang07, CheWubMor11}.  For freely propagating waveguide modes, a side-coupled or embedded quantum system of interest couples both to forward- and backward-propagating modes, although not necessarily symmetrically \cite{LeKRau14}.  A single-mode approximation depends on the nature of the input field; i.e. it may apply when the input light is prepared in a symmetric combination of forward- and backward-propagating guided modes.    

In the following sections, we abstract away the physical details, realizing that the relevance of our theory relies on the degree to which single-, two-, or finite multi-mode approximations apply in any physical situation.  We work from an idealized standpoint where coupling other than to the modes of interest can be fully suppressed, and we set $\gamma_\perp = 0$.  With that in mind, we point out that using the tools of \srf{sec_2modefock}, the effects of these ancillary modes can be readily included.

	\subsection{Interactions with Fock states: two-level atom}

The non-classical temporal correlations in Fock states along with the fact that they are modified during interactions generates rich and interesting dynamics which arise even for the relatively simple case of a two-level atom.  Consider such an atom with ground and excited states $\ket{g}$ and $\ket{e}$ separated by transition frequency $\omega_0$ as it interacts with a continuous-mode field, assumed to be strongly coupled to the atom in both spatial and polarization degrees of freedom.  The system and field couple via the interaction-picture Hamiltonian in \erf{Eq::IntHamRWA} with $\hat{c} = \op{g}{e}$ and coupling strength $\kappa(\omega_0)$ as in \erf{Eq::ModeCouplingStrength}.  When the atom is initially in the ground state and the input field is a single-photon Fock state with resonant carrier frequency ($\omega_c = \omega_0$), the joint state can be written in the single-excitation subspace as
	\begin{align} \label{Eq::JointStateOnePhoton}
		\ket{\psi(t)} = c_e(t) \ket{e} \otimes \ket{0} + \ket{g} \otimes \int d\omega \, c_{g,\omega}(t) \hat{b}\dg(\omega) \ket{0}.
	\end{align}
For an off-resonant carrier frequency $\omega_c$, the resulting detuning appears in an additional system Hamiltonian $\hat{H}_{\rm sys}$\footnote{For example, a two-level atom with bare Hamiltonian in the Schr\"{o}dinger picture, $\hat{H}_{\rm sys} = \smallfrac{\omega_0}{2} \hat{\sigma}_z$, has a remaining bare Hamiltonian in the interaction picture, $\hat{H}_{\rm sys} = -\smallfrac{\Delta}{2} \hat{\sigma}_z$, where $\Delta = \omega_c - \omega_0$.}.  The equations of motion for the probability amplitudes follow from the Hamiltonian,
	\begin{align}
		\dot{c}_{g,\omega}(t) &= \kappa(\omega) c_e(t) e^{i (\omega - \omega_0) t},  \\
		\dot{c}_e(t) &= -  \int d\omega \, \kappa(\omega) c_{g,\omega}(t) e^{-i (\omega - \omega_0) t},
	\end{align}
and, within the Markov approximation, can be integrated to find the formal solutions,
	\begin{align}
		c_{g,\omega}(t)  = & c_{g,\omega}(0) + \kappa(\omega_0) \int_0^t dt' c_{e}(0) e^{i(\omega - \omega_0) t'} e^{-\frac{\Gamma}{2}t'} \label{Eq::GroundAmp}  \\
		& + |\kappa(\omega_0)|^2 \int_0^t dt'  \int_0^{t'} dt'' e^{i(\omega - \omega_0) t'}e^{-\frac{\Gamma}{2}(t'-t'')} \int d\omega' \, c_{g,\omega'}(0) e^{-i(\omega' - \omega_0)t''} , \nn \\
		c_e(t) = & c_e(0)e^{- \frac{\Gamma}{2}t} + \kappa(\omega_0) \int_0^t dt' e^{-\frac{\Gamma}{2}(t-t')} \int d\omega \, c_{g,\omega}(0) e^{-i (\omega - \omega_0) t'} , \label{Eq::ExcitedAmp}
	\end{align}
where $\Gamma$ is the total decay rate.  

For an atom in the ground state interacting with a quasi-monochro-matic, single-photon Fock state, [\erf{Eq::SinglePhotonFreqDomain}], the initial conditions are $c_e(0) = 0$ and $c_{g,\omega}(0) = \xi(\omega).$  Recognizing the inverse Fourier transform, we rewrite the coefficient in terms of the time-domain wave packet,
	\begin{align}
		c_{g,\omega}(t)  = & \xi(\omega)+ 2 \pi|\kappa(\omega_0)|^2 \int_0^t dt'  \int_0^{t'} dt'' e^{i (\omega - \omega_0) t'} e^{-\frac{\Gamma}{2}(t'-t'')} \xi(t'') , \nn \\
		c_e(t)  = & \sqrt{2\pi} \kappa(\omega_0) e^{ - \frac{\Gamma}{2}t} \int_0^t dt' \ e^{ \frac{\Gamma}{2} t'} \xi(t') .
	\end{align}
Using these solutions in the expression for the joint state, \erf{Eq::JointStateOnePhoton}, the probability of finding the atom in the excited state at time $t$ is 
	\begin{align} \label{Eq::AnalyticProb}
		\mathbbm{P}_e(t) = 2 \pi |\kappa(\omega_0)|^2 e^{-\Gamma t} \bigg| \int_0^t dt' \, \xi(t') e^{ \frac{\Gamma}{2} t' } \bigg|^2.
	\end{align}
In the input-output formalism, note that the coupling rate to the guided modes, [\erf{Eq::KappaGamma}], is $\gamma = 2 \pi |\kappa(\omega_0)|^2.$

	\subsubsection{Excitation with an $N$-photon wave packet}

The method in the previous section can be extended to higher photon number with a solution given by Julio Gea-Banacloche \cite{BanaclochePrivate}. The details are given in Appendix \ref{Appendix::NPhotonExcitation}; the result is summarized here.  While not restricted to the single-excitation subspace, the joint state at any can time can still generally be written as
	\begin{equation}
		\ket{\psi(t)} = \ket e \otimes \ket{\phi_e(t)} + \ket g \otimes \ket{\phi_g(t)},
		\label{e7}
	\end{equation}
where $\ket{\phi_e(t)}$ and $\ket{\phi_g(t)}$ are photonic wave functions to be determined.  Their equations of motion, Eqs. (\ref{ApEq::PhotonicEOMs}), follow from the white-noise Hamiltonian, \erf{Eq::WhiteNoiseDipHam}.  When the atom is initially in the ground state and the input field is an $N$-photon Fock state, \erf{Eq::ContModeFockState}, the analytic solution for the excited photon wave function, derived in Appendix \ref{Appendix::NPhotonExcitation}, is
	\begin{align}
		\ket{\phi_e(t)} &= -\sqrt{\gamma N} \int_0^t dt_1 e^{-\frac{\gamma}{2}(t-t_1)}  \xi(t_1)\,  \ket{N-1_\xi}\label{e13} \\
		&\! \!+\gamma^{3/2} \sqrt{N(N-1)} \int_0^t \!\! dt_1\!  \int_0^{t_1} \!\! dt_2 \!  \int_0^{t_2} \!\! dt_3 \, e^{-\gamma(t-t_1)} e^{-\gamma(t_2-t_3)} \xi(t_1)  \xi(t_3) \hat{b}^\dagger(t_2)  \ket{N-2_\xi} \cr
		&\! \!-\gamma^{5/2}\sqrt{N(N-1)(N-2)}  \int_0^t dt_1 \int_0^{t_1} dt_2 \int_0^{t_2} dt_3 \int_0^{t_3} dt_4 \int_0^{t_4} dt_5 \nn \\
		&\qquad \times e^{-\gamma(t-t_1)} e^{-\gamma(t_2-t_3)} e^{-\gamma(t_4-t_5)} \xi(t_1)  \xi(t_3) \xi(t_5)  \hat{b}^\dagger(t_2) \hat{b}^\dagger(t_4) \ket{N-3_\xi} + \dots \nn
	\end{align}
The expression for the ground photonic wave function $\ket{\psi_g(t)}$ can be found by plugging this solution into (\ref{e8b}) and integrating, thus giving the full joint state.  The beauty of the analytic solution is that, in principle, any expectation values can be calculated; for instance, the excitation probability at any time is $\mathbbm{P}_e(t) = \ip{\phi_e(t)}{\phi_e(t)}$.  In practice, however, this may prove challenging due to complicated time-ordered integrals.

	\subsection{Tracing over the field: master equations}

While it can be useful to have an analytic solution for the full joint state of the system and field, \erf{e13} involves quite a few nested, time-ordered integrals and can be difficult to integrate, even numerically.  The expression results from the fact that the quantum state of the field is modified during the interaction, and Fock states of different photon numbers couple in via the exchange of excitations between the atom and field.  When the system itself is the primary object of interest, we may benefit from using a master equation approach, where the field is traced out.  The resulting equation of motion for the reduced system state, $\hat{\varrho}$, includes the effects of the field which may manifest in coherent driving, and to the extent that the system and field become entangled, decoherence.  

To introduce the machinery used in the derivation of the master equations for Fock-state input fields, we begin with the much simpler case of a system driven by a field in vacuum.  The technique rests on the foundation of quantum stochastic calculus, reviewed in \srf{SEC::QWNL}. The time evolution of an arbitrary system operator, $\hat{X}(0) = \hat{X} \otimes \hat{I}_{\rm field}$, is given by the It\={o} Langevin equation [\erf{Eq::dXAppendix}],
	\begin{align} \label{Eq::ItoLangevin}
		d\hat{X} =&\Big( \smallfrac{i}{\hbar}[\hat{H}_{\rm sys}, \hat{X}] + \mathcal{L}_L\dg[\hat{X} ]\Big) dt +[\hat{L}\dg,\hat{X}]\hat{S} dB_{t} + \hat{S}\dg[\hat{X},\hat{L}] dB\dg_{t}+ (\hat{S}\dg \hat{X}\hat{S}-\hat{X}) d\Lambda_{t}.
	\end{align}
To find the master equation, we first take vacuum expectations of \erf{Eq::ItoLangevin} using the following notation (to be explained in detail in \srf{sec_fock}):  $\mathbbm{E}_{0,0}[d\hat{X}]=\Tr_{\rm sys + field} \big[ (\hat{\rho}_{\rm sys} \otimes \op{0}{0}) d\hat{X} \big]$. The action of the quantum noise increments on vacuum,
	\begin{subequations}
	\begin{align} \label{Eq::VacuumRelations}
		dB_t\ket{ 0}&= \int_t^{t+dt} ds \, \hat{b}(s) \ket{0} = 0,\\
		d\Lambda_t\ket{0}&= \int_t^{t+dt} ds \, \hat{b}\dg (s)\hat{b}(s) \ket{0} = 0,
	\end{align}
	\end{subequations}
reveals that all of the terms in \erf{Eq::ItoLangevin} involving $dB_t$, $dB_t\dg$, and $d\Lambda_t$ vanish.  The equation of motion then becomes,
	\begin{align} 
		\frac{d}{dt} \mathbbm{E}_{0,0}[d\hat{X}] =&\Tr_{\rm sys + field} \big[ (\hat{\rho}_{\rm sys} \otimes \op{0}{0}) \smallfrac{i}{\hbar} [\hat{H}_{\rm sys}, \hat{X}(t)] + (\hat{\rho}_{\rm sys} \otimes \op{0}{0}) \mathcal{L}_L\dg[\hat{X}(t) ] \big] \nn\\
		=&\Tr_{\rm sys} \Big[ \hat{\varrho}_{0,0}(0) \big( \smallfrac{i}{\hbar} [\hat{H}_{\rm sys}, \hat{U}\dg(t)\hat{X}\hat{U}(t)] +  \mathcal{L}_L\dg[\hat{U}\dg(t) \hat{X} \hat{U}(t) ] \big) \Big] \\
		=&\Tr_{\rm sys} \Big[ \hat{X} \big( \smallfrac{-i}{\hbar} [\hat{H}_{\rm sys}, \hat{\varrho}_{0,0}(t) ]  +  \mathcal{L}_L[\hat{\varrho}_{0,0}(t) ] \big) \Big].  \label{Eq::VacdX}
	\end{align}
From the first line to the second, we performed a trace over the field. In doing so we have defined the \emph{reduced} density matrix for the system,
	\begin{align}
		\hat{\varrho}_{0,0}(t) \equiv \Tr_{\rm field} \big[ U(t) \hat{\rho}_{\rm sys} \otimes \op{0}{0} U\dg(t) \big].
	\end{align}

This step is critical to the derivation and to the understanding of the Fock-state master equations derived later in this chapter since, as we will see, interactions with an $N$-photon field can be described by a coupling of systems prepared in different initial states.  Although superfluous in this example, the subscripts label the initial state of the field, vacuum -- labeled 0,0 -- in this case.  Using the cyclic property of the trace in \erf{Eq::VacdX} we obtain the vacuum master equation,
	\begin{align}\label{Eq::vacME}
		\frac{d}{dt} \hat{\varrho}_{0,0}(t) = - \frac{i}{\hbar} \big[\hat{H}_{\rm sys}, \hat{\varrho}_{0,0}(t) \big] + \mathcal{L}_L \big[ \hat{\varrho}_{0,0}(t) \big],
	\end{align}
where the action of Lindblad superoperator in the \sch{} picture is
	\begin{align}  \label{Eq::LindbladSuperO}
		\mathcal{L}_L[ \hat{\varrho} ] \equiv L \hat{\varrho} L\dg - \smallfrac{1}{2} \left( L\dg L \hat{\varrho} + \hat{\varrho} L\dg L \right).
	\end{align}
The master equation evolution of the reduced state in \erf{Eq::vacME} describes a system interacting at each time $t$ with fresh vacuum that is uncorrelated with the system.  The Lindblad dissipator describes the decoherence introduced when excitations present in the atom decay back into the field.  The reduced system dynamics can be expressed as a completely-positive map, and the resulting master equation is closed in the initial conditions.  We will see that for $N$-photon states, this is not the case.



\section{Fock-state master equations} \label{sec_fock}

In this section we derive master equations for a quantum system interacting with a field prepared in a Fock state [\erf{Eq::ContModeFockState}].  The derivation is performed in the Heisenberg picture where the time-dependent operators evolve according to \erf{Eq::ItoLangevin}. Similar to what was done for vacuum in the previous section, we take expectations with respect to input Fock states and then convert to the \sch{} picture for the master equation. To facilitate the derivation we first introduce notation convenient for representing expectations with respect to a Fock states. It should be noted that our method is a generalization to $N$-photon states of a method introduced in Refs.~\cite{GJNphoton, GJNCgen} for a single photon.
	
Before taking expectations, we need the action of the quantum noise increments that appear in \erf{Eq::ItoLangevin} on the input Fock states.  Using \erf{Eq::bAction} we obtain the general actions of the quantum noise increments on Fock states \cite{TeamAwesome, SonXi13},
	\begin{subequations}
	\begin{align}\label{Eq::IncrementActions}
		dB_{t}\ket{n_{\xi}}& =dt \sqrt{n} \xi(t) \ket{n-1_\xi} \\
		 d\Lambda_{t}\ket{n_{\xi}}& = dB\dg_t \sqrt{n} \xi(t) \ket{n-1_\xi}.
	\end{align}		
	\end{subequations}
The composition rules for the quantum noise increments, expressed in \erf{Eq::ItoTable}, are generally modified for non-vacuum fields \cite{GardinerBook,WisMilBook}.  However, the It\={o} table for Fock states is identical to that for vacuum \cite{GJNphoton, GJNCgen, TeamAwesome, CarJam12, SonXi13}.

Assuming no correlations before the interaction, the initial joint state is described by the product state 
	\begin{align}\label{Eq::initState}
		\hat{\rho}(0) = \hat{\rho}_{\rm sys} \otimes \op{N_\xi}{N_\xi}, 
	\end{align}
with the system in the state $\hat{\rho}_{\rm sys}$ and the field in the Fock state $\ket{N_\xi}$. At any time, the reduced state of the system is found by tracing over the field,
	\begin{align}
		\hat{\varrho}_{N,N}(t) \equiv \Tr_{\rm field} \big[ \hat{U}(t) \hat{\rho}_{\rm sys} \otimes \op{N_\xi}{N_\xi} \hat{U}\dg(t) \big],
	\end{align}
where, as above, we use the subscripts $N,N$ to label the reduced system state by the initial field state.  Using the Hilbert-Schmidt inner product for operators $\hat{A}$ and $\hat{B}$,
	\begin{align} \label{Eq::HSInnerP}
		\expt{\hat{A}|\hat{B}} \equiv \Tr [\hat{A}\dg \hat{B}],
	\end{align}
one can take expectations with respect to system and/or field states\footnote{The dagger in the HS inner product is important when considering general operators $\hat{A}$ and $\hat{B}$.  It does not typically appear when taking expectation values, as the physical density matrix is Hermitian; $\hat{\rho} = \hat{\rho}\dg$. }. For the following derivation we make use of the \emph{asymmetric expectation value} \cite{GJNCgen}, defined using \erf{Eq::HSInnerP},
	\begin{align}\label{Eq::GenExpectation1}
		\mathbbm{E}_{m,n}[\hat{\mathcal{O}}] \equiv \mbox{Tr}_{\rm sys + field} \left[ \left( \hat{\rho}_{\rm sys} \otimes \op{m_\xi}{n_\xi} \right) \dg \hat{\mathcal{O}} \right],
	\end{align}
where $\hat{\mathcal{O}}$ can be a joint operator on the system and field and is not necessarily separable\footnote{For a pure initial system state, $\hat{\rho}_{\rm sys} = \op{\psi}{\psi}$, the asymmetric expectation in \erf{Eq::GenExpectation1} above is $\bra{m_\xi} \bra{\psi} \hat{\mathcal{O}} \ket{\psi} \ket{n_\xi}$, and the ordering of the subscripts suddenly makes sense.}.  We use a convention where capital letters, $\ket{N_\xi}$ denote the number of photons in the input field. Lowercase letters, that is, $\ket{n_\xi}$ where $n = \{0,...,N-1\}$, label ``reference'' Fock states to which the system couples.  The asymmetric expectation in \erf{Eq::GenExpectation1} is the link between the Heisenberg picture where the trace over input field states is performed and the \sch{} picture master equation for the reduced quantum state.  

After some interaction time the system and field become entangled. The expectation value in \erf{Eq::GenExpectation1} is
	\begin{align}\label{Eq::GenExpectation1}
		\mathbbm{E}_{m,n} \big[\hat{\mathcal{O}}(t) \big] & = \mbox{Tr}_{\rm sys} \big[ \hat{\varrho}_{m,n}\dg(t) \hat{\mathcal{O}} \big],
	\end{align}
where we have defined a set of generalized density operators, $\hat{\varrho}_{m,n}$ -- first introduced in Ref.~\cite{Zoolander} -- by tracing over the field,
	\begin{align} \label{Eq::GeneralDensityOps}
		\hat{\varrho}_{m,n}(t) \equiv \Tr_{\rm field} \big[ \hat{U}(t) \hat{\rho}_{\rm sys} \otimes \op{m_\xi}{n_\xi} \hat{U}\dg(t) \big].
	\end{align}
Again, the subscripts $m,n$ refer to the initial field state.  Such generalized density operators were also used in Refs. \cite{GJNphoton,GJNCgen, CarJam12} for a single photon. Further interpretation and understanding of these generalized density operators is given in \srf{sec_fockgen}.

As the trace in \erf{Eq::GenExpectation1} is over both system and field, it gives a $c$-number expectation value.  Using the partial trace we can also define an asymmetric partial expectation over the field alone which results in an operator.  We define this operation with the notation\footnote{The symbol $\varpi$ was used in Refs.~\cite{GJNphoton,GJNCgen}. Our definition is different.},
	\begin{align} \label{Eq::GenExpectation2}
		\varpi_{m,n}\big( \hat{\mathcal{O}} \big) &\equiv \Tr_{\rm field} \big[ ( \hat{I}_{\rm sys}\otimes  \op{m_\xi}{n_\xi} ) \dg \hat{\mathcal{O}}  \big]\!.
	\end{align}
The derivation of the Fock-state master equations in the following section is based on the It\={o} Langevin equations of motion for system operators.  In this picture, the state remains separable and the expectations will always have the form of \erf{Eq::GenExpectation1} and \erf{Eq::GenExpectation2}.

	\subsection{Fock-state master equations }\label{sec_fockgen}
%

We are now equipped to derive the Fock-state master equations. To extract the \sch-picture master equations, we take equation of motion for a system operator, \erf{Eq::ItoLangevin}, making use of the asymmetric expectation in \erf{Eq::GeneralDensityOps}: $\mathbbm{E}_{m,n}[ X(t) ] =   \Tr_{\rm sys }[ \hat{\varrho}\dg_{m,n}(t) X]$. Then, using the cyclic property of the trace, we can write down the master equations for the system state:
	\begin{align} \label{Eq::SingleModeMESch}
 		\frac{d}{dt} \hat{\varrho}_{m,n}(t) & = - \smallfrac{i}{\hbar}[\hat{H}_{\rm sys}, \hat{\varrho}_{m,n} ] + \mathcal{L}_L [\hat{\varrho}_{m,n} ]   \\
 		& \!+\! \sqrt{m} \xi(t) [S \hat{\varrho}_{m-1,n}, \hat{L}^\dagger] \!+\!  \sqrt{n} \xi^*\!(t) [\hat{L},\hat{\varrho}_{m,n-1} S^\dagger  ] \nonumber \\
		& \!+\! \sqrt{mn} |\xi(t)|^2\!\left(S \hat{\varrho}_{m-1,n-1} S^\dagger - \hat{\varrho}_{m-1,n-1}  \right) \nonumber\!.
	\end{align}
This set of coupled differential equations is the main result of this section. The initial conditions for these equations follow from \erf{Eq::GeneralDensityOps},
	\begin{align} \label{Eq::RhomnInitCond}
		\hat{\varrho}_{m,n}(0) = \Tr_{\rm field} \big[ \hat{\rho}_{\rm sys} \otimes \op{m_\xi}{n_\xi} \big] = \hat{\rho}_{\rm sys} \delta_{m,n}.
	\end{align}
That is, the diagonal equations $\hat{\varrho}_{n,n}$ should be initialized with the initial system state $\hat{\rho}_{\rm sys}$, while the off-diagonal equations should be initialized to zero. In order to calculate expectation values of system operators for an $N$-photon Fock state one needs only the top-level density operator that describes the physical state, $\hat{\varrho}_{N,N}$.  However, extracting $\hat{\varrho}_{N,N}(t)$ requires propagating all equations between 0 and $N$ to which it is coupled. We note some special cases of \erf{Eq::SingleModeMESch} were derived previously in Refs.~\cite{Zoolander,GJNphoton,GJNCgen}; however, little intuition or physical interpretation was given to these equations.

The master equations in \erf{Eq::SingleModeMESch} require further explanation. The diagonal terms, $\hat{\varrho}_{n,n}$, are valid state matrices describing the evolution of the system interacting with an $n$-photon Fock state for $n \in \{ 0, \dots, N \}$.  For example, when $N=0$ we recover the vacuum master equation, \erf{Eq::vacME}.  From the actions of the quantum noise increments on vacuum, \erf{Eq::VacuumRelations}, we see that the vacuum master equation is the only closed-form equation in \erf{Eq::SingleModeMESch}.  For $N \geq 1$, the diagonal equations couple ``downward'' towards the vacuum master equation via the off-diagonal equations $\hat{\varrho}_{m,n}$ where $m \neq n$.  These off-diagonal operators are non-Hermitian of trace-class zero \cite{Zoolander}; consequently they are not valid state matrices but do satisfy $\hat{\varrho}_{m,n}=\hat{\varrho}_{n,m}\dg$.

The fact that the equations couple downward means that we need only consider a \emph{finite} set of equations, which can be integrated numerically and in some cases, analytically.  For a field in an $N$-photon Fock state there are $(N+1)^2$ equations.  From the symmetry $\hat{\varrho}_{n,m}=\hat{\varrho}_{m,n}\dg$, the number of independent coupled equations reduces to $\half(N+1)(N+2)$. 

Finally, we comment on the physical interpretation of these equations. Absorption of a photon by the system significantly changes a field prepared in a Fock state, so its dynamics are non-Markovian \cite{Zoolander,GJNphoton}.  This necessitates propagating a set of coupled master equations.  In contrast, for coherent states, photons can be removed while leaving the field state unchanged and a single master equation suffices.  Before the wave packet has interacted with the system $\xi(t)$ is zero and only the top level equation $\hat{\varrho}_{N,N}$ contributes to the evolution of the system.  In other words, the system evolves solely under the terms on the first line of \erf{Eq::SingleModeMESch}, which describe evolution from an external Hamiltonian and decay due to coupling to the vacuum. When the wave packet begins to interact with the system $\xi(t)$ becomes nonzero, and the other coupled equations contribute to the evolution of the system. Then, the information flow propagates \emph{upwards} from $\hat{\varrho}_{0,0}$ to $\hat{\varrho}_{N,N}$. 
	
So far we have discussed the dynamics of the system before and during the interaction. The last physically important observation is related to the correlation between the system and the outgoing field during and {\em after} the interaction. Consider the case where $\xi(t)$ is bimodal.  When the temporal spacing between the peaks is much greater than the characteristic decay time of the system and since $\xi(t)$ is zero at these intermediate times, the coherence between the first peak of the wave packet and the system is lost before the second peak begins to interact. Thus only the top-level equation must be propagated at these times, and the only nonzero terms describe external Hamiltonian drive and decay into the vacuum.  When the temporal spacing between the two peaks is on the order of the system decay time or shorter,  then the initial temporal coherence between the peaks can affect the system.
	
From \erf{Eq::GenExpectation2}, we can take the partial trace over Fock states for an arbitrary system operator $\hat{\mathcal{O}} = \hat{X} \otimes \hat{I}_{\rm field}$, whose equation of motion is given by \erf{Eq::ItoLangevin}. Doing so yields the Heisenberg-picture master equations:
	\begin{align} \label{Eq::SingleModeMEHei}
		 \frac{ d }{dt} \varpi_{ m,n} \big( \hat{X}(t) \big) = & \frac{i}{\hbar}\varpi_{m,n}\big( [ \hat{H}_{\rm sys}, \hat{X} ] \big) + \varpi_{m,n}\big(\mathcal{L}\dg_L [ \hat{X} ] \big) \\
          	&+ \nn \sqrt{m} \xi^*(t) \varpi_{m-1,n}\big( S\dg [\hat{X}, \hat{L}]\big) + \sqrt{n} \xi(t) \varpi_{m,n-1}\big([\hat{L}\dg,\hat{X}]S\big) \\
          	&+ \nn \sqrt{ m n } |\xi(t)|^{2} \varpi_{m-1,n-1}\big(S\dg \hat{X}S-\hat{X}\big). 
	\end{align}
In order to find solutions in the Heisenberg picture, one needs to find not only $\varpi_{ m,n} \big( \hat{X}(t) \big)$ for all $\{m,n\}$, but all other operators to which $\hat{X}$ couples as well.  In many cases, solving \erf{Eq::SingleModeMESch} for the reduced state in the \sch-picture provides easier access to expectation values.  However, in situations where the full quantum state is hard to determine or requires massive resources\footnote{Consider a large ensemble of $N$ qubits, whose Hilbert space dimension scales as $2^N$.}, Heisenberg-picture equations can prove useful.

	\subsection{Output field expectation values }\label{Sec::outputfield}

In addition to expectation values of system observables, we may also be interested in features of the output field.  Consider a field observable $\hat{Y}(t)$ with initial condition $\hat{Y}(t_0) = \hat{I}_{\rm sys} \otimes \hat{Y}$.   We insert the It\=o Langevin equation of motion for $\hat{Y}$ into the asymmetric expectations. Using \erf{Eq::GenExpectation2} for the partial trace $\varpi_{m,n}( \hat{Y}(t) )$, the result is operator-valued Heisenberg master equations. We focus here on expectation values, $\mathbbm{E}_{m,n}[\hat{Y}(t)]$, which are found by tracing over the system as well, as in \erf{Eq::GenExpectation1}.   For two field quantities of interest -- photon flux and field quadratures -- we produce a set of coupled differential equations similar in form to \erf{Eq::SingleModeMESch}.  The initial conditions are $\varpi_{m,n}(\hat{Y}(t_0)) = \bra{m_\xi} \hat{Y} \ket{n_\xi} \hat{I}_{\rm sys}$ and similarly $\mathbbm{E}_{m,n}[\hat{Y}(t_0)] = \bra{m_\xi} \hat{Y} \ket{n_\xi}$.  Since the wave packet is assumed at the initial time to be far from the system (which defines $t_0$ in the observables) these expectations vanish at the initial time.  
		
The photon flux is given by $d\Lambda_t$, which counts the number of photons in the field in the infinitessimal time increment $t$ to $t+ dt$.  The rules of It\={o} calculus are used to find the equation of motion for the output photon flux [\erf{Eq::dLambdaAppendix}],
	\begin{align}
		d\Lambda_t^{\rm out} =   \hat{L}\dg \hat{L} dt  + \hat{L}\dg \hat{S} dB_t + \hat{S}\dg \hat{L} dB\dg_t + d\Lambda_t.
	\end{align}
Taking expectations over Fock states using \erf{Eq::GenExpectation2} yields an equation for the mean photon flux operator in terms of remaining system operators, 
	\begin{align}
		\frac{d}{dt} \varpi_{m,n} \big( \Lambda_t^{\rm out} \big) = &\varpi_{m,n} \big( \hat{L}\dg \hat{L} \big)   +  \sqrt{n} \xi(t)\varpi_{m,n-1} \big( \hat{L}\dg \hat{S} \big) \label{Eq::VarPiLambda} \\
		&+  \sqrt{m} \xi^*(t)\varpi_{m-1,n} \big( \hat{S}\dg \hat{L} \big)  + \sqrt{mn}  |\xi(t)|^2  \varpi_{m-1,n-1}\big( \hat{I}_{\rm sys} \big) \nn.
	\end{align}
Tracing over the system yields an equation for the mean photon flux\footnote{Note that in the final term in \erf{Eq::VarPiLambda} and in \erf{Eq::SingleModeFieldME_lambda}, the off-diagonal expectations vanish.  }, 
	\begin{align} \label{Eq::SingleModeFieldME_lambda}
		\nn \frac{d}{dt} \emn{m,n} \big[\Lambda^{\rm out}_t \big] = & \emn{m,n} \big[ \hat{L}\dg \hat{L} \big] + \sqrt{n} \xi(t) \emn{m,n-1} \big[ \hat{L}\dg \hat{S} \big]    \\
 		& + \sqrt{m} \xi^*(t) \emn{m-1,n} \big[  \hat{S}\dg \hat{L} \big] + \sqrt{mn} |\xi(t)|^2 \mathbbm{E}_{m-1,n-1} \big[ \hat{I}_{\rm sys} \big] .
	\end{align}
	
For an input Fock state with $N$ photons, the top-level solution to this equation, $\mathbbm{E}_{N,N}[\Lambda_t^{\rm out}(t)]$, gives the integrated mean photon number up to time $t$. If one can solve \erf{Eq::SingleModeMESch} for the generalized density operators, the most direct route to finding the expectation value of field operators is to use $\hat{\varrho}_{m,n}(t)$ to directly calculate the asymmetric expectation values, \erf{Eq::GenExpectation1}, that appear in \erf{Eq::SingleModeFieldME_lambda}.  In this case, one need only calculate the physical observable found from the top-level equation,
	\begin{align} \label{Eq::TopLevel_lambda}
		\nn \frac{d}{dt} \emn{N,N} \big[\Lambda^{\rm out}_t \big] = & \emn{N,N} \big[ \hat{L}\dg \hat{L} \big] + \sqrt{N} \xi(t) \emn{N,N-1} \big[ \hat{L}\dg \hat{S} \big]    \\
 		& + \sqrt{N} \xi^*(t) \emn{N-1,N} \big[  \hat{S}\dg \hat{L} \big] +N |\xi(t)|^2 .
	\end{align}
This equation has the following interpretations: the final term is the photon flux from the unperturbed free Fock state [\erf{Eq::FockPhotonProb}], the first term arises from the field radiated due to an excitation in the system, and the two middle terms describe interference between absorbed and emitted photons.  

We can also calculate the output quadratures.  A Hermitian field quadrature $Z_t$ measurable via homodyne detection is described by  
	\begin{align}
		Z_t= e^{i \phi} B_t +e^{-i\phi} B\dg_t.
	\end{align}
Following the same prescription, the equation of motion for the quadrature after the interaction is given in \erf{Eq::dBAppendix}, 
	\begin{align}
		dZ_t^{\rm out}  &= e^{i\phi}dB_t^{\rm out} + e^{-i\phi}dB_t^{\dagger \rm out} \nonumber \\
		&= e^{i\phi}(\hat{L} dt + \hat{S} dB_t) + e^{-i\phi}(\hat{L}\dg dt + \hat{S}\dg dB_t\dg) .
	\end{align}
Taking expectations over Fock states using \erf{Eq::GenExpectation1} gives the mean homodyne current,
	\begin{align} \label{Eq::SingleModeFieldME_B}
		\frac{d}{dt} \emn{m,n} \big[Z_t^{\rm out}(t) \big] = & \emn{m,n} \big[e^{i \phi} \hat{L} + e^{-i \phi} \hat{L}\dg \big] \\
		&+ e^{i \phi} \sqrt{n} \xi(t) \emn{m,n-1}\big[ \hat{S} \big] + e^{-i \phi} \sqrt{m} \xi^*(t) \emn{m-1,n}\big[ \hat{S}\dg \big] \nonumber.
	\end{align}

	\subsection{System correlation functions: quantum regression theorem for Fock-state input}

One method for calculating correlation functions for system operators is using the quantum regression theorem \cite{Lax63}, reviewed in Appendix \ref{Appendix::QRT}.  Since the Fock-state master equation formalism requires the use of a whole set of auxiliary generalized density operators, it is not clear how one should apply the quantum regression theorem.  However, Gheri \emph{et al.} had already understood (and possibly worked out) the relevant equations for the case of a single photon in 1999, hinted at in Ref. \cite{Zoolander}.  The results in this subsection first appeared in the supplemental material of Ref. \cite{SatJoh14}\footnote{The notation therein differs slightly from that in this dissertation.}, with a derivation in the Schr\"{o}dinger picture, which is, in fact, much more involved.  Here we present the Heisenberg-picture derivation, which follows in a straightforward manner from the techniques in \srf{sec_fockgen}.
	
For a system interacting with an $N$-photon Fock state, we wish to calculate two-time correlations between system operators $\hat{A} \otimes \hat{I}_{\rm field}$ and $\hat{B}\otimes \hat{I}_{\rm field}$, for times $t$ and $t+\tau$,
	\begin{align}
	 	\expt{\hat{A}(t)\hat{B}(t+\tau)} & = \Tr_{\rm sys+field}[\hat{A}(t) \hat{B}(t+\tau) \hat{\rho}_{\rm tot}(0)] \\
		& = \Tr_{\rm sys} \Big[  \hat{B} \hat{\mathcal{A}}_{N,N}(t, t+\tau) \Big].
	 \end{align}
Using the joint initial state of system and field in \erf{Eq::initState}, we define a two-time operator,
	\begin{align}
		\hat{\mathcal{A}}_{N,N}(t, t+\tau) \equiv & \Tr_{\rm field} \big[ \hat{U}(t, t+ \tau)  \hat{\rho}_{\rm tot}(t) \hat{A} \hat{U}\dg(t+ \tau,t)  \big] 
 \big] \\
 		= & \Tr_{\rm field} \big[ \hat{U}(t, t+ \tau) \hat{U}(t,0)  \hat{\rho}_{\rm sys} \otimes \op{N_\xi}{N_\xi} \hat{U}\dg(t,0) \hat{A} \hat{U}\dg(t, t+ \tau)  \big] 
 \big] 
	\end{align}
subject to the boundary condition ($\tau=0$),
	\begin{align}
		\hat{\mathcal{A}}_{N,N}(t, t) 
		= & \hat{\varrho}_{N,N}(t) \hat{A} . \label{Eq::QRTInitCond}
	\end{align}
	
The equation of motion for $\hat{\mathcal{A}}_{N,N}(t, t+ \tau)$ can be derived in the Heisenberg picture using the same technique as in \srf{sec_fockgen}.  That is, from time $t$ to $t' \equiv t+\tau$, $\hat{\mathcal{A}}_{N,N}(t, t')$ is nothing more than a system operator evolving just as if it were a density operator, \erf{Eq::SingleModeMESch}, with initial condition ($t' = t$) given by \erf{Eq::QRTInitCond}.  Just as for the Fock-state master equations, we see that the physical operator $\hat{\mathcal{A}}_{N,N}(t, t')$, couples to auxiliary operators $\hat{\mathcal{A}}_{m,n}(t, t')$, defined
	\begin{align}
		\hat{\mathcal{A}}_{m,n}(t, t') \equiv \Tr_{\rm field} \Big[ \hat{U}(t,t')  \hat{U}(0, t) \hat{\rho}_{\rm sys} \otimes \op{m_\xi}{n_\xi} \hat{U}\dg (0, t) \hat{A} \hat{U}\dg(t,t') \Big].  
	\end{align}
subject to the boundary conditions,
	\begin{align} 
		\hat{\mathcal{A}}_{m,n}(t, t) = \hat{\varrho}_{m,n}(t) \hat{A}. \label{Eq::InitCondQRTGen}
	\end{align}
The equations of motion for $\hat{\mathcal{A}}_{m,n}(t, t')$ from $t$ to $t'$ are 
	\begin{align} \label{Eq::Amn}
		\frac{d}{dt'}\hat{\mathcal{A}}_{m,n}(t,t')  = & -\frac{i}{\hbar}[\hat{H}_{\rm sys}, \hat{\mathcal{A}}_{m,n}(t, t')] + \mathcal{L}_L \big[ \hat{\mathcal{A}}_{m,n}(t, t') \big] \\
			& +  \sqrt{m} \xi(t') \big[\hat{S} \hat{\mathcal{A}}_{m-1,n}(t, t'), L\dg \big] + dt \sqrt{n} \xi^*(t') \big[L, \hat{\mathcal{A}}_{m,n-1}(t, t') \hat{S}\dg \big] \nn \\
			& +   \sqrt{mn} |\xi(t')|^2 \left( \hat{S} \hat{\mathcal{A}}_{m-1,n-1}(t, t') \hat{S}\dg - \hat{\mathcal{A}}_{m-1,n-1}(t, t') \right). \nn 
	\end{align}

To perform calculations, follows these steps:  first, find all the generalized density operators, $\hat{\varrho}_{m,n}(t)$, at time $t$ and use them to calculate the boundary conditions, \erf{Eq::InitCondQRTGen}.  Second, evolve all of the operators $\hat{\mathcal{A}}_{m,n}(t, t')$ from time $t$ to $t'$.  Finally, at time $t'$ take the trace of the physical operator, $\hat{\mathcal{A}}_{N,N}(t, t')$, with system operator $\hat{B}$.	Since Fock states are defined within a temporal envelope, in general there is no steady state, and the correlation functions and related spectra are time dependent.

	\subsection{Superpositions and mixtures of Fock states}\label{sec_fockinput}

The previous Fock-state results can be generalized to field states described by an arbitrary superpositions and/or mixtures of Fock states in the same temporal wave packet.  As the Fock states span the full Hilbert space, they form a basis for arbitrary states in the wave packet $\xi(t)$,
	\begin{align} \label{Eq::CominationStates}
		\hat{\rho}_{\mathrm{field}} = \sum_{m,n=0}^\infty c_{m,n} \op{m_{\xi}}{n_{\xi}}.
	\end{align}
The coefficients are constrained by the requirements of valid quantum states:  $\hat{\rho} _{\mathrm{field}} \geq 0$, $\mathrm{Tr}[\hat{\rho} _{\mathrm{field}}]=1$, and $\hat{\rho} _{\mathrm{field}\phantom{\dg}}=\hat{\rho} _{\mathrm{field}}\dg$. 

When the input field is described by \erf{Eq::CominationStates} the reduced system state is a superposition over the generalized density matrices in \erf{Eq::GeneralDensityOps}, weighted by the appropriate coefficients,
\begin{equation} \label{Eq::gen_me}
	\hat{\varrho}_{\rm phys} (t)=\sum_{m,n} c_{m,n} \hat{\varrho}_{m,n}(t),
\end{equation}
where $\hat{\varrho} _{m,n}(t)$ are the solutions to the master equations\footnote{Here we use a slightly different convention from Ref. \cite{TeamAwesome} so that the combination of generalized density operators, rather than the combination of asymmetric expectations, uses the unstarred coefficients.}.  Generating the full, physical density operator for an arbitrary field requires combining the appropriate solutions from the hierarchy of coupled equations in \erf{Eq::SingleModeMESch} with associated weights $c_{m,n}$.  This does not change the initial conditions, 

The Heisenberg master equation is found in the same manner,
\begin{equation} \label{Eq::GeneralHEM}
	\varpi_{\rm phys} (t)=\sum_{m,n}c^*_{m,n} \varpi_{m,n}(t).
\end{equation}
Finally, the physical expectation value of a system operator $\hat{X}$ is given by
	\begin{align}
		 \mathbbm{E}_{\rm phys}[\hat{X}(t)] &={\rm Tr}_{\rm sys+field}\big[ \hat{\varrho}\dg_{\rm phys}(t) \hat{X} \big] \\
	&=\sum_{m,n} c^*_{m,n} \mathbbm{E}_{m,n}\big[\hat{X}(t)\big] \label{Eq::GenExpectations}.
\end{align}
This technique also applies to the output field quantities in \srf{Sec::outputfield}.  Note that the definition of the Hilbert-Schmidt inner product, \erf{Eq::HSInnerP}, gives rise to the conjugate coefficients in Eq. (\ref{Eq::gen_me}) but not in Eqs. (\ref{Eq::GeneralHEM}, \ref{Eq::GenExpectations}).  The field quantities behave in the same way.  For example, the output photon flux is given by
	\begin{align} \label{Eq::Superposition_lambda}
		\mathbbm{E}_{\rm phys}\big[\Lambda^{\rm out}_t \big] = \sum_{m,n} c^*_{m,n} \mathbbm{E}_{m,n}\big[\Lambda^{\rm out}_t \big].
	\end{align}

The same procedure applies to the quantum regression theorem for system driven by superpositions and mixtures of Fock states.  In the context of the quantum regression theorem, the solutions to \erf{Eq::Amn}, $\hat{\mathcal{A}}_{m,n}(t,t')$, are combined just as the \sch-picture states in \erf{Eq::gen_me}; that is, using the unstarred coefficients $c_{m,n}$.

	\subsection{Displaced Fock states} \label{SEC::DisplacedFockStates}
	
The Mollow transformation can be used to coherently displace a single frequency mode of the field;  $\hat{D}_\omega\dg(\alpha) \hat{b}(\omega) \hat{D}_\omega(\alpha) = \hat{b}(\omega) + \alpha(\omega).$  Thus, the quantum white noises transform according to
	\begin{align}
		\hat{b}(t)& \rightarrow  \frac{1}{\sqrt{2\pi}} \int_{-\infty}^{\infty} d\omega \, D_\omega \dg(\alpha) \hat{b}(\omega) D_\omega(\alpha) e^{-i \omega t} \nn \\
			& = \frac{1}{\sqrt{2\pi}} \int_{-\infty}^{\infty} d\omega \, \left( \hat{b}(\omega) + \alpha(\omega) \right) e^{-i \omega t} \nn \\
			& = \hat{b}(t) + \alpha(t). \label{Eq::DisplacedQuantumNoise}
	\end{align}
We see that the time domain fields are displaced by a coherent-state wave packet $\alpha(t) = (2\pi)^{-1/2}\int_{-\infty}^\infty d\omega \alpha(\omega) e^{-i \omega t}$, which must also satisfy the quasi-monochromatic, slowly-varying envelope conditions.  The photon flux of the coherent-state wave packet is $|\alpha(t)|^2$, and the total average photon number in the pulse is given by
	\begin{align} \label{Eq::CohPhotonNumber}
		\int dt |\alpha(t)|^2 = \bar{n}.
	\end{align}
In contrast to Fock-state wave packets, the coherent field is not required to have finite photon number - for example, a continuous coherent pump - it is only the flux that must be finite and \erf{Eq::CohPhotonNumber} may be infinite.  Using the displaced white noise operators, \erf{Eq::DisplacedQuantumNoise}, the fundamental quantum noises become
	\begin{align}
		B_t  \rightarrow \int_0^t dt' \big( \hat{b}(t') + \alpha(t') \big)  \quad \quad B_t\dg  \rightarrow \int_0^t dt' \big( \hat{b}\dg(t') + \alpha^*(t') \big) \\
		\Lambda_t  \rightarrow \int_0^t dt' \big(  \hat{b}\dg(t')\hat{b}(t') + \alpha^*(t')\hat{b}(t') + \alpha(t')\hat{b}\dg(t') + |\alpha(t')|^2 \big).
	\end{align}
The displaced quantum noise increments are then,
	\begin{subequations} \label{Eq::DisplacedIncrements}
	\begin{align}
		dB_t & \rightarrow dB_t  + \alpha(t)dt \\
		dB_t\dg & \rightarrow dB_t\dg + \alpha^*(t)dt \\
		d\Lambda_t & \rightarrow 	d\Lambda_t + \alpha^*(t)dB_t + \alpha(t)dB_t\dg + |\alpha(t)|^2 dt.
	\end{align}
	\end{subequations}
Using these quantum noise increments in \erf{Eq::dXAppendix} and 
taking expectations with respect to Fock states using the methods in \srf{sec_fock} gives the displaced Fock-state master equation,
	\begin{align}
 		\frac{d}{dt} \hat{\varrho}_{m,n}(t) & = - \smallfrac{i}{\hbar}[\hat{H}_{\rm sys}, \hat{\varrho}_{m,n} ] + \mathcal{L}_L [\hat{\varrho}_{m,n} ]  \label{Eq::DisplacedFockME} \\
		& \!+\! \alpha(t) [S \hat{\varrho}_{m,n}, L^\dagger] \!+\!\alpha^*\!(t) [L,\hat{\varrho}_{m,n} S^\dagger  ]	\!+\! |\alpha(t)|^2\!\left(S \hat{\varrho}_{m,n} S^\dagger - \hat{\varrho}_{m,n}  \right) \nonumber\! \\
		& \!+\! \sqrt{m} \alpha^*(t) \xi(t) \!\left(S \hat{\varrho}_{m-1,n} S^\dagger - \hat{\varrho}_{m-1,n}  \right) 			 \!+\! \sqrt{n} \alpha(t) \xi^*(t) \!\left(S \hat{\varrho}_{m,n-1} S^\dagger - \hat{\varrho}_{m,n-1}  \right)\nn \\
 		& \!+\! \sqrt{m} \xi(t) [S \hat{\varrho}_{m-1,n}, L^\dagger] \!+\!  \sqrt{n} \xi^*\!(t) [L,\hat{\varrho}_{m,n-1} S^\dagger  ] \nonumber \\
		& \!+\! \sqrt{mn} |\xi(t)|^2\!\left(S \hat{\varrho}_{m-1,n-1} S^\dagger - \hat{\varrho}_{m-1,n-1}  \right) \nonumber\!.
	\end{align}
The first two lines originate entirely from driving by the coherent field\footnote{For $m,n = 0,0$, \erf{Eq::DisplacedFockME} is the standard coherent state master equation. }.  The third line describes processes, mediated by the scattering operator $\hat{S}$, where photons are scattered between the Fock and coherent states.   The two final lines are those dynamics from \erf{Eq::SingleModeMESch} that result from the Fock state field.  Because the displacement is independent of the Fock states, the envelopes $\alpha(t)$ and $\xi(t)$ are not required to be the same shape.  Indeed, by setting $\alpha(t)$ constant, one can model continuous coherent pumping punctuated by interaction with a Fock-state wave packet.  Equation (\ref{Eq::DisplacedFockME}) can also be found by using a modified It\={o} table for coherent states \cite{GardinerBook} and then taking expectations with respect to Fock states.

For completeness, we take expectations of the \emph{output} displaced quantum noise increments with respect to Fock states to find the output photon flux,
	\begin{align}
		\frac{d}{dt} \mathbbm{E}_{m,n} \big[ \Lambda^{\rm out}_t \big] =& \emn{m,n} \big[ \hat{L}\dg \hat{L} \big] + \sqrt{n} \xi(t) \emn{m,n-1} \big[ \hat{L}\dg \hat{S} \big]  + \sqrt{m} \xi^*(t) \emn{m-1,n} \big[  \hat{S}\dg \hat{L} \big]  \nn \\
		& + \alpha^*(t) \big( \mathbbm{E}_{m,n} \big[\hat{L}\big] + \sqrt{n} \xi(t) \mathbbm{E}_{m,n-1} \big[ \hat{S} \big] \big) \nn \\
		&+ \alpha(t) \big( \mathbbm{E}_{m,n} \big[\hat{L}\dg\big] +  \sqrt{m} \xi^*(t) \mathbbm{E}_{m-1,n} \big[ \hat{S}\dg \big] \big) \nn\\
		& + |\alpha(t)|^2 \mathbbm{E}_{m,n} \big[ \hat{I}_{\rm sys} \big] + \sqrt{mn} |\xi(t)|^2 \mathbbm{E}_{m-1,n-1} \big[ \hat{I}_{\rm sys} \big]. \nn
	\end{align}
and output quadratures,
	\begin{align} 
		\frac{d}{dt} \emn{m,n} \big[Z_t^{\rm out}(t) \big] = & \emn{m,n} \big[e^{i \phi} \hat{L} + e^{-i \phi} \hat{L}\dg \big] \\
		&+ e^{i \phi} \sqrt{n} \xi(t) \emn{m,n-1}\big[ \hat{S} \big] + e^{-i \phi} \sqrt{m} \xi^*(t) \emn{m-1,n}\big[ \hat{S}\dg \big] \nonumber \\
		& + \big( e^{i \phi} \alpha(t) + e^{-i \phi} \alpha^*(t) \big)  \mathbbm{E}_{m,n} \big[ \hat{I}_{\rm sys} \big]. \nn
	\end{align}

\section{Examples using the Fock-state master equations} \label{Sec::2LevelExcitation}	

In this section, we present a variety of examples to illustrate the use and breadth of the Fock-state master equations.  In \srf{SEC::2photonME}, we examine the form of the master equation for the simple case of a two-photon Fock state. Next in \srf{SEC::1n2photons} we numerically examine a two-level atom interacting on a dipole transition with a wave packet prepared with at most two photons. First, we reproduce the single-photon excitation results from prior studies, then we broaden these results to include two photons and output field quantities. Finally in \srf{Sec::LargePhotonNumbers} we present a numerical study for large-photon-number Fock states. This allows us to explore the relationship between excitation probability, bandwidth, interaction time, and photon number. For photon numbers $N \gg 1$, we identify a region of average strong coupling. 

To use the formalism developed in the previous section, the underlying physical interaction is assumed to be single mode, as discussed in \srf{SEC::SingleModeInteractions}.

\subsection{Two-photon Fock-state master equations}\label{SEC::2photonME}

It is instructive to examine the form of the master equation for the simple case of interaction with a two-photon Fock state where both photons are created in the same temporal wave packet $\xi(t)$, $\ket{\psi}_{ \mathrm{field}}= \ket{2_{\xi}}$. From \erf{Eq::SingleModeMESch}, the two-photon Fock state master equations are,
	\begin{subequations}\label{Eq::twophoton_eg}
	\begin{align}
 		\dot{\hat{\varrho}}_{2,2}(t) = &- \smallfrac{i}{\hbar}[\hat{H}_{\rm sys}, \hat{\varrho}_{2,2} ] + \mathcal{L}_L[ \hat{\varrho}_{2,2}]  + \sqrt{2} \xi(t) [\hat{S} \hat{\varrho}_{1,2}, \hat{L}^\dagger] \label{rho22} +  \sqrt{2} \xi^*(t) [\hat{L},  \hat{\varrho}_{2,1} \hat{S}^\dagger ] \\
		& \nn+ 2 |\xi(t)|^2 \big( \hat{S} \hat{\varrho}_{1,1} \hat{S}^\dagger - \hat{\varrho}_{1,1}  \big) \nn \\
	\dot{\hat{\varrho}}_{2,1}(t) = &- \smallfrac{i}{\hbar}[\hat{H}_{\rm sys}, \hat{\varrho}_{2,1} ] + \mathcal{L}_L[  \hat{\varrho}_{2,1}] + \sqrt{2} \xi(t) [\hat{S} \hat{\varrho}_{1,1}, \hat{L}^\dagger] + \xi^*(t) [\hat{L}, \hat{\varrho}_{2,0} \hat{S}^\dagger ]  \label{rho21} \\
		& \nn + \sqrt{2} |\xi(t)|^2\big( \hat{S} \hat{\varrho}_{1,0} \hat{S}^\dagger - \hat{\varrho}_{1,0} \big)\\
	\dot{\hat{\varrho}}_{2,0}(t) = &- \smallfrac{i}{\hbar}[\hat{H}_{\rm sys}, \hat{\varrho}_{2,0} ] + \mathcal{L}_L[  \hat{\varrho}_{2,0} ]
 		+ \sqrt{2} \xi(t) [\hat{S} \hat{\varrho}_{1,0}, \hat{L}^\dagger] \\ 	
	\dot{\hat{\varrho}}_{1,1}(t) = &- \smallfrac{i}{\hbar}[\hat{H}_{\rm sys}, \hat{\varrho}_{1,1} ] +\mathcal{L}_L[  \hat{\varrho}_{1,1} ]  + \xi(t) [\hat{S} \hat{\varrho}_{0,1}, \hat{L}^\dagger] + \xi^*(t) [\hat{L}, \hat{\varrho}_{1,0} \hat{S}^\dagger ]  \label{rho11} \\
		& \nn + |\xi(t)|^2\big( \hat{S} \hat{\varrho}_{0,0} \hat{S}^\dagger - \hat{\varrho}_{0,0} \big)
\\
	\dot{\hat{\varrho}}_{1,0}(t) = &- \smallfrac{i}{\hbar}[\hat{H}_{\rm sys}, \hat{\varrho}_{1,0} ] +\mathcal{L}_L[  \hat{\varrho}_{1,0} ] + \xi(t) [\hat{S} \hat{\varrho}_{0,0}, \hat{L}^\dagger] \label{rho10}\\
	\dot{\hat{\varrho}}_{0,0}(t) = &- \smallfrac{i}{\hbar}[\hat{H}_{\rm sys}, \hat{\varrho}_{0,0} ] +\mathcal{L}_L[  \hat{\varrho}_{0,0} ] \label{rho00}
	\end{align}
	\end{subequations}
with the initial conditions:
	\begin{align}
		\hat{\varrho}_{2,2}(0) &= \hat{\varrho}_{1,1}(0) = \hat{\varrho}_{0,0}(0) = \hat{\rho}_{\rm sys} \\
		\hat{\varrho}_{2,1}(0) &= \hat{\varrho}_{2,0}(0) = \hat{\varrho}_{1,0}(0) = 0.
	\end{align}
Similar equations to Eqs.~(\ref{Eq::twophoton_eg}) were originally derived in Ref.~\cite[Equations 71 (a)-(f)]{Zoolander} for a two-level atom but without the scattering operator $\hat{S}$.  Later, Ref.~\cite{GJNphoton} generalized to an arbitrary quantum system for single photon input. Then Ref.~\cite{GJNCgen} showed how to propagate these equations for superpositions and mixtures of one photon and vacuum.

Now suppose the input field is in a superposition of one and two photons, $\ket{\psi}_{\mathrm{field}} = c_1 \ket{1_\xi} + c_2 \ket{2_\xi}$ with $ |c_1|^2 + |c_2|^2 = 1$. From \erf{Eq::gen_me} we combine the solutions to the master equations, \erf{Eq::twophoton_eg}, to get the physical reduced system state,
	\begin{align}  \label{Eq::Varrho20}
		\hat{\varrho}_{\rm sys} (t) =& |c_1|^2 \hat{\varrho}_{1,1}(t) + |c_2|^2\hat{\varrho}_{2,2}(t) + c_1 c_2^* \, \hat{\varrho}_{1,2}(t) + c_1^* c_2 \, \hat{\varrho}_{2,1}(t). 
	\end{align}
Notice that the last two terms of \erf{Eq::Varrho20} originate in the coherences of the input field.  It is interesting that the ``off-diagonal,'' traceless generalized density operators (e.g. $\hat{\varrho}_{1,2}$) contribute to the calculation of physical quantities, albeit in Hermitian combinations. They are required to calculate the interference between the one- and two-photon sectors.  Had the field been a pure Fock state or a statistical mixture of one and two photons, these terms would not appear.  Output field quantities are calculated in the same fashion as \erf{Eq::Varrho20}. For example, the mean photon flux is,
	\begin{align} \label{Eq::TwoPhotonLambda}
 		\mathbbm{E}_{\rm phys} [\Lambda_{t}^{\rm out}] &=  |c_1|^2 \mathbbm{E}_{1,1} [\Lambda_t^{\rm out}] + |c_2|^2 \mathbbm{E}_{2,2} [\Lambda_t^{\rm out}]  + c_1^* c_2 \mathbbm{E}_{1,2}[\Lambda_t^{\rm out}] + c_1 c_2^* \mathbbm{E}_{2,1} [\Lambda_t^{\rm out}] ,
	 \end{align}
where \erf{Eq::GenExpectations} is used to calculate $ \mathbbm{E}_{\rm phys} [\cdot]$\footnote{In Ref. \cite{TeamAwesome} there is an error in the coefficients on the interference terms in \erf{Eq::Varrho20} and \erf{Eq::TwoPhotonLambda}.}.

	\subsection{Excitation of a two-level atom with few-photon wave packets}\label{SEC::1n2photons}

Efficient photon absorption is important for information transfer from a flying to a stationary qubit. In this section we analyze this problem with a study of the excitation probability and output field quantities for Fock states interacting with a two-level system.  This problem has been studied before in much detail for a single photon \cite{StoAlbLeu07,WanSheSca10,StoAlbLeu10}. Our intention is to make a direct connection to established results and then to extend those results to higher photon numbers. 

We specialize to a wave packet prepared with up to two photons interacting on a dipole transition with a two-level atom initially in the ground state.  We focus on two square-normalized wave packets.  The first is a rising exponential,
	\begin{align}
		\xi_{\rm rexp}(t)=&\sqrt{\Delta_\omega} \exp{\left[\frac{\Delta_\omega}{2} (t - t_a )\right] } \label{Eq::rexp_xi} ,
	\end{align}
which is known to be optimal for single-photon excitation with parameter $\Delta_\omega = 1$.  The second is a Gaussian as defined in Ref. \cite{WanSheSca10}, 
	\begin{align}
		\xi_{\rm gau}(t)=&\left( \frac{\Delta_\omega^2}{2 \pi} \right)^{1/4} \exp{\left[ -\frac{\Delta_\omega^2}{4} (t - t_a )^ 2\right] } . \label{Eq::gau_xi} 
	\end{align}
For both wave packets, the peak arrives at time $t_a$, which choose to be $0$ for simulations.  For Gaussian wave packets the simple relationship between bandwidth $\Delta_\omega$ and temporal width enables us to explore the tradeoff between interaction time and spectral support around resonance\footnote{As defined in \erf{Eq::gau_xi}, the variance of $|\xi(t)|^2$ is $1/\Delta_\omega^2$ and of $\xi(t)$ is $\sigma_T^2 = 2/\Delta_\omega^2$.  The variance of $|\xi(\omega)|^2$ is $\Delta_\omega^2$ and of $\xi(\omega)$ is $\sigma_\omega^2 = \Delta_\omega^2/2$.  This parameterization of a Gaussian was chosen to aid comparison with previous studies.}. The $(S,L,H)$ triple used in the Fock-state master equations is: $\hat{H}_{\rm sys}=-\hbar\smallfrac{\Delta_0}{2} \big( \op{e}{e} - \op{g}{g}\big)$, $\hat{L}= \sqrt{\gamma} \op{g}{e}$, $\hat{S}=\hat{I}_{\rm sys}$, where $\Delta_0 = \omega_c - \omega_0 = 0$. The atom is assumed to be perfectly coupled to the mode of the input field with a rate is chosen to be $\gamma = \Gamma = 1$.  To study the excitation probability we numerically integrate the master equations (\ref{rho22})--(\ref{rho00}) for a resonant carrier frequency, $\Delta_0 = 0$.  Then, for a given input field state we calculate the excitation probability,
	\begin{align}
		\mathbbm{P}_e(t) & = \Tr \big[ \hat{\varrho}_{\rm sys}(t) \op{e}{e} \big], 
	\end{align} 
where $\hat{\varrho}_{\rm sys}(t)$ is given by \erf{Eq::gen_me}.

	\begin{figure}[!h]
	\centering
	\includegraphics[scale=0.33]{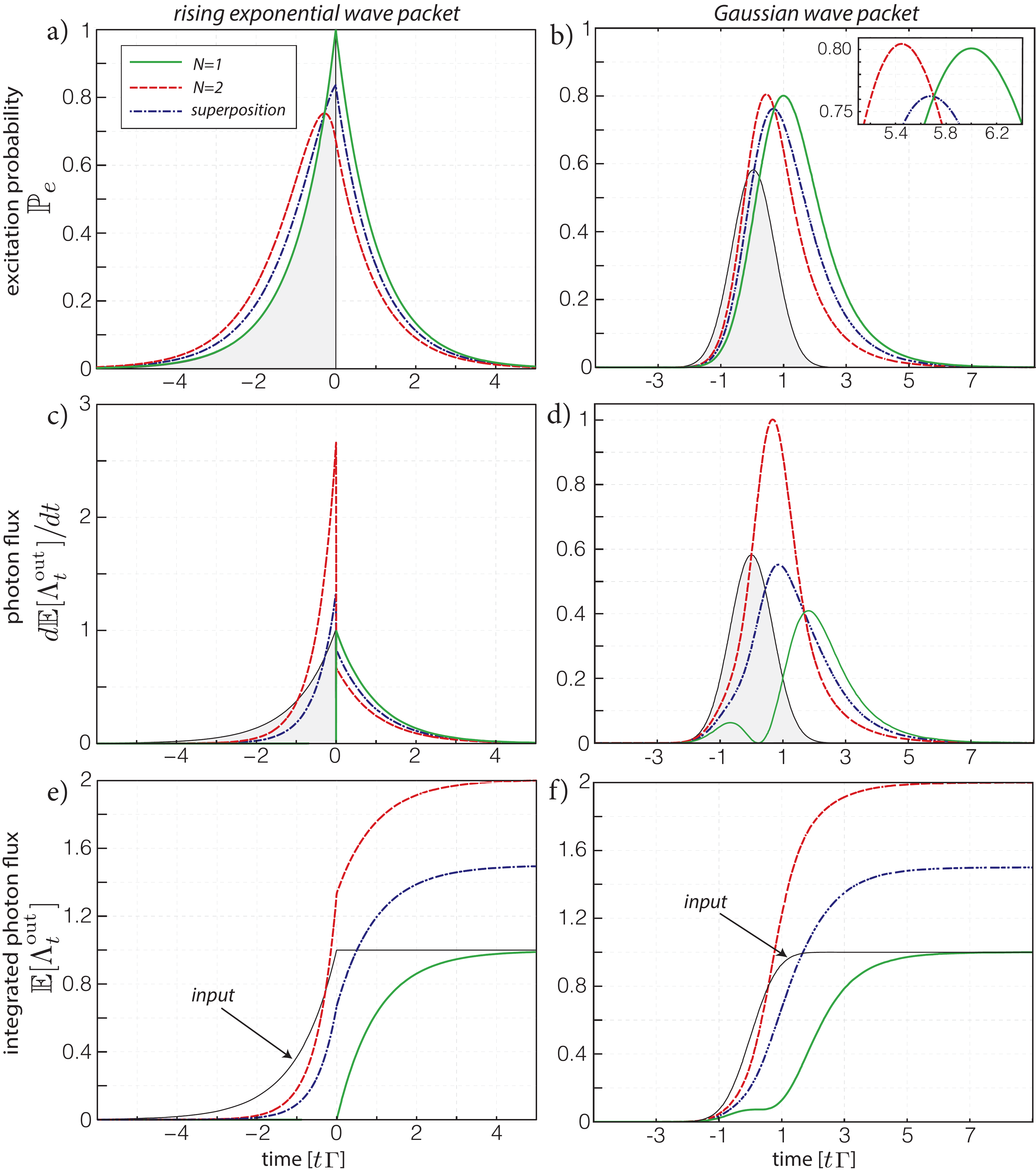} 
		\caption[Excitation and photon flux dynamics for a two-level atom interacting with Fock states]{Dynamics for a two-level atom interacting with rising exponential (first column) and Gaussian (second column) wave packets in three field states: a single-photon Fock state (solid green), a two-photon Fock state (dashed red), and an equal superposition of one and two photons (dash-dot blue).  Both wave packets are chosen to be optimal for a single photon: $\Delta_\omega/\Gamma = 1$ for the rising exponential and $\Delta_\omega/\Gamma = 1.46$ for the Gaussian.  The input wave packet, $|\xi(t)|^2$, is shown in black filled grey. a-b) Excitation probability. c-d) Output photon flux. e-f) Integrated output photon flux.  For comparison the integrated input single-photon flux is shown.  } \label{Fig::2PhotonCompare}
	\end{figure}

Figures \ref{Fig::2PhotonCompare}(a)-(b) present the excitation probability for a two-level atom interacting with rising exponential \erf{Eq::rexp_xi} and Gaussian wave packets \erf{Eq::gau_xi} prepared in three different states: (i) a single-photon Fock state, (ii) a two-photon Fock state, and (iii) an equal superposition of one and two photons; $c_1 = c_2 = 1/\sqrt{2}$ in \erf{Eq::Varrho20}. In the simulations we use a bandwidth known to be optimal for a single-photon wave packets: $\Delta_\omega/\Gamma = 1$ for the rising exponential and $\Delta_\omega/\Gamma = 1.46$ for the Gaussian \cite{StoAlbLeu07}.  As the optimally shaped rising exponential is the time-reversed shape of a decaying excited atom, it can lead to full excitation as seen in \ref{Fig::2PhotonCompare}(a).  A second photon interferes with the excitation, and the maximum is reduced.  For the Gaussian wave packet, the maximum excitation probability is found to be $\mathbbm{P}_e^{\rm max} \approx 0.801$ for $N=1$ as found in other works \cite{StoAlbLeu07,WanSheSca10,StoAlbLeu10}.  Putting a second photon in the wave packet slightly increases this to $\mathbbm{P}_e^{\rm max} \approx 0.805$; however, we see in \srf{Sec::LargePhotonNumbers} that this is not universal behavior for all bandwidths and photon numbers.

In Figs. \ref{Fig::2PhotonCompare}(c)-(d) we plot the mean photon flux of the output field, $d\mathbbm{E}_{\rm phys}[\Lambda_t^{\rm out}]/dt$, after interaction with the atom.  For a single photon, we see a drastic change in the output photon flux for both wave packets.  The rising exponential is completely absorbed by the atom, with re-radiation from atomic decay interfering destructively with the incoming field.  For the Gaussian wave packet, the absorption is not complete; however, the ``double-hump" in the photon flux indicates a period where destructive interference is playing an important role.  The first hump (to the left) is the attenuated input wave packet and the second (to the right) is the re-radiation of the excitation back into the field.  For two photons, however, much of the wave packet travels through the atom undisturbed, since a two-level atom can absorb at most one photon.  The related integrated mean photon flux (total integrated photon number), $\mathbbm{E}_{\rm phys}[\Lambda_t^{\rm out}]$, is plotted in Figs. \ref{Fig::2PhotonCompare}(e)-(f).  For these pure one- and two-photon Fock states there exist a definite number of excitations, whereas for the superposition there is not. Regardless, any excitation induced in the atom through absorption of a photon eventually decays back into the field\footnote{Since the interaction is single mode, ``transmission" and ``reflection" describe the same process.  Scattering between two modes is treated later in \srf{Sec::2modeExample}.}.  This is shown in Fig. \ref{Fig::2PhotonCompare}(c) where the integrated mean photon flux for long times approaches the number of initial excitations, $\{1, 1.5, 2\}$.  During the absorption of the single-photon wave packet, the integrated intensity is zero for the rising exponential and flattens out for the Gaussian since the photon has been transferred to an atomic excitation and arrives only later after decay.  

For a single-photon wave packet, the Schr\"{o}dinger equation can be solved analytically for the excitation probability [\erf{Eq::AnalyticProb}]. The simulations in \frf{Fig::2PhotonCompare} agree with the analytic expression in \erf{Eq::AnalyticProb}.  For larger photon number, the excitation probability can likewise be calculated using embedded time integrals, [\erf{e13}], but the system of Fock-state master equations developed here is easier to integrate numerically \cite{BanaclochePrivate} and, being in the \sch{} picture (as compared to the Heisenberg-picture results in Refs. \cite{DomHorRit02,WanSheSca10,WanSca12}), it gives access to all system expectation values.

	 \subsubsection{System-field entanglement}

	\begin{figure}[!h]
	\centering
	\includegraphics[scale = 0.33]{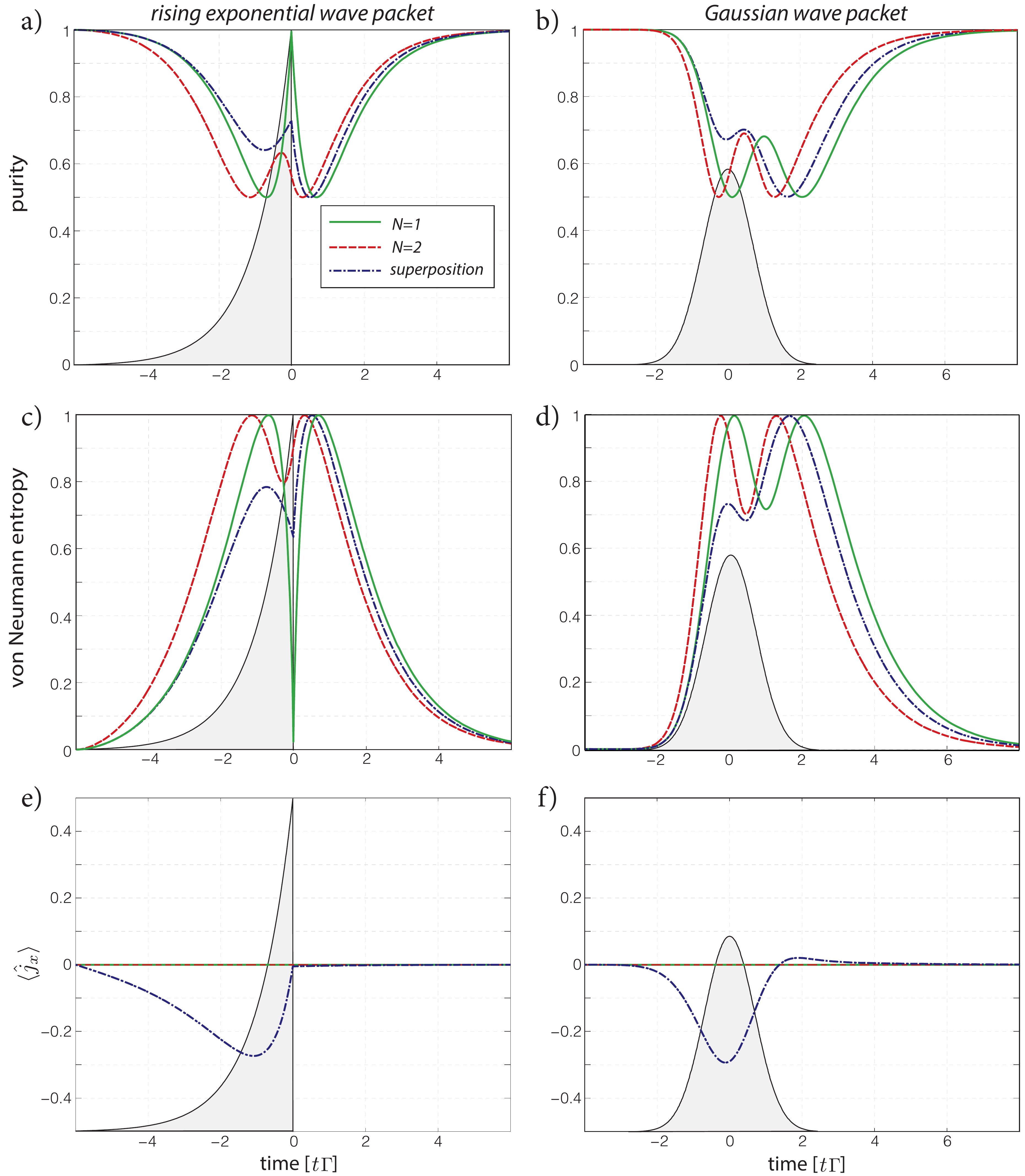}
		\caption[Excitation of a two-level atom with Fock states - dynamics]{Dynamics of the purity, von Neumann entropy, and $\expt{\hat{j}_x}$ Bloch component for a two-level system interacting with rising exponential (first column) and Gaussian (second column) wave packets in three states: a single-photon Fock state (solid green), a two-photon Fock state (dashed red), and an equal superposition of one and two photons (dash-dot blue).  The wave packets, $|\xi(t)|^2$, are shown in black filled grey.  Wave packet parameters and field states are those used in \frf{Fig::2PhotonCompare} (a)-(b) Purity.  (c)-(d) Von Neumann entropy. (e)-(f) $\expt{\hat{j}_x}$ Bloch vector component with $|\xi(t)|^2$ superimposed for clarity.} \label{Fig::PurityEntropy} 
	\end{figure} 
 
As the atom and field interact they become entangled, and at intermediate times we expect to see signatures of this entanglement in the reduced state of the atom. The purity, $\Tr [ \hat{\varrho}_{\rm phys}^2]$, can be used as a witness for system-field entanglement \cite{RosFan11} when the system is prepared in a pure state.  A more direct entanglement measure is the von Neumann entropy, $- \Tr \big[ \hat{\varrho}_{\rm phys} \log_2 ( \hat{\varrho}_{\rm phys} ) \big]$ \cite{NielsenChuang07}.   In \frf{Fig::PurityEntropy} we compare the purity and von Neumann entropy for a two-level atom interacting with rising exponential and Gaussian wave packets prepared with a single-photon, with two photons, and with an equal superposition of one and two photons using the same parameters as in the previous section.  As pure Fock states have no associated phase, the excitation dynamics (see \frf{Fig::2PhotonCompare}) drive the state directly through the center of the Bloch sphere.  When the excitation probability is 0.5, the reduced state is fully mixed \frf{Fig::PurityEntropy} (a)-(b) and the atom is maximally entangled with the field as confirmed by the von Neumann entropy in \frf{Fig::PurityEntropy} (c)-(d).  Similar results were found for single-photon input in Ref. \cite{DaeShe13}.  For the optimal rising exponential wave packet, at the moment of complete absorption of the photon the joint system-field is described by a product state and the entanglement vanishes.  It may seem counterintuitive that the superposition state does not achieve maximum entanglement at 0.5 excitation probability.  This can be understood by looking at the components of the Bloch vector, $\expt{ \hat{j}_i(t) } = \Tr [\hat{\varrho}_{\rm phys}(t) \hat{j}_i] $. The superposition coefficients set a relative phase, which, in our example, manifests as a non-zero $\expt{ \hat{j}_x }$ component of the Bloch sphere \frf{Fig::PurityEntropy} (e)-(f).

Here we have only shown some results relating to system-field entanglement; one may also be interested in measures of initial system-field correlations at each time $t$, which are expected for systems interacting with Fock states.  Further studies are needed to quantify the degree of the non-Markovianity in the Fock-state system dynamics.  A connection between witnesses for non-Markovianity as well as initial system-field correlations considered in Ref. \cite{RodAsp12} could be useful in this endeavor.

	\subsection{Excitation for large photon numbers} \label{Sec::LargePhotonNumbers}

In this section we expand the numerical study of excitation probability to larger photon numbers.  Since perfect excitation can be achieved with a rising exponential pulse with but a single photon, we focus here on Gaussian wave packets, [\erf{Eq::gau_xi}], prepared Fock states with photon number $N\ge1$.

		\subsubsection{Scaling}
	
For small bandwidths ($\Delta_\omega/\Gamma \ll 1$), see the left side of \frf{Fig::NPhotonScaling}(a), one would expect a high probability of excitation from the substantial spectral support near the transition frequency of the atom. However, the long temporal extent of the wave packet means the photon flux over the relevant interaction time scale $\tau = 1/\Gamma$ is too small to significantly excite the atom \cite{WanSheSca10}. A complementary way of understanding this is that the dissipative terms in the master equations [terms on the first line of \erf{Eq::SingleModeMESch}] prevail over the coherent coupling (terms on the other lines). By extending the analysis in Ref.~\cite{Zoolander}, we find a recursive scaling of the excitation probability for very long temporal wave packets: $\mathbbm{P}_{e}^{\rm max}\approx P_{N}$, where $P_{N} = N P_{1}(1-2P_{N-1})$  with  $P_{1}= 4\max |\xi(t)|^{2}$. 
	
In the other asymptotic regime where bandwidths are large ($\Delta_\omega/\Gamma \gg 1$), see the right side of \frf{Fig::NPhotonScaling}(a), the maximum excitation probability is small even for large photon numbers. This is due to the wave packet being so short that its bandwidth is spread over frequencies far from the atomic resonance. We numerically find the asymptotic scaling $\mathbbm{P}_{e}^{\rm max}= 5 N\Gamma/\Delta_\omega$ for $\Delta_\omega/\Gamma \in [10^3,10^7]$ with $R^2=1$ for photon numbers $N \in \{1,\dots,10\}$. 
	
At intermediate bandwidths, we note several interesting features.  First, the maximum excitation probabilities are not universally ordered by photon number and adding photons to the field can, in fact, decrease $\mathbbm{P}_e^{\rm max}$.  Indeed, there exists a bandwidth region in \frf{Fig::NPhotonScaling} where a single photon in the wave packet is optimal for excitation, $\Delta_\omega/\Gamma \in [.5,1.4]$. In a related study, \cite{WanSca12}, it was found that excitation with coherent states of increasing mean photon number do not exhibit crossings in the excitation curves; adding more photons always increases the maximum excitation probability.  Second, for each photon number there exists an optimal bandwidth for excitation.  In \frf{Fig::NPhotonScaling} (b) we have plotted the absolute maximum of $\mathbbm{P}_{e}$ (maximized over $t$ and $\Delta_\omega/\Gamma$) as a function of the number of photons. We find excellent agreement ($R^2 = 1$) by fitting to the model $\mathbbm{P}_e^{\rm max}(N) = 1-aN^{-b}$ over the range $N \in \{10,\dots,40\}$  with coefficients (95\% confidence):  $a = 0.2694  (0.2678, 0.271), \,\,  b = 0.973  (0.9709, 0.975)$. The conclusion is that the absolute maximum of $\mathbbm{P}_{e}$ does monotonically increase with $N$, but with diminishing returns.
	
	\begin{figure}[!t]
	\centering
	\includegraphics[scale=0.7]{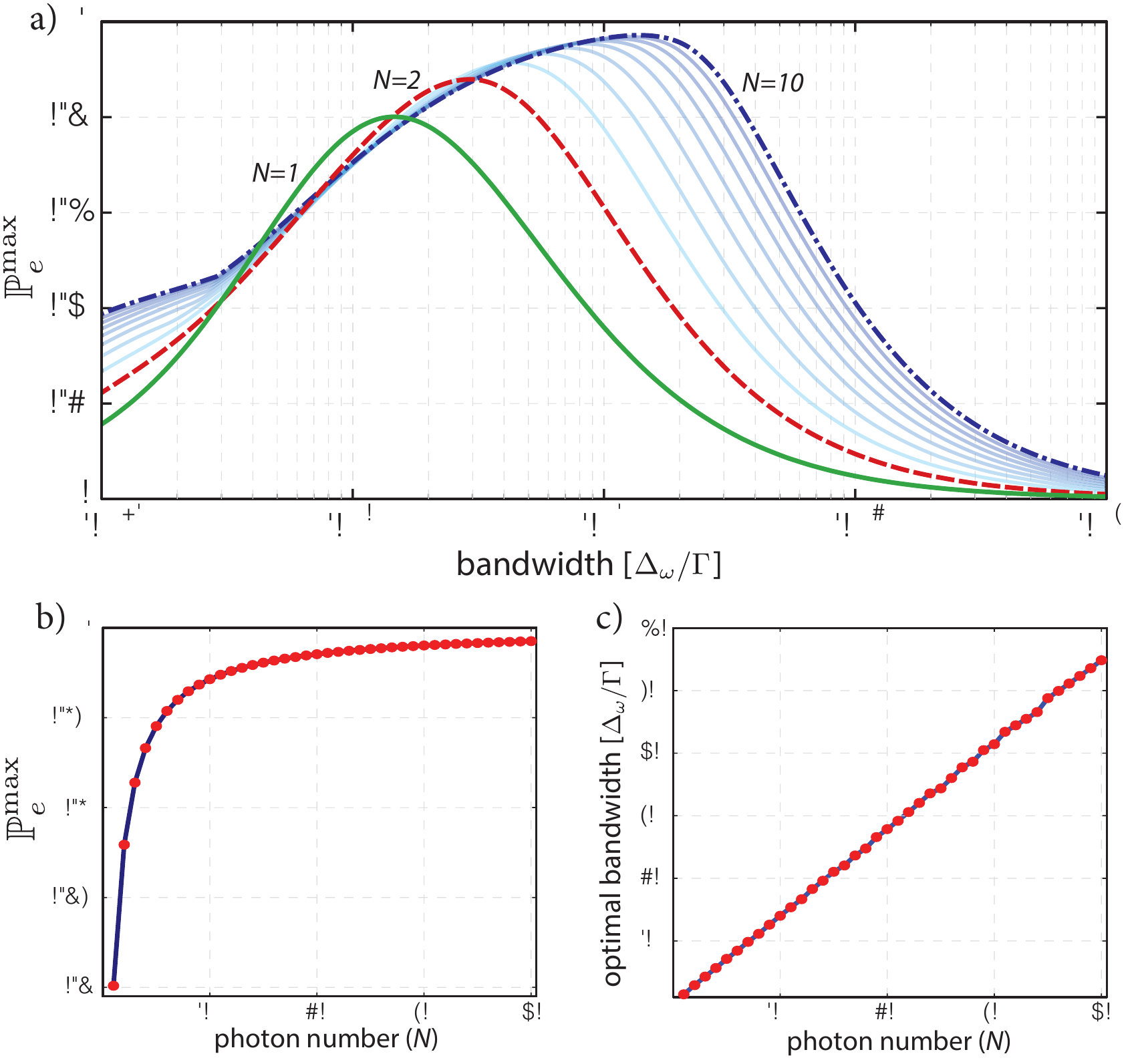}
		\caption[Excitation of a two-level atom with Fock states - scaling]{(a) Maximum excitation probability $\mathbbm{P}_e^{\rm max}$ of a two-level atom interacting with Gaussian wave packets of bandwidth $\Delta_\omega/\Gamma$ for photon numbers $N \in \{1,\dots 10\}$. Small (large) bandwidths correspond to long (short) temporal wave packets. (b) Scaling of $\mathbbm{P}_e^{\rm max}$ with photon number (red circles).  The fit shown is $\mathbbm{P}_e^{\rm max}(N) = 1-0.269\,N^{ -0.973}$ (blue line). (c ) Scaling of $\mathbbm{P}_e^{\rm max}$ with optimal bandwidth for each photon number $N$ (red circles).  The fit is $\Delta_\omega^{\rm opt} (N)/\Gamma = { 1.45} N^{0.987}$.
		} \label{Fig::NPhotonScaling} 
	\end{figure}
	
In \frf{Fig::NPhotonScaling} (c) we investigate the optimal bandwidth for excitation for each photon number $N$. Fitting to the model $\Delta_\omega^{\rm max}(N)/\Gamma = aN^b$ gives $a=1.447(1.418, 1.476)$ and $b=0.9869(0.981, 0.9928)$ with 95\% confidence and $R^2=0.9998$.  Thus, to achieve this scaling for photon number $N$, the optimal bandwidth of the wave packet is $\Delta_\omega^{\rm opt} (N)/\Gamma \approx { 1.45} N^{0.987}$.  Thus, the optimal width seems to be proportional to the single-photon optimal bandwidth, $\Delta_\omega^{\rm opt} (N)/\Gamma \approx  1.46 N$.

			\subsubsection{Excitation dynamics}

\begin{figure}[!h]
	\centering
	\includegraphics[scale = 0.575]{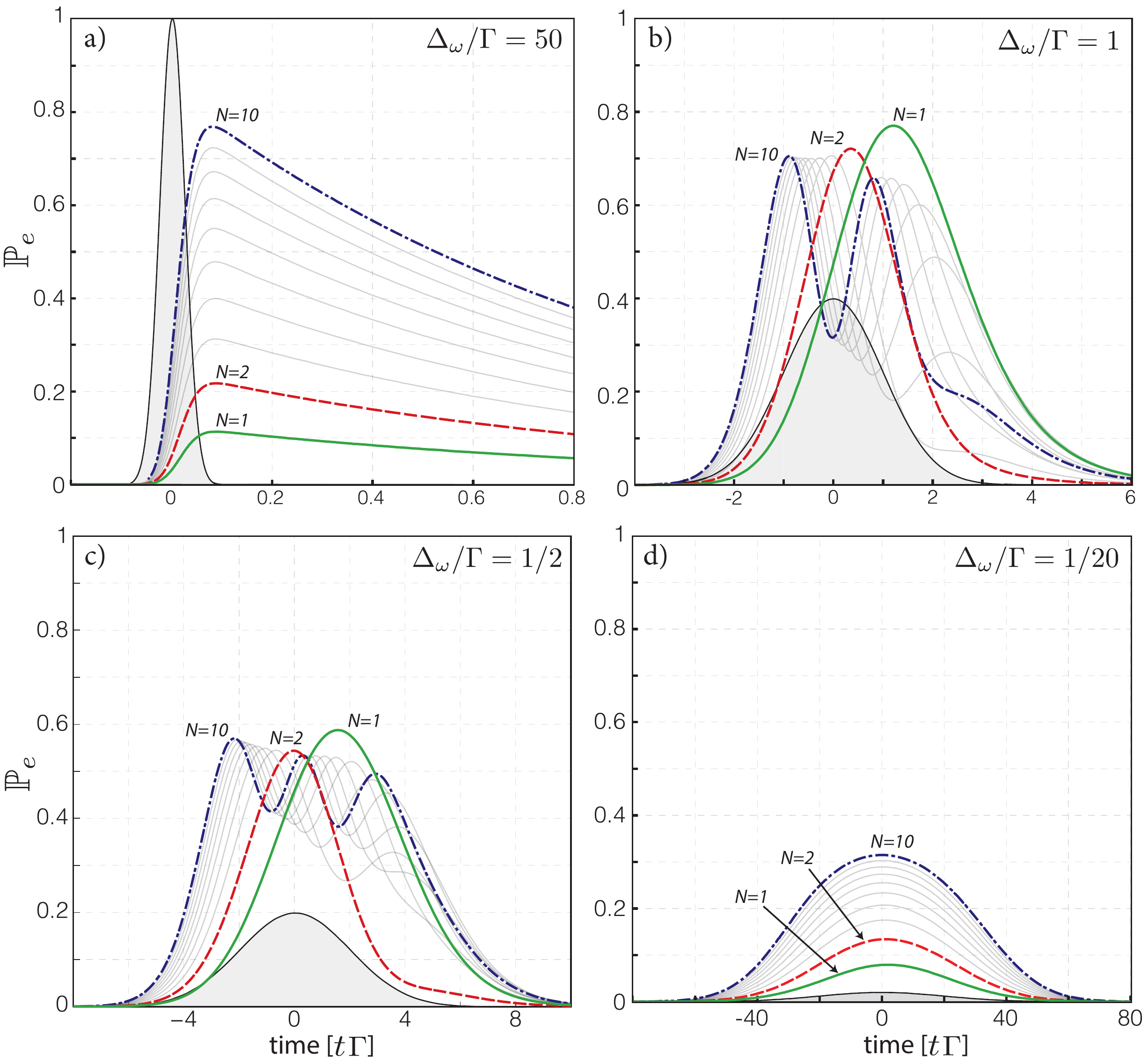}
		\caption[Excitation of a two-level atom with Fock states - dynamics]{Excitation probability $\mathbbm{P}_e$ of a two-level atom interacting with Gaussian wave packets of bandwidth $\Delta_\omega/\Gamma = \{ 50, 1, 1/2, 1/20 \}$ prepared with $N \in \{1, \dots, 10\}$ photons. Highlighted are $N=1$ (solid green), $N=2$ (dashed red), and $N=10$ (dash-dot blue).  The input photon flux $|\xi(t)|^2$ is plotted in black filled grey (normalized in (a) for clarity).  (a) Behavior of short temporal wave packets (large bandwidths) shows $\mathbbm{P}_e$ is ordered by photon number.  (b)-(c) For intermediate bandwidths, we see damped Rabi oscillations, discussed in \srf{SEC::StrongCoupling}.  Note that $\mathbbm{P}_e$ is not necessarily ordered at any time. (d) Behavior of long temporal wave packets (small bandwidths) where $\mathbbm{P}_e$ is again ordered. Note the different time scales in (a)-(d). 
		} \label{Fig::NPhotonBandwidthComp} 
	\end{figure}

Finally we illustrate the excitation probability dynamics.  Figure (\ref{Fig::NPhotonBandwidthComp}) shows $\mathbbm{P}_e$ for bandwidths $\Delta_\omega/\Gamma \in \{ 50 , 1, 1/2, 1/20  \}$, chosen to illustrate three types of behavior.  In each subplot (a)-(d), excitation curves are plotted for photon numbers $N \in \{1,\dots,10\}$.  

In \frf{Fig::NPhotonBandwidthComp}(a) a short pulse quickly excites the atom, which then decays into vacuum with rate $\Gamma$ after the wave packet leaves the interaction region.  Larger photon number corresponds directly to larger maximum excitation. In the intermediate bandwidth regime, $\Delta_\omega/\Gamma \approx 1$, excitations can be coherently exchanged between the atom and field, leading to oscillations in the excitation probabilities.  This continues until the wave packet leaves the interaction region as shown in \frf{Fig::NPhotonBandwidthComp}(b)-(c).  Similar damped Rabi oscillations were observed for large-photon-number coherent state wave packets in Ref.~\cite[Fig. 5]{WanSheSca10}.   For a single photon in the field, these oscillations are never seen due to the tradeoff between spectral bandwidth and photon density \cite{DomHorRit02, SilDeu03}. At the chosen bandwidth $\Delta_\omega/\Gamma = 1$, a single photon achieves the highest maximum excitation with maximum excitation falling off roughly with photon number in agreement with \frf{Fig::NPhotonScaling}.  Finally, in \frf{Fig::NPhotonBandwidthComp}(d) we see that an atom interacting with a long wave packet is excited and then decays well within the wave packet envelope and the $\mathbbm{P}_e(t)$ curves are nearly symmetric around the peak of the wave packet for all photon numbers $N = \{1,\dots,10 \}$.

		\subsection{Strong coupling with Fock states} \label{SEC::StrongCoupling}
		
	\begin{figure}[!h]
	\begin{center}
	\centering
	\includegraphics[scale = 0.83]{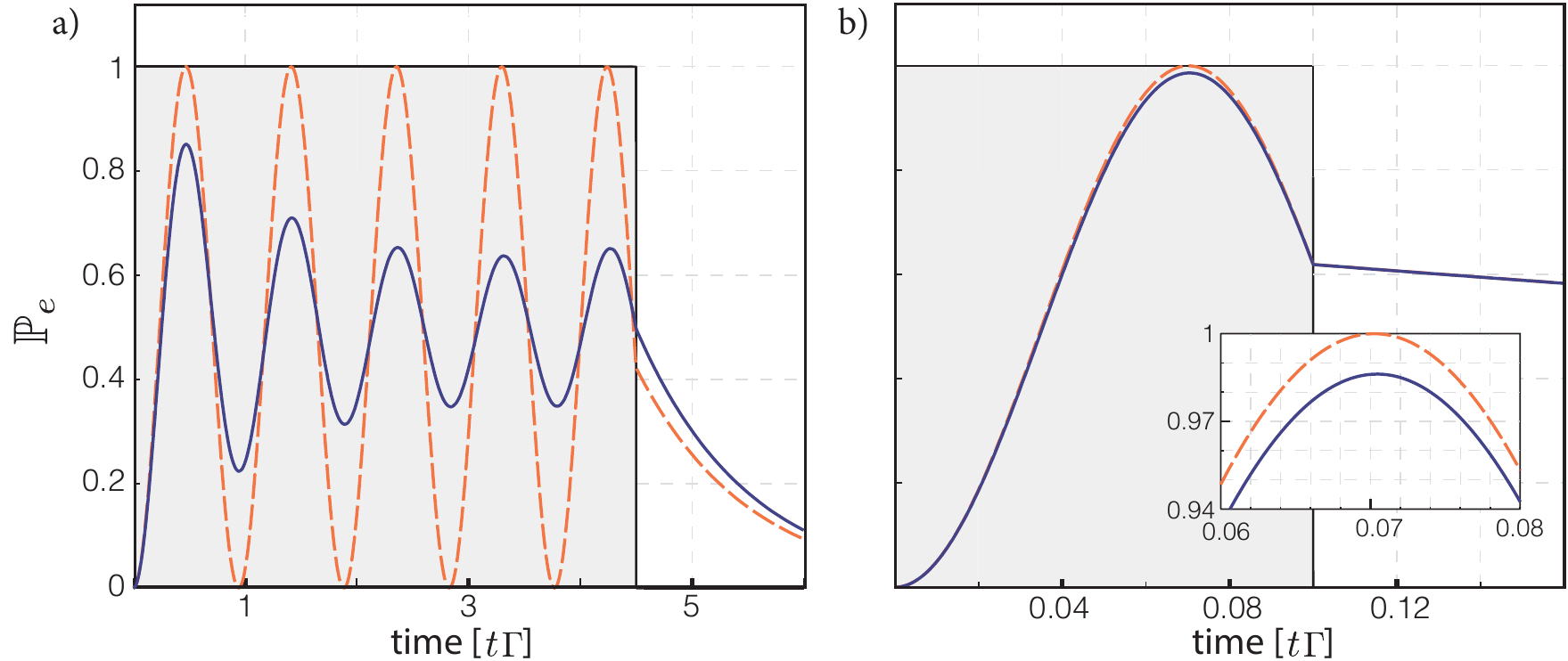}
		\caption[Damped Rabi oscillations for large photon number excitation]{Comparison of the numerically-calculated (dark blue) and analytically-predicted (dashed orange) Rabi oscillations for rectangular wave packets (normalized for clarity) with $N=50$ photons.  (a) Wave packet length $t_{\rm max}$ large compared to $1/\Gamma$. (b) Wave packet length approaching the limit $t_{\rm max} \ll 1/\Gamma$. We see increasing agreement between prediction and numerics.} \label{Fig::IvanMatch} 
	\end{center}
	\end{figure}
	
The damped Rabi oscillations seen in \frf{Fig::NPhotonBandwidthComp}(b) suggest the existence of a regime where coherent processes dominate over dissipation, known in cavity QED as the strong coupling regime. The authors of Ref.~\cite{SilDeu03} defined a strong coupling parameter (for {\em very} short rectangular wave packets): $\sqrt{N}g_{\rm eff} \gg \Gamma$ where $g_{\rm eff} = \xi(t) \sqrt{\gamma_g}$. Specifically the wave packet was taken to be a rectangular pulse; $\xi(t)= 1/\sqrt{t_{\rm max}}$ for times $t \le  t_{\rm max}\ll 1/\Gamma$ and zero otherwise. In this limit they showed that full Rabi oscillations for $N$ photons occur at frequency $\omega_R=g_{\rm eff}\sqrt{N}$.  In  \frf{Fig::IvanMatch} we compare their analytically-predicted excitation oscillations with our numerical calculations for $N=50$ photons.  In (a), the wave packet is long compared to $1/\Gamma$ and, while the oscillation frequencies match, the amplitudes do not due to dissipation.  For short wave packets, as seen in (b), coherent coupling prevails over dissipation, we see excellent agreement with the predicted frequency (in our parameters: $\omega_R=2 \xi(t)\sqrt{\gamma_g N}$) and good agreement with the predicted amplitude.

For non-rectangular pulses the frequency of the Rabi oscillations is time-dependent as seen in \frf{Fig::NPhotonBandwidthComp}(b). The time variation of the wave packet $\xi(t)$ must be accounted for in order to define a general strong coupling parameter.  To achieve strong coupling, the coherent coupling rate into the guided modes $\sqrt{N\gamma_g}|\xi(t)|$ must dominate the total relaxation rate $\Gamma$.  We can immediately define the condition for instantaneous strong coupling:  $\sqrt{N\gamma_g}|\xi(t)| / \Gamma \gg 1$.   However, in order to see interesting dynamics such as a complete Rabi oscillation, the coupling must remain strong over a characteristic timescale $\tau$. From this argument we define an {\em average} strong coupling parameter,
	\begin{align}\label{Eq::strongcoupling}
		\frac{\sqrt{N\gamma_g}}{\tau \Gamma}\int_{t_s-\tau/2}^{t_s+\tau/2}dt  \,  |\xi(t)|\gg 1 \,\,\, \forall \, t_s.
	\end{align}
If, for any wave packet $\xi(t)$, there is a value of $t_s$ such that \erf{Eq::strongcoupling} is much greater than one, then average strong coupling has been achieved over the time window $\tau$.  

	\begin{figure}[!t]
	\begin{center}
	\includegraphics[scale=0.75]{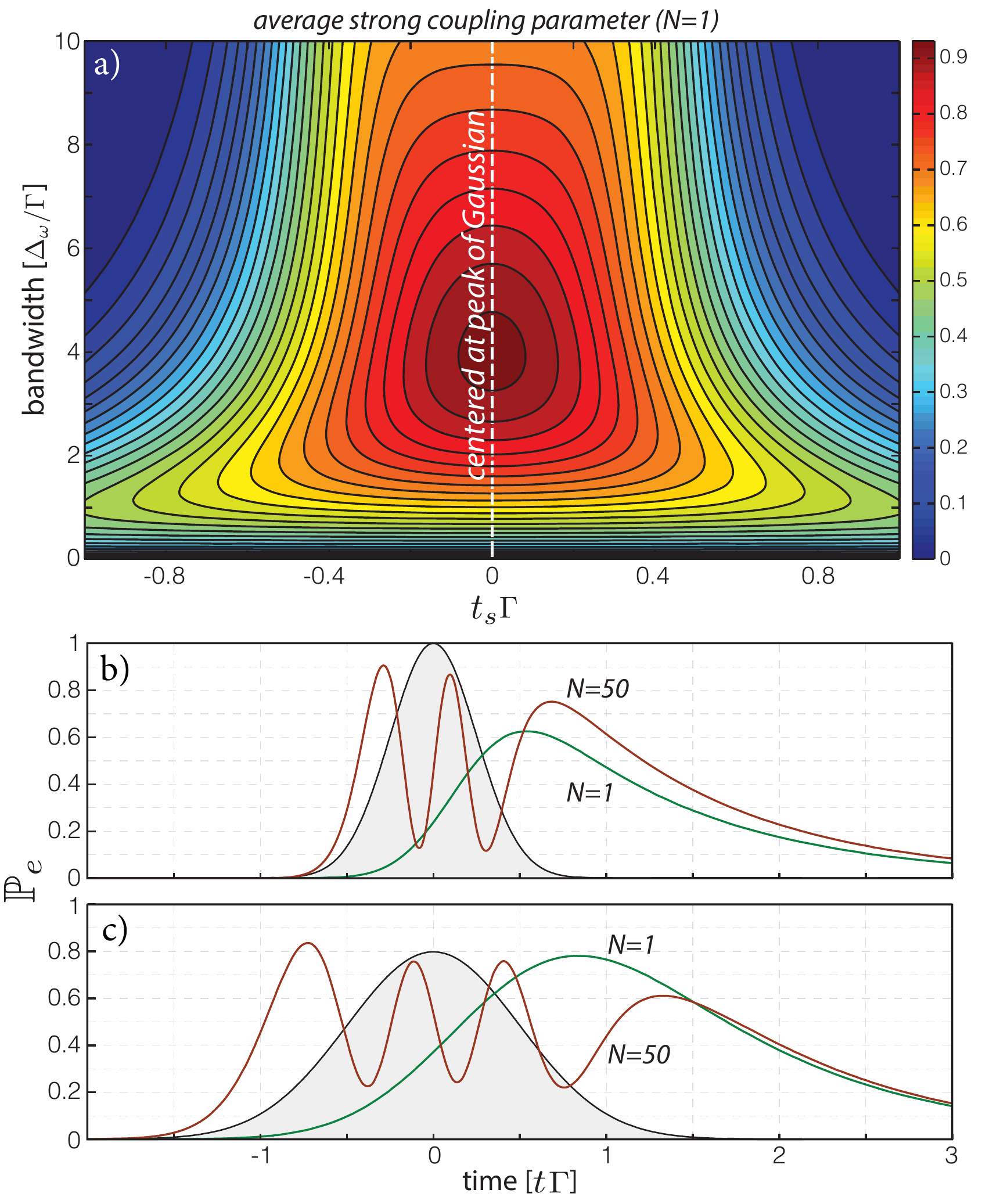}
		\caption[Average strong coupling for Fock-state excitation]{(a) Contour plot of the average strong coupling parameter for a Gaussian wave packet prepared with a single-photon as a function of center of the time window ($t_s$) and bandwidth $\Delta_\omega/\Gamma$ (where $\tau =1/\Gamma$).  (b) and (c): Excitation probability of a two-level atom interacting with a wave packet of bandwidths $\Delta_\omega/\Gamma = 4$ for (b) and $\Delta_\omega/\Gamma = 2$ for (c). Only $N=1$ and $N=50$ photons are shown. The input photon flux $|\xi(t)|^2$ is shown in black filled grey.} \label{Fig::StrongCoupling} 
	\end{center}
	\end{figure}

A natural choice for $\tau$ is the characteristic decay time of the atom, $1/\Gamma$.  In \frf{Fig::StrongCoupling}(a) we present a contour plot of the average strong coupling parameter for Gaussian wave packets prepared in a single-photon Fock state ($N=1$).  Ideal coupling to the guided mode is assumed, $\gamma_g = \Gamma = 1$.  For any bandwidth, maximum coupling occurs when the time window is centered at the Gaussian peak (indicated by the vertical, dashed white line) as expected, and the strongest coupling is achieved for $\Delta_\omega/\Gamma = 4.$  Note that although the average strong coupling parameter for a single photon never exceeds one, for larger photon numbers the $\sqrt{N}$ factor can lead to significant coupling. In \frf{Fig::StrongCoupling}(b) the excitation probability dynamics are shown for an optimal bandwidth $\Delta_\omega/\Gamma = 4$ wave packet.   We see the appearance of damped Rabi oscillations when the wave packet has $N=50$ photons that are completely absent when only a single photon is in the field.  For comparison, a wave packet of bandwidth $\Delta_\omega/\Gamma = 2$ is shown in \frf{Fig::StrongCoupling}(c).  Even at this bandwidth, damped Rabi oscillations appear for $N=50$ photons, albeit with reduced contrast and frequency.

	\subsection{Fock approximation to field states}    
      
Since the Fock states, $\ket{n_\xi}$, form a complete basis, they can be used to construct any state in the temporal mode $\xi(t)$.  This allows the study of quantum systems interacting with arbitrary field states using appropriately weighted solutions to the Fock-state master equations.  In the Fock basis, states may require an infinite number of terms, such as coherent states, thermal states, and squeezed states, but often can be well approximated with a finite number.  In this section we illustrate this with an truncated Fock-state approximation to a coherent state wave packet and compare the exact and approximate dynamics as it interacts with a two-level atom.

		\subsubsection{Example: Fock-state approximation to a coherent state wave packet}
	
	\begin{figure}[!t]
	\centering
	\includegraphics[scale=0.55]{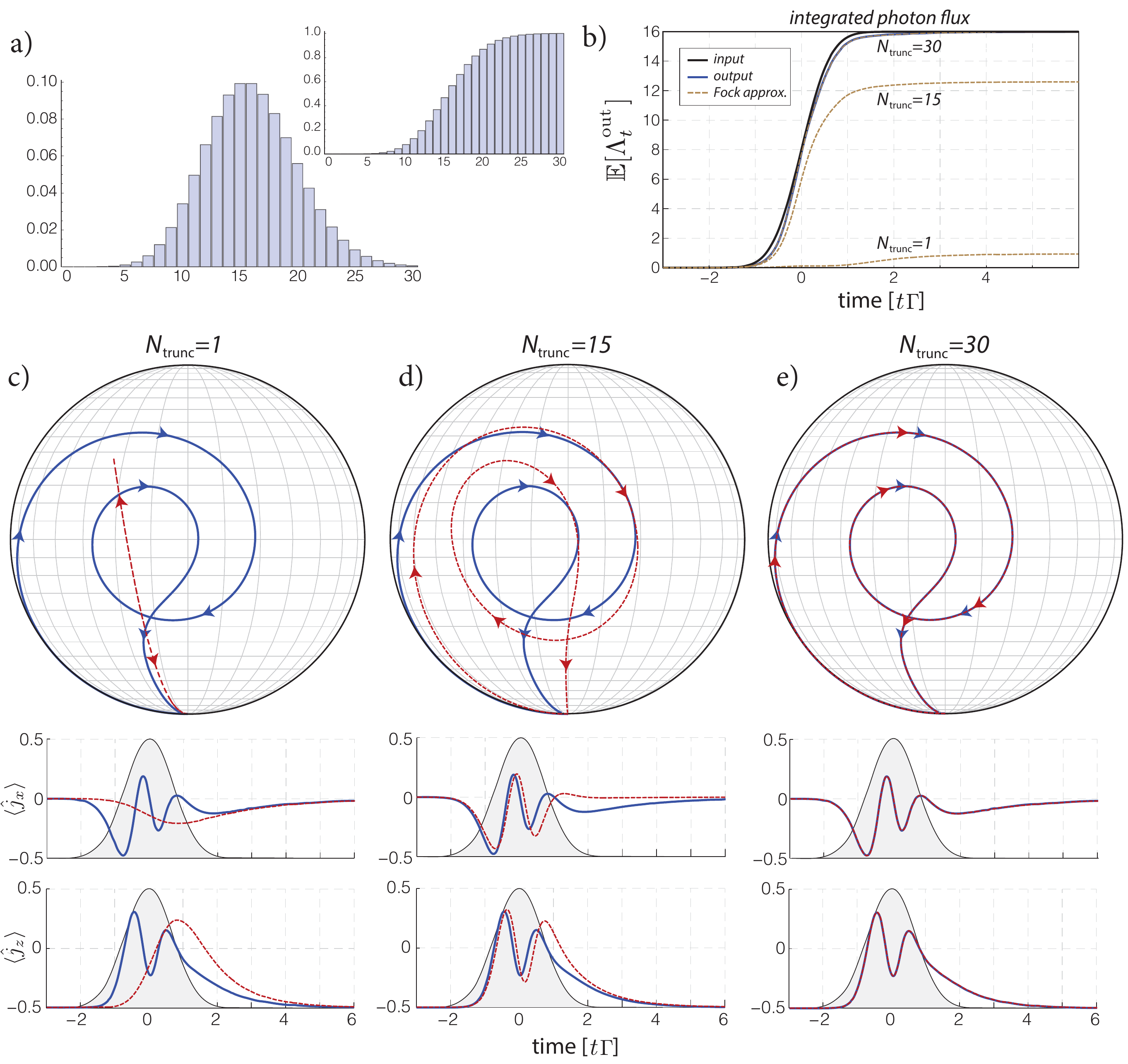}
	 \caption[Fock-state approximations to a coherent state]{Fock-state approximations to a coherent state in a Gaussian wave packet with $\Delta_\omega/\Gamma = 2$ interacting with a two-level atom.  a) Photon number distribution (probabilities) for an $\bar{n}=16$ coherent state.  Subplot shows the cumulative probabilities.  b) Coherent state input and output photon flux and output flux for Fock approximations with $N_{\rm trunc} \in \{1, 15, 30\}$. The $N_{\rm trunc}=30$ and exact curves overlap.  c)-e) Bloch vector trajectory in the $x-z$ plane and dynamics of its components for coherent state (blue) and Fock approximations (red dashed). Arrowheads indicate the trajectory on the Bloch sphere.  Below, the dynamics of the $\expt{ \hat{j}_x}$ and $\expt{ \hat{j}_z}$ components are shown ($\expt{ \hat{j}_y(t)} = 0$). For reference, the wave packet, $|\xi(t)|^2$, is shown normalized (black line filled grey).} \label{Fig::CohStateApprox}
	\end{figure}
	
A continuous-mode coherent state with total average photon number $\bar{n}$ in the wave packet $\alpha(t) = \sqrt{ \bar{n} } \, \xi(t)$ \cite{LoudenBook}, can be expanded in the Fock basis as
	\begin{align}
		\ket{\alpha (t)} = e^{- \bar{n}^2/2} \sum_n \frac{ \bar{n}^n}{ \sqrt{n!} } \ket{n_\xi}.
	\end{align}
Thus, the coefficients for the representation of $\op{\alpha (t)}{\alpha (t)}$ in \erf{Eq::CominationStates} are
	\begin{align} \label{Eq::CoherentCoefficients}
		c_{m,n} = e^{-\bar{n}^2} \frac{\bar{n}\,^{m+n}}{\sqrt{m!n!}}.
	\end{align}
For a given mean $\bar{n}$, one can find a suitable, finite approximation by truncating the Fock expansion at a desired degree of accuracy depending on the photon number distribution.  

In \frf{Fig::CohStateApprox} we plot the Bloch sphere representation and components of the pseudo-spin of a two-level atom interacting with a coherent-state Gaussian wave packet with $\Delta_\omega = 2$ and mean photon number $\bar{n} = 16$.  The exact dynamics for the reduced state, $\hat{\varrho}_\alpha(t)$, can be found from the coherent-state master equation, given by \erf{Eq::DisplacedFockME} with $m,n=0$.  The Bloch vector components are calculated with $\expt{\hat{j}_i(t)} = \Tr [\hat{\varrho}_\alpha(t) \hat{j}_i ]$ where $i \in \{ x,y,z\}$ and are shown (blue lines) in \frf{Fig::CohStateApprox}(c).  The exact dynamics are compared to a Fock-state approximation using a finite number of terms, with Bloch sphere components calculated with $\expt{\hat{j}_i(t)} = \Tr [\hat{\varrho}_{\rm total}(t) \hat{j}_i ]$.  The approximate state $\hat{\rho}_{\rm total}(t)$ is composed of the solutions $\hat{\varrho}_{m,n}(t)$ weighted by the coefficients \erf{Eq::CoherentCoefficients}.  For $\bar{n} = 16$, the photon number distribution and cumulative probabilities that determine the diagonal coefficients, \erf{Eq::CoherentCoefficients}, are shown in Figure \ref{Fig::CohStateApprox}(a) shows the photon number distribution that determines the diagonal coefficients in \erf{Eq::CoherentCoefficients}.  In this example, the approximations are truncated at $N_{\rm trunc} \in \{1,15,30\}$, which have cumulative probabilities in the photon number distribution of $\{ <\!\!.01, .467, .999 \}$ (see \frf{Fig::CohStateApprox}(a) subplot), chosen to illustrate poor, average, and excellent agreement.  For these truncations, the total integrated photon flux is shown in \frf{Fig::CohStateApprox}(b) as it converges to $\bar{n}$ for $N_{\rm trunc}=30$.

Whereas a Fock state has an indeterminate phase, a coherent state's phase breaks the symmetry in the $\expt{\hat{j}_x}$ and $\expt{\hat{j}_y}$ Bloch sphere components and leads to coherent rotations in around the $\expt{\hat{j}_y}$ axis (Rabi flopping).  Meanwhile, coupling to the reservoir via the Lindblad terms drives decoherence and the Bloch vector spirals toward the fully mixed state at the origin, as seen in the exact dynamics (blue curves) in \frf{Fig::CohStateApprox}(c).  Eventually, the wave packet exits the interaction region and the atom state decays back to the ground state.  In \frf{Fig::CohStateApprox} the Fock approximation approaches the coherent state dynamics for the Bloch vector and the output photon flux when the truncation is large enough.

	\subsection{Higher-dimensional systems: cavity QED}
	
In the previous examples, we have chosen as our system a two-level atom for its simplicity.  In physical applications one often probes systems with more complex internal structures; as an example we briefly study dynamics for a richer system consisting of a two-level atom placed in an optical cavity.  The addition of the cavity removes the direct coupling of the atom to the input fields, replacing it with an indirect cavity-mediated coupled.  Now, rather than just two internal levels, the atom-cavity system has infinitely many; and in some sense this is more interesting as such a system can store multiple excitations as they arrive in the form of Fock states.  

The total Hamiltonian has the form of \erf{Eq::TotalHam}.  The system includes both the atom and cavity, and thus the system Hamiltonian has three parts,
	\begin{align}
		\hat{H}_{\rm sys} &= \hat{H}_{\rm atom} + \hat{H}_{\rm cav} + \hat{H}_{\rm JC}.
	\end{align}
The bare atomic Hamiltonian with transition frequency $\omega_0$ and the bare cavity Hamiltonian with cavity frequency $\omega_{\rm cav}$,
	\begin{align}
			\hat{H}_{\rm atom} & =  \frac{\hbar \omega_{0}}{2} \big( \op{e}{e} -  \op{g}{g} \big) \\
			\hat{H}_{\rm cav} & = \hbar \omega_{\rm cav} \hat{a}\dg \hat{a}.
	\end{align}
where $\hat{a}$ and $\hat{a}\dg$ are cavity annihilation and creation operators, not to be confused with continuous-mode field operators.  The Jaynes-Cummings interaction that describes coherent transfer of excitations between the atom and cavity is given by
	\begin{align}
		\hat{H}_{\rm JC} = \hbar g \big( \hat{a} + \hat{a}\dg \big) \big( \op{e}{g} + \op{g}{e} \big),
	\end{align}	
with interaction strength $g$.  The bare Hamiltonian of the free field is given by \erf{Eq::ContModeFieldHam}, and the continuous-mode field interacts directly with the cavity (and only indirectly with the atom) via \erf{Eq::GardinerHam} with $\hat{c} \rightarrow \hat{a}$ and $\kappa(\omega_0)$ related to the cavity decay rate.  We move into an interaction picture with respect to the carrier frequency of the input Fock-state wave packet, $\omega_c$, with the choice,
	\begin{align}
		\hat{H}_0 = \hbar \omega_c \big( \op{e}{e} -  \op{g}{g} \big) + \hbar \omega_c \hat{a}\dg \hat{a}+ \int_0^\infty d\omega \, \hbar\omega \, \hat{b}(\omega)\dg \hat{b}(\omega).
	\end{align}
In this interaction picture the $(S,L,H)$ parameters that go into the Fock-state master equations are
	\begin{align}
		\hat{H}_{\rm sys} & =  -\hbar \Delta_0 \op{e}{e} - \hbar \Delta_{\rm cav} \hat{a}\dg \hat{a}  + \hbar g \big( \hat{a} \op{e}{g} + \hat{a}\dg  \op{g}{e} \big), \\
		\hat{L} & = \sqrt{\gamma} \hat{a},  \\
		\hat{S} & = \hat{I}_{\rm sys},
	\end{align}	
where $\Delta_0 = \omega_c - \omega_0$ is the detuning of the atom, $\Delta_{\rm cav} = \omega_c - \omega_{\rm cav}$ is the detuning of the cavity,  and the cavity coupling rate is $\gamma = 2 \pi |\kappa(\omega_0)|^2.$  In this system we consider perfect matching of the input field into the cavity, and thus the cavity coupling rate is equal to the total rate, $\gamma = \Gamma$.  We also assume that the atom interacts only with the cavity mode and we ignore side coupling to other modes of the free field.

	\subsubsection{Dynamics for Fock-state input}

	\begin{figure}[!h]
	\centering
	\includegraphics[scale=0.55]{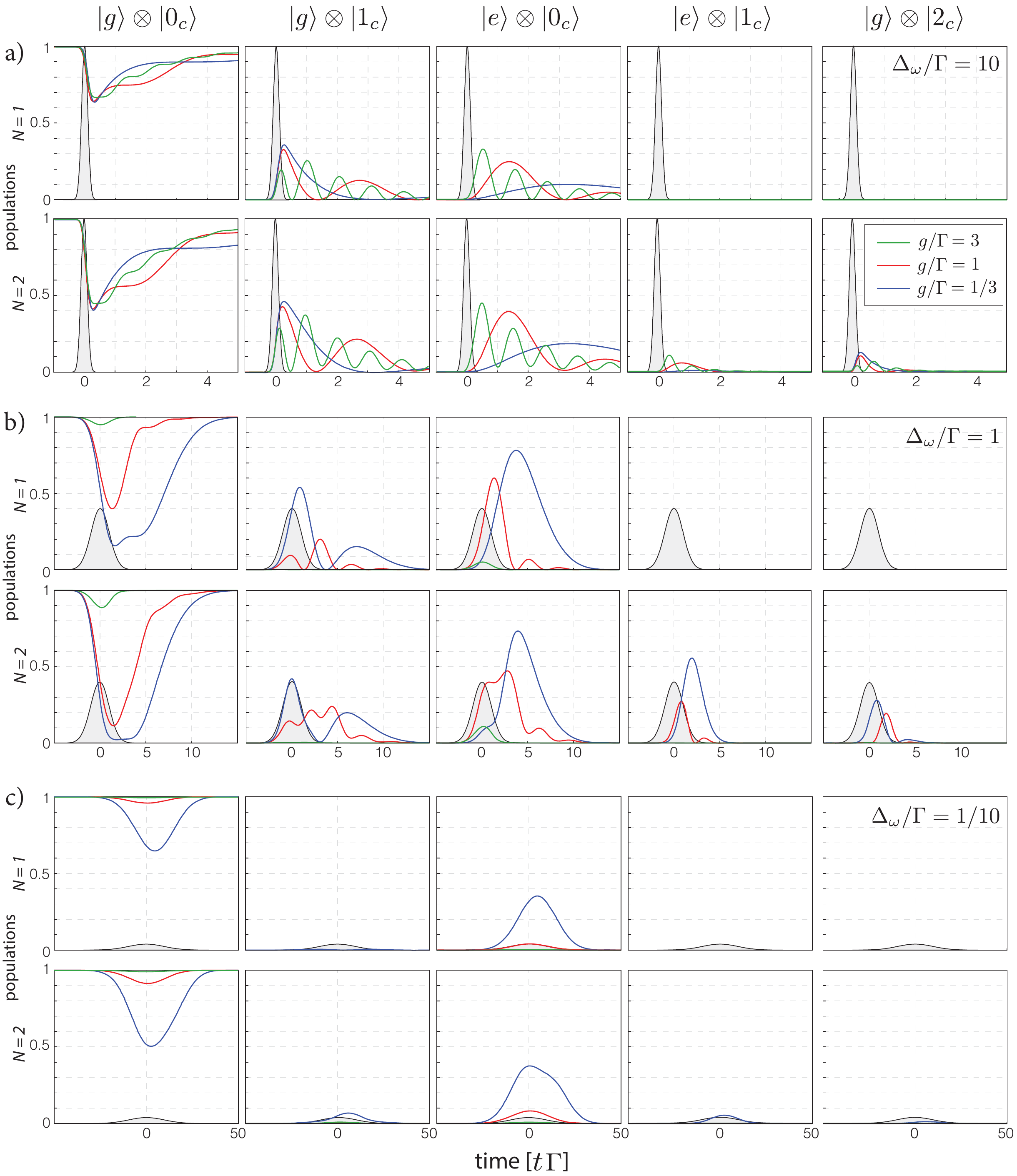}
	 \caption[]{Basis-state populations (columns) for a cavity QED system interacting with a Fock state in a Gaussian wave packet of bandwidth $\Delta_\omega$.  For each bandwidth, the first row is single-photon input and the second is two-photon input.  The wave packet $|\xi(t)|^2$ is indicated in black, filled grey.  The dynamics are shown for three values of the parameter, $\gamma/g \in \{ 1/3, 1, 3\}$.  a) $\Delta_\omega/\Gamma = 10$ (here $\xi(t)$ has been normalized for clarity), b) $\Delta_\omega/\Gamma = 1$, c) $\Delta_\omega/\Gamma = 1/10$.} \label{Fig::CavityQEDdynamics}
	\end{figure}
	
With this model the dynamics of the cavity QED system can be explored numerically as it is probed by Fock states.  
Varying the relative strengths of the cavity coupling rate and the atom-cavity coupling rate $g$ allows us to consider strong coupling ($g/\Gamma > 1$), weak coupling ($g/\Gamma < 1$), and intermediate coupling ($g/\Gamma \approx 1$) regimes.  We consider Fock states in Gaussian wave packets given by \erf{} with frequency bandwidth $\Delta_\omega$ and focus on wave packets with $N \in \{1,2\}$ photons.  The Fock-state master equations are numerically integrated and then the populations in the tensor product basis states, $\ket{a} \otimes \ket{n_c}$, are calculated, where $\ket{a}$ is an atomic state, either $\ket{g}$ or $\ket{e}$, and $\ket{n_c}$ is a Fock state in the cavity.  Since we put at most two photons into the system and it begins in the ground state, we truncate the system's Hilbert space to include up to two excitations.  
	
In \frf{Fig::CavityQEDdynamics} we plot the dynamics of the populations in the basis states for three bandwidths $\Delta_\omega/\Gamma \in \{10, 1, 1/10\}$ for one and two photons.  For each bandwidth, we simulate the dynamics for three values of the parameter, $g/\Gamma \in \{3,1,1/3\}$.  When the wave packet is prepared with only a single photon, the two-excitation subspace remains unoccupied as seen in the first row of \frf{Fig::CavityQEDdynamics}(a), (b), and (c).  Adding a second photon accesses these states.  In \frf{Fig::CavityQEDdynamics} (a) we consider short wave packets ($\Delta_\omega/\Gamma = 10$).  When the cavity-atom coupling is large enough ($g/\Gamma > 1$), then the excitations can be coherently transferred between the atom and cavity - Rabi flopping - many times before decaying back out of the cavity.  Lower coupling rates transfer the excitation more slowly.  For intermediate bandwidths ($\Delta_\omega/\Gamma = 1$), a high cavity-atom coupling rate prevents significant excitation.  For long wave packets ($\Delta_\omega/\Gamma \gg 1$), the excitations are spread out temporally, and only for relatively weak coupling ($g/\Gamma < 1$) does the system become moderately excited, but even then we see no oscillations.

	\subsubsection{Optimal atomic excitation for a single photon input}

For the case of a two-level atom without a cavity to mediate interactions with the input field, the maximum excitation for single-photon in a Gaussian wave packet was found in \srf{SEC::1n2photons} to be $\mathbbm{P}_e^{\rm max} = 0.801$.  With the addition of the cavity this can not only be improved.  This section is based on the ideas and conclusions of Ref. \cite{JohFio11}, where the authors consider an ```artificial atom" consisting of a quantum dot in an optical nanocavity.  For this system, application of a large electric field can decouple the quantum dot from the cavity to allow storage and release of an input photon.  We reproduce their results using the Fock-state formalism developed in this dissertation.  

	\begin{figure}[!t]
	\centering
	\includegraphics[scale=0.60]{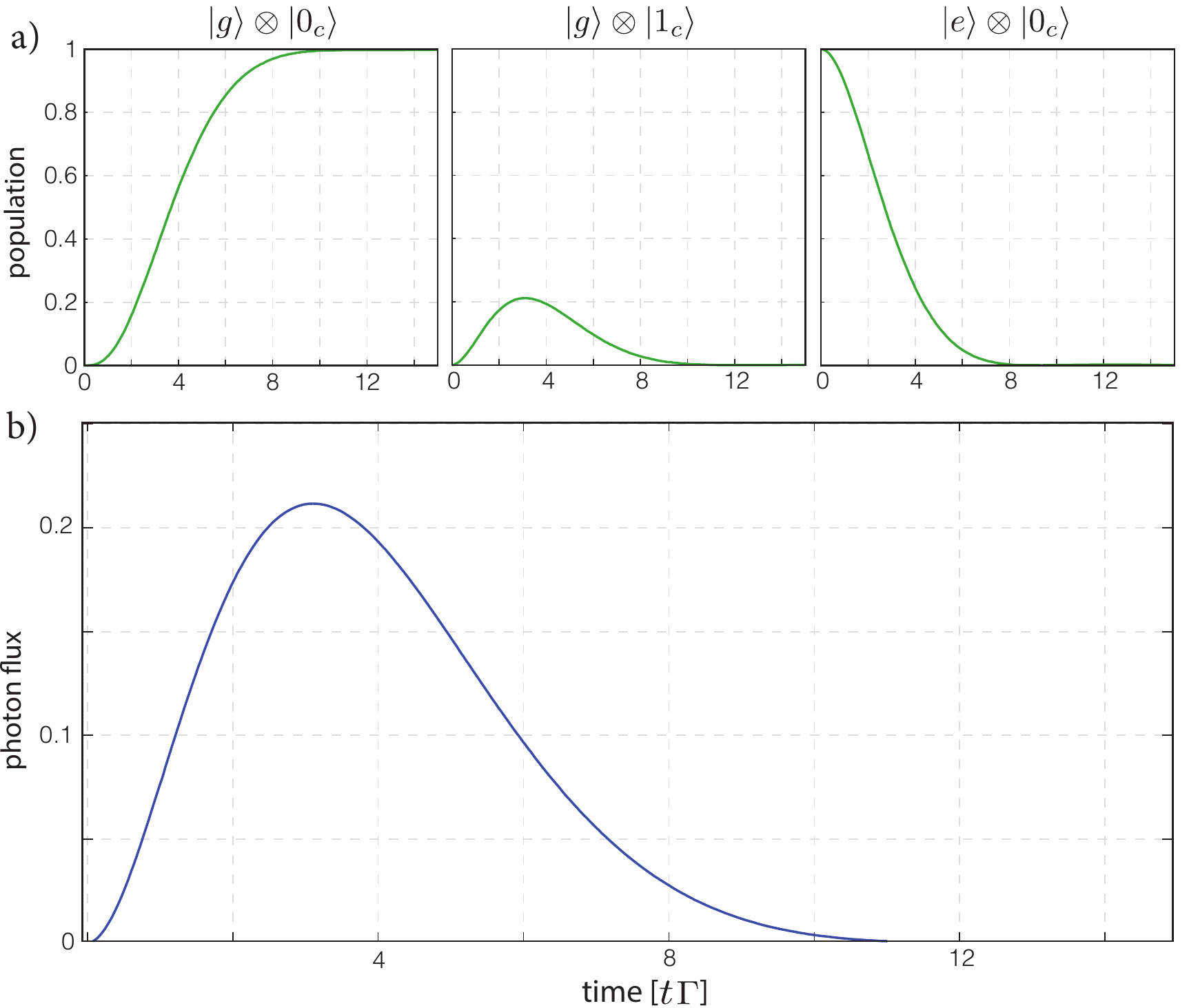}
	 \caption[Decay of a cavity QED system]{Decay dynamics when the atom is prepared in the excited state for $g/\Gamma = 0.9/\sqrt{2\pi}$.  a) Basis-state populations.  b) Output photon flux showing the shape of the emitted single-photon wave packet.} \label{Fig::CavityQED_photondecay}
	\end{figure}
	
	\begin{figure}[!h]
	\centering
	\includegraphics[scale=0.75]{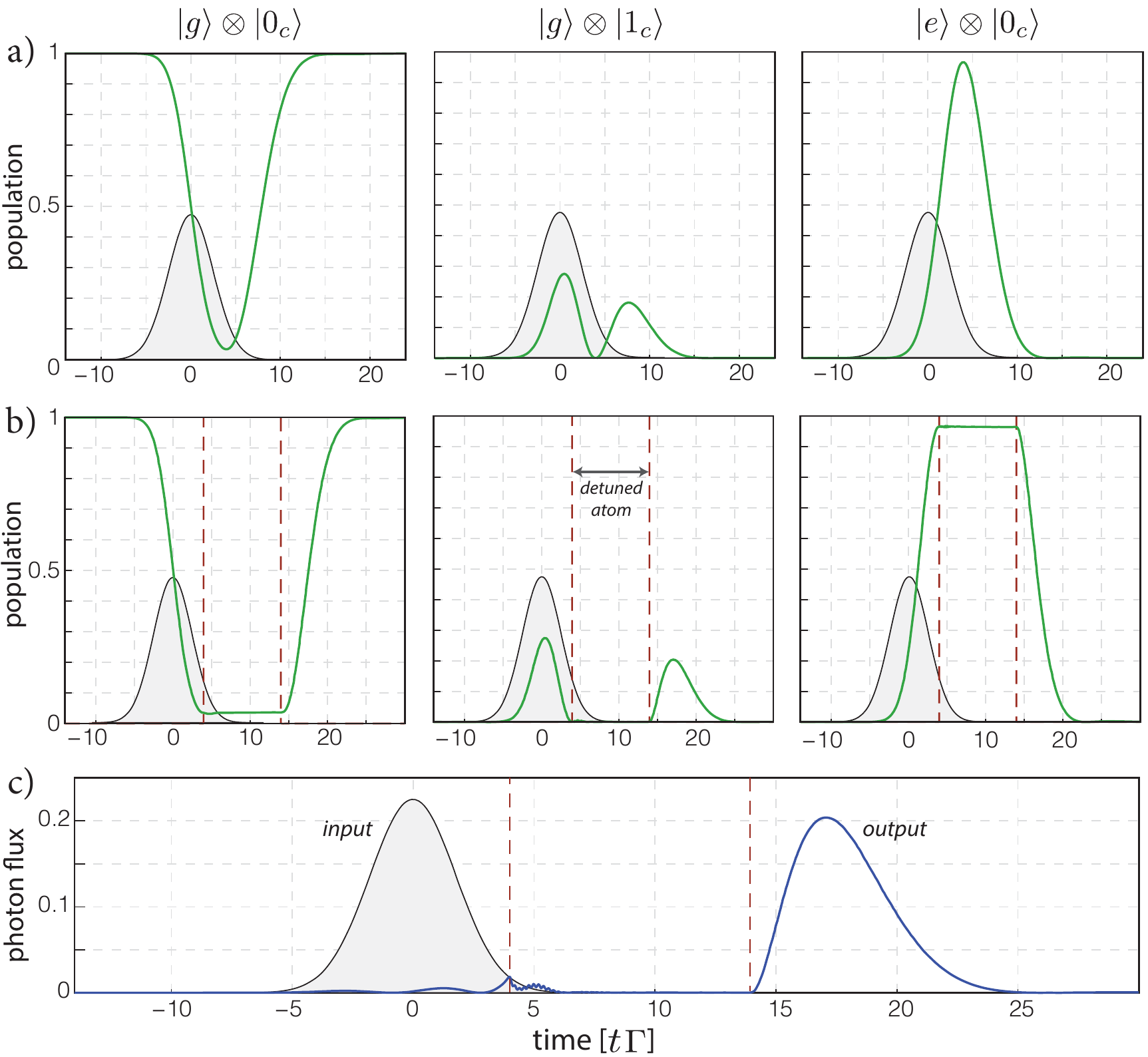}
	 \caption[Exciting an atom in a cavity with a single-photon Gaussian wave packet]{Exciting the atom with an optimal single-photon Gaussian wave packet $|\xi(t)|^2$, shown in black, filled grey.  a) Basis-state population dynamics.  In the third column, the atom achieves excitation probability of 0.966.   b) Protocol for storing the excitation: once full excitation is achieved, the atom is detuned off resonance.  When it is brought back to resonance with the cavity, the photon is released.  c) Input and output photon flux for the stored photon. } \label{Fig::CavityQED_photonstorage}
	\end{figure}
	
A fundamental consideration when exciting a system with a single photon is the overlap between the system's natural temporal mode for decay.  An excited two-level generates a photon with a decaying exponential wave packet, and excitation with a Gaussian wave packet is limited by the overlap of the  temporal modes.  When a cavity is placed around the atom, the temporal mode of decay can have a much higher overlap with a Gaussian as seen qualitatively in \frf{Fig::CavityQED_photondecay}.  This overlap depends highly on the relative atom-cavity coupling rate $g/\Gamma$.  In the strong coupling regime ($g/\Gamma > 1$) the excitation coherently oscillates between the atom and cavity before decaying out, while in the weak coupling regime ($g/\Gamma < 1$) the excitation decays slowly.  In both cases, the output wave packet is poorly matched to a Gaussian (further details can be found in Ref. \cite{JohFio11}).  For well-chosen parameter, $g/\Gamma = 0.9/\sqrt{2\pi}$, the output wave packet is shown in \frf{Fig::CavityQED_photondecay}(b).  This wave packet has excellent overlap (0.97) with a Gaussian wave packet of spectral width $\Delta_\omega/\Gamma = 1/\sqrt{\pi}$ \cite{JohFio11}.  Excitation with this optimal wave packet is shown in \frf{Fig::CavityQED_photonstorage}(a).  The achieved excitation of 0.966 is far beyond the maximum for an atom without a cavity.  
	
This near-perfect excitation of the atom can be used to store the input photon \cite{JohFio11}.  The atomic excited state can be maintained by rapidly switching on an external electric field to detune the atom away from resonance with the cavity, as seen in \frf{Fig::CavityQED_photonstorage}(b).   Any portion of the excitation that resides as a cavity photon leaks out, while the atomic excitation persists.  Tuning the two-level atom back on resonance releases the photon back into the field,  \frf{Fig::CavityQED_photonstorage}(c).  This idea serves as the foundation for the proposal of a photonic phase gate in Ref. \cite{JohFio12}.  For a more complete model, one would need to include the atom's coupling to the field out the side of the cavity, which would deco here the atom, reduce the output photon flux, and ultimately limit the gate fidelity.

\section{Multi-mode Fock-state master equations } \label{sec_2modefock}

In this section we derive the master equations for a system interacting with an arbitrary combination of continuous-mode Fock states in multiple modes (spatial or polarization).  With this generalization, as we will see in \srf{}, multi-photon states with correlations across modes can be treated.  This includes manifestly orthogonal modes such as transverse spatial and polarization modes as well as temporal modes, which are inescapably intertwined with longitudinal spatial modes (and frequency modes via Fourier transform).   In just two modes, this allows one to consider wave packets scattering off of atoms or addressing multiple dipole transitions, for instance.  The analysis and formalism for multiple field modes is conceptually identical to but algebraically more complicated than the single-mode case.  Because the procedure to derive the multi-mode Fock-state master equation is exactly the same as for the single-mode case, we omit some of the details.  

The evolution of a system operator interacting with multiple continuous-mode fields is given by the multi-mode It\={o} Langevin equation, [\erf{Eq::dXTwoMode}],
	\begin{align}\label{2mode_flow}
		d\hat{X} = &\Big(\smallfrac{i}{\hbar} \big[\hat{H}_{\rm sys}, \hat{X} \big] + \sum_i \mathcal{L}_{L_i}[\hat{X}] \Big) dt  + \sum_{i,j} [\hat{L}_i\dg,\hat{X}]\hat{S}_{ij} dB_{j}  \\
		&+\sum_{i,j} \hat{S}_{ij}\dg[\hat{X},\hat{L}_i] dB_j\dg  + \sum_{i,j} \Big( \sum_k \hat{S}_{ki}\dg \hat{X}\hat{S}_{kj}-\delta_{ij}\hat{X} \Big) d\Lambda_{ij}, \nn 
	\end{align}
where the modes are labeled by the subscripts $\{i,j,k\}$.  
%
We consider interactions with a multi-mode Fock state, where for each mode $i$ there is temporal wave packet $\xi_i(t)$ with $N_i$ photons.  The multi-mode Fock state can be written,
	\begin{align} \label{Eq::multi-modeFockState}
		 \ket{{N_1}_{\xi_1}} \otimes \ket{ {N_2}_{\xi_2} } \otimes \dots =\frac{1}{\sqrt{N_1! N_2! ...}} \big [ B_1\dg(\xi_1) \big ]^{N_1} \big[ B_2\dg(\xi_2) \big]^{N_2} \cdots \ket{0},
	\end{align}	
where the temporal mode creation operators, $B_i\dg(\cdot)$, are defined in \erf{eqBdef}.  To simplify the notation, we henceforth drop the tensor products and write the state in \erf{Eq::multi-modeFockState} as $ \ket{ N_1,N_2,...}$, where strict ordering within the ket labels the modes.

The action of the quantum noise increments in each mode,
	\begin{align}\label{eq:bla}
		&dB_i \ket{ n_1,n_2,...} = dt \sqrt{n_i} \xi_i(t) \ket{ n_1, n_2, ... ,n_i-1,... } \\
		&d\Lambda_{ij}   \ket{ n_1,n_2,...} = dB_i \dg \sqrt{n_j} \xi_j(t) \ket{ n_1, n_2, ... ,n_j-1,... },
	\end{align}
couples together multi-mode Fock states, just as in the single-mode case.  Thus we follow the same prescription and define the multi-mode generalized density operators with somewhat inelegant notation,
	\begin{align}
		\hat{\varrho}_{m_1,m_2,...| n_1,n_2...}(t) \equiv \Tr_{\rm field} \big[ \hat{U}(t) \hat{\rho}_{\rm sys} \otimes \op{ m_1,m_2,...}{n_1,n_2,...} \hat{U}\dg(t) \big], \label{Eq::multi-modeRhoMN}
	\end{align}
in order to avoid confusion\footnote{This notation differs from that of Ref. \cite{TeamAwesome}, wherein the subscripts were grouped by mode.  I prefer this notation because it has some advantages.  For example, taking a conjugate transpose is simply making the replacements $m_i \leftrightarrow n_i$ for all $i$.}.  Within the subscript, the order of the indices on either side of the divider ``$|$" labels the mode, and the value of each index is the numbers of photons in that mode.  By taking the trace of \erf{2mode_flow} with the multi-mode Fock states defined in \erf{Eq::multi-modeFockState}, we find the coupled master equations for these generalized density operators,
	\begin{align} \label{Eq::multi-modeME}
		\frac{d}{dt} & \hat{\varrho}_{m_1,...| n_1,...}(t) = \\
		&- \frac{i}{\hbar}[ \hat{H}_{\rm sys}, \hat{\varrho}_{m_1,...| n_1,...}] +  \sum_ i \mathcal{L}_{L_i} \big[  \hat{\varrho}_{m_1,...| n_1,...} \big] \nn \\
		&+ \sum_{i,j} \sqrt{m_j} \xi_j(t) \big[\hat{S}_{ij} \hat{\varrho}_{m_1,...,m_{j}-1,...| n_1,...}, \hat{L}_i\dg \big] \nn \\
		&+ \sum_{i,j} \sqrt{n_j} \xi^*_j(t) \big[\hat{L}_i, \hat{\varrho}_{m_1,...| n_1,...n_{j} - 1, ...} \hat{S}\dg_{ij} \big] \nn \\
		&+ \sum_{i,j} \sqrt{m_{i} n_{j}} \xi_i(t) \xi^*_{j}(t) \Big( \sum_k \hat{S}_{ki} \hat{\varrho}_{m_1,...,m_{i}-1,...| n_1,...n_{j} - 1, ...} \hat{S}_{kj}\dg -  \hat{\varrho}_{m_1,...,m_{i}-1,...| n_1,...n_{j} - 1, ...} \Big) \nn.
	\end{align}
Again, the sum over $k$ in the final line is only over terms involving subscripts $k$.  The initial conditions follow from \erf{Eq::multi-modeRhoMN},
	\begin{align} \label{Eq::multi-modeRhoMNInitCond}
		\hat{\varrho}_{m_1,m_2,...| n_1,n_2...}(0)  = \hat{\rho}_{\rm sys} \delta_{m_1,n_1} \delta_{m_2,n_2}, \dots
	\end{align}
To solve a multi-mode master equation with $N_i$ photons in the $i^{th}$ mode, we need to propagate $ \Pi_{i}(N_i+1)^2$ coupled equations. As in the single-mode case the symmetries in the generalized density operators, $\hat{\varrho}_{n_1,...| m_1,...} = \hat{\varrho}_{m_1,...| n_1,...}\dg$, reduce the number of independent equations to $ \smallfrac{1}{4}\Pi_i (N_i+1)(N_i+2)$.  With these generalized density operators we define the asymmetric expectations of system operators,
	\begin{align} \label{Eq::multi-modeExpectation}
		\mathbbm{E}_{m_1,m_2,...|n_1,n_2,...}[\hat{X}(t)] \equiv \Tr_{\rm sys} \big[ \hat{\varrho}_{m_1,...| n_1,...}\dg (t) \hat{X}\big],
	\end{align} 
which are used to calculate expectation values.

	\subsection{Multi-mode output field expectation values}\label{}

By taking asymmetric expectations of output field operators, we can calculate expectation values of photon flux and quadrature current in multiple modes.  The number of photons scattered from mode $j$ into mode $i$ in the interval $t$ to $t+d t$ is given by $d\Lambda_{ij}^{\rm out}$.  The diagonal elements $\Lambda_{ii}^{\rm out}$ give the photon flux in mode $i$.  The output relation for $d\Lambda_{ij}^{\rm out}$, given in \erf{Eq::LambdaOutTwoMode}, is
	\begin{align} \label{Eq::dLambdaTwoMode}
		d\Lambda_{ij}^{\rm out} =  \hat{L}_i\dg \hat{L}_j dt + \sum_k \hat{L}\dg_i \hat{S}_{jk} dB_k +  \sum_k \hat{S}\dg_{ik} \hat{L}_j dB_k +  \sum_{k,l} \hat{S}\dg_{ik} \hat{S}_{jl} d\Lambda_{kl}.
	\end{align}
Taking asymmetric expectations with respect to Fock states gives
	\begin{align} \label{Eq::multi-modeFieldME_lambda}
		\frac{d}{dt} & \mathbbm{E}_{m_1...|n_1...} \big[\Lambda_{ij}^{\rm out}(t) \big] = \mathbbm{E}_{m_1...|n_1...} \big[\hat{L}_i\dg \hat{L}_j \big] \\
		& + \sum_{k} \sqrt{m_k} \xi_k^*(t) \mathbbm{E}_{m_1...m_k-1,...|n_1...} \big[ \hat{S}_{ik}\dg \hat{L}_{j} \big]   + \sum_{k}  \sqrt{n_k} \xi_k(t)  \mathbb{E}_{m_1...|n_1...n_k-1,...} \big[ \hat{L}_i\dg \hat{S}_{jk} \big], \nonumber \\
		& + \sum_{k,l} \sqrt{m_k n_{l}} \xi_k^*(t) \xi_{l}(t) \mathbbm{E}_{m_1...m_k-1,...|n_1...n_{k'}-1,...} \big[  \hat{S}\dg_{ik}\hat{S}_{jl} \big] \nn.
	\end{align}
The output quantum noise increment in mode $i$ is given by \erf{Eq::BOutTwoMode},
	\begin{align} \label{Eq::multi-modeBOut}
 		dB_i^{\rm out} =  \hat{L}_i dt + \sum_j \hat{S}_{ij} dB_j.
	\end{align}
Taking asymmetric expectations with respect to Fock states gives
	\begin{align} \label{Eq::TwoModeFieldME}
		\frac{d}{dt} \mathbbm{E}_{m_1...|n_1...} \big[ dB_i^{\rm out} \big] =&\mathbbm{E}_{m_1...|n_1...} \big[ \hat{L}_i ]  + \sum_j \sqrt{n_j} \xi_j(t) \mathbbm{E}_{m_1...|n_1...n_j-1...} \big[\hat{S}_{ij}   \big]. 
	\end{align}
Physically observable quadratures are given by Hermitian combinations of $B_i^{\rm out}$ and $B_j^{\rm out \dagger}$.  For example, when the modes are two transverse polarizations orthogonal to the propagation direction, such combinations are generated in a laboratory setting with wave plates and polarizing beamsplitters.

	\subsection{Superpositions and mixtures of Fock states in multiple modes}\label{TwoModeCombinations}

One would like to describe system interacting with field states that, in general, are not multi-mode Fock states.  However, as Fock states form a basis, and thus by proper combination of the solution to the multi-mode master equations in \erf{Eq::multi-modeME}, such situations can be treated.  Consider the input field state
	\begin{align} \label{Eq::2ModeGeneralField}
		\hat{\rho}_{\mathrm{field}}&= \sum_{m_1,m_2,...}\sum_{n_1,n_2,...} c_{m_1,m_2,...|n_1,n_2,...} \op{ m_1,m_2,...}{ n_1,n_2,...}	\end{align}
As before, the coefficients, $c_{m_1,m_2,...|n_1,n_2,...}$, are constrained by the requirements of valid quantum states.  When the input field is described by \erf{Eq::2ModeGeneralField}, the total system state is given by 
	\begin{equation}\label{Eq::gen_me2Mode}
		\hat{\varrho}_{\rm phys} (t)= \sum_{m_1,m_2,...}\sum_{n_1,n_2,...} c_{m_1,m_2,...|n_1,n_2,...}  \hat{\varrho}_{m_1,m_2,...|n_1,n_2,...}(t),
	\end{equation}
The composition rule for system expectation values is given by 
	\begin{equation}\label{Eq::EXPgen_me2Mode}
		\mathbbm{E}_{\rm phys}[\hat{X}(t)] = \sum_{m_1,m_2,...}\sum_{n_1,n_2,...} c^*_{m_1,m_2,...|n_1,n_2,...}  \mathbbm{E}_{m_1,m_2,...|n_1,n_2,...}[ \hat{X}(t)].
	\end{equation}
where the asymmetric expectation value is defined in \erf{Eq::multi-modeExpectation}.  As before, the conjugate coefficients in \erf{Eq::gen_me2Mode} come from the conjugate transpose in the Hilbert-Schmidt inner product, \erf{Eq::HSInnerP}.  This technique also applies to the field quantities, such as the output photon flux,
	\begin{equation}\label{Eq::LambdaCombmulti-mode}
		\mathbbm{E}_{\rm phys}[\hat{\Lambda}_{ij}^{\rm out}(t)] = \sum_{m_1,m_2,...}\sum_{n_1,n_2,...} c^*_{m_1,m_2,...|n_1,n_2,...}  \mathbbm{E}_{m_1,m_2,...|n_1,n_2,...}[ \hat{\Lambda}_{ij}^{\rm out}(t) ].
	\end{equation}

	\subsection{Example: two-mode Fock-state master equations} \label{2msec_fockgen}

The formalism developed above is by no means transparent or easy to use at first glance.  In this section, we provide the details for two modes, in hopes that the simplest nontrivial example will help guide an understanding of the multi-mode formalism.  Here, we examine the photon flux of the transmitted and reflected fields when Fock states are incident on a two-level atom \cite{DroHavBuz00, DomHorRit02, CheWubMor11,ZheGauBar10,SheFan07b,ZhoGonSun08,LonSchBus09,SheFan05,Roy10}.  In this setting, we have two spatial modes -- the forward- and backward-propagating fields -- as in a tightly-confined waveguide QED setting \cite{SheFan07b,ZheGauBar10}.  

Consider the case where photons in mode one are prepared in a temporal wave packet $\xi(t)$ and those in mode two are in the wave packet $\eta(t)$.  A two-mode Fock state with $N_1$ photons in mode one and $N_2$ photons in mode two is,
	\begin{align}
		\ket{{N_1}_\xi} \otimes \ket{ {N_2}_\eta } = \ket{N_1, N_2} = \frac{1}{\sqrt{N_1! N_2!}} \big[ B_1\dg(\xi) \big]^{N_1} \big[ B_2\dg(\eta) \big]^{N_2} \ket{0} .
	\end{align}	
We specialize the It\={o} Langevin equation,  \erf{2mode_flow}, and the output photon flux,  \erf{Eq::dLambdaTwoMode}, to two modes by restricting the indices to run over the mode labels $\{1,2 \}$. The two-mode generalized density operators are,
	\begin{align}
		\hat{\varrho}_{m_1,m_2|n_1,n_2}(t) \equiv & \Tr_{\rm field} \Big[ \hat{U}(t) \big(\hat{\rho}_{\rm sys}\! \otimes\op{{m_1}_\xi}{{n_1}_\xi} \otimes \op{ {m_2}_\eta }{ {n_2}_\eta }  \big) \hat{U}\dg(t)  \Big] \\
		= &  \Tr_{\rm field} \Big[ \hat{U}(t) \big(\hat{\rho}_{\rm sys}\! \otimes \op{m_1,m_2}{ n_1, n_2 } \big) \hat{U}\dg(t)  \Big]. 
	\end{align}
The labels $\{m_1,n_1\}$ refer to mode one and $\{m_2,n_2\}$ to mode two.  With these we can find asymmetric expectations for two-mode Fock states,  
	\begin{align} \label{Eq::TwoModeGeneralDensityOps}
		\mathbbm{E}_{m_1,m_2|n_1,n_2}[ \hat{X}(t) ] 
		& \equiv \!\Tr_{\rm sys} \big[ \hat{\varrho}_{m_1,m_2|n_1,n_2}\dg(t)  \hat{X} \big].
	\end{align}
The actions of the quantum noise increments on two-mode Fock states are
	\begin{subequations}
	\begin{align}
		dB_1\ket{n_1,n_2} & = \,dt\sqrt{n_1} \xi(t) \ket{n_1-1, n_2},\\
		dB_2 \ket{n_1,n_2} & = \,dt\sqrt{n_2} \eta(t) \ket{n_1, n_2-1},\\
		d\Lambda_{11} \ket{n_1,n_2} & = dB_1\dg \sqrt{n_1} \xi(t) \ket{n_1-1, n_2},\\
		d\Lambda_{22} \ket{n_1,n_2} & = dB_2\dg \sqrt{n_2} \xi(t) \ket{n_1, n_2 -1},\\
		d\Lambda_{12} \ket{n_1,n_2} & = dB_1\dg \sqrt{n_2} \eta(t) \ket{n_1, n_2-1} \\
		d\Lambda_{21} \ket{n_1,n_2} & = dB_2\dg \sqrt{n_1} \eta(t) \ket{n_1-1, n_2}.
	\end{align}
	\end{subequations}

Assuming only linear, dipole coupling for brevity ($\hat{S}_{ij} = \delta_{i,j} \hat{I}_{\rm sys}$), the two-mode Fock-state master equations then follow from \erf{Eq::multi-modeME},
	\begin{align} \label{Eq::TwoModeME} 
		 \frac{d}{dt} \hat{\varrho}&_{m_1,m_2|n_1,n_2}  = - \frac{i}{\hbar}[ \hat{H}_{\rm sys},\hat{\varrho}_{m_1,m_2|n_1,n_2}] + \sum_{i} \mathcal{L}_{L_i}[ \hat{\varrho}_{m_1,m_2|n_1,n_2}]   \\ 
                  &  \nn+ \sqrt{m_1} \xi(t) [\hat{\varrho}_{m_1-1,m_2|n_1,n_2}, \hat{L}_1\dg  ] + \sqrt{n_1} \xi^*(t) [\hat{L}_1, \hat{\varrho}_{m_1,m_2|n_1-1,n_2}  ] \\
                  & + \sqrt{m_2} \eta(t) [ \hat{\varrho}_{m_1,m_2-1|n_1,n_2}, \hat{L}_2\dg ]  + \sqrt{n_2} \eta^*(t)[  \hat{L}_2, \hat{\varrho}_{m_1,m_2|n_1,n_2-1} ] \nn .
	\end{align}
According to \erf{Eq::multi-modeRhoMNInitCond}, the diagonal generalized density operators are initialized to the system state; that is, 
	\begin{align}
		 \hat{\varrho}_{m_1,m_2|n_1,n_2}(0) &= \bigg\{ \begin{array}{cl}
 \hat{\rho}_{\rm sys}  & \quad \mbox{if }  m_1 = n_1  \mbox{ and }   m_2 = n_2  \\
0 & \quad  \mbox{if } m_1 \neq n_1  \mbox{  or } m_2 \neq n_2.  \end{array}
	\end{align}
To solve a two-mode master equation with $N_1$ photons in mode one and $N_2$ photons in mode two, $\hat{\rho}_{\rm field} = \op{N_1, N_2}{N_1,N_2}$, we need to propagate $(N_1+1)^2\times(N_2+1)^2$ coupled equations. 

	\subsubsection{Two-mode output photon flux}\label{2msec_fockgen_output}

The output equations for field observables in two modes are significantly more complicated than the single-mode case because one can consider linear combinations of the modes. Thus, there is a continuum of possible of output photon fluxes and field quadratures. Here we focus on photon fluxes that are diagonal in the modes.  More complicated output observables that combine both modes can be obtained using beam splitter relations -- effectively, a change of basis -- as described in Ref.~\cite{GouJam09}.

From \erf{Eq::multi-modeFieldME_lambda}, the mean photon flux in mode one is governed by the equation,
	\begin{align} \label{Eq::TwoModeFieldME_lambda}
		\frac{d}{dt}  \mathbbm{E}_{m_1,m_2|n_1,n_2} [\Lambda_{11}^{\rm out}(t)] = 
		& \mathbbm{E}_{m_1,m_2|n_1,n_2}[\hat{L}_1\dg \hat{L}_1] \\
		& + \sqrt{m_1} \xi^*(t) \mathbbm{E}_{m_1-1,m_2|n_1,n_2} [ \hat{L}_{1} ]  + \sqrt{n_1} \xi(t) \mathbbm{E}_{m_1,m_2|n_1-1,n_2}[ \hat{L}_1\dg ]  \nonumber \\
		& + \sqrt{m_1n_1} |\xi(t)|^2 \delta_{m,n}. \nonumber 
	\end{align}
The equation for mode two follows similarly. 

\subsubsection{Fock states in a Gaussian wave packet scattering from a two-level atom} \label{Sec::2modeExample}

As above we specialize to a resonant Gaussian wave packet $\xi(t)$ described by \erf{Eq::gau_xi} interacting with a two-level atom in two propagating modes. The master equation parameters we use are again those for dipole coupling without external Hamiltonian drive: $\hat{H}_{\rm sys}=0$, $\hat{L}_i= \sqrt{\gamma_i} \op{g}{e}$, $\hat{S}_{11}=\hat{S}_{22}=\hat{I}_{\rm sys}$, $\hat{S}_{12}=\hat{S}_{21}=0$ for $i\neq j$, and the coupling rate is chosen to be symmetric in the modes $\gamma_1 = \gamma_2 = 1/2$.  The forward-propagating field, mode 1, is prepared in a Fock state with $N \in \{1,\dots,5\}$ photons while the backward mode is initially in vacuum; that is, $\ket{\psi_{\rm field}} = \ket{N, 0}$.


	\begin{figure}[!t]
	\begin{center}
	\includegraphics[scale=0.8]{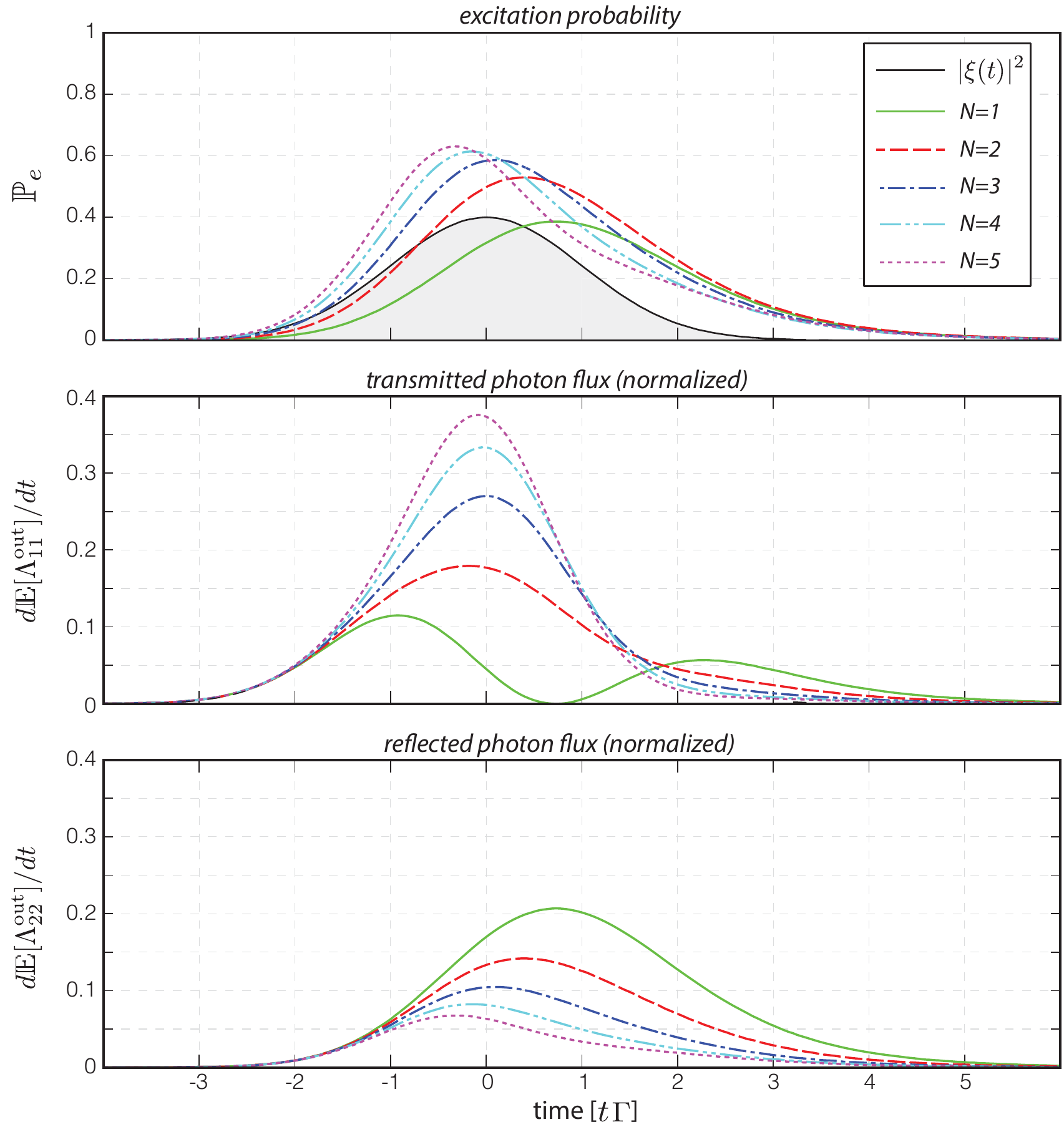}
		\caption[Fock states scattering from a two-level atom - dynamics]{Scattering of a Gaussian wave packet of bandwidth $\Delta_\omega/\Gamma=1$ from a two-level atom.  The wave packet $|\xi(t)|^2$ (black filled grey) is prepared with $N \in \{1,\dots,5 \}$ photons. (a) Excitation probability.  Photon flux of the transmitted (b) and reflected (c) fields, normalized to input photon number.} \label{Fig::2m_reflect}
	\end{center}
	\end{figure}


	\begin{figure}[t]
	\begin{center}
	\includegraphics[scale=0.8]{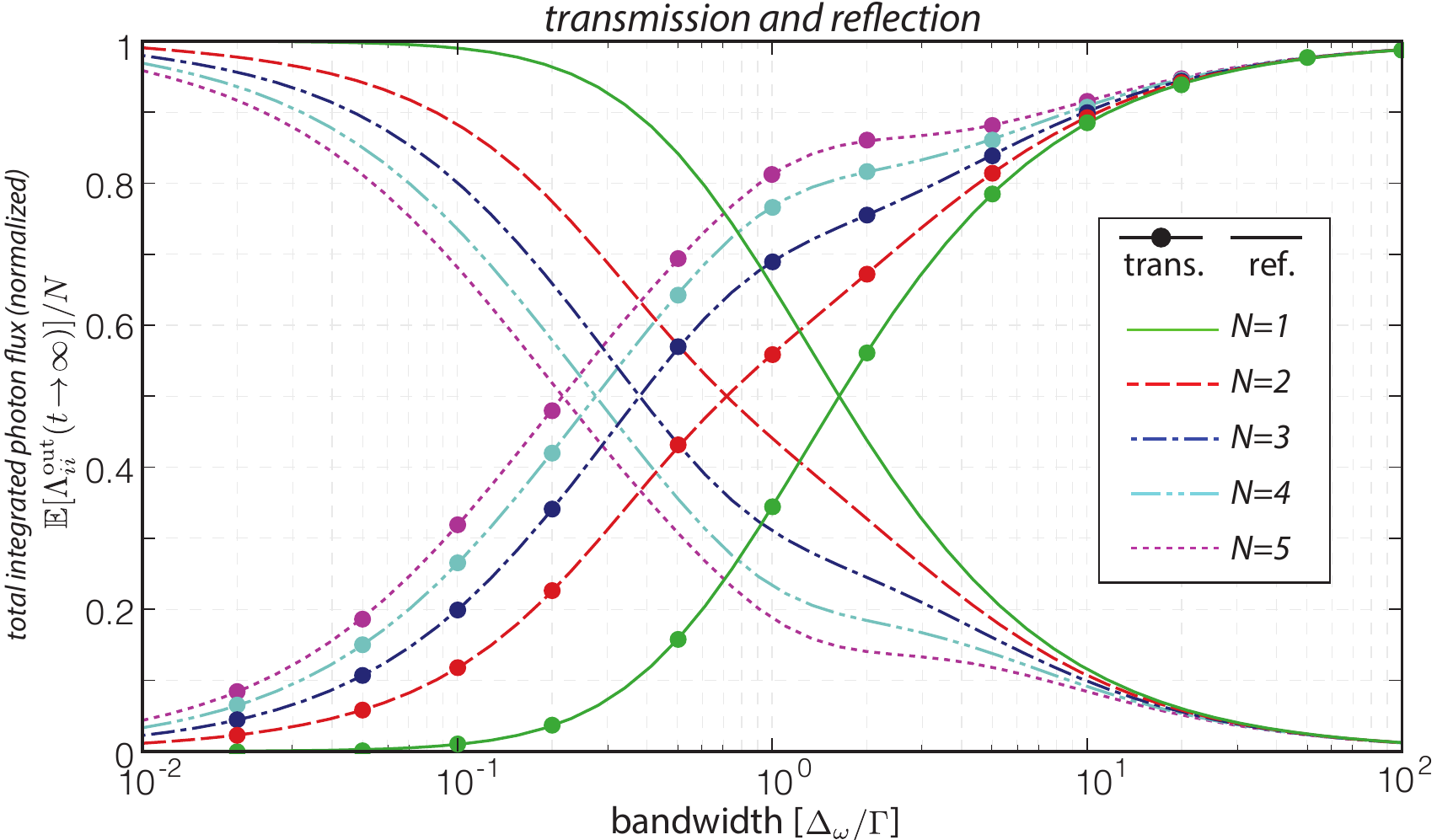}
		\caption[Fock states scattering from a two-level atom - transmission and reflection]{Normalized transmission and reflection for Gaussian wave packets,  prepared with $N\in \{1,\dots,5\}$ photons, with bandwidths $\Delta_\omega/\Gamma$ scattering from a two-level atom.  The left (right) side represents long (short) temporal wave packets. For larger photon number, note the increased transmission at intermediate bandwidths. } \label{Fig::poindexter}
	\end{center}
	\end{figure}
	
In \frf{Fig::2m_reflect}(a) we plot the excitation probability $\mathbbm{P}_e$ for a two-level atom interacting with a wave packet with bandwidth $\Delta_\omega/\Gamma=1$. The photon flux of the transmitted, \erf{Eq::TwoModeFieldME_lambda}, and reflected fields is plotted in Figures \ref{Fig::2m_reflect}(b) and (c), normalized to the number of input photons $N$. We first examine the single-photon input state (solid green curves). While absorbing the photon, the atom has a substantial $\mathbbm{P}_e$. The two peaks in the transmitted flux correspond to the attenuated input wave packet and the contribution from remission into the forward mode~\cite{DroHavBuz00}.  Notice the dip between the peaks occurs when there is a large atomic excitation. Consequently this dip in the transmitted photon flux is due to atomic absorption and destructive interference with the incoming wave packet \cite{DroHavBuz00,LonSchBus09,CheWubMor11,SheFan05}. Energy from the field that is not absorbed is scattered into the backward mode through the reemission process~\cite{DroHavBuz00}.  For $N>1$, we see that the excitation probability is comparable to that for a single photon, but the relative transmitted and reflected photon fluxes are quite different. In particular the ratio of transmitted to reflected flux increases with $N$.

In order to understand this phenomena we consider the normalized transmitted and reflected photon numbers in the long-time limit ($\mathbbm{E}[\Lambda_{11}]$ and $\mathbbm{E}[\Lambda_{22}]$) at different bandwidths \cite{DroHavBuz00,ZheGauBar10}. In \frf{Fig::poindexter} we explore this issue numerically. Recall that the reflection process involves absorption and then reemission into the backward mode. Thus one would expect reflection to dominate for small bandwidth wave packets, which is indeed what is seen in the left hand side of \frf{Fig::poindexter}. In the large bandwidth limit very little of the wave packet is near resonance with the atomic transition so no absorption occurs and the wave packet is transmitted. The bump in the $N>1$ transmission and reflection curves is a consequence of an effective photon-photon interaction \cite{DeuChi92, SheFan07b,ZheGauBar10}.  By calculating the scattering eigenstates, Zheng et al. found ``multi-photon bound states"~\cite{ZheGauBar10} which can increase transmission in that bandwidth region. They also considered coherences between photons scattered between the forward and backward modes, which can be done in our formalism with $\Lambda_{12}$ and $\Lambda_{21}$.

\section{General $N$-photon master equations } \label{Sec::GeneralWavepackets}

In many experimental settings multiple photons are not created in Fock states. Fock states are a subset of more general $N$-photon states, which have a definite number of photons but an arbitrary SDF $\tilde\psi$.  Indeed, a quantum tomography protocol for characterizing the SDF was recently proposed \cite{RohdeSep06} and implemented \cite{WasKolRob07}.  The techniques developed earlier in this chapter can be applied to general $N$-photon states by first decomposing the temporal mode function in a basis of temporal modes\footnote{Master equations similar to those in this section were derived in Ref.~\cite{Zoolander} for two photons without the terms involving $d\Lambda_t$ or $\hat{S}$.   Also, the master equations and conditional quantum filters for $N$-photon states were recently derived using a non-Markovian embedding \cite{SonXi13}.}.

	\subsection{$N$-photon states}
	
	In a single spatial and polarization mode, a general $N$-photon state is 
	\begin{align}\label{Eq::Genphotyfreq}
		\ket{\psi_N} = \frac{1}{\sqrt{\mathcal{N}}} \int d\omega_1 & \dots d\omega_N \, \tilde\psi(\omega_1,\dots,\omega_N) b\dg(\omega_1)\dots b\dg(\omega_N) \ket{0}.
	\end{align}  
Again we assume quasi-monochromatic wave packets such that the Fourier transform of the SDF, $\psi = \mathcal{F}[\tilde\psi]$, is a slowly-varying envelope with respect to the carrier frequency.  Then, in the time domain a general $N$-photon state can be written as
	\begin{align}\label{Eq::Genphoty}
	\begin{split}
		\ket{\psi_N} = \frac{1}{\sqrt{\mathcal{N}}}\int dt_1 & \dots dt_N \, \psi(t_1,\dots,t_N) b\dg(t_1)\dots b\dg(t_N) \ket{0},
		\end{split}
	\end{align} 
where $\mathcal{N}$ is a normalization factor that depends on the permutation symmetry of $\psi(t_1,\dots,t_N)$\footnote{Although the temporal envelope (or in the frequency domain the SDF) is not required to be permutation symmetric, the quantum state is.  Permutation of the indices in the full $N$-photon state, [\erf{Eq::Genphotyfreq}], leaves the state invariant thanks to the fact that the creation operators commute and to the integration over all $\omega_i$.  For this reason, one may choose to symmetrize $\tilde\psi(\omega_1,\dots,\omega_N)$ beforehand; here we follow Refs. \cite{Ou06, Roh07, Ou08} and do not assume permutation symmetry.  Within those publications, the goal was often to describe the distinguishability of the photons in a generalized Hong-Ou-Mandel interferometer, which can be directly related to the unsymmetrized $\tilde\psi(\omega_1,\dots,\omega_N)$ as we use it.}.  To describe $N$-photon states we make use of the occupation number representation developed by Rohde et al.~\cite{Roh07}, which we review and modify slightly in Appendix \ref{Appendix::NPhotonStates}.  Using \erf{AEq::NPhotonState}, the $N$-photon state in \erf{Eq::Genphoty} can be written in a basis of Fock states defined on orthogonal temporal modes,
	\begin{align} \label{Eq::GeneralNPhotonState}
		\ket{\psi_N} =& \sum_{n_1,n_2,...} c_{n_1, n_2,...} \ket{ {n_1}_{\xi_1} } \ket{ {n_2}_{\xi_2} } \dots \rightarrow \sum_{n_1,n_2,...} c_{n_1, n_2,...} \ket{ {n_1}, {n_2}, ... }, 
	\end{align}
where $\ket{ {n_i}_{\xi_i} }$ is a normalized Fock state described by \erf{Eq::ContModeFockState} with $n_i$ photons in temporal mode $\xi_i(t)$.  We have essentially projected the input state into a basis of unentangled Fock states.  The initial field state that enters the multi-mode Fock state formalism is
	\begin{align} \label{Eq::GeneralNPhotonDensOp}
		\op{\psi_N}{\psi_N}  = \sum_{m_1,m_2,...} \sum_{n_1,n_2,...} c_{m_1,m_2,...|n_1, n_2,...} \op{m_1,m_2,...}{ {n_1}, {n_2}, ... }, 
	\end{align}
with coefficients that arise from products of the coefficients in \erf{Eq::GeneralNPhotonState} when the outer product is taken. 

The occupation number representation relies on a set of temporal modes to represent the states.  Thus, everything follows exactly the multi-mode Fock-state formalism from \srf{sec_2modefock}, with one notable difference.  The quantum noise increments are not indexed by these modes, and each acts on Fock states in \emph{every} temporal mode.  This can be seen when we write down the action of the quantum noise increments on the basis Fock states in \erf{Eq::GeneralNPhotonState} \cite{TeamAwesome,SonXi13}:
	\begin{align}\label{eq:bla}	
		&dB_t \ket{ n_1,n_2,...} = dt \sum_i \sqrt{n_i} \xi_i(t) \ket{ n_1, n_2, ... ,n_i-1,... } \\
		&d\Lambda_t   \ket{ n_1,n_2,...} = dB_t\dg \sum_i \sqrt{n_i} \xi_i(t) \ket{ n_1, n_2, ... ,n_i-1,... }
	\end{align}
The action of the quantum noise increments leads to a superposition of multi-mode Fock states with one fewer photons, weighted by the value of the temporal mode envelopes at time $t$.  The generalized density operators are precisely the multi-mode operators in \erf{Eq::multi-modeRhoMN},
	\begin{align}
		\hat{\varrho}_{m_1,m_2,...| n_1,n_2...}(t) \equiv \Tr_{\rm field} \big[ \hat{U}(t) \hat{\rho}_{\rm sys} \otimes \op{ m_1,m_2,...}{n_1,n_2,...} \hat{U}\dg(t) \big], \label{Eq::multi-modeRhoMN}
	\end{align}
with a physical state given by the proper superposition,
	\begin{align} \label{Eq::NPhotonPhysicalState}
		\hat{\varrho}_{\rm phys}(t) = \sum_{m_1,m_2,...} \sum_{n_1,n_2,...} c_{m_1,m_2,...|n_1,n_2,...} \,	\hat{\varrho}_{m_1,m_2,...| n_1,n_2...}(t) 
	\end{align}
using the coefficients from \erf{Eq::GeneralNPhotonDensOp}.  The master equations are solved with initial conditions given in \erf{Eq::multi-modeRhoMNInitCond}, and the solutions recombined via \erf{Eq::NPhotonPhysicalState} in order to calculate expectation values.  The equations for the output field quantities follow in the same way.

The succinct summary is that, once the input field state is represented in a basis unentangled Fock states over a set of temporal modes, the $N$-photon master equations can be found simply from \erf{Eq::multi-modeME} with $\hat{L}_i \rightarrow \hat{L}$ and $\hat{S}_{ij} \rightarrow \delta_{i,j} \hat{S}$.  Each master equations couples to a set of equations enumerated by the indices $\{m_1,...,n_1,...\}$.  The total number of equations required to describe such a state depends on the overlap of the initial wave packet $\psi(t_1,...t_N)$ with the particular choice of basis.  In order for this formalism to be useful, a particular set of modes in which to decompose an $N$-photon state much be chosen wisely to reduce as much as possible the number of equations one is required to track.

	\subsection{Non-orthogonal, factorized temporal envelopes}
	
While the occupation number representation gives us a method to deal with arbitrary field states by projecting them into a basis of unentangled Fock states, the applicability of this technique relies heavily on the existence of a convenient basis of temporal modes.  Imagine the case of two single-photon wave packets following one another with a time delay $\tau$.  As long as $\tau$ is much larger than the temporal width of the wave packets, they are effectively orthogonal and the state is well described by two temporal modes.  As $\tau$ is shortened, the wave packets begin to overlap. In order to use the occupation number representation, one must find a suitable basis (with the smallest number of necessary basis functions) on which to project these modes, which may not be an easy task.  However, this particular case can be treated directly.  For situations when the input photons are not entangled, one may proceed directly without resorting to projection onto an orthogonal basis\footnote{The technique in this subsection is both inspired the ideas in Ref. \cite{SonXi13}.  That paper is but an onslaught of equations that, to be honest, I could not follow, so I proceeded to derive similar equations myself.  }.  
	
In this situation, the input field's temporal envelope factorizes with respect to the photon labels and can be written \cite{OzdImo02},
	\begin{align}\label{}
		\psi(t_1,\dots,t_N)  = \xi_{t_1}(t_1) \xi_{t_1}(t_2) \dots \xi_{t_N}(t_N).
	\end{align} 
If we gather all the terms for which the wavepacket is that same, $\xi_{t_i}(t_i) = \xi_{t_j}(t_j)$, then the field state can be written,
	\begin{align}\label{Eq::NonOrthogonalState}
		\ket{\psi_N} = & \frac{1}{\sqrt{\mathcal{N}}} \big[ B\dg(\xi_1) \big]^{n_1} \big[ B\dg(\xi_2) \big]^{n_2}  \dots \big[ B\dg(\xi_k) \big]^{n_k} \ket{0} \nn \\
		\equiv & \ket{\psi^N_{n_1,n_2,...n_k}}.
	\end{align} 
This representation of the state reveals that there are $m_i$ photons in wave packet $\xi_i(t)$, within $k$ different wave packets (temporal modes), with total number of photons given by the superscript, $N = \sum_i n_i$\footnote{The superscript on the state in \erf{Eq::NonOrthogonalState} is somewhat superfluous, as one can simply add up all the $n_i$ to find the total photon number $N$, which is why it does not appear in the definition of the generalized density matrices, \erf{Eq::NonorthogonalGDM}.  The superscript does help with notation when calculating the inner product in \erf{Eq::NonorthogonalInnerProduct}.}.  The actions of the increments are
	\begin{align}
		dB_t\ket{\psi^N_{n_1,...}} = & \frac{dt}{\sqrt{\mathcal{N}}} \sum_{i=n_1}^{n_k} n_i \xi_i(t) \big[ B\dg(\xi_1) \big]^{n_1} \dots \big[ B\dg(\xi_i) \big]^{n_i-1} \dots \big[ B\dg(\xi_k) \big]^{n_k} \ket{0} \nn \\
		 = & \frac{dt}{\sqrt{\mathcal{N}}} \sum_{i=n_1}^{n_k} n_i \xi_i(t) \ket{\psi^{N-1}_{n_1,...n_i-1,...}}, \\
		 d\Lambda_t\ket{\psi^N_{n_1,...}} = &\frac{dB_t\dg}{\sqrt{\mathcal{N}}} \sum_{i=n_1}^{n_k} n_i \xi_i(t) \ket{\psi^{N-1}_{n_1,...n_i-1,...}}.
	\end{align}

Following the prescription within this chapter, we define generalized density matrices with respect to the states in \erf{Eq::NonOrthogonalState},
	\begin{align} \label{Eq::NonorthogonalGDM}
		\hat{\varrho}_{m_1,...|n_1,...}(t) \equiv \frac{1}{\mathcal{N}}\Tr_{\rm field} \big[ \hat{U}(t) \hat{\rho}_0 \otimes \op{\psi^M_{m_1,...}}{\psi^N_{n_1,...} } \hat{U}\dg(t) \big]. 
	\end{align}
The master equations for these generalized density matrices are
	\begin{align} \label{Eq::MENonOrthogonalState}
		\frac{d}{dt} & \hat{\varrho}_{m_1,...| n_1,...}(t) = \\
		&- \frac{i}{\hbar}[ \hat{H}_{\rm sys}, \hat{\varrho}_{m_1,...| n_1,...}] + \mathcal{L}_{L} \big[  \hat{\varrho}_{m_1,...| n_1,...} \big] \nn \\
		&+ \sum_{j} m_j \xi_j(t) \big[\hat{S} \hat{\varrho}_{m_1,...,m_{j}-1,...| n_1,...}, \hat{L}\dg \big] \nn \\
		&+ \sum_{j} n_j \xi^*_j(t) \big[\hat{L}, \hat{\varrho}_{m_1,...| n_1,...n_{i - 1}, ...} \hat{S}\dg \big] \nn \\
		&+ \sum_{i,j} m_{i} n_{j} \xi_i(t) \xi^*_{j}(t) \Big( \hat{S} \hat{\varrho}_{m_1,...,m_{i}-1,...| n_1,...n_{j} - 1, ...} \hat{S}\dg -  \hat{\varrho}_{m_1,...,m_{i}-1,...| n_1,...n_{j - 1}, ...} \Big) \nn.
	\end{align}
The sums run over the indices regardless of the total number of photons.  These $N$-photon master equations are very nearly the same as those in \srf{}, other than a conspicuous lack of square roots in the prefactors in each term.  Whereas in the previous sections we had the convenience of representing our coupled field states in terms of orthogonal Fock states, here we do not.  The square roots arose from the normalization factors on Fock states; here the normalization appears in the definition of the generalized density matrices, \erf{Eq::NonorthogonalGDM}.  

The major difference arises in the initial conditions, 
	\begin{align}
		\hat{\rho}_{m_1,...|n_1,...}(0) = \frac{1}{\mathcal{N}} \Tr_{\rm field} \big[ \hat{\rho}_0 \otimes \op{\psi^M_{m_1,...}}{\psi^N_{n_1,...}} \big] = \hat{\rho}_0  \frac{\ip{\psi^N_{n_1,...}}{\psi^M_{m_1,...}}}{\mathcal{N}} ,
	\end{align}
which are modified by the nonorthogonality of the wave packets.  For multi-mode Fock states, the inner products lead to $\delta$-functions in the subscripts (mode labels), but here they must be computed.  It is true that the sum of the values of the subscripts on each state must be equal, as they give the number of creation/annihilation operators.  The inner products are then,
	\begin{align} \label{Eq::NonorthogonalInnerProduct}
		\ip{\psi^N_{n_1,...}}{\psi^M_{m_1,...}} = \delta_{M,N} \bra{0} \big[ B(\xi_1) \big]^{n_1} \dots \big[ B(\xi_k) \big]^{n_k} \big[ B\dg(\xi_1) \big]^{m_1} \dots \big[ B\dg(\xi_k) \big]^{m_k}  \ket{0},
	\end{align}
where the evaluation may be simplified\footnote{``May be" as in ``it is possible that", rather than ``I know how and will show you."} with the commutation relation for the wave packet creation/annihilation operators,
	\begin{align}
		\big[ B(\xi_i), B\dg(\xi_j)\big] = \int dt \, \xi_i^*(t) \xi_j(t).
	\end{align}

Taking asymmetric expectations in the standard way, [\erf{Eq::multi-modeExpectation}], gives an output photon flux,
	\begin{align} 
		\frac{d}{dt} & \mathbbm{E}_{m_1...|n_1...} \big[\Lambda_t^{\rm out}(t) \big] = \mathbbm{E}_{m_1...|n_1...} \big[\hat{L}\dg \hat{L}\big] \\
		& + \sum_{i} m_i \xi_i^*(t) \mathbbm{E}_{m_1...m_i-1,...|n_1...} \big[ \hat{S}\dg \hat{L} \big]   + \sum_{i} {n_i} \xi_i(t)  \mathbb{E}_{m_1...|n_1...n_i-1,...} \big[ \hat{L}\dg \hat{S} \big], \nonumber \\
		& + \sum_{i,j} {m_i n_{j}} \xi_i^*(t) \xi_{j}(t) \mathbbm{E}_{m_1...m_i-1,...|n_1...n_{j}-1,...} \big[  \hat{S}\dg \hat{S} \big] \nn.
	\end{align}
and an expectation value for the output quantum noise,
	\begin{align} 
		\frac{d}{dt} \mathbbm{E}_{m_1...|n_1...} \big[ dB_t^{\rm out} \big] =&\mathbbm{E}_{m_1...|n_1...} \big[ \hat{L} ]  + \sum_i {n_i} \xi_i(t) \mathbbm{E}_{m_1...|n_1...n_i-1...} \big[\hat{S}   \big]. 
	\end{align}
Because of the non-orthogonality, one must take care when taking expectation values.

Superpositions and mixtures of input states of the form \erf{Eq::NonOrthogonalState}, are composed just as in \erf{Eq::gen_me2Mode}.  The rules for calculating physical observables are found in \srf{TwoModeCombinations}.

	\subsection{Example: two-photon state in two non-orthogonal wave packets}
	
To illustrate the $N$-photon master equation formalism, we examine the case of a system interacting with a two-photon state $\ket{\psi_2}$, whose temporal function factorizes, $\psi(t_1, t_2) = \xi(t_1)\eta(t_2)$.  Using \erf{Eq::NonOrthogonalState} the state can be written\footnote{If $\xi(t)$ and $\eta(t)$ are orthogonal, then \erf{ApEq::ExampleTwoPhoton} can be written $\ket{\psi_2} = \ket{1,1}$.},
	\begin{align} 
		\ket{\psi_2} = &\frac{1}{\sqrt{\mathcal{N}}} \int dt_1 \int dt_2 \,\xi(t_1) \eta(t_2) \hat{b}\dg(t_1) \hat{b}\dg(t_2) \ket{0} \nn \\
			= & \frac{1}{\sqrt{\mathcal{N}}}B\dg(\xi) B\dg(\eta) \ket{0} \nn \\
			\equiv &  \frac{1}{\sqrt{\mathcal{N}}} \ket{\psi_{1,1}}. \label{Eq::ExampleTwoPhoton}
	\end{align}
The normalization is given by \erf{Eq::NormEx}, 
	\begin{align}
		\mathcal{N} = & \ 1 + \bigg| \int dt \, \xi(t)\eta^*(t) \bigg|^2,
	\end{align}
and depends on the degree to which the wave packets are orthogonal.  
	
	\begin{figure}[!t]
	\begin{center}
	\includegraphics[scale=0.7]{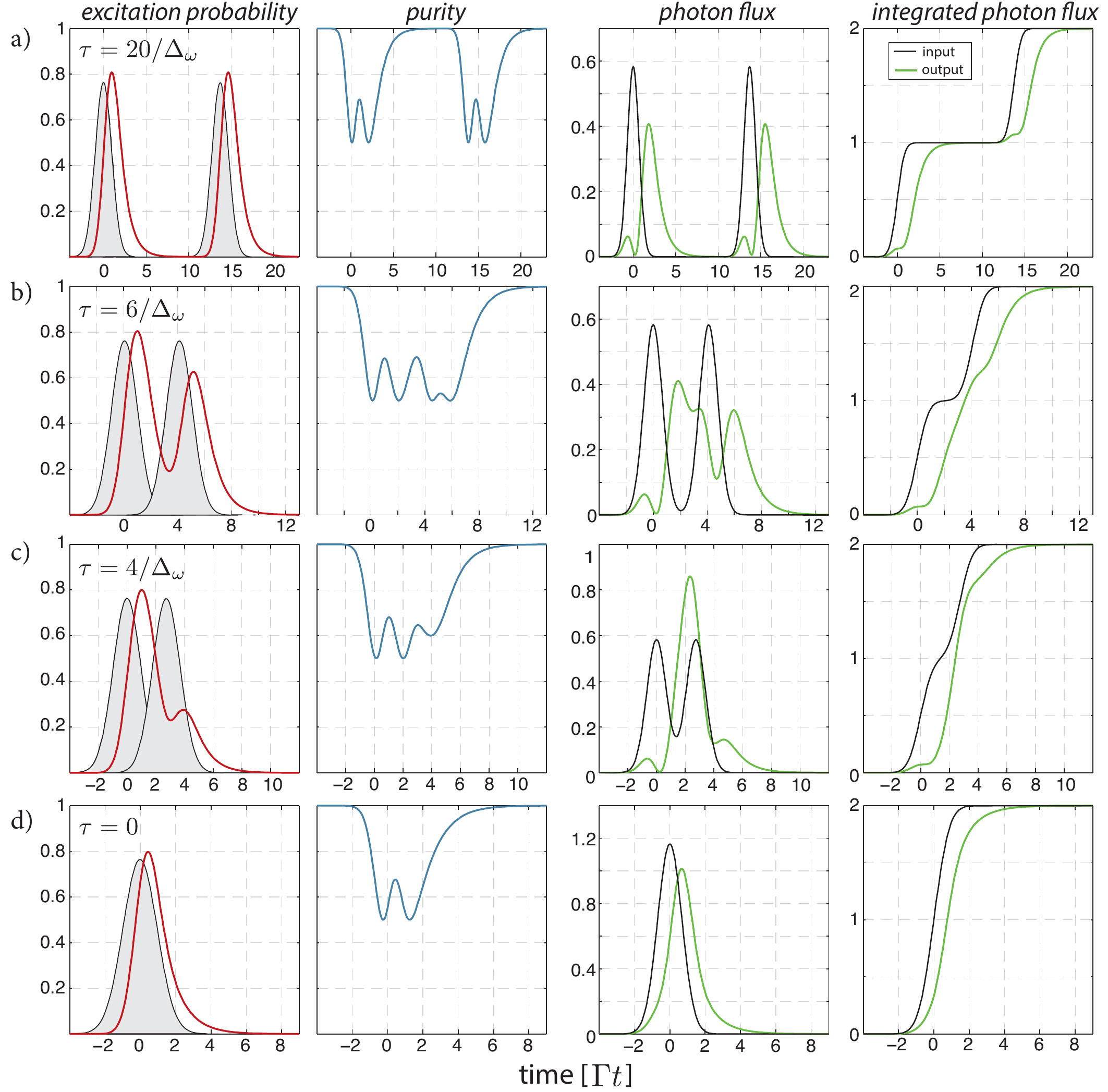}
		\caption[Two single-photon pulses exciting a two-level atom]{Excitation dynamics, reduced state purity, photon flux, and integrated photon flux for a two-level atom interacting with the general two-photon state.  The input field state, described by \erf{Eq::ExampleTwoPhoton}, is composed of two unentangled, Gaussian, single-photon wave packets with spectral width $\Delta_\omega/\Gamma = 1.46$ and varying delay time between the peaks, $\tau$.  The rows correspond to a different delays between the pulses: a) $\tau = 20/\Delta_\omega$, b) $\tau = 4/\Delta_\omega$, c) $\tau = 6/\Delta_\omega$, d) $\tau = 0$.} \label{Fig::2photonpulses}
	\end{center}
	\end{figure}

The physical reduced density operator is then 
	\begin{align}
		\hat{\varrho}_{\rm phys}(t) = \frac{1}{\mathcal{N}} \Tr_{\rm field} \big[ \hat{U}(t) \rho_0 \otimes \op{\psi_{1,1}}{\psi_{1,1}} \hat{U}\dg(t) \big] \equiv \hat{\varrho}_{1,1|1,1}(t).
	\end{align}
For simplicity we consider only dipole-type interactions, $\hat{S} \rightarrow \hat{I}_{\rm sys}$ in \erf{Eq::MENonOrthogonalState}, to give the master equations,
	\begin{align} \label{Eq::TwoModeME} 
		 \frac{d}{dt} \hat{\varrho}_{m_1,m_2|n_1,n_2}  = & - \smallfrac{i}{\hbar} \big[ \hat{H}_{\rm sys}, \hat{\varrho}_{m_1,m_2|n_1,n_2} \big] + \mathcal{L}_{L}[ \hat{\varrho}_{m_1,m_2|n_1,n_2}]   \\ 
                  &  \nn+ {m_1} \xi(t) [\hat{\varrho}_{m_1-1,m_2|n_1,n_2}, \hat{L}\dg  ] + {n_1} \xi^*(t)  [\hat{L}, \hat{\varrho}_{m_1,m_2|n_1-1,n_2} \dg  ] \\
                  & + {m_2} \eta(t)  [ \hat{\varrho}_{m_1,m_2-1|n_1,n_2}, \hat{L}\dg ]  + {n_2} \eta^*(t)[  \hat{L}, \hat{\varrho}_{m_1,m_2|n_1,n_2-1} \dg ]\nn.
	\end{align}
The initial conditions are 
	\begin{align}
		 \hat{\varrho}_{1,1|1,1}(0) &= \hat{\rho}_{\rm sys},  \\
		  \hat{\varrho}_{1,0|0,1}(0) &= \frac{\ip{\psi_{0,1}}{\psi_{1,0}}}{\mathcal{N}}\hat{\rho}_{\rm sys} \\
		   \hat{\varrho}_{0,1|1,0}(0) &= \frac{\ip{\psi_{1,0}}{\psi_{0,1}}}{\mathcal{N}}\hat{\rho}_{\rm sys} \\
		 \hat{\varrho}_{1,0|1,0}(0) &= \hat{\varrho}_{0,1|0,1}(0) = \hat{\varrho}_{0,0|0,0}(0) =\frac{1}{\mathcal{N}}\hat{\rho}_{\rm sys}, 
	\end{align}
with all others initialized to zero.  The inner products are evaluated simply,
	\begin{align}
		 \ip{\psi_{0,1}}{\psi_{1,0}} = \ip{\psi_{1,0}}{\psi_{0,1}}^* = \int dt \, \xi(t) \eta^*(t).
	\end{align}
For orthogonal wave packets this inner product vanishes and gives normalization $\mathcal{N} = 1$, and for identical wave packets (Fock states) the inner product is 1 and we find $\mathcal{N} = 2$ for the normalization\footnote{Taking the Fock state limit, $\xi(t) = \eta(t)$, can be confusing because the off-diagonal equations are initialized to $\hat{\rho}_{\rm sys}$.  This is in fact correct, but to make the proper comparison it is important to realize that in this limit the subscripts collapse to a single subscript, as for the Fock-state case in \srf{sec_fock}.}.
	
These situations are illustrated in Fig. \ref{Fig::2photonpulses}, which plot various quantities for a two-level atom probed by two consecutive Gaussian single-photon pulses each described by \erf{Eq::gau_xi} for several choices of delay time, $\tau$.  The atom interacts on a dipole transition with parameters $\hat{H}_{\rm sys} = 0$, $\hat{L} = \sqrt{\gamma} \op{g}{e}$, and $\hat{S} = \hat{I}_{\rm sys}$ in a single mode, $\gamma = \Gamma$.  The system interacts with the photon in the first wave packet and remains entangled with the field until the excitation is released through decay.  The case of large $\tau$ is shown in \ref{Fig::2photonpulses}(a).  We see in the first column that, before the second photon arrives, the atom returns to the ground state.  The atom's purity, used as a witness for system-field entanglement (see \srf{}), is shown in the second column and likewise shows that the atom returns to the pure ground state before the second photon arrives.  Each of the two excitation curves in \frf{Fig::2photonpulses}(a) is identical to the single-photon excitation in \frf{Fig::2PhotonCompare}(b), and similarly the photon fluxes, third and fourth column, are those in \frf{Fig::2PhotonCompare}(d) and (f).  When the second photon arrives before the atom has fully decayed, \frf{Fig::2photonpulses}(b)-(d), then this entanglement and the resulting interference between excitation from the first and second photon affect the system's dynamics as well as the output photon flux, as seen in Figs. \ref{Fig::2photonpulses}(b)-(d).  When the pulses completely overlap, the field is a two-photon Fock state, and the dynamics agree with the results from \srf{SEC::1n2photons}, as seen in the two-photon curves in Fig. \ref{Fig::2PhotonCompare}(b), (d), and (f).

	\subsection{Multi-mode multi-photon master equations }

Finally, one may wish to consider input fields in combinations (superpositions and/or mixtures) of different $N$-photon states in multiple modes, similar to the analysis of Ref. \cite{SonXi13}.  With the techniques from this chapter, one may first represent the initial field state in a basis of Fock states both over spatial/polarization modes and over temporal modes.  From there, the generalized density matrices are those of \erf{Eq::multi-modeME}.  One need only take care to properly identify the coupling operators for each mode $\hat{L}_i$ and $\hat{S}_{ij}$, since, for all temporal modes within a single spatial/ polarization mode, these coupling operators will be the same.  Such a situation follows in a straightforward way but is quite complicated and, as such, is not explicitly included in this dissertation.

	\chapter{Dispersive atom-light interaction} \label{Ch::DispersiveInt}
%


We now shift our attention to the study of an atomic ensemble interacting dispersively with a paraxial electric field.  Entanglement generated between the collective atomic state and the propagating modes of the field makes atomic ensembles a rich and promising platform for quantum information processing.  The entangling atom-light interaction is produced by detuning the probe laser such that the excited states are negligibly populated. Two critical consequences are that (i) the light effectively couples to the robust and controllable atomic ground-state spins, and (ii) decoherence from spontaneous photon scattering is reduced.  An interface based on a dispersive atom-light interaction has been extensively studied in the literature; our goal is to treat, in a fully quantum fashion, the spatial effects for a paraxial probe laser that is inhomogeneous in both amplitude and phase across the ensemble.  Before navigating the complicated problem of a multi-atom interaction with spatially varying electric fields, treated in Chapter \ref{Ch::3DInterface}, we lay the groundwork with a careful analysis for a single multi-level alkali atom interacting with a monochromatic probe laser of frequency $\omega_0$.  In this chapter, we review the dispersive atom-light interaction for a single, multi-level alkali atom closely following Ref. \cite{DeuJes10}. 

Following Refs. \cite{HammererThesis, VasSor12}, we partition the quantized electric field into paraxial modes and nonparaxial, diffuse modes, 
	\begin{align} \label{Eq::ModeDecomp}
		\hat{\mathbf{E}}^{(+)}(\mathbf{r}, t) = \hat{\mathbf{E}}_{\rm fwd}^{(+)}(\mathbf{r}, t) + \hat{\mathbf{E}}_{\rm diff}^{(+)}(\mathbf{r}, t).
	\end{align}	
This decomposition is motivated by the geometry we wish to consider -- photon scattering of a paraxial laser beam by an extended atomic ensemble.  For this analysis we assume the positive frequency component of the paraxial electric field operator in the spatial mode of the laser is
	\begin{align} \label{Eq::SimpleField}
		\hat{\mathbf{E}}_{\rm fwd}^{(+)}(\mathbf{r}) = \sqrt{\frac{2 \pi \hbar \omega_0}{A c}}  \mathcal{U}(\mathbf{r}_\perp, z) \big( \mathbf{e}_x \hat{a}_x(z,t) +  \mathbf{e}_y \hat{a}_y(z,t) \big) e^{i {k}_0 z },
	\end{align}
and all other spatial modes have been traced out and give rise to decoherence and loss.  The dimensionless mode function $\mathcal{U}(\mathbf{r}_\perp, z)$ plays no role in the simplified analysis in this chapter and will summarily be ignored until Chapter \ref{Ch::3DInterface}, when it emerges as a central complicating factor.  
	
The dispersive light shift interaction describes an atom's effective response to an electric field when the excited states can be adiabatically eliminated from the dynamics.  This reduced description applies when the saturation parameter is small,
	\begin{align} \label{Eq::SatParameter}
		s = \frac{ \Omega^2/4}{\Delta^2 + \Gamma^2/4} \ll 1,  
	\end{align}
where $\Omega$ is the Rabi frequency, $\Gamma$ is the spontaneous emission rate, and $\Delta = \omega_0 - \omega_{eg}$ is the probe detuning from the atomic resonance frequency $\omega_{eg}$. As the saturation parameter is proportional to the ratio of the Rabi frequency to the detuning, $\Omega/\Delta$, physically this allows condition \erf{Eq::SatParameter} to be satisfied even for large driving power (Rabi frequency) so long as the probe laser is far enough detuned.  In this limit, the atom-light coupling is described by the dispersive light-shift interaction,
	\begin{align} \label{Eq::LightShiftHam}
		\hat{H}_{\rm eff} & = - \hat{\mathbf{E}}_{\rm fwd}^{(-)}(\mathbf{r}_A ) \cdot \tensor{\alpha} \cdot \hat{\mathbf{E}}_{\rm fwd}^{(+)}(\mathbf{r}_A ), 
	\end{align}
where the electric field operator in \erf{Eq::SimpleField} is evaluated at the atomic position $\mathbf{r}_A$.  This coupling gives rise to the Faraday interaction, which serves as the basis for the QND squeezing protocol in Chapter \ref{Ch::SpinSqueezing}.  In the next subsection we review the irreducible decomposition of the atomic polarizability tensor, $\tensor{\alpha}$.  	
	
\section{Atomic polarizability tensor}
	
The atomic polarizability tensor that appears in \erf{Eq::LightShiftHam} is the dyad of two vector dipole operators and connects between ground states through the excited states.  The atom is probed by a laser detuned near a fine structure transition with excited electronic angular momentum $j' = l' \pm s$ for $l'=1$ ($P$ orbital) and the single spin of the valence electron, $s=1/2$.  Two hyperfine ground states emerge from the coupling of the nuclear spin ($i$ = 7/2 in ${}^{133}$Cs) to the total electronic angular momentum, $\hat{\mbf{j}} = \hat{\mbf{l}} + \hat{\mbf{s}}$, to form total hyperfine angular momentum $\hat{\mbf{f}} = \hat{\mbf{i}} + \hat{\mbf{j}}$.  A similar hyperfine splitting occurs in each of the excited $P$ states.  Each hyperfine state has $2f + 1$ magnetic sublevels, where $f = i + j$.  
	
Due to the massive fine-structure splitting between the first excited $P_{1/2}$ and $P_{3/2}$ states ($\Delta \omega \approx 2\pi \times 7000$ THz in ${}^{133}$Cs), one need only consider a single $j \rightarrow j'$ transition, e.g. the D2 line ($j' = 3/2$).  It can be shown that the interaction Hamiltonian is effectively block-diagonal in the hyperfine ground states \cite{DeuJes10}, $\tensor{\alpha} = \sum_f \tensor{\alpha}(f)$, where 
	\begin{align} \label{Eq::PTensorGen}
		\tensor{\alpha}(f) = -\frac{1}{\hbar} \sum_{f'} \frac{ \hat{\mathbf{d}}_{f f'} \hat{\mathbf{d}}_{f' f}^\dagger}{\Delta_{f, f'} + i \Gamma/2 }. 
 	\end{align}
Within the effective interaction, \erf{Eq::LightShiftHam}, the real and imaginary parts of the complex polarizability tensor comprise a non-Hermitian Hamiltonian. The real part drives coherent, entangling evolution between the atom and paraxial field, and the imaginary part leads to decoherence and loss. 

First, we set the convention for the atomic dipole operators on the $j \rightarrow j'$ transition.  The operator that raises the atom from the ground hyperfine state $f$ to the excited state $f'$ is
	\begin{align}
		\hat{ \mathbf{d} }_{f' f}^\dagger  & = \sum_q \sum_{m, m'}  \bra{f' m'} d_{f',f}^q \ket{f m} \op{f' m'}{f m} \mathbf{e}_q^* \\
		& = \bra{ f' }| d | \ket{ f } \sum_q \sum_{m, m'}  \braket{f m; 1 q}{f' m'} \op{f' m'}{f m} \mathbf{e}_q^*
	\end{align}
where we used the Wigner-Eckart theorem to pull out the reduced matrix element $\bra{ f' }| d | \ket{ f } $.  It can be further simplified with another application of the Wigner-Eckart theorem,
	\begin{align}
		\bra{ f' }| d | \ket{ f } = \bra{ j' }| d | \ket{ j } o^{j' f'}_{j f},
	\end{align}
in terms of a reduced matrix element involving the $j \rightarrow j'$ transitions and a relative oscillator strength,
	\begin{align} \label{Eq::OscStrength}
		o^{j' f'}_{j f} \equiv (-1)^{f'+i + j' + 1} \sqrt{ (2 j'+1) (2f + 1) } 
			\left\{ 
				\begin{array}{ccc}
					f' & i & j' \\
					j & 1 & f
				\end{array}
			\right\} .
	\end{align}
that determines the spontaneous decay branching ratios on allowed dipole transitions; $\Gamma_{j' f' \rightarrow j f} / \Gamma_{j'\rightarrow j} = |o^{j' f'}_{j f}|^2$ \cite{DeuJes10}.	
	
This allows us to factor out characteristic units from the dipole raising operator and define dimensionless dipole operators,
	\begin{align}
		\hat{ \mathbf{D}}_{f' f}^\dagger & = \frac{ \hat{ \mathbf{d} }_{f' f}^\dagger }{\bra{ j' }| d | \ket{ j } } \\
		& = \sum_q \sum_{m, m'} \mathbf{e}_q^* o^{j' f'}_{j f} \braket{f' m'}{f m; 1 q} \op{f' m'}{f m}. 
	\end{align}
Finally, we write the detuning from a particular hyperfine transition as
	\begin{align}
		\Delta_{f,f'} = \Delta + \delta_{f'},
	\end{align}
where we have factored out the detuning relative to the largest hyperfine excited state,
	\begin{align} \label{Eq::DetuningChoice}
		\Delta \equiv \Delta_{f,f'_{\rm max}} = \omega_0 - ( \omega_{f'_{\rm max}} - \omega_f),
	\end{align} 
and $\delta_{f f'} \equiv \Delta_{f,f'} - \Delta$ is the residual detuning for the other hyperfine excited states.  All the units are collected in a characteristic polarizability,
	\begin{align} \label{Eq::alpha0}
		\alpha_0(\Delta) & =  -\frac{|  \bra{ j' }| d | \ket{ j } |^2}{\hbar \Delta } = - \frac{3 \lambda_{j' j}^3}{32 \pi^3} \frac{ \Gamma }{ \Delta }.
	\end{align}
The wavelength of the transition $\lambda_{j' j}$ and spontaneous emission rate $\Gamma$ are defined with respect to the fine-structure splitting $j' \rightarrow j$; that is,
	\begin{align}
		\Gamma = \frac{1}{\hbar} \frac{4}{3} \frac{\omega_{j j'}^3}{c^3} | \bra{ j' }| d | \ket{ j } |^2.
	\end{align}
One could define a slightly different characteristic polarizability using a different detuning in \erf{Eq::DetuningChoice}, such as the fine structure splitting on the $j \rightarrow j'$ transition.

The atomic polarizability tensor in \erf{Eq::PTensorGen} can then be written in terms of the dimensionless dipole operators and the characteristic polarizability,
	\begin{align} \label{Eq::AtomicPolarizabilityF}
		\tensor{\alpha}(f) =  \alpha_0(\Delta) \sum_{f'} \frac{ \hat{\mathbf{D}}_{f f'} \hat{\mathbf{D}}_{f' f}^\dagger}{ 1 + \delta_{f'} / \Delta + i \Gamma/(2\Delta)}.
	\end{align}
For ${}^{133}$Cs with ground hyperfine manifolds $f = \{3,4\}$, this becomes
	\begin{align}
		\tensor{\alpha} & =  \tensor{\alpha}(3) + \tensor{\alpha}(4) \nn \\
		& = \alpha_0(\Delta_{3}) \sum_{f'} \frac{ \hat{\mathbf{D}}_{3 f'} \hat{\mathbf{D}}_{F' 3}^\dagger}{ 1 + \delta_{3, f'} / \Delta_{3} + i \Gamma/(2\Delta_3) } +  \alpha_0(\Delta_{4}) \sum_{F'} \frac{ \hat{\mathbf{D}}_{4 f'} \hat{\mathbf{D}}_{f' 4}^\dagger}{ 1 + \delta_{4, f'} / \Delta_{4} + i \Gamma/(2\Delta_4) }.
	\end{align}
I have explicitly labeled the detunings by the ground state hyperfine level $f$ in order to emphasize that the detunings are different on each of the two terms.  Unless the laser detuning is chosen such that $\Delta_3$ and $\Delta_4$ are of the same order, one of the two terms will dominant the dynamics and the other can be ignored.

	\subsection{Irreducible representation of the atomic polarizability tensor}
		
The atomic polarizability tensor, being a dyad of two vector operators, can be decomposed into rank-0, rank-1, and rank-2 irreducible tensor components \cite{StocktonThesis, HammererThesis, GerMab06, DeuJes10, LeKRau13}.  The terms in the Hamiltonian can be written in a Cartesian basis, useful for describing atomic interaction with linearly polarized light.  The dimensionless Cartesian $ij$-component of the atomic polarizability tensor in the ground hyperfine state $f$ is defined,
	\begin{align} \label{Eq::alphaij}
		\hat{\alpha}_{ij}(f) \equiv \mathbf{e}_i \cdot \hat{\mathbf{D}}_{f f'} \hat{\mathbf{D}}_{f' f}^\dagger \cdot \mathbf{e}_j.
	\end{align}	
It can be shown that the block-diagonal terms have a basis-independent form whose Cartesian $ij$-components are \cite{DeuJes10},
	\begin{align} \label{Eq::IrreducibleDecomp}
		\hat{\alpha}_{ij} (f) = C_{j' f f'}^{(0)} \delta_{ij} \hat{ I } + i C_{j' f f'}^{(1)} \varepsilon_{ijk} \hat{f}_k + C_{j' f f'}^{(2)} \left(\frac{1}{2} \big(\hat{f}_i \hat{f}_j + \hat{f}_j \hat{f}_i \big) - \frac{1}{3} \delta_{ij} \hat{ \mathbf{f} } \cdot \hat{ \mathbf{f} }  \right).  
	\end{align}
The $\hat{f}_i$ are dimensionless hyperfine spin operators satisfying
	\begin{align}
		[\hat{f}_i, \hat{f}_j] = i \varepsilon_{ijk} \hat{f}_k,
	\end{align}
with total angular momentum $f$ for each ground hyperfine manifold.  The tensor coefficients are \cite{DeuJes10}
		\begin{align}
			C_{j' f f'}^{(0)} &= (-1)^{3f - f' + 1} \frac{1}{ \sqrt{3} } \frac{2f'+1}{\sqrt{2f+1}}  \left\{ \begin{array}{ccc} f& 1 & f' \\ 1 & f & 0 \end{array} \right\} |o^{j' f'}_{j f}|^2, \\
			C_{j' f f'}^{(1)} &= (-1)^{3f - f' } \frac{3}{ \sqrt{2} } \frac{2f'+1}{ \sqrt{f(f+1)(2f+1)} }  \left\{ \begin{array}{ccc} f & 1 & f' \\ 1 & f & 1 \end{array} \right\} |o^{j' f'}_{j f}|^2, \\
			C_{j' f f'}^{(2)} &= (-1)^{3f - f' } \frac{\sqrt{30} (2f'+1)}{\sqrt{ f(f+1)(2f+1)(2f-1)(2f+3))} }  \left\{ \begin{array}{ccc} f & 1 & f' \\ 1 & f & 2 \end{array} \right\} |o^{j' f'}_{j f}|^2 .
		\end{align}	
These coefficients depend on the fine-structure quantum numbers $\{j, j'\}$ through the relative oscillator strengths, $o^{j' f'}_{j f}$, given in \erf{Eq::OscStrength}.  Note that the notation in \erf{Eq::alphaij} is different from that for the full polarizability tensor, \erf{Eq::AtomicPolarizabilityF}, (and different from that in Ref. \cite{DeuJes10}) in that it contains no units, detunings, or sums over excited states.  Instead its intention is to isolate the irreducible tensor components.   

	When the detuning is also large compared to the excited hyperfine splitting the detuning becomes independent of $f'$ (essentially $\delta_{f'}/\Delta \rightarrow 0$).  Then, the sums over the tensor coefficients in Eqs. (\ref{Eq::GenPolarizability}) can be done explicitly to yield,
	\begin{align}
		C_{j' f}^{(0)} &\equiv \sum_{f'} C_{j' f f'}^{(0)} =   \frac{2^{j'-1/2}}{3} \label{Eq::ScalarCoefSum} \\ 
		C_{j' f}^{(1)} &\equiv  \sum_{f'} C_{j' f f'}^{(1)} = (-1)^{j'-1/2} \frac{g_f}{3} \\
		C_{j' f}^{(2)} &\equiv  \sum_{f'} C_{j' f f'}^{(2)} = 0. \label{Eq::Rank2Sum}
	\end{align}
The Land\'{e} g-factor depends on the ground state manifold:  $g_f = 1/f_{\uparrow}$ for $f_{\uparrow} = i + 1/2$ and $g_f = -1/f_{\uparrow}$ for $f_{\downarrow} = i - 1/2$.  In this far-detuned limit, we see that the $C_{j' f f'}^{(2)}$ coefficients sum to zero and the rank-2 terms in \erf{Eq::GenPolarizability} vanish,
	\begin{equation} \label{Eq::GenPolarizability}
		\hat{\alpha}_{ij} (f) \rightarrow  C_{j' f}^{(0)} \delta_{ij} \hat{ I } + i C_{j' f }^{(1)} \varepsilon_{ijk} \hat{f}_k  .
	\end{equation}
This reflects the fact that in the absence of hyperfine resolution, the nuclear spin is decoupled and the ground state angular momentum is given by the total electronic angular momentum, $j=1/2$.

\section{Interaction with a quantum field}

	The coherent interaction can be written in a useful form involving the coupling of the \emph{Stokes vector}, which describes the field's polarization, to the angular momentum of the atom.  The operator components of the Stokes vector are
	\begin{subequations} \label{Eq::StokesOps}
	\begin{align}
		 \hat{s}_0(z,t) = & \frac{1}{2} \Big(  \hat{a}^\dagger_{x}(z,t)  \hat{a}_{x}(z,t) + \hat{a}^\dagger_{y}(z,t)  \hat{a}_{y}(z,t) \Big) \\
		\hat{s}_1(z,t) = & \frac{1}{2} \Big(  \hat{a}^\dagger_{x}(z,t)  \hat{a}_{x}(z,t) - \hat{a}^\dagger_{y}(z,t)  \hat{a}_{y}(z,t) \Big) \\
		\hat{s}_2(z,t) = & \frac{1}{2} \Big(  \hat{a}^\dagger_{x}(z,t)  \hat{a}_{y}(z,t) + \hat{a}^\dagger_{y}(z,t)  \hat{a}_{x}(z,t) \Big) \\
		\hat{s}_3(z,t) = & \frac{1}{2i} \Big(  \hat{a}^\dagger_{x}(z,t)  \hat{a}_{y}(z,t) - \hat{a}^\dagger_{y}(z,t)  \hat{a}_{x}(z,t) \Big), 
	\end{align}
	\end{subequations}
and, from \erf{Eq::ParaxialFieldCommutator}, they satisfy the unequal-$t$, unequal-$z$ commutation relations of an effective angular momentum, 
	\begin{align}
		\big[ \hat{s}_i(z,t), \hat{s}_j(z',t') \big] = i \varepsilon_{ijk} \hat{s}_k(z,t) \delta (t-t' - (z-z')/c ).
	\end{align}  
Using these definitions in the real part of the light shift interaction, \erf{Eq::LightShiftHam}, gives a coherent Hamiltonian of the form,
	\begin{align} \label{Eq::StokesHam}
		\hat{H}_{\rm coh} = \hbar \chi_0 \sum_{F'} \frac{1}{1+ \delta_{f'}/\Delta} \Big( & \hat{A}_0 \hat{s}_0(z_A,t) +\hat{A}_1 \hat{s}_1(z_A,t)  \\
		&+\hat{A}_2 \hat{s}_2(z_A,t)  +\hat{A}_3 \hat{s}_3(z_A,t)  \Big), \nn
	\end{align}
where the atomic operators that couple to the Stokes components at the atom's position $z_A$ are
	\begin{subequations} \label{Eq::StokesSpinOps}
	\begin{align}
		\hat{A}_0 = & C^{(0)}_{j'f f'} \hat{I} + C^{(2)}_{J'FF'} \left( \frac{3 \hat{f}_z^{2} - \hat{\bf{f}} \cdot  \hat{\mbf{f}} }{6} \right) \\
		\hat{A}_1 = & C^{(2)}_{j'f f'} \left( \frac{\hat{f}_x^{2} - \hat{f}_y^{2} }{2} \right) \\
		\hat{A}_2 = & C^{(2)}_{j'f f'} \left( \frac{\hat{f}_x \hat{f}_y + \hat{f}_y \hat{f}_x }{2} \right) \\
		\hat{A}_3 = & - C^{(1)}_{j'f f'} \hat{f}_z.
	\end{align}
	\end{subequations}
The dimensionless coupling constant,
	\begin{align} \label{Eq::FaradayAngle}
		\chi_0 = - \left(\frac{4\pi \omega_0}{A c} \right) \alpha_0(\Delta) = \Big( \frac{\sigma_0}{A} \Big)\Big( \frac{\Gamma}{2 \Delta} \Big),
	\end{align} 
is a measure of the strength of the light-matter interaction.  It is proportional to the detuning, $\Delta^{-1}$, but more importantly to the ratio of the resonant atomic cross section, $\sigma_0 = 3 \lambda_{j j'}^2/2 \pi$, to the transverse mode area $A$.  This ratio sets the strength of the single-atom coupling, as it roughly describes the amount of light scattered from the atom back into the probe mode.  The form of the real part of the interaction, \erf{Eq::StokesHam}, assumes we are working in a regime where the detuning is much greater than the linewidth, $\Delta_{f,f'} \gg \Gamma$, for all excited hyperfine states.  Note that the sign of $\hat{A}_3$ is opposite that found in Ref. \cite{DeuJes10}.  

Although we ultimately focus on the mapping of atomic spin noise to the field via the interaction Hamiltonian in \erf{Eq::StokesHam}, it is also useful for the study of the polarization dynamics.  The state of a macroscopically prepared field (a coherent state) can be represented by a Stokes vector whose position on the Poincar\'{e} sphere describes the polarization state of the field.  Through interaction with a multilevel atom, the field's polarization undergoes a rotation on the Poincar\'{e} sphere which depends on the spin state of the atom through the operators in \erf{Eq::StokesSpinOps}.  A particular effect of interest is Faraday rotation, where an input linear polarization experiences a rotation proportional to the atomic polarization along the direction of propagation, $\Theta \propto \chi_0 \expt{ \hat{f}_z }$ \cite{DeuJes10}.

	\subsection{Coherent driving field}

In many cases, such as the generation of spin squeezing in Chapter \ref{Ch::SpinSqueezing}, we are specifically interested in probing an atomic ensemble with a coherent laser.  We can make a Mollow transformation on the paraxial field that displaces the mode of the laser into a large amplitude coherent state,
	\begin{align} \label{Eq::MollowTrans}
		\hat{\mathbf{E}}^{(+)}_{\rm fwd}(\mathbf{r}) \rightarrow \mathcal{E}^{(+)}_L(\mathbf{r}) \vec{\epsilon}_L + \hat{\mathbf{E}}^{(+)}_{\rm fwd}(\mathbf{r}),
	\end{align}
where $\mathcal{E}^{(+)}_L(\mathbf{r}) = \mathcal{E}^{(+)}_0 \mathcal{U}(\mathbf{r})e^{i k_0 z} $ is a classical paraxial laser field with polarization $\vec{\epsilon}_L$\footnote{Note that $\mathcal{E}^{(+)}_0$ is the positive-frequency amplitude, which is twice the real amplitude. This distinction is important when comparing our results to the literature.}. As was done in \erf{Eq::SimpleField}, we henceforth bury the explicit appearance of the spatial mode.  The remaining quantum field operator, $\hat{\mathbf{E}}_{\rm fwd}(\mathbf{r})$, describes the underlying quantum fluctuations. 

Making this transformation on the effective Hamiltonian, \erf{Eq::LightShiftHam}, we can decompose the coherent (real) part of the interaction into a portion that couples the atom to the quantum paraxial electric field,
	\begin{align}
		\hat{H}_{\rm int} = & - \alpha_0(\Delta) \sum_{f'} \frac{ \hat{\mathbf{E}}^{(-)}_{\rm fwd}(\mathbf{r}_A) \cdot \hat{\mathbf{D}}_{f f'} \hat{\mathbf{D}}_{f' f}^\dagger \cdot \vec{\epsilon}_L \mathcal{E}^{(+)}_L(\mathbf{r}_A)   }{ 1 + \delta_{f'}/\Delta  }  + \mbox{H.c.},
	\end{align}
and a portion that does not couple the field, but drives coherent dynamics in the internal spin state of the atom,	
	\begin{align}
		\hat{H}_{\rm LS} 
			= & \hbar \frac{\Omega(\mathbf{r}_A)}{4 \Delta} \sum_{f'} \frac{  \vec{\epsilon}_L^* \cdot \hat{\mathbf{D}}_{f f'} \hat{\mathbf{D}}_{f' f}^\dagger  \cdot  \vec{\epsilon}_L }{1 + \delta_{f'}/\Delta }.
	\end{align}
The effective Rabi frequency,
	\begin{align} \label{Eq::EffRabiFreq}
		\Omega(\mathbf{r}) = \frac{\langle j' || d || j \rangle 2 \mathcal{E}^{(+)}_0}{\hbar},
	\end{align}
has been defined in terms of the positive-frequency amplitude of the classical probe, $\mathcal{E}^{(+)}_0$.  The terms that describe scattering between quantum fluctuations (fields initially in vacuum) have been ignored.   

In the irreducible representation of the atomic polarizability, \erf{Eq::GenPolarizability}, an effective the interaction describes coupling of the atomic operators to the field quadratures with polarization orthogonal to that of the laser.  Of specific interest to us is the Faraday interaction, which we isolate with a choice of $\vec{\epsilon}_L = \mathbf{e}_x$.  Neglecting the rank-2 terms, in this case the interaction Hamiltonian becomes
	\begin{align}
		\hat{H}_{\rm int} & = \alpha_0(\Delta) \mathcal{E}^{(+)}_L (\mathbf{r}_A) \sqrt{\frac{4 \pi \hbar \omega_0}{Ac}} \sum_{f'} \frac{ C^{(1)}_{j'f f'} }{1 + \delta_{f'}/\Delta }  \hat{f}_z \hat{P}(z_A,t) \\
			& = - \hbar \chi_0\sqrt{\frac{\dot{N}_L}{2}} \sum_{f'} \frac{ C^{(1)}_{j'f f'} }{1 + \delta_{f'}/\Delta }  \hat{f}_z \hat{P}(z_A,t) 
	\end{align}
and the state-independent residual light shift is
	\begin{align}
		\hat{H}_{\rm LS} = & \hbar \frac{\Omega(\mathbf{r}_A)}{4 \Delta} \sum_{f'} \frac{  C^{(0)}_{j'f f'} }{1 + \delta_{f'}/\Delta } \hat{I}.
	\end{align}
As expected we see a coupling of the laser to the field quadrature, $\hat{P}(z,t) =  -i(\hat{a}_y(z,t) - \hat{a}\dg_y(z,t))/\sqrt{2}$, via the $z$-component of the atom's hyperfine spin.  A rotation of polarization on the Poincar\'{e} sphere becomes, in this linearized regime, a translation in quadrature phase space.  In the next chapter, we develop this further for multiple atoms, including a detailed description of the effects of the ignored rank-2 components of the interaction.

	\section{ Single-atom master equation }

In addition to the coherent interaction which generates entanglement between the atom and the paraxial field, light is scattered into other field modes and carries away information and leads to decoherence.  Keeping track of this decoherence is critical, as it sets limits on potential uses for the light-atom interface.  The effects of this decoherence can be included in a master equation description of the atom-light dynamics,	
	\begin{align} \label{Eq::StandarME}
		\frac{d \hat{\rho} }{dt} = -\frac{i}{\hbar} \big[ \hat{H}_{\rm eff} \hat{\rho} - \hat{\rho} \hat{H}_{\rm eff}\dg \big] +  \Gamma \sum_{q} \hat{W}_q \hat{\rho} \hat{W}_q\dg.
	\end{align}
This involves a non-Hermitian Hamiltonian that drives loss and a ``feeding term" that describes incoherent repopulation of the atomic ground states when a photon is scattered.  The feeding term is comprised of \emph{jump operators},
	\begin{align} \label{Eq::GenJumpOG}
		\hat{W}_q &=  \sum_{f'} \frac{ \bra{ j' }| d | \ket{ j }/\hbar }{ \Delta_{f'} + i\Gamma/2 } \left( \mathbf{e}_q^* \cdot \hat{\mathbf{D}}_{f f'} \right) \left( \hat{\mathbf{D}}_{f'f}^\dagger \cdot  \hat{\mathbf{E}}^{(+)}_{\rm fwd}(\mathbf{r}_A)    \right), 
	\end{align}
where $q$ sums over all spherical basis components.  This differs from the polarization indices of the paraxial field operator, which sum only over allowed polarizations orthogonal to the direction of propagation.  Since the quantized field operators used in this dissertation have units of $\sqrt{\text{photon flux}}$, the jump operators have a slightly different form than those found in Ref. \cite{DeuJes10}.  

The effective non-Hermitian Hamiltonian is given by the dispersive interaction, \erf{Eq::LightShiftHam}, which can be divided into coherent (Hermitian) and loss (anti-Hermitian) terms,
	\begin{align} \label{Eq::nonHHam}
		\hat{H}_{\rm eff} = \hat{H}_{\rm coh} + \hat{H}_{\rm loss}.
 	\end{align}	
The coherent portion of the effective Hamiltonian, \erf{Eq::StokesHam}, was the focus of the previous section.  The loss Hamiltonian arises from the imaginary part of atomic polarizability tensor.  Expanding  \erf{Eq::AtomicPolarizabilityF} and keeping terms to order $1/\Delta^2$,
	\begin{align} \label{Eq::AtomicPolBreakdown}
		\tensor{\alpha}(f) = \alpha_0(\Delta) \sum_{f'} \frac{ \hat{\mathbf{D}}_{f f'} \hat{\mathbf{D}}_{f' f}^\dagger}{1 + \delta_{f'}/\Delta} - i \alpha_0 (\Delta)  \frac{ \Gamma}{ 2 \Delta }  \sum_{f'} \hat{\mathbf{D}}_{f f'} \hat{\mathbf{D}}_{f' f}^\dagger . 
	\end{align}
Using the Mollow transformation, \erf{Eq::MollowTrans}, the anti-Hermitian part of the effective Hamiltonian that drives loss is	
	\begin{align} \label{Eq::LossHamiltonian}
		 \hat{H}_{\rm loss} = - i \hbar \frac{\gamma_s(\mathbf{r}_A)}{2} \sum_{f'}  \vec{\epsilon}_L^* \cdot \hat{\mathbf{D}}_{f f'} \hat{\mathbf{D}}_{f' f}^\dagger \cdot  \vec{\epsilon}_L ,
	\end{align}
where we have written it in terms of the \emph{photon scattering rate},
	\begin{align}
		\gamma_s(\mathbf{r}_A) = \frac{\Omega^2(\mathbf{r}_A) }{4 \Delta^2} \Gamma .
	\end{align}
In the same limit, to order $1/\Delta^2$, the jump operators in \erf{Eq::GenJumpOG} become,
	\begin{align} \label{Eq::GenJump}
		\hat{W}_q & =  \frac{ \Omega(\mathbf{r}_A) }{ 2 \Delta } \sum_{f'}  \left( \mathbf{e}_q^* \cdot \hat{\mathbf{D}}_{f f'} \right) \left( \hat{\mathbf{D}}_{f'f}^\dagger \cdot  \vec{\epsilon}_L    \right) .
	\end{align}	
Higher-order corrections as well as a justification of the omitted terms are presented in Appendix \ref{Appendix::MECorrections}.  

The master equation \erf{Eq::StandarME} can be divided up conveniently into the coherent and incoherent pieces,
	\begin{align} \label{Eq::MEbreakdown}
		\frac{d \hat{\rho}}{dt} =  - \frac{i}{\hbar}  [\hat{H}_{\rm coh}, \hat{\rho}] + \frac{d \hat{\rho}}{dt} \Big|_{\rm dec}.
	\end{align}
The terms that drive loss and decoherence are proportional to the rate of incoherently scattered photons,
	\begin{align} \label{Eq::MEDec}
		\frac{d \hat{\rho}}{dt} \Big|_{\rm dec} =&\,  \frac{\gamma_s(\mathbf{r}_A)}{2} \bigg\{ \sum_{f'}  \vec{\epsilon}_L^* \cdot \hat{\mathbf{D}}_{f f'} \hat{\mathbf{D}}_{f' f}^\dagger \cdot  \vec{\epsilon}_L, \hat{\rho} \bigg\}_+ \\
			& + \gamma_s(\mathbf{r}_A) \sum_q \bigg( \sum_{f'}  \mathbf{e}_q^* \cdot \hat{\mathbf{D}}_{f f'} \hat{\mathbf{D}}_{f'f}^\dagger \cdot  \vec{\epsilon}_L   \bigg) \hat{\rho} \bigg( \sum_{f'} \vec{\epsilon}^*_L  \cdot \hat{\mathbf{D}}_{f f'} \hat{\mathbf{D}}_{f'f}^\dagger \cdot \mathbf{e}_q  \bigg). \nn
	\end{align}
The anti-commutator term on the first line results from the loss Hamiltonian in \erf{Eq::LossHamiltonian}, and the second line is the feeding term using the jump operators in \erf{Eq::GenJump}.  

For a probe laser linearly polarized along $\mathbf{e}_x$, we can input the irreducible decomposition of the atomic polarizability tensor into \erf{Eq::MEDec} to find the loss Hamiltonian,
	\begin{align}
		\hat{H}_{\rm loss} = - i \hbar \frac{ \gamma_s(\mathbf{r}_A)  }{2} C^{(0)}_{j'f} \hat{I},
	\end{align}
and the feeding terms,
	\begin{align}
		\Gamma \sum_{q} \hat{W}_q\hat{\rho} \hat{W}_q^{\dagger} = \gamma_s(\mathbf{r}_A) \left( |C_{j'f}^{(0)}|^2 \hat{\rho} + |C_{j'f}^{(1)}|^2 \big( \hat{f}_z \hat{\rho} \hat{f}_z + \hat{f}_y \hat{\rho} \hat{f}_y \big) \right).
	\end{align}
This gives a master equation,
	\begin{align} \label{Eq::XPolarizedME}
		\frac{d \hat{\rho}}{dt} \Big|_{\rm dec} =  - \gamma_s(\mathbf{r}_A)  \left( C_{j'f}^{(0)} - |C_{j'f}^{(0)}|^2  \right)\hat{ \rho} + \gamma_s(\mathbf{r}_A) |C^{(1)}_{j'f}|^2  \left(\hat{f}_z\hat{ \rho} \hat{f}_z +  \hat{f}_y \rho \hat{f}_y \right).
	\end{align}
For an atom driven on an $S_{1/2} \rightarrow P_j$ transition (such as ${}^{133}$Cs) the coefficients are,
	\begin{align} \label{Eq::CsCoefficients}
		|C_{j'f}^{(0)}|^2 - C_{j'f}^{(0)} & = - \frac{2}{9} \\
		|C_{j'f}^{(1)}|^2 & = \frac{g^2_f}{9}.
	\end{align} 
Defining the quantization axis along $\mathbf{e}_x$, $\hat{f}_y = (\hat{f}_+ + \hat{f}_-)/2$ and $\hat{f}_z = (\hat{f}_+ - \hat{f}_-)/2i$, and focusing only on the incoherent terms, we arrive at the standard optical pumping equation \cite{DalCoh89},
	\begin{align}
		\frac{d \hat{\rho}}{dt} \Big|_{\rm dec} = - \gamma_s(\mathbf{r}_A) \frac{2}{9}\hat{ \rho} + \gamma_s(\mathbf{r}_A) \frac{g_f^2}{18}  \left(\hat{f}_+ \hat{\rho} \hat{f}_- +  \hat{f}_- \hat{\rho }\hat{f}_+ \right).
	\end{align}

	\chapter{Three-dimensional light-matter interface for atomic ensembles} \label{Ch::3DInterface}
%

\section{Introduction }	
	
\begin{figure}
	\centering
		\includegraphics[scale=0.8]{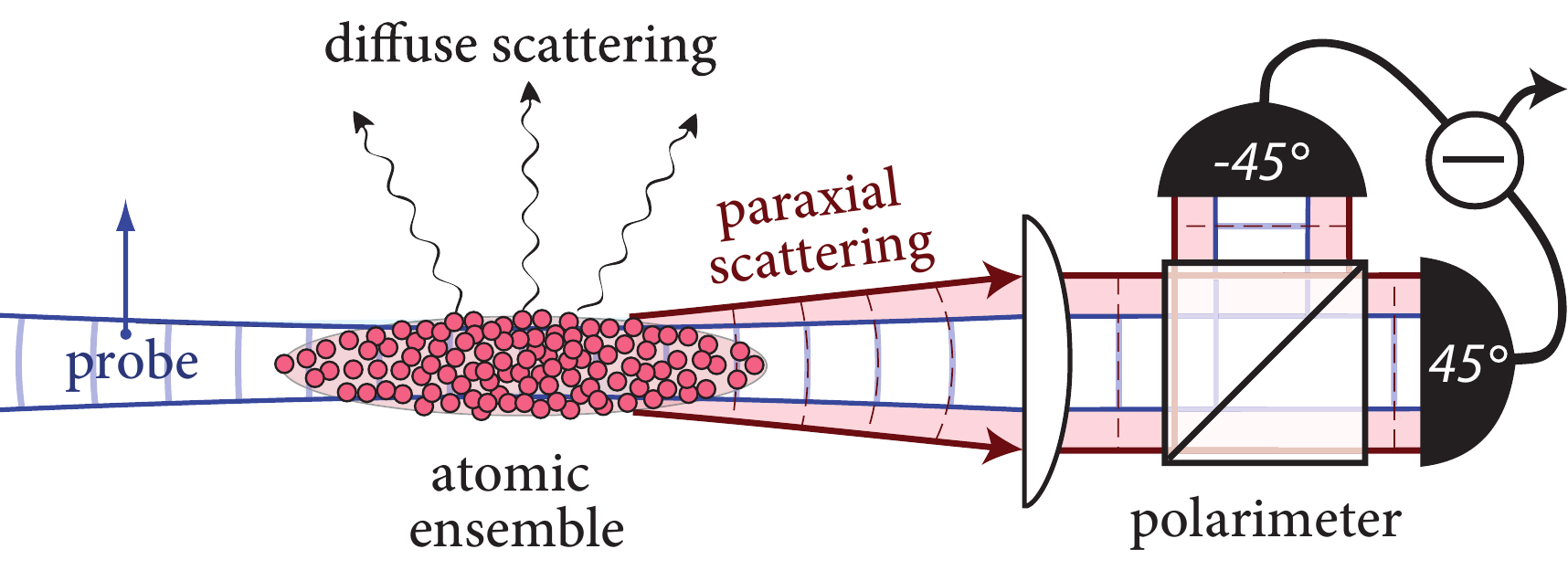} 
       		 \caption[Schematic for a paraxial laser interacting with an atomic ensemble]{ Schematic of a linearly polarized laser probe with a Gaussian spatial mode (solid blue lines entering from the left of the cloud) interacting dispersively with a cold, trapped atomic ensemble.  The light that is indistinguishably scattered by the average atomic density distribution defines the radiated paraxial mode (solid red arrows emanating from the cloud to the right).  The spatial overlap of the collectively scattered field and the probe, which determines the strength of the atom-light interface, depends highly on geometry.  Density fluctuations drive diffuse scattering out of the paraxial modes which leads to local decoherence of the ensemble.} \label{Fig::Schematic}   	
	\end{figure}

In this chapter, we present a three-dimensional model for the interaction of a paraxial probe laser with a spatially extended ensemble of spin-$f$ atoms, depicted in Fig. \ref{Fig::Schematic}.  When driven by an off-resonant laser field such that the excited state probability is small, the atoms and light interact dispersively as studied in Chapter \ref{Ch::DispersiveInt}.  In a rigorous field-theoretic analysis, S$\o$rensen and  S$\o$rensen showed that the mean-field effect of the light interacting with an atomic ensemble gives rise to an index of refraction of the gas due to the spatially-averaged local density of the atoms \cite{SorSor08}.  Diffuse scattering into 4$\pi$ arises from the random positions of the point atomic scatterers and is equivalent to local spontaneous emission.  This leads to attenuation of the incident wave and optical pumping of the atomic state, accounted for by the imaginary part of the polarizability according to the optical theorem.  
		
A number of novel features emerge from the fact that atoms experience position-dependent probe amplitudes and phases.  First, probe photons are coherently scattered into many paraxial modes, similar to the diffraction effects considered semiclassically in Ref. \cite{MulPol05}.  The resulting interaction between that atoms and and paraxial quantum field is inherently multi-mode \cite{KupPol05} and can be decomposed in a set of collective atomic \emph{spin waves}, each of which couples to a different transverse mode of the paraxial field.  Second, photons diffusely scattered out of the probe result in decoherence that dictates the limitations for any quantum information task.  This process occurs locally according to the spatially varying probe intensity experienced by individual atoms.  The relative rate at which coherent, collective evolution dominates over local decoherence depends on the resonant optical density (OD) of the ensemble,
	\begin{align} \label{Eq::ODdef}
		\mbox{OD} = N \frac{\sigma_0}{A},
	\end{align}
 a concept which we generalize for multiple spatial modes.  The OD is routinely used as a figure-of-merit for strong coupling in atomic ensembles.  The ratio of the single-atom scattering cross section on resonance, $\sigma_0 = 3 \lambda_0^2/2 \pi$, to the transverse beam area, $A$, describes the coupling strength to a single atom, which is multiplied by the total number of atoms $N$.  While the OD describes the rate of extinction of resonant light as it traverses a medium, in the far detuned, dispersive regime, the atoms are transparent to the laser, and the OD serves to characterize the strong coupling.  

After the interaction, the forward-scattered paraxial light is continuously measured with a balanced polarimeter.  We spend some time discussing this measurement using a classical model of the paraxially scattered light before including the quantum effects.  Balanced polarimetry selects only light scattered into the spatial mode of the probe.  Light in all other paraxial modes is lost, which leads to an additional, \emph{collective} form of decoherence. The stochastic master equation presented in this chapter describes all the relevant physical features including the conditional dynamics of collective atomic state from measurement backaction, collective decoherence from unmeasured paraxial light, and local decoherence from diffuse photon scattering.

\section{Multi-atom dispersive light-matter interaction }

The mean-field, spatially averaged atomic density, which plays the role of the index of refraction in the classical theory, appears as coherent radiation by a collective atomic observable in the quantum theory.  The coupling of collective atomic observables to paraxial modes thus describes the coherent atom-photon light-shift interaction, mediated by the Hermitian part of the atomic polarizability operator.  The diffuse modes, in contrast, couple to the density fluctuations in the ensemble due to the discrete atomic positions and thus act locally on each atom \cite{SorSor08}.  In the usual Born-Markov approximation, tracing over these modes leads to decoherence and is described by the anti-Hermitian part of the atomic polarizability \cite{DeuJes10}. In this section we first derive a multi-mode generalization of the tensor polarizability interaction, reviewed in Chapter \ref{Ch::DispersiveInt}, that coherently entangles the atomic ensemble and the paraxial quantum field. 

Quantization of the paraxial electromagnetic field was discussed in \srf{Sec::ParaxialQuantization}.  The paraxial field operator is decomposed into a set of transverse spatial modes $\{ u_i(\mathbf{r}_\perp, z)\}$, which are orthonormal in a longitudinal plane.  The positive-frequency component of the electric field restricted to the paraxial subspace is
	\begin{align}
		\hat{\mbf{E}}^{(+)}_{\rm fwd}(\mbf{r},t) & = \sum_{i,\lambda}  \sqrt{\frac{2 \pi \hbar \omega_0}{c A}}  \mbf{e}_\lambda \,   u_{i}(\rperp,z) \hat{a}_{i,\lambda}(z,t)   e^{i(k_0 z - \omega_0 t)},
	\end{align}
where $A$ is the quantization area.  Relevant spatial symmetries in a particular problem under consideration motivates a specific choice of modes -- Laguerre-Gauss, Hermite-Gauss, etc.

For weak excitation the atomic response is linear in the field, and the excited atomic states can be adiabatically eliminated.  The effective interaction governing the coupling of the quantized paraxial modes with the atomic ensemble is given by the non-Hermitian Hamiltonian,
	\begin{align} \label{Eq::ScatteringInteraction}
		\hat{H}_{\text{eff}} = - \sum_{n=1}^N \hat{\mathbf{E}}^{(-)}_{\rm fwd}(\mbf{r}_n, t) \cdot \hat{\tensor{\alpha}} \phantom{}^{(n)} \cdot \hat{\mathbf{E}}^{(+)}_{\rm fwd}(\mbf{r}_n, t).
	\end{align}
The index $n$ labels atoms at respective positions $\mathbf{r}_n$, each with dynamic tensor polarizability $\hat{\tensor{\alpha}}\phantom{}^{(n)}$.  Upon substitution of the irreducible components given in Eq. (\ref{Eq::IrreducibleDecomp}), we find scalar (rank-0), vector (rank-1), and tensor (rank-2) contributions to the interaction.  Defining multi-mode generalizations of the Stokes operators in \erf{Eq::StokesOps}, akin to those in Ref. \cite{SorSor08},
	\begin{subequations} \label{Eq::MultimodeStokesOps}
	\begin{align}
		 \hat{s}^{ij}_0(z,t) = & \frac{1}{2}  \left(  \hat{a}^\dagger_{i,x}(z,t)  \hat{a}_{j,x}(z,t) + \hat{a}^\dagger_{i,y}(z,t)  \hat{a}_{i,y}(z,t) \right) \\
		\hat{s}^{ij}_1(z,t) = & \frac{1}{2}\left(  \hat{a}^\dagger_{i,x}(z,t)  \hat{a}_{j,x}(z,t) -  \hat{a}^\dagger_{i,y}(z,t)  \hat{a}_{i,y}(z,t) \right) \\
		\hat{s}^{ij}_2(z,t) = & \frac{1}{2}  \left( \hat{a}^\dagger_{i,x}(z,t)  \hat{a}_{j,y}(z,t) +  \hat{a}^\dagger_{i,y}(z,t)  \hat{a}_{i,x}(z,t) \right) \\
		\hat{s}^{ij}_3(z,t) = & \frac{1}{2i} \left( \hat{a}^\dagger_{i,x}(z,t)  \hat{a}_{j,y}(z,t) -  \hat{a}^\dagger_{i,y}(z,t)  \hat{a}_{i,x}(z,t) \right),
	\end{align}
	\end{subequations}
the interaction can be written as a multi-mode version of the single atom interaction in \erf{Eq::StokesHam}:
	\begin{align} \label{Eq::MultimodeHamQuantum}
		\hat{H}_{\text{coh}} = &  \hbar \chi_0  \sum_{i,j} \sum_{n=1}^N \sum_{f'} \frac{u_i^*(\mathbf{r}_{\perp n}, z_n) u_j(\mathbf{r}_{\perp n}, z_n)}{1+\delta_{f'}/\Delta} \\
		& \times \left( \hat{s}^{ij}_0(z_n,t) \hat{A}_0^{(n)} + \hat{s}^{ij}_1(z_n,t) \hat{A}_1^{(n)} + \hat{s}^{ij}_2(z_n,t) \hat{A}_2^{(n)} + \hat{s}^{ij}_3(z_n,t) \hat{A}_3^{(n)} \right) \nn.
	\end{align}
Here, $\hat{A}_i^{(n)}$ are the components of the polarizability tensor for the $n^{th}$ atom at position $\mathbf{r}_n$ which couple to the Stokes components in \erf{Eq::MultimodeStokesOps}:
	\begin{subequations}
	\begin{align}
		\hat{A}_0^{(n)} = & 2 C^{(0)}_{j'ff'} \hat{I}^{(n)} + C^{(2)}_{j'ff'} \left( \frac{3 \hat{f}_z^{(n)2} - \hat{\mbf{f}}^{(n)} \cdot  \hat{\mbf{f}}^{(n)} }{6} \right) \\
		\hat{A}_1^{(n)} = & C^{(2)}_{j'ff'} \left( \frac{\hat{f}_x^{(n)2} - \hat{f}_y^{(n)2} }{2} \right) \\
		\hat{A}_2^{(n)} = & C^{(2)}_{j'ff'} \left( \frac{\hat{f}_x^{(n)} \hat{f}_y^{(n)} + \hat{f}_y^{(n)} \hat{f}_x^{(n)} }{2} \right) \\
		\hat{A}_3^{(n)} = & -C^{(1)}_{j'ff'} \hat{f}_z^{(n)}.
	\end{align}
	\end{subequations}
The interaction has been written in terms of the dimensionless coupling strength $\chi_0$, given in \erf{Eq::FaradayAngle}.  


	\subsection{Coherent driving field}
	
	The multi-mode light-matter interaction can be useful in the fully quantum form, \erf{Eq::MultimodeHamQuantum}, for instance when studying input fields at the single photon level.  For our purposes, we wish to model the interaction with a driving laser by promoting the appropriate paraxial modes using the Mollow transformation.  We take the spatial mode of the laser as the fundamental mode with label $i = 0$.  The laser also has a well-defined polarization, and thus only one of its polarization components is promoted to a classical field.  This is important because we consider scattering of probe photons into orthogonal photons in the same spatial mode.  As it will be useful in the next chapter, we consider a probe with linear polarization along $\mathbf{e}_x$.  We make the Mollow transformation, equivalent to \erf{Eq::MollowTrans},
	\begin{align} \label{Eq::MultimodeMollow}
		\hat{a}_{0,x}(z,t) \rightarrow \sqrt{\dot{N}_L} + \hat{a}_{0,x}(z,t),
	\end{align}
where the average photon flux of the laser is related to its peak intensity $I_0$ via the relation $\dot{N}_L = A I_0 / \hbar \omega_0 $.  This gives a classical coherent complex amplitude for the laser,
	\begin{align} \label{Eq::ClassicalXField}
		\vec{ \mathcal{E}}^{(+)}_L(\mathbf{r},t) = \mathcal{E}^{(+)}_0 u_0(\mathbf{r}_\perp, z, t) e^{i(k_0 z - \omega_0 t)} \mathbf{e}_x,
	\end{align}
with positive-frequency electric field amplitude
	\begin{align}
		\mathcal{E}^{(+)}_0  = \sqrt{ \frac{ 2 \pi \hbar \omega \dot{N}_L }{Ac} }.
	\end{align}
When the description of the probe laser requires more than one mode, the generalization is straightforward using \erf{Eq::MultimodeMollow} for the photon flux in each mode.  
	
	We now rewrite the multi-mode interaction Hamiltonian with the transformation, \erf{Eq::MultimodeMollow}, neglecting all higher order terms that describe scattering between modes initially in vacuum.  The resulting linearized interaction is
	\begin{eqnarray} \label{Eq::MultimodeHamQuantum2}
		\hat{H}_{\rm coh}  &=&  \hbar \frac{ \chi_0 \sqrt{\dot{N}_L } }{2}  \sum_{n=1}^N \sum_{f'}  \sum_{i} \Bigg\{  \frac{ u_0^*(\mathbf{r}_{\perp n}, z_n) u_i(\mathbf{r}_{\perp n}, z_n) }{1 + \delta_{f'}/ \Delta}  \\
	& \quad & \times \bigg[ i  C^{(1)}_{j'ff'} \hat{f}_z^{(n)} + \smallfrac{1}{2} C_{j'ff'}^{(2)} \Big( \hat{f}_x^{(n)} \hat{f}_y^{(n)} + \hat{f}_y^{(n)} \hat{f}_x^{(n)} \Big) \bigg] \hat{a}_{i,y}(z_n,t)  + \rm{ H.c.} \Bigg\} \nn \\
	& + & \hbar  \frac{ \chi_0 \sqrt{\dot{N}_L } }{2}  \sum_{n=1}^N \sum_{f'}  \sum_{i} \Bigg\{  \frac{ u_0^*(\mathbf{r}_{\perp n}, z_n) u_i(\mathbf{r}_{\perp n}, z_n) }{1 + \delta_{f'}/ \Delta} \nn  \\
	& \quad & \times \bigg[ C^{(0)}_{j'ff'} \hat{I}^{(n)} + C_{j'ff'}^{(2)} \Big( \hat{f}_x^{(n)2}- \smallfrac{1}{3} \hat{\mbf{f}}^{(n)} \cdot \hat{\mbf{f}}^{(n)} \Big)  \bigg] \hat{a}_{i,x}(z_n,t)  + \rm{ H.c.} \Bigg\} \nn \\
	& + & \hbar \frac{ \chi_0 \dot{N}_L }{2}  \sum_{n=1}^N \sum_{f'}  \frac{ | u_0(\mathbf{r}_{\perp n}, z_n) |^2 }{1 + \delta_{f'}/ \Delta} \bigg[ C^{(0)}_{j'ff'} \hat{I}^{(n)} + C_{j'ff'}^{(2)} \Big( \hat{f}_x^{(n)2}- \smallfrac{1}{3} \hat{\mbf{f}}^{(n)} \cdot \hat{\mbf{f}}^{(n)} \Big) \bigg]. \nonumber 
	\end{eqnarray}  
This expression is quite a beast at the moment, but we will trim it down to useful size and form.  For now, let us examine the physical interpretation of this interaction.  The terms in the first two lines describe scattering of $x$-polarized probe photons into $y$-polarized photons in all spatial modes mediated by the vector and tensor components of the atomic polarizability.  The most useful for our purposes is the effective Faraday interaction from the rank-1 vector part of the polarizability.  The terms in the third and fourth lines describe scattering of $x$-polarized probe photons back into $x$-polarized photons in all modes mediated by the scalar and tensor components of the atomic polarizability.  The terms in the final line describe the dynamics induced on the internal states of the atoms due to the classical electric field, which do not generate atom-light entanglement.

\section{Isolating the multi-mode Faraday interaction}

	The linearized interaction in \erf{Eq::MultimodeHamQuantum2} that coherently entangles the paraxial field and the atomic ensemble is not only unwieldy, but contains rank-2 tensor terms which spoil the multi-mode Faraday interaction we hope to isolate.  In a single mode, the Faraday interaction rotates the linear polarization of the probe by an amount proportional to the collective atomic magnetization.  In the spatial mode of the probe, the interaction is analogous, and we find rotation of the macroscopic polarization mediated by the collective magnetization of the atoms that couple to its spatial mode.  An additional effect involves the scattering of probe photons into other spatial modes.  Since these modes are initially in vacuum, this cannot be thought of as a rotation of the polarization in these modes.  In the next chapter, we will analyze mode matching to maximize the portion of the scattered light into the spatial mode of the probe.  For the moment, we simply want to isolate the multi-mode Faraday interaction such that the input quadrature of light and the collective angular momentum of the atomic cloud are QND observables.  

	\subsection{Averaging out the tensor light shift dynamics} \label{Sec::BFieldAveraging}
	
In some cases, the deleterious effects of the rank-2 component of the tensor polarizability on the QND measurement can be removed via dynamical decoupling~\cite{KosMit10}.  Here we take a different approach.  Application of a uniform bias magnetic field along the $z$-direction not only sets the quantization axis along the direction of propagation, but will also benefit by helping to isolate the Faraday interaction.  The Hamiltonian for a magnetic field of strength $B$ is
	\begin{align}
		\hat{H}_B = \hbar \Omega_B \sum_{n=1}^N \hat{f}_z^{(n)},
	\end{align}
with Larmor frequency $\Omega_B = g_F \mu_B B$.  In a frame rotating at the Larmor frequency, the effective Hamiltonian -- both the Hermitian and anti-Hermitian components -- and the jump operators are transformed,
	\begin{align}
		\hat{H}_{\rm eff} \rightarrow \exp \bigg( i \Omega_0 t \sum_n \hat{f}_z^{(n)} \bigg) \hat{H}_{\rm eff} \exp \bigg( -i \Omega_0 t \sum_n \hat{f}_z^{(n)} \bigg).
	\end{align}	
Under this transformation, the hyperfine spin operators for each atom become
	\begin{subequations} \label{Eq::BiasTransformations}
	\begin{align} 
		\hat{f}_x & \rightarrow \hat{f}_x  \cos \Omega_B t +  \hat{f}_y \sin\Omega_B t, \\
		\hat{f}_y & \rightarrow -\hat{f}_x  \sin \Omega_B t +  \hat{f}_y \cos \Omega_B t, \\
		\hat{f}_z & \rightarrow \hat{f}_z.
	\end{align}
	\end{subequations}
The second-order spin operators which couple to the Stokes components in \erf{Eq::MultimodeHamQuantum} (and linearly to the field operators in \erf{Eq::MultimodeHamQuantum2}) similarly transform.
When the bias field is strong compared to the rank-2 component of the interaction, i.e. $\Omega_0 \gg C^{(2)}_{j'ff'} \chi_0 (\delta_{F'}/\Delta)$, then the rotating wave approximation can be made and the rapidly oscillating terms can be ignored, as their effects average out.  This is equivalent to making the substitutions $\sin^2 \Omega_B t = \cos^2 \Omega_B t \rightarrow \frac{1}{2}$ and $\sin \Omega_B t \cos \Omega_B t = 0$.  Then we see that the cycle-averaged operators become, 
	\begin{subequations} \label{Eq::BiasTransformations}
	\begin{align}
		\hat{f}_x^2 & \rightarrow  \frac{1}{2} \left( \hat{\mathbf{f}} \cdot \hat{\mathbf{f}}   -  \hat{f}_z^2 \right) \\
		\hat{f}_y^2 & \rightarrow  \frac{1}{2} \left( \hat{\mathbf{f}} \cdot \hat{\mathbf{f}}   -  \hat{f}_z^2 \right) \\
		 \hat{f}_x\hat{f}_y + \hat{f}_y\hat{f}_x  & \rightarrow 0.
	\end{align}
	\end{subequations}
The result is that the coupling to the $\hat{s}^{ij}_1(z,t)$ and $\hat{s}^{ij}_2(z,t)$ Stokes components averages to zero, i.e. $\hat{A}_1^{(n)} \rightarrow 0$ and $\hat{A}_2^{(n)} \rightarrow  0.$  Thus, the induced birefringence on the probe does not spoil the desired Faraday interaction.  

The third and fourth lines of \erf{Eq::MultimodeHamQuantum2} must now be understood.  Since the rank-2 tensor coefficients sum to zero, [\erf{Eq::Rank2Sum}], the only nonvanishing terms that generate nontrivial atom-light entanglement by scattering $x$-polarized photons are of order in $1/\Delta^{2}$.  But, unlike the unentangling residual light shift, \erf{Eq::ResidualLightShift}, they are proportional only to the square root of the photon flux.  Finally, in the spatial mode of the probe, these fluctuations are small compared to the laser amplitude.  For these reasons, they will be neglected. 

The result is that \erf{Eq::MultimodeHamQuantum2} becomes
	\begin{align} \label{Eq::HCoh}
		\hat{H}_{\rm coh}  =  \hat{H}_{\rm int} + \hat{H}_{\rm LS},
	\end{align}	
where the coherent interaction is decomposed into the multi-mode Faraday Hamiltonian,
	\begin{equation} \label{Eq::HamAlmostFaraday}
		\hat{H}_{\rm int} = -\hbar \frac{ \chi^{(1)} \sqrt{ \dot{N}_L } }{2}  \sum_{n=1}^N \sum_{i} \Big[ i u_0^*(\mathbf{r}_{\perp n}, z_n) u_i(\mathbf{r}_{\perp n}, z_n) \hat{a}_{i,y}(z_n,t)  + {\rm H.c.} \Big] \hat{f}_z^{(n)} ,  
	\end{equation}
and a remaining light shift that acts on each atom's internal hyperfine spin, 
	\begin{align} \label{Eq::ResidualLightShift}
		\hat{H}_{\rm LS} =  \hbar \frac{  \chi^{(2)}  \dot{N}_L   }{2} \sum_{n=1}^N | u_0(\mathbf{r}_{\perp n}, z_n) |^2  \Big(  \hat{f}_z^{(n)2}- \smallfrac{1}{3} \hat{\mbf{f}}^{(n)} \cdot \hat{\mbf{f}}^{(n)}  \Big) .  
	\end{align}  
We have defined parameters which group together the dimensionless constant $\chi_0$ in \erf{Eq::FaradayAngle} with the remaining hyperfine excited state detunings and the tensor coefficients,
	\begin{align} \label{Eq::NewChi}
		\chi^{(i)} \equiv \chi_0 \sum_{f'} \frac{ C_{j' f f'}^{(i)} }{1 + \delta_{f'}/\Delta}.
	\end{align} 
Technically, these coefficients should be inside the sum over atom indices $n$, but the since the detuning across the ensemble is identical, it can be treated as a constant and factored out.

	\subsection{Longitudinal coarse graining and collective spin waves}

The Hamiltonian in \erf{Eq::HamAlmostFaraday} describes a multi-atom interaction that scatters $x$-polarized probe photons into $y$-polarized photons.  However, its form is not yet that of the traditional Faraday Hamiltonian, nor is it particularly obvious how to identify the collective atomic observables which couple to the scattered light.  Here, we massage the interaction until it yields a description that is easily recognized as the multi-mode extension of the standard Faraday Hamiltonian.  
	
We begin by noting that \erf{Eq::HamAlmostFaraday} describes the collective coupling of all atoms located at longitudinal plane to the field operators at that plane.  We coarse grain the interaction using longitudinal slices of thickness $\delta z$ at longitudinal coordinate $z_k$, where $k$ indexes the slices.  At each longitudinal slice $z_k$ we perform the sum over atoms within that slice and define $z$-local, collective spin-wave operators,
	\begin{align} \label{Eq::LocalSpinWave}
		\hat{F}_z^{i}(z_k) \equiv  \sum_{n_k=1}^{N_k} \beta_i(\mathbf{r}_{\perp n_k}, z_{n_k}) \hat{f}_z^{(n)}.
	\end{align}
The subscript $k$ in the sum indicates that only atoms in longitudinal slice $z_k$ are included.  Each local spin wave includes contributions from every atom in longitudinal plane $z_k$ with a spatial weighting $\beta_{i}(\mbf{r}_\perp, z)$ that depends upon the atom's position relative to the transverse mode functions,
	\begin{align} \label{Eq::Beta}
		\beta_{i}(\mbf{r}_\perp, z)  \equiv u^*_{i}(\mbf{r}_\perp, z) u_{0}(\mbf{r}_\perp, z). 
	\end{align}
To reiterate, each of the operators defined in \erf{Eq::LocalSpinWave} is the collective spin wave at longitudinal plane $z_k$ that couples to the field in transverse mode $i$ at that plane.  Using the definitions in \erf{Eq::LocalSpinWave} and \erf{Eq::Beta}, the Hamiltonian in \erf{Eq::HamAlmostFaraday} can be written, 
	\begin{align} \label{Eq::HamAlmostFaraday2}
		\hat{H}_{\rm int} = i \hbar \frac{ \Cstrength  }{2}\sum_{k} \hat{F}_z^{i\dagger}(z_k) \hat{a}_{i,y}(z_n,t)  - \hat{F}_z^{i}(z_k) \hat{a}_{i,y}\dg(z_n,t),
	\end{align}
where $k$ sums over longitudinal slices.  The interaction has been expressed in terms of the \emph{measurement strength},
	\begin{align} \label{Eq::kappa}
		\kappa \equiv \big| \chi^{(1)} \big|^2 \dot{N}_L,  
	\end{align}
with $ \chi^{(1)}$ given in \erf{Eq::NewChi}.

As we have linearized the interaction around the classical mean of the input probe, we can describe the $y$-polarized quantum field in terms of quadratures for each mode,
	\begin{subequations}\label{Eq::LocalQuadatures}
	\begin{align} 
		\hat{X}_{i}(z, t) \equiv&  \frac{1}{\sqrt{2 }} \left( \hat{a}_{i,y}(z, t) + \hat{a}_{i,y}^\dagger(z, t) \right) \\
		\hat{P}_{i}(z, t) \equiv & \frac{1}{i \sqrt{2 } } \left( \hat{a}_{i,y}(z, t) - \hat{a}_{i,y}^\dagger(z, t) \right),
	\end{align}
	\end{subequations}
and from \erf{Eq::ParaxialFieldCommutator} they satisfy the commutation relation,
	\begin{align}
		\big[ \hat{X}_{i}(z, t), \hat{P}_{j}(z', t') \big] = i  \delta_{i,j} \delta(t-t' - (z-z')/c) .
	\end{align}
Finally, we divide the z-local spin waves into real and imaginary parts,
	\begin{align} \label{Eq::FReIm}
		\hat{F}_z^i(z_k) = \mbox{Re} \big\{ \hat{F}_{z}^i(z_k) \big\} + i \, \mbox{Im} \big\{ \hat{F}_{z}^i(z_k) \big\} ,
	\end{align}
that arise from the spatial coefficients, \erf{Eq::Beta}.  The multi-mode Faraday Hamiltonian can now be written in the pleasing form, 
	\begin{equation} \label{Eq::MultiModeFARADAY}
		\hat{H}_{\rm int} =  \hbar\sqrt{ \frac{ \kappa }{2} }  \sum_k \sum_i \left(  \mbox{Im} \big\{ \hat{F}_{z}^i (z_k) \big\} \hat{X}_{i}(z_k,t) -\mbox{Re} \big\{ \hat{F}_{i}^z(z_k) \big\} \hat{P}_{i}(z_k,t) \right) . 
	\end{equation} 
where $\hat{X}_{i}(z_k,t)$ and $\hat{P}_{i}(z_k,t)$ are coarse-grained, $z$-local quadratures.

	\subsection{Heisenberg-picture dynamics} \label{Sec::HeisenbergDynamics}
	
The Heisenberg equation of motion for a $y$-polarized traveling wave mode interacting with the atomic media in the presence of the probe field follows from \erf{Eq::HamAlmostFaraday2}:
	\begin{equation} \label{Eq::aoutDiffEq}
		\left(\frac{\partial}{\partial t} +c \frac{\partial}{\partial z}\right) \! \hat{a}_{i,y}(z,t) = -\frac{\Cstrength}{2} \sum_{n=1}^N \beta_{i} (\mbf{r}_{\perp n}, z_n) \hat{f}_z^{(n)}  (t) \delta(z-z_n).
	\end{equation} 
The formal solution in the Born approximation is
	\begin{equation} \label{Eq::aoutSolution}
		\hat{a}_{i,y}(z,t) = \, \hat{a}_{i,y}(0,t-z/c) - \frac{\Cstrength}{2} \sum_{n=1}^N \beta_{i} (\mbf{r}_{\perp n}, z_n) \hat{f}_z^{(n)} (t-(z-z_n)/c) \Theta(z-z_n),
	\end{equation}
where $ \Theta(z)$ is the Heaviside step function.  This solution can be verified by plugging \erf{Eq::aoutSolution} into \erf{Eq::aoutDiffEq}.  The time derivative of any quantity proportional to the retarded time, $\tau = t - z/c$, vanishes, and the $z$-derivative of the Heaviside function yields the spatial $\delta$-function.

Neglecting the time it takes light to propagate across the sample, $(z-z_n)/c \rightarrow z/c$, the mode amplitude at a detector plane in the far field, $z_D$, is
	\begin{equation} \label{Eq::OutputMode}
		\hat{a}_{i,y}(z_D,t) =\hat{a}_{i,y}(0,t-z_D/c) - \frac{\Cstrength}{2} \hat{F}_z^{i}(t - z_D/c) . 
	\end{equation} 
The first term describes the vacuum fluctuations of the free, paraxial quantum field and the second term is the ``source term" that arises from scattering off the atoms.  The collective atomic spin wave that couples to the traveling paraxial mode $i$ is the sum over the $z$-local spin waves given in \erf{Eq::LocalSpinWave},
	\begin{align} \label{Eq::SpinWave}
		\hat{F}_z^{i} & \equiv \sum_k \hat{F}_z^{i}(z_k) = \sum_{n=1}^N \beta_{i} (\mbf{r}_{\perp n}, z_n) \hat{f}_z^{(n)},
	\end{align}
As the light travels through the ensemble, it becomes locally entangled with the spin waves at every longitudinal plane, which in the far field are indistinguishable.  

From the Heisenberg equation for the output modes, \erf{Eq::OutputMode}, and the definitions in \erf{Eq::LocalQuadatures}, the output quadratures at the detector plane are,
	\begin{subequations} \label{Eq::ScatQuadratures}
	\begin{align} 
		\hat{X}_{i}(z_D,t) = &\hat{X}_{i,y}(0,t-z_D/c) - \sqrt{\frac{ \CstrengthSq}{2} } \mbox{Re} \big\{ \hat{F}_z^{i}(t-z_D/c) \big\} , \label{Eq::QuantumXOut} \\
		\hat{P}_{i}(z_D,t) = &\hat{P}_{i,y}(0,t-z_D/c) -  \sqrt{\frac{ \CstrengthSq}{2} } \mbox{Im} \big\{ \hat{F}_z^{i}(t-z_D/c) \big\} . 
	\end{align}
	\end{subequations}
In those modes for which the spin wave is purely real, such as the spatial mode of the probe, the $\hat{P}_{i,y}$ quadrature undergoes no Hamiltonian evolution and is a constant of motion.  

When considering measurements of the output quadratures, the appearance of the retarded time in the spin wave is a consequence of the detector being spatially separated from the atomic cloud.  In a laboratory experiment, a typical separation of 30 cm leads to a time delay of $z_D/c \approx$ 1 ns, which is much slower than the collective or local spin dynamics. Thus, we can reasonably conclude that the replacement $ t - z/c \rightarrow t$ can be made without consequence.  An alternative is to keep track of the fact that measurements at time $t$ are correlated with spin dynamics at a previous time. This could matter, for instance, if one were interested in doing feedback control for rapid spin dynamics.  

Here we note a connection to the formalism of input-output theory presented in \srf{} and used in Chapter \ref{Ch::NPhotonME}.  It is precisely the first-Born-approximation scattering solutions for the field operators [\erf{Eq::aoutSolution}] and quadratures [\erf{Eq::ScatQuadratures}] that one refers to as the \emph{output} with the free-field solution as the \emph{input}.  Translated to the language of input-output theory, \erf{Eq::aoutSolution} becomes,
	\begin{align} \label{Eq::aoutInputOutput}
		\hat{a}_{i,y}^{\rm out}(t) = \, &\hat{a}_{i,y}^{\rm in}(t) - \frac{\Cstrength}{2} \hat{F}_z^i(t).
	\end{align}

The $z$-component of the spin waves are themselves constants of motion since they commute with the full Hamiltonian,
	\begin{align} \label{Eq::FzCommutes}
		[ \hat{F}_z^{i}, \hat{H}_{\rm int} + \hat{H}_{\rm LS} ] = 0,
	\end{align}	
and thus are QND observables.  The $x$- and $y$-components of the $z$-local spin waves, defined 
	\begin{align} \label{Eq::OrthogonalSpinWaves}
		\hat{F}_x^{i} & \equiv \sum_{n=1}^N \beta_{i} (\mbf{r}_{\perp n}, z_n) \hat{f}_x^{(n)} = \sum_{k} \hat{F}_x^{i}(z_k) \\
		\hat{F}_y^{i} & \equiv \sum_{n=1}^N \beta_{i} (\mbf{r}_{\perp n}, z_n) \hat{f}_y^{(n)} = \sum_{k} \hat{F}_y^{i}(z_k),
	\end{align}
commute with neither the Faraday Hamiltonian [\erf{Eq::MultiModeFARADAY}] nor the residual light-shift [\erf{Eq::ResidualLightShift}].  The $z$-local, coarse-grained spin-wave components, $\hat{F}_x^{i}(z_k)$ and $\hat{F}_y^{i}(z_k)$, are defined just as in \erf{Eq::LocalSpinWave}. Thus, each will experience both collective, entangling dynamics and local dynamics due to light shifts on the spin of each atom.

\section{Stochastic master equation for continuous balanced polarimetry measurements} \label{Sec::SME}

The Faraday Hamiltonian, \erf{Eq::MultiModeFARADAY}, is an entangling interaction between the atomic spin waves and the paraxial quadratures of the field.  When the light is measured with a polarimeter, quantum backaction leads to stochastic evolution of the atomic state, conditioned on the random measurement results.  A complete description of the dynamics can be described by a stochastic master equation (SME), which accounts for this stochastic evolution of the collective atomic state.  If all of the scattered light is measured, then all of the information about the atoms imprinted on the field is recovered, and the SME preserves purity.  As discussed in the previous section, diffuse scattering propels information into nonparaxial modes which, never impinging on a detector, give rise to decoherence that is local at the level of individual atoms.  This is because, in principle, this scattering is distinguishable.  With a fine enough microscope, one could determine which atom emitted a diffusely scattered photon.  Additionally, we show in this section that in a multi-mode description, collectively scattered light -- that is, paraxial light scattered indistinguishably from the ensemble at large -- can lead to further decoherence that acts collectively on the ensemble.  

While the description of the measurement process requires a fully quantum treatment to understand the role of quantum fluctuations, much can be understood using a semiclassical treatment.  The mean polarimetry signal can be described with classical fields, and from this description we can identify the collective atomic spin wave that couples to the measurement.  We begin with a classical description of balanced polarimetry, move to a semiclassical description where the tensor susceptibility arises from alkali atoms with internal hyperfine structure, and then complete the description with fully quantum fields, which will be required for the derivation of the SME and ultimately for the generation of spin squeezing.

	\subsection{Balanced polarimetry} \label{Sec::BalancedPolarimetry}
		
In balanced polarimetry, the output light, composed of a superposition of input and paraxially scattered fields, is collected and fed into a polarizing beam splitter (PBS) oriented at $45^\circ$ to the input linear polarization of the probe.  The light exiting each output port of the PBS is sent to its own detector, and the signals are subtracted.  This performs an effective homodyne detection of the light scattered into the orthogonal polarization, in the spatial mode of the probe.  Some details are discussed here, more information about continuous polarimetry can be found in Refs. \cite{SmiJes03, StocktonThesis, DeuJes10}, among others.   
	
	For an $x$-polarized probe, the PBS acts as a transformation into the diagonal/anti-diagonal polarization basis, $\{ \mathbf{e}_d$, $\mathbf{e}_{\bar{d}}\}$,
	\begin{subequations}
	\begin{align}
		\mathbf{e}_x &= \frac{1}{\sqrt{2}} ( \mathbf{e}_d -  \mathbf{e}_{\bar{d}}  ) \\
		\mathbf{e}_y &= \frac{1}{\sqrt{2}} ( \mathbf{e}_d +  \mathbf{e}_{\bar{d}}  ). 
	\end{align}
	\end{subequations}
At a single point on each of the two detectors, the intensity is proportional to the square of the field amplitude:		
	\begin{subequations} \label{Eq::DiagBasisIntensity}
	\begin{align}  
		I_d(\mathbf{r}_{\perp}, z_{D_1}) & \propto  |\mathcal{E}_{ \text{out,} x}|^2 + |\mathcal{E}_{ \text{out,} y}|^2 + \mathcal{E}_{ \text{out,} x}^* \mathcal{E}_{ \text{out,} y} + \mathcal{E}_{ \text{out,} x} \mathcal{E}_{ \text{out,} y}^*  \\
		I_{\bar{d}}(\mathbf{r}_{\perp}, z_{D_2}) & \propto |\mathcal{E}_{ \text{out,} x}|^2 + |\mathcal{E}_{ \text{out,} y}|^2 - \mathcal{E}_{ \text{out,} x}^* \mathcal{E}_{ \text{out,} y} -  \mathcal{E}_{ \text{out,} y}^* \mathcal{E}_{ \text{out,} x}. 
	\end{align}
	\end{subequations}
The measurement current from each detector results from integration of the intensity over the detector surface.  When all of the paraxial light is collected, this isolates the spatial mode overlap of the $x$- and $y$-polarized output fields in the interference terms in \erf{Eq::DiagBasisIntensity}.  Intuitively, this makes sense as one can only see interference for indistinguishable processes, in this context meaning that only light in the same spatial mode interferes.  

Each of the measurement currents contains information about the scattered light, and by transitive property about the atoms.  However, the signal of interest is hiding beneath the dominant term that arises from the intensity of the probe at each detector.  The idea behind balanced polarimetry is to isolate the interference terms by subtracting the signals from the two detectors,
	\begin{align}  \label{Eq::HomodyneSignal}
		\mathcal{M} & \propto \int_{D_1} d^2 \mathbf{r}_\perp I_d (\mathbf{r}_\perp, z_{D}) - \int_{D_2} d^2 \mathbf{r}_\perp I_{ \bar{d} } (\mathbf{r}_\perp, z_{D} ) \nonumber \\
		    & \propto \int_{D} d^2 \mathbf{r}_\perp \text{Re} \left\{\mathcal{E}_{ \text{out,} x}^*(\mathbf{r}_\perp, z_{D}) \, \mathcal{E}_{ \text{out,} y}(\mathbf{r}_\perp, z_{D}) \right\}.
	\end{align}	
 It is balanced because the orientation of the PBS is chosen such that, in the absence of any atoms, an equal amount of probe light falls on each detector, and thus the subtracted measurement current has zero mean and is dominated solely by shot-noise fluctuations, which only appear in a quantum treatment.   

Through interaction with the atoms, light will be scattered into both the $x$ and $y$ polarizations.  The amplitude of the scattered $x$-polarized light is much smaller than that of the incident probe and can be ignored.  In this case, the $x$-polarized output field at the detector can be replaced with the input, probe field, $\mathcal{E}_{ \text{out,} x}(\mathbf{r}_\perp, z_{D}) \rightarrow \mathcal{E}_{L}(\mathbf{r}_\perp, z_{D})$.  The $y$-polarized field is composed entirely of scattered light:  $\mathcal{E}_{ \text{out,} y}(\mathbf{r}_\perp, z_{D}) \rightarrow \mathcal{E}_{ \text{scat,} y}(\mathbf{r}_\perp, z_{D})$.  We can then write the polarimetry signal as
	\begin{align}  \label{Eq::HomodyneSignal1}
		\mathcal{M} & \propto \int_{D} d^2 \mathbf{r}_\perp \text{Re} \big\{ \mathcal{E}_L^*(\mathbf{r}_\perp, z_{D}) \mathcal{E}_{ \text{scat,} y}(\mathbf{r}_\perp, z_{D}) \big\}. 
	\end{align}	
The transverse integration isolates only the portion of the $y$-polarized, scattered field that is in the spatial mode of the probe.  This is seen very clearly if one uses a mode decomposition of the paraxial field.  For an input probe prepared in spatial mode $u_0(\mathbf{r}_\perp, z)$, as in \erf{Eq::ClassicalXField}, the orthogonality of the spatial modes explicitly shows that the measurement current is proportional to light scattered back into mode $u_0(\mathbf{r}_\perp, z)$.

Using $\lambda/2$ and $\lambda/4$ waveplates along with a PBS, one can measure in any polarization basis and can in principle measure any Stokes component.  That is, using probe light prepared with arbitrary polarization, $\vec{\epsilon}_L \propto \mathbf{e}_x + e^{i \phi} \mathbf{e}_y$, one can always engineer a balanced polarimetry measurement to detect scattered, orthogonally polarized photons using the probe as a local oscillator.  Since these polarization component couple differently to atomic operators, this could be an avenue to study interesting light-matter interactions and generate non-classical collective spin states.

	\subsubsection{Classical description of paraxially scattered fields and measurement} \label{Sec::ClassScatParaxF}
	
Consider an incident quasi-monochromatic, paraxial laser beam with frequency $\omega_0$ and complex amplitude, $\mbf{E}_L(\mbf{r}_\perp, z) = \vec{\mathcal{E}}_L (\rperp , z) e^{ik_0 z}$.  The mean incident field is described by the electric field envelope $\vec{\mathcal{E}}_L (\rperp , z,t)= \mathcal{A}(t-z/c) \vec{\mathcal{U}}_{\rm in} (\rperp, z)$, where $\mathcal{A}(t)$ is the temporal pulse envelope and $\vec{\mathcal{U}}_{\rm in} (\rperp, z)$ is the spatial envelope of the laser.  The laser scatters off of an dielectric medium with a tensor susceptibility, $\tensor{\chi}(\rperp, z)$.  At the level of the mean scattered electric field, an atomic ensemble can be treated as a smooth, spatially extended dielectric scatterer.  The total spatial field $\vec{\mathcal{U}}(\rperp , z)$ satisfies the paraxial wave equation, \erf{paraxEq}, whose solution describes the coherent interference of the input and scattered fields.  In the first Born approximation, i.e. for dilute samples where multiple scattering is negligible, the total field is given in \erf{Eq::ClassicalScattering} as
	\begin{equation} \label{Eq::ClassicalVecScattering}
		\vec{\mathcal{U}}(\rperp , z) =  \vec{\mathcal{U}}_{\rm in} (\rperp,z) + i 2 \pi k_0  \int_{-\infty}^z  dz'   \int  d^2 \rperp' K(\rperp  - \rperp', z-z')  \tensor{\chi}( \rperp', z')  \cdot \vec{\mathcal{U}}_{\rm in} (\rperp',z') , 
	\end{equation}
where $K(\rperp-\rperp', z-z')$ is the paraxial propagator, \erf{Eq::ParaxialPropagator}.  

The solution for a paraxial field scattered from a single point dipole at position $\mbf{r}'$ is found by setting $\tensor{\chi}(\mbf{r}) = \tensor{\alpha} \, \delta^{(3)}(\mbf{r}-\mbf{r}') $. However, in such a situation much of the scattered field is into nonparaxial directions, as in \frf{Fig::ModeMatching}, and the use of the paraxial propagator will fail to describe the off-axis field.  Roughly, the degree to which a finite dielectric scatterer radiates paraxially is determined by its Fresnel number $\mathcal{F} = \sigma_\perp^2/\lambda_0 \sigma_z$, where $\sigma_\perp$  and $\sigma_z$ are the transverse and longitudinal extents of the dielectric, respectively. For transversely extended clouds where $\mathcal{F} \gg 1$, scattered light is not only paraxial but is reradiated perfectly back into the spatial mode of the probe.  In this case the light-matter interface is well described as single-mode \cite{SorSor08}.  However, a dielectric with a small Fresnel number can also scatter predominantly paraxial light if the ratio of longitudinal extent to the wavelength of light, $\sigma_z/\lambda_0$ is large.  A wave incident on a dielectric scatterer that extends over many wavelengths, $\sigma_z/\lambda_0 \gg 1$, creates a phased array of induced dipoles that scatter preferentially into the forward direction\footnote{This concept is used for single-photon input to create and detect so-called timed Dicke states \cite{ScuWod06}.}.  
Indeed, for the longitudinally extended atomic clouds often used in experiments with dipole trapped atoms, the scattering can be overwhelmingly paraxial even with a Fresnel number much less than one\footnote{Consider the experimental set up in Ref. \cite{KosMit11}. A cloud of $10^6$ $^{87}$Rb atoms with dimensions $\sigma_\perp = 20$ $\mu$m and $\sigma_z = 3000$ $\mu$m is probed on the D2-line with light of wavelength  $\lambda_0 = 780$ nm.  The Fresnel number for this situation is $\mathcal{F} \approx 1.7 \times 10^{-4}$.}.  However, as the scattering is not entirely into spatial mode of the probe \cite{SorSor08}, the first step is understanding how geometry of the probe and ensemble affects the mode-matching.  Henceforth, we consider geometries where the scattering is entirely paraxial.  
	
We wish to isolate the Faraday effect that arises from the scattering of linearly polarized input light into the orthogonal linear polarization, governed by the off-diagonal element of the dielectric susceptibility matrix, $\chi_{yx} = \mathbf{e}_y \cdot  \tensor{\chi} \cdot \mathbf{e}_x$.  The diagonal matrix elements of the average susceptibility (in the basis of the laser polarization) describe the amount of scattering from one polarization back to the same polarization.  The real part of the diagonal elements gives rise to an index of refraction and the imaginary part describes attenuation and loss, resulting in a distortion of the wavefront of the beam.  We can neglect these effects for dilute gases, though they are easily accounted for.  

	\begin{figure}[!t]
	\centering
    		\includegraphics[scale=1.4]{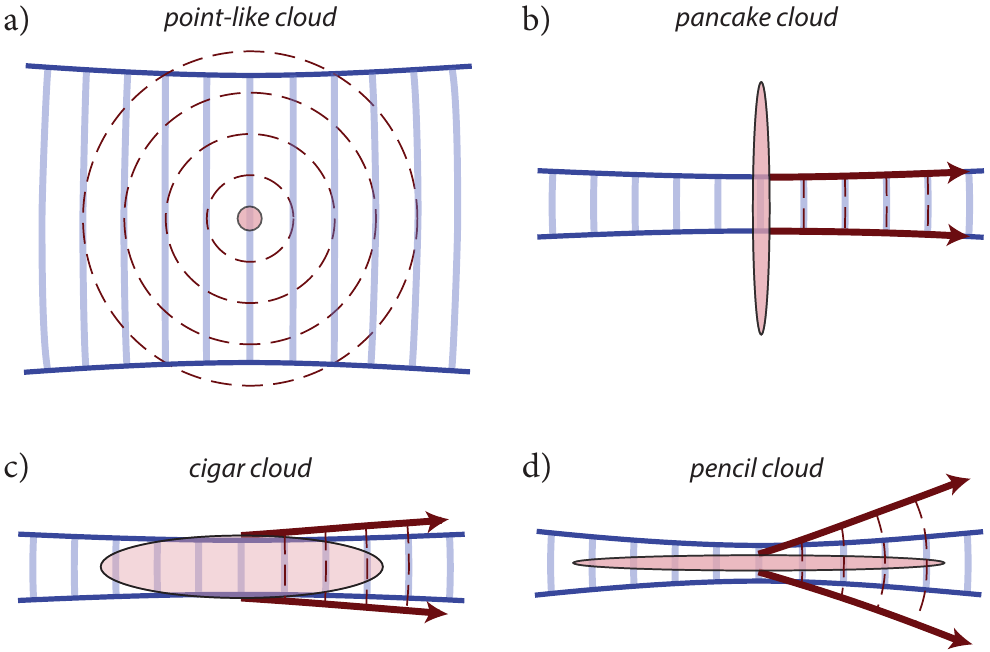}
       		 \caption[Radiation patterns for various geometries]{Cartoons of the scattered modes for various atomic cloud and beam geometries.  The spatial profile of the probe mode is indicated with solid blue lines and that of the field scattered by a given dielectric distribution is indicated by solid red lines with arrows. a) A point-like atomic ensemble scatters light isotropically.  b) A ``pancake"-shaped cloud at a fixed $z$-plane radiates nearly perfectly into the probe mode.  Extended clouds can radiate into the probe mode well, as seen in c) for a ``cigar"-shaped cloud, or poorly, as seen in d) for a ``pencil"-shaped cloud. }  \label{Fig::ModeMatching}     	
	\end{figure}

We specify the probe as the input field with $x$-polarization, $ \vec{\mathcal{U}}_{\rm in} (\rperp,z) = \mathcal{U}_x(\mathbf{r}_\perp,z) \mathbf{e}_x$, which scatters $y$-polarized light according to \erf{Eq::ClassicalVecScattering},
	\begin{equation} \label{Eq::yScattering}
		\mathcal{U}_y(\rperp, z) =  i 2 \pi k_0  \int_{-\infty}^z  dz'   \int  d^2 \rperp' K(\rperp  - \rperp', z-z')  \chi_{yx}( \rperp', z')  \cdot \mathcal{U}_{x} (\rperp',z').
	\end{equation}
When the average susceptibility results from a rarefied cloud of atoms and the light is detuned from resonance, the scattered field is quite weak compared to the input coherent laser probe. Thus, Faraday rotation is measured with a balanced polarimeter at position $z_D$ in the far field as discussed in \srf{Sec::BalancedPolarimetry}.  The mean signal $\mathcal{M}$ is proportional to the classical $\mathcal{S}_2$ Stokes vector, $\mathcal{U}_x^* \mathcal{U}_y +\mathcal{U}_y^* \mathcal{U}_x$, integrated across the detector surface [\erf{Eq::HomodyneSignal1}].  Using the solution for  $ \mathcal{U}_y(\rperp, z_D)$, \erf{Eq::yScattering}, and the properties of the propagator, Eq. (\ref{backprop}), we find 
	\begin{align} 
		\mathcal{M} \propto & \int_{-\infty}^{z_D} \! \! dz' \! \int \! d^2 \rperp' \text{Re}\big\{ i \chi_{yx}(\rperp', z')\big\}  \mathcal{U}_x (\rperp',z') \! \int \! d^2 \rperp  \, \mathcal{U}_x^* (\rperp, z_D) K(\rperp-\rperp', z_D-z') \nn \\
		=& -\int d^3 \mbf{r}' \, \text{Im}\big\{ \chi_{yx}(\rperp', z')\big\} |\mathcal{U}_x (\rperp',z')|^2 . \label{Eq::ClassicalMeasurementSignal}
	\end{align}
The measured signal is proportional to the local value of the susceptibility component $\text{Im}\left\{ \chi_{yx}(\rperp, z)\right\}$ integrated over the average positions of the scatterers, weighted by the local field intensity $|\mathcal{U}_x (\rperp,z)|^2$.

	\begin{figure}[!t]
	\centering
    		\includegraphics[scale=0.6]{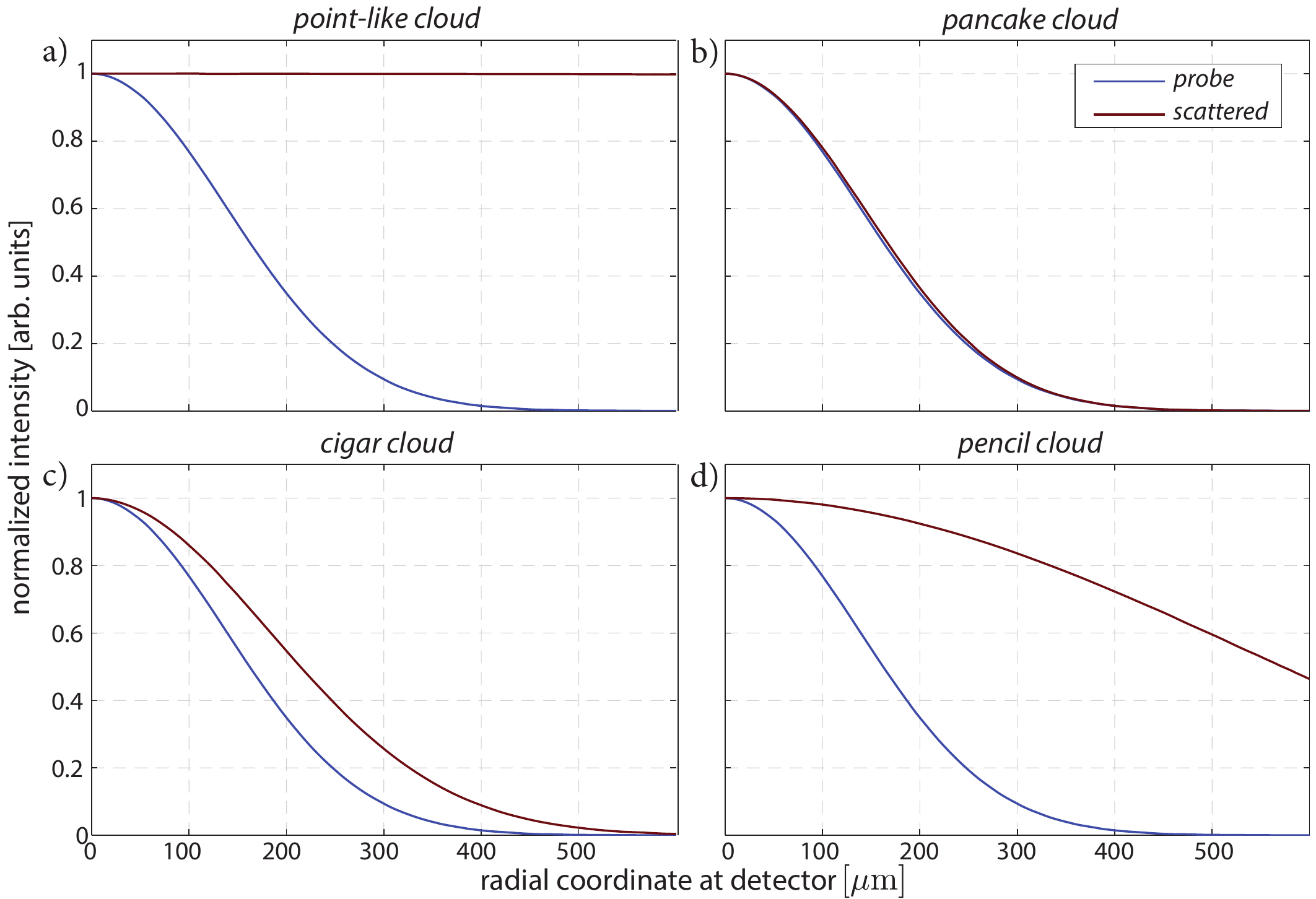}
       		 \caption[Far-field intensity for various geometries]{Normalized mean intensity as a function of the radial coordinate at the detector plane for a three-dimensional Gaussian cloud of scatterers described by \erf{Eq::AtomicDistribution}.  The probe is a TEM$_{00}$ beam with a waist $w_0 = 50$ $\mu$m and the detector plane is in the far field at $z_D = 5$ cm.  The radial profile of the probe intensity (blue lines) is Gaussian, and the radial profile of the scattered field is shown in red.  a) In the far field, a point-like cloud ($\sigma_\perp = 1$ $\mu$m, $\sigma_z = 1$ $\mu$m) radiates a spherical wave that approximates a plane wave in the paraxial regime.  b) A ``pancake" cloud ($\sigma_\perp = 1000$ $\mu$m, $\sigma_z = 10$ $\mu$m) radiates nearly perfectly into the probe mode.  Longitudinally extended clouds can radiate into the probe mode well, as seen in c) for a ``cigar" cloud ($\sigma_\perp = 80$ $\mu$m, $\sigma_z = 1000$ $\mu$m), or poorly, as seen in d), for a ``pencil" cloud ($\sigma_\perp = 20$ $\mu$m, $\sigma_z = 1000$ $\mu$m). }  \label{Fig::ModeMatching2}     
	\end{figure}

We can acquire some physical intuition from this classical model. When the squared probe amplitude, $ |\mathcal{U}_x (\mbf{r}_{\perp}, z)|^2$, is constant over the atomic ensemble then the coupling is symmetric.  Geometrically, this is approximately achieved when the beam waist at the focus, $w_0$, is much larger than the transverse extent of the cloud and the length of the cloud is short compared to twice the Rayleigh range, $z_R = k_0 w_0^2/ 2$.    The mean-field radiation pattern of such a cloud described by \erf{Eq::yScattering}, however, has poor overlap with the probe.  The asymptotic limit is a point-like scatterer, which scatters a spherical wave as depicted in Fig. \ref{Fig::ModeMatching}(a).  The end result is that the polarimeter detects only a small fraction of the signal photons.  On the other hand, perfect mode matching is achieved for atoms confined as a uniform dielectric sheet at a fixed $z$-plane, similar to the geometry shown in Fig. \ref{Fig::ModeMatching}(b).  Indeed, a uniform dielectric slab of extent much larger than the beam waist achieves perfect mode matching, but one cannot achieve such an dielectric distribution with high OD using cold atomic gases.  Further, for a finite number of atoms, the realizable OD [\erf{Eq::ODdef}] is low in this configuration, since the majority of the atoms are outside the beam's focus and experience little electric field.  An more realistic intermediate geometry where the ensemble is pencil-shaped allows for reasonable mode matching while maintaining a high OD, as depicted in Figs. \ref{Fig::ModeMatching}(c-d).  To quantify these tradeoffs, it is clear that the concept of optical density in \erf{Eq::ODdef} requires some notion of spatial dependence to describe the effective number of atoms addressed by a particular laser and atomic cloud geometry, which we will return to in \srf{Sec::QuantumMeasurement}.

These concepts can be reinforced with a numerical calculation of the spatial profile of the scattered amplitude and intensity.  Consider a cylindrically symmetric cloud of scatterers with a mean density described by a three-dimensional Gaussian distribution,
	\begin{align} \label{Eq::AtomicDistribution}
		\eta(\mathbf{r}) = \eta_0 \exp \left( - 2\frac{\rho^2}{\sigma_\perp^2}  - 2\frac{z^2}{\sigma_z^2} \right),
	\end{align}
where $\sigma_\perp^2$ and $\sigma_z^2$ are the transverse and longitudinal $e^{-2}$ variances, $\eta_0$ is the peak density, and the total atom number is found by integrating over the cloud, $N = \int d^3 \mathbf{r}\, \eta(\mathbf{r})$.  The cloud is probed by a TEM$_{00}$ laser with waist at the focus of $w_0 = 50$ $\mu$m.  In the far field, $z_D = 5$ mm, the probe and scattered electric fields are calculated for the scattering geometries in \frf{Fig::ModeMatching}.

		\subsubsection{Semi-classical scattering from an ensemble of alkali atoms using a mode decomposition of the paraxial field}
		
We now specify the scatterers as alkali atoms with hyperfine atomic structure, and we decompose the paraxial field into a set of orthonormal modes.  For a dilute ensemble of $N$ cold atoms at fixed positions $\mbf{r}_n$, the dielectric susceptibility of the gas is
	\begin{equation}
		\tensor{\chi}(\mbf{r}) = \sum_{n=1}^N \expects{ \hat{\tensor{\alpha}}\phantom{}^{(n)} } \, \delta^{(3)}(\mbf{r}-\mbf{r}_n),
	\end{equation}
where $\hat{\tensor{\alpha}}\phantom{}^{(n)}$ is the the dynamic polarizability tensor operator for the $n^{th}$ atom. We consider here alkali atoms restricted to a subspace defined by a total (hyperfine) angular momentum $f$ as described in Chapter \ref{Ch::DispersiveInt}. In terms of the total angular momentum operator for each atom, $\hat{\mbf{f}}^{(n)}$, the polarizability operator can be decomposed into its irreducible tensor components, \erf{Eq::IrreducibleDecomp}.  The effect of the rank-2 tensor component complicates both the collective coupling of the atoms to the probe as well as the internal spin dynamics. Alternatively, as shown in \srf{Sec::BFieldAveraging}, by applying a strong bias magnetic field in the direction of the probe's propagation, the birefringent effect on the probe arising from the coupling to the atoms via the rank-2 tensor term averages to zero \cite{NorDeu12}.  The residual effect of the rank-2 component is a nonlinear dynamics on the internal spin state of each atom~\cite{SmiJes03}, given in \erf{Eq::ResidualLightShift}, which does not affect the the QND measurement under consideration here.  

The paraxial field is expanded in a a set of orthonormal spatial modes, with the probe described by the fundamental mode, $u_0(\mathbf{r}_\perp, z)$.  The scattered field amplitude radiated by the induced dipoles is
	\begin{align} \label{Eq::DipoleScattering}
		\vec{\mathcal{U}}_{\rm scat}(\mbf{r}_\perp, z)&= i 2 \pi k_0 \mathcal{E}_0 \sum_{n=1}^N \big[ \expt{ \talphan } \cdot \mathbf{e}_x \big]_\perp  u_{0}(\mathbf{r}_{\perp n}, z_n) K(\mathbf{r}_\perp - \mathbf{r}_{\perp n}, z - z_n), 
	\end{align}
where the subscript on the square brackets, $\perp$, denotes the component of the dipole transverse to the direction of observation.  This is a consequence of free-space paraxial scattering; the electric field vector must vanish along the direction of propagation.   Thus, the means scattered paraxial electric field is determined by the $\hat{\alpha}_{xx}$ and  $\hat{\alpha}_{yx}$ components of the polarizability tensor, but not by $\hat{\alpha}_{zx}$. 

Because the scattered radiation in general is not mode-matched with the Gaussian laser beam, the light is scattered into all paraxial modes.  In the far field, $z \gg z'$, we can separate $ \mbf{E}_{\rm scat}(\mbf{r})$ into a portion forward-scattered into the spatial mode of the probe with amplitude $\vec{\Upsilon}\peakprobe e^{ik_0 z}$, and a portion scattered into all other spatial modes, $\mbf{E}'_{\rm scat}(\mbf{r})$.  The total field $\mbf{E}_{\rm out}(\mbf{r})  =\mbf{E}_L(\mbf{r}) + \mbf{E}_{\rm scat}(\mbf{r})$, takes the form
	\begin{align} 
		\mbf{E}_{\rm out}(\mbf{r})  &=\mbf{E}_{\rm out}(\mbf{r})  =\mbf{E}_L(\mbf{r}) + \mbf{E}_{\rm scat}(\mbf{r}) \nn \\
		& = \big[  \mathbf{e}_x + \vec{\Upsilon} \big] \peakprobe u_{0}(\rperp,z) e^{ik_0 z}+ \mbf{E}'_{\rm scat}(\mbf{r}).
	\end{align}
The amplitude of the scattered field into the probe mode is given by
	\begin{align} 
 		\vec{\Upsilon} \peakprobe e^{ik_0 z} \equiv & \int \frac{d^2\rperp}{A} u_{0}^*(\rperp,z) \mbf{E}_{\rm scat}(\mbf{r}) \nn \\
		= & i  \frac{2\pi k_0}{A} \peakprobe \sum_{n=1}^N \big[ \expt{ \talphan } \cdot \mathbf{e}_x  \big]_\perp   \beta_{0}(\mathbf{r}_{\perp n},z_n)   e^{ik_0(z-z_n)}. \label{Eq::ClasFaraday},
 	\end{align} 
where the square of the probe's spatial mode, $\beta_{0}(\mathbf{r}_{\perp n},z_n)$, is defined in \erf{Eq::Beta}.
The amplitude of the scattered light in the probe mode at a distant observation plane $z$ is proportional to the square of the probe's spatial function at the atomic positions. 
  
Equation (\ref{Eq::ClasFaraday}) reveals the key physical effects on the light which could be measured in the far field.  The component of $\vec{\Upsilon}$ along the laser polarization $\mathbf{e}_x$ gives rise to the scalar index of refraction and attenuation.  The component of $\vec{\Upsilon}$ along $\mathbf{e}_y$ (component orthogonal to the probe) gives rise to a rotation of the polarization on the Poincar\'{e} sphere -- Faraday rotation and birefringence.  Written in terms of the components of the tensor polarizability, the total field is
 	\begin{align} \label{Eq::OutputField}
 		& \mbf{E}_{\rm out} (\mbf{r}) = \left[ \left(1+i\delta \phi-\frac{a}{2}\right) \mbf{e}_x + \left(\frac{\chi + i\nu}{2}\right) \mbf{e}_y \right]  \peakprobe u_{0}(\rperp,z) e^{ik_0 z} + \mbf{E}'_{\rm scat}(\mbf{r}), 
	\end{align}
 where,
 	\begin{subequations} \label{Eq::FundModeShifts}
 	\begin{align}
 		\delta \phi &= \left( \frac{2 \pi k_0}{A} \right) \sum_{n=1}^N  \beta_{0}(\mathbf{r}_{\perp n},z_n)   \text{Re} \left\{ \expt{ \talphanOp_{xx} } \right\}  \label{Eq::ScatteringPhaseShift} \\
 		a &= \left( \frac{4 \pi k_0}{A} \right)  \sum_{n=1}^N  \beta_{0}(\mathbf{r}_{\perp n},z_n)   \text{Im} \left\{ \expt{ \talphanOp_{xx} } \right\}  \\
 		\chi &=-\left( \frac{4 \pi k_0}{A} \right)  \sum_{n=1}^N  \beta_{0}(\mathbf{r}_{\perp n},z_n)  \text{Im} \left\{ \expt{ \talphanOp_{yx} } \right\} \label{Eq::ClassicalFaradayRotation} \\
 		\nu &=\left( \frac{4 \pi k_0}{A} \right)  \sum_{n=1}^N  \beta_{0}(\mathbf{r}_{\perp n},z_n)  \text{Re} \left\{ \expt{ \talphanOp_{yx} } \right\} 
 	\end{align}
	\end{subequations}
 are respectively: $\phi$ is the index of refraction phase shift, $a$ is the Beer's law attenuation coefficient\footnote{In this model, attenuation of the field in the spatial mode of the probe is caused by scattering into other modes, including off-axis scattering, rather than dissipative processes such as heating.}, $\chi$ is the rotation angle of the Stokes vector corresponding to the Faraday effect, and $\nu$ is the corresponding angle for birefringence, with the polarizability matrix elements denoted as $ \hat{\alpha}_{ij} = \mathbf{e}_i\cdot \tensor{\alpha} \cdot \mathbf{e}_j$ in the $x$-$y$ basis.  An alternate and insightful classical derivation of the attenuation and phase shift can be found in Ref. \cite{TanVul11}.  
 
Balanced polarimetry is sensitive to Faraday rotation, \erf{Eq::ClassicalFaradayRotation}.  Retaining only the vector component of the off-diagonal component of the atomic polarizability tensor, \erf{Eq::GenPolarizability}, describes a pure Faraday interaction.  Substituting $\text{Im} \big\{  \expt{ \hat{ \tensor{\alpha} }^{(n)}_{yx} } \big\} \propto  \expects{ \hat{f}^{(n)}_z }$ into the formula for the mean polarimetry signal, \erf{Eq::ClassicalMeasurementSignal}, yields
	\begin{align}
		\mathcal{M} & \propto \sum_{n=1}^N   \beta_{0} (\mbf{r}_{n}) \expect{ \hat{f}^{(n)}_z } = \expt{ \hat{F}_z^{0} }  \label{Eq::FundamentalSpinWave}
	\end{align}
which is the spin wave given in \erf{Eq::SpinWave}.  We see that for paraxial beams, the polarimeter measures the effective collective spin determined by the inhomogeneous weighting, \erf{Eq::Beta}, of the atomic spin operators by the spatial mode of the beam.  The spin wave is stationary because it is coupled to the forward-scattered light, where the absorbed and emitted modes are the same. Physically, this spin wave is the collective observable that radiates indistinguishably into the probe mode and is effectively selected by the homodyne measurement of the polarimeter. In a plane wave, homogeneous, one-dimensional description, the measured observable is proportional to the symmetric collective spin, $\mathcal{M} \propto \sum_i  \expects{ \hat{f}^{(i)}_z }  =  \expects{ \hat{F}_z } $.  

The remaining paraxial field scattered into modes other than the probe can also be decomposed in terms of the modes,  
	\begin{align}
		\mathbf{E}'_{\rm scat}(\mathbf{r}_\perp, z) = \sum_{i \neq 0} \vec{c}_{i} u_{i}(\mathbf{r}_\perp, z) e^{ik_0z},
	\end{align} 
with vector overlap coefficients found using the scattering solution and the properties of the modes and of the propagator, \erf{Eq::EssentialProperties},
	\begin{align}
		\vec{c}_{i} 
			=& \, i \frac{2 \pi k_0}{A} \mathcal{E}_0 \sum_{n=1}^N  \big[ \expt{ \talphanOp }  \cdot \mathbf{e}_x \big]_\perp \beta_{i}(\mathbf{r}_{\perp n}, z_n). \label{Eq::HigherModeClassicalScat}
	\end{align}
Similar to the physical quantities in \erf{Eq::FundModeShifts}, such a decomposition allows one to characterize the phase and polarization of light scattered from the probe mode into any higher order transverse mode.  The real and imaginary parts of the vector component of the off-diagonal term in the susceptibility allow us to identify the spin wave that scatters into each of the higher-order transverse modes $i \neq 0$, $\hat{F}_z^{i}$, defined in \erf{Eq::SpinWave}. The higher order spin waves are different from the fundamental spin wave \erf{Eq::FundamentalSpinWave} which couples to the probe mode and is measured in the polarimeter.  In fact, these higher order spin waves couple to paraxial radiation in modes that are orthogonal to the probe (by construction), and as such are not measured in the balanced polarimeter. 
		
To characterize the atomic geometry with respect to the mode decomposition, we introduce an effective atom number that describes the fraction of atoms scattering probe light into mode $i$,  
	\begin{equation} \label{Eq::N1eff}
		N^{(1)}_{\eff, i}  = \sum_{n=1}^N u^*_{i}(\mbf{r}_n) u_{0}(\mbf{r}_n) = \sum_{n=1}^N \beta_{i}(\mbf{r}_n).  \\ 
	\end{equation}
In the continuum limit, the mean atomic density distribution is described by a continuous function, $\eta(\mbf{r})$, often determined by the geometry of a trapping potential.  The density distribution is normalized so that integrating over the cloud yields the total atom number,
	\begin{equation}
		\int d^3\mbf{r} \, \eta(\mbf{r}) = N.
	\end{equation}
Then, the sum in \erf{Eq::N1eff} becomes an integral and the effective atom number can be written,
	\begin{equation} \label{Eq::N1effCont}
		N^{(1)}_{\eff, i} = \int d^3\mbf{r} \,  \eta(\mbf{r}) \beta_{i}(\mbf{r}).
	\end{equation} 

Optimizing laser and atomic cloud geometry to maximize the mean scattered field in the fundamental mode is related to maximizing $N^{(1)}_{\eff}$ (in the fundamental mode, $u_0(\mathbf{r})$, the label on the effective atom numbers will be dropped).  The quantities in \erf{Eq::FundModeShifts} that describe the phase shift, attenuation, Faraday rotation, and birefringence can all be expressed in terms of $N^{(1)}_{\eff}$ when the particles are identically prepared.  For alkali atoms probed by a laser detuned far enough that the rank-2 components and the imaginary part of the polarizability can be ignored, the attenuation and birefringence are effectively zero.  The phase shift and the Faraday rotation angle in the fundamental mode, 
	\begin{subequations} 
 	\begin{align}
 		\delta \phi = & \frac{1}{2} \left(\frac{\Gamma}{2 \Delta } \right) C_{j' f}^{(0)} \, \text{OD}^{(1)}_\eff ,\label{Eq::PhaseShiftOD1} \\
 		\chi =&  - \left(\frac{\Gamma}{2 \Delta } \right) C_{j' f }^{(1)}  \, \text{OD}^{(1)}_\eff  \expt{\hat{f}_z}, \label{Eq::RotationAngleOD1}
	\end{align}
	\end{subequations}
can be expressed in terms of the effective resonant optical density\footnote{Using a resonant quantity in off-resonant situations may seem odd.  The goal is to identify a geometric quantity, independent of detuning or the atomic state, which characterizes the measurable quantities.  In the literature, one often encounters a detuning-dependent optical density, OD$(\Delta) = N \frac{\sigma_0(\Delta)}{A}$, where $\sigma_0(\Delta)$ is the detuning-dependent, single-atom scattering cross section.} seen by the probe,
	\begin{align} \label{Eq::ODeff1}
		\text{OD}_{\rm eff}^{(1)} \equiv N_\eff^{(1)} \frac{\sigma_0}{A} .
	\end{align}
This generalizes the standard definition of the OD, \erf{Eq::ODdef}, to include spatial variations of a probe that address an effective atom number $N_\eff^{(1)}$.  The equality in \erf{Eq::RotationAngleOD1} holds only when each atom's expectation value $\expt{f}_z$ is the same, such as for a spin coherent state.    Closer to resonance, the attenuation and birefringence will become non-negligible, and they, too, can be expressed in terms of $\text{OD}_{\rm eff}^{(1)}$.  

The maximum phase shift for a single alkali atom occurs when it is placed at the focus of the Gaussian probe, $\delta \phi^{\rm max} =  \frac{\Gamma}{4 \Delta} \frac{\sigma_0}{A}C_{j' f}^{(0)}$.  The tighter the focus, the more the probe and scattered light overlap in the far field\footnote{The intensity seen by the atom is also larger for small beam areas, and so the scattered field is larger.  This does not increase the phase shift, however, as it is scaled by the electric field amplitude of the probe.}. The atom scatters a spherical wave that interferes with the Gaussian beam, which also approaches a spherical wave in the far field with an additional $\pi/2$ Guoy phase.  In \frf{Fig::ScatteringPhaseShift}, the total phase shift, \erf{Eq::PhaseShiftOD1}, is numerically calculated for different cloud geometries at fixed peak atomic density, $\eta_0 = 5 \times 10^{11}$ atoms/cm$^3$.  The clouds are probed by light far detuned from resonance, $\Delta/\Gamma = 200$, with scalar polarizability coefficient given by \erf{Eq::ScalarCoefSum}; $C^{(0)}_{J' F} = 2/3$.  In \frf{Fig::ScatteringPhaseShift}(a), we see that the cigar and pencil geometries make best use of the available atoms, with the pancake and pencil geometries producing negligible phase shifts.  Although the total phase shift for the point-like geometry may be small due to the minuscule effective atom number, we see in \frf{Fig::ScatteringPhaseShift}(b) that the phase shift per atom (normalized to the maximum single-atom phase shift) can be quite high.  This is due to the fact that for a point-like geometry, the atoms are localized where the probe field amplitude is the largest.  Both the cigar and pencil geometries have large relative contributions from their constituent atoms, whereas the pancake does not.  
	
	\begin{figure}[!t]
	\centering
    		\includegraphics[scale=0.85]{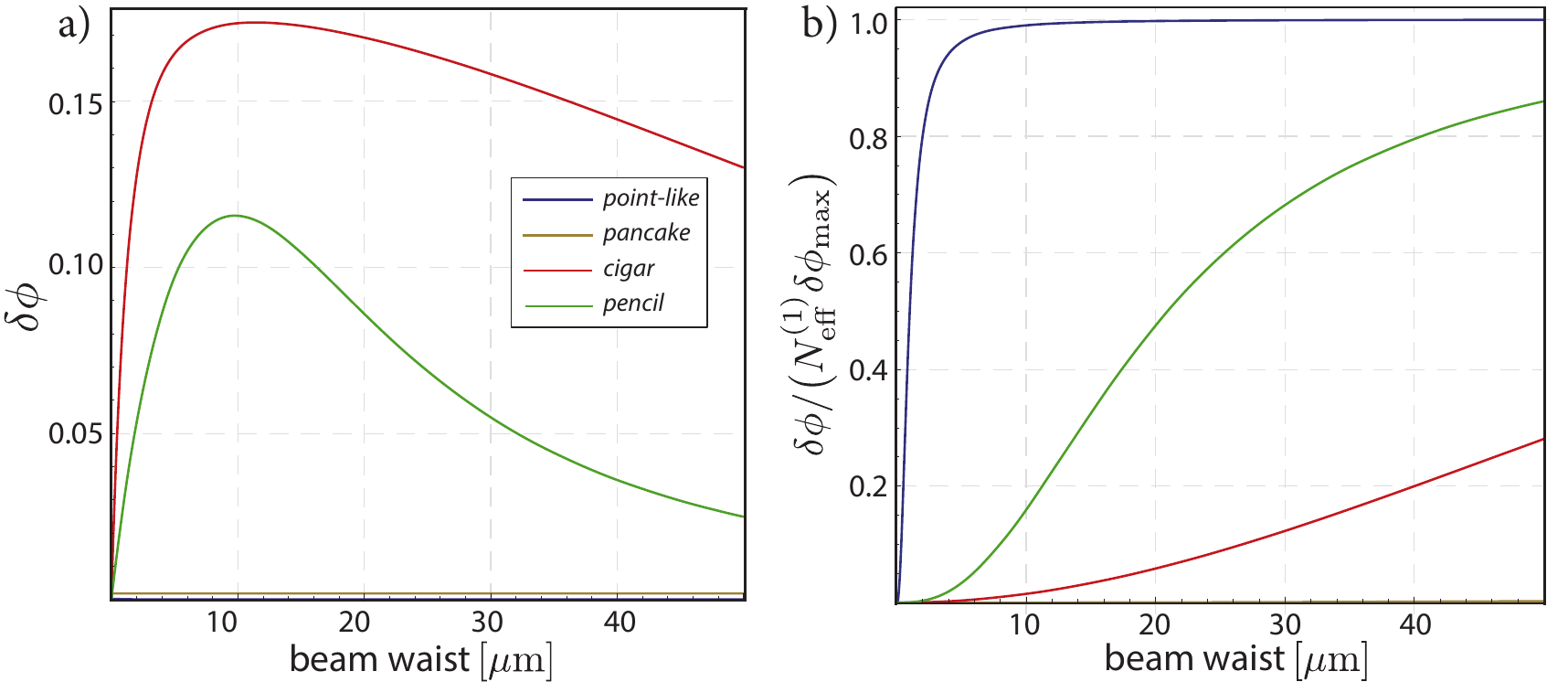}
       		 \caption[Scattering phase shift for spatially-extended atomic clouds]{Scattering phase shift in the probe mode calculated for different cloud geometries -- those from \frf{Fig::ModeMatching2} -- for fixed peak atomic density, as a function of probe waist.  a) Total phase shift.  b) Phase shift per atom, normalized to the maximum single atom phase shift.}  \label{Fig::ScatteringPhaseShift}     
	\end{figure}

Note that for symmetric coupling, when $\beta_i(\mathbf{r}) \rightarrow 1$ in \erf{Eq::N1eff} and \erf{Eq::N1effCont}, then the effective atom number in any mode approaches the total atom number, $N^{(1)}_{\eff,i} \rightarrow N$.  This indicates that every atom in the ensemble contributes to the scattered field in that mode.  There is no requirement that the atom numbers in each mode sum to the total atom number $N$.  This is seen by considering a cloud of atoms localized at the focus of a very wide beam, as in Fig. \ref{Fig::ModeMatching}.  Although $\beta_0 \rightarrow 1$, indicating that all atoms scatter into the fundamental mode, the isotropic radiation pattern reveals that they also scatter into many other modes.  Using a Laguerre-Gauss mode decomposition given in Appendix \ref{Appendix::LGModes}, it can be shown that such a cloud scatters equally into all $l=0$ modes since $\beta_{p0}(\mathbf{r}) \rightarrow 1$ for all $p$\footnote{The $l=0$ generalized Laguerre polynomials evaluated at the origin are $L_p^0(0) = 1.$}.

	\subsubsection{Quantum description of measurement} \label{Sec::QuantumMeasurement}

In addition to maximizing the signal by mode-matching the pattern of scattered light with that of the probe mode, which can be done using the semiclassical description in the previous section, we are also be interested in the fluctuations and noise. There are two fundamental quantum effects: (i) the polarimeter has finite shot-noise resolution, and (ii) atoms scatter photons diffusely into all directions (spontaneous emission).  We will deal with spontaneous emission and optical pumping in \srf{Sec::Decoherence}.  Here, we investigate the unavoidable quantum fluctuations of the field and, in addition, we characterize how the quantum fluctuations of the \emph{atoms} manifest in the measurement signal.  For the generation of spin squeezing, the mean polarimetry signal is zero and these fluctuations play a critical role.  

In a fully quantum model, balanced polarimetry can be interpreted as a measurement of the 2-component of the Stokes operator in \erf{Eq::MultimodeStokesOps}, $\hat{s}^{ii}_2(z,t)$, for a probe with spatial mode $u_i(\mathbf{r}_\perp, z)$.  We have considered the specific case of a large amplitude, $x$-polarized probe field in spatial mode $u_0(\mathbf{r}_\perp, z)$ that gives the linearization, $\hat{s}^{00}_2(z,t) \rightarrow \sqrt{ \dot{N}_L/2 } \hat{X}_{0}(z,t)$.  Using the multi-mode Faraday interaction in the previous section, the output field operators are given by \erf{Eq::OutputMode}, and the measurement current is generated by the scattered $\hat{a}_{0,y}(z,t)$ component.  

The measured quadrature at the detector plane, found from \erf{Eq::QuantumXOut}, is proportional to the \emph{fundamental spin wave}, $\hat{F}^{0}_z$ --  the same result as from the semiclassical calculation, \erf{Eq::FundamentalSpinWave}.  The polarimeter signal is determined by the measurement operator\footnote{The measurement operator $\hat{\mathcal{M}}$ is exactly the output Stokes operator $\hat{s}_2^{00}(z_D,t)$ integrated over the measurement time.}
	\begin{align}\label{Eq::Measurement}
		\hat{\mathcal{M}}(t) 
		 &= \sqrt{\frac{\dot{N}_L}{2} } \int_0^t dt' \hat{X}_{0}(0,t') - \sqrt{ \frac{\dot{N}_L \CstrengthSq}{4}}  \int_0^t dt' \hat{F}^{0}_z(t').
	\end{align}
The statistical fluctuations in the integrated signal for a collection of experiments is given by
	\begin{align}\label{Eq::MeasurementFluctuations}
		\Delta \mathcal{M}^2 
		&= \frac{\dot{N}_L t}{4}  + \frac{\dot{N}_L \CstrengthSq}{4} \int_0^t dt' \int_0^t dt'' \big\langle \Delta \hat{F}^{0}_z (t') \Delta \hat{F}^{0}_z (t'') \big\rangle \\
		&\equiv \Delta \mathcal{M}_{\rm SN}^2  + \Delta \mathcal{M}_{\rm PN}^2, \label{Eq::NoiseBreakdown}
	\end{align}
where $\dot{N}_L t$ is the total average number of probe photons.  The fully quantum theory explicitly includes the additional vacuum shot noise entering the polarimeter, $\langle \Delta X_{0}(0,t)\Delta X_{0}(0,t') \rangle=\delta(t-t')/2$,  that leads to the shot-noise variance, $\Delta \mathcal{M}_{\rm SN}^2$.  The additional fluctuations in the signal, $\Delta \mathcal{M}_{\rm PN}^2$, come from the shot-to-shot uncertainty in $z$-projections of the collective spin wave.  

When diffuse scattering is neglected $\hat{F}^0_z$ is a QND observable [\erf{Eq::FzCommutes}] and, for vacuum input, the mean integrated signal, 
	\begin{equation}\label{Eq::Measurement}
		\expt{ \hat{\mathcal{M}}(t) } 
			=- \sqrt{ \frac{\dot{N}_L \CstrengthSq}{4}} t \expt{ \hat{F}^{0}_z(0) }.
	\end{equation}
and projection-noise fluctuations,
	\begin{equation}\label{Eq::MeasurementPN}
		\Delta \mathcal{M}_{\rm PN}^2 
			= \frac{\dot{N}_L \CstrengthSq}{4} t^2  \big( \Delta  \hat{F}^{0}_z (0)\big)^2,
	\end{equation}
are entirely determined by the initial collective atomic state.  Increasing the projection-noise contribution to the fluctuations in the measurement signal is the key to achieving QND spin squeezing, as we will see in Chapter \ref{Ch::SpinSqueezing}.  In terms of its atomic constituents, the fundamental spin wave variance,
	\begin{align}\label{Eq::FundamentalVariance}
		\left( \Delta F^{0}_z \right)^2 =  & \sum_{n=1}^N  \big[ \beta_{0}(\mbf{r}_n) \big]^2 \expect{ \big(\Delta \hat{f}_z^{(n)})^2}  + \sum_{m\neq n}  \beta_{0}(\mbf{r}_m) \beta_{0}(\mbf{r}_n )\expect{\Delta \hat{f}_z^{(m)}\Delta \hat{f}_z^{(n)}},
	\end{align}
consists of single-atom fluctuations (first term) as well as atomic correlations (second term).  In experimental realizations, the atoms are initialized by optically pumping into a spin coherent state (SCS); a product state devoid of atomic correlations.  To maximize the projection noise contribution, the SCS is oriented orthogonal to the probe's propagation direction.  This results in a spin wave variance in mode $i$,
	\begin{align}\label{Eq::PNvariance}
		\left( \Delta F^{i}_z \right)_{\rm SCS}^2 & = \sum_{n=1}^N \big[\beta_i(\mathbf{r}_n) \big]^2 \expt{ \hat{f}_z^{(n)2}}  = \frac{f}{2} N^{(2)}_{\eff,i} ,
	\end{align}
where we have introduced another effective atom number, 
	\begin{align} \label{Eq::N2eff}
		N^{(2)}_{\eff,i} & \equiv \sum_{n=1}^N |\beta_{i}(\mbf{r}_n)|^2  \rightarrow \int d^3\mbf{r} \,  \eta(\mbf{r}) |\beta_{i}(\mbf{r})|^2,
	\end{align}
similar to $N^{(1)}_{\eff}$ in \erf{Eq::N1eff}\footnote{Again, we drop the label for the fundamental mode $i = 0$.}.  In the fundamental probe mode, $N^{(2)}_{\eff}$ describes the geometric dependence of the projection noise contribution in the measured signal.  To maximize $\Delta \mathcal{M}_{\rm PN}^2$ one maximizes $N^{(2)}_{\eff}$.

The coupling strength $\xi$ that sets the degree of attainable entanglement between the atoms and photons is the ratio of the projection noise variance to the shot-noise resolution \cite{DeuJes10}.  Using Eqs. (\ref{Eq::kappa}) and (\ref{Eq::PNvariance}) we find
	\begin{equation}\label{Eq::CouplingStrength}
		\xi = \frac{\Delta \mathcal{M}_{\rm PN}^2}{\Delta \mathcal{M}_{\rm SN}^2} = \kappa t \big( \Delta F_z^0 \big)^2 = \mbox{OD}^{(2)}_\eff \, \frac{ g_f^2 }{18} \Big| \sum_{f' }1 + \delta_{f'}/\Delta \Big|^{-2} \gamma_0 t, 
	\end{equation}	
where the unit-oscillator-strength photon scattering rate at the peak intensity is
	\begin{align} \label{Eq::PeakScat}
		\gamma_0 
		= \dot{N}_L  \left( \frac{ \sigma_0}{A} \right)  \left( \frac{\Gamma}{2 \Delta} \right)^2.
	\end{align}
In \erf{Eq::CouplingStrength} we have defined another effective resonant optical density as it relates to projection-noise fluctuations of the signal, 
	\begin{align} \label{Eq::ODeff}
		\mbox{OD}^{(2)}_\eff \equiv N^{(2)}_\eff \frac{\sigma_0}{A}.
	\end{align}  	
The key to achieving a large OD$^{(2)}_\eff$ is choosing an atomic and beam geometry that addresses a large number of atoms to maximize $N_\eff^{(2)}$ while keeping the mode area $A$ small. Note that $\mbox{OD}^{(2)}_\eff$ is different from $\mbox{OD}^{(1)}_\eff$, defined in \erf{Eq::ODeff1}.  In a one-dimensional description a single optical density, \erf{Eq::ODdef}, is associated both with the coupling strength and with the phase shift, Beer's law attenuation, Faraday rotation, and birefringence.  In the three-dimensional case different effective atom numbers are associated with these effects, and the geometric couplings are not the same.

Thus far, we have analyzed the polarimetry measurements, including the fluctuations in the signal from shot noise and atomic projection noise.  In the following section, we present a stochastic master equation that describes how the state of the atoms is modified based on the projective polarimetry measurements of the field.  Further, while \erf{Eq::CouplingStrength} implies an ever increasing coupling strength with integration time $t$, we have neglected so far the decoherence that limits the total useful  integration time and the strength of the atom-light interface.  In \srf{Sec::Decoherence} we treat these effects a first-principles perspective, including spatial variations in the scattering rate which drives local decoherence.  


	\subsection{Stochastic master equation for balanced polarimetry}
	
We have learned from the previous section that balanced polarimetry selects light scattered only into the spatial mode of the probe, and thus effectively measures only the inhomogeneous spin wave coupled to that light.  Continuous measurement of this light reveals the stochastic dynamics of the spins in the ensembles through the random measurement current.  The paraxially scattered light orthogonal to the probe mode goes undetected and carries away any information imprinted on it by other spin waves.  This ultimately leads to decoherence in the atomic ensemble.  In this section we derive the stochastic master equation (SME) in the Schr\"{o}dinger picture that quantitatively describes these processes.

The expectation value in \erf{Eq::Measurement} and the related projection noise, [\erf{Eq::MeasurementPN}], are statistical statements about an ensemble of measurement records, not a single experiment.  Within a single experiment, the conditional expectation value $\expt{\hat{F}_z^0(t)}_c$ changes stochastically as a function of the measurement record, \emph{even when $\hat{F}_z$ is a} QND \emph{observable}.  Only after the system has been ``projected" into an eigenstate in the long-time limit \cite{StoMab04, BouJam07}, does its expectation value $\expt{\hat{F}_z}$ remain constant in time.  For vacuum input, the classical measurement record from a single shot of an experiment is determined by the conditional $z$-projection of the atoms in the fundamental mode,  $\expt{ \hat{F}^{0}_z(t) }_c$, with additive Gaussian noise,
	\begin{align} \label{Eq::SingleShotExperiment}
		\mathcal{M}(t) = \sqrt{ \frac{ \dot{N}_L  }{4} } \int_0^t  dW  - \sqrt{ \frac{\dot{N}_L \CstrengthSq}{4}} \int_0^t dt' \expt{ \hat{F}^{0}_z(t') }_c,
	\end{align}
where $dW$ is a Wiener increment satisfying $\mathbbm{E}[dW] = 0$ and $dW^2 = dt$.  The infinitesimal increments of the continuous polarimetry measurement are 
	\begin{align} \label{Eq::dM}
		d \mathcal{M} = \sqrt{ \frac{ \dot{N}_L  }{4} } dW - \sqrt{ \frac{\dot{N}_L \CstrengthSq}{4}} \expt{ \hat{F}^{0}_z(t) }_c dt.
	\end{align}
The underlying signal in \erf{Eq::SingleShotExperiment} - rightmost term - is the Faraday rotation angle, equivalent to \erf{Eq::ClassicalFaradayRotation} multiplied by the total number of photons in the integration time\footnote{Also equivalent to \erf{Eq::RotationAngleOD1} when the collective atomic state is uncorrelated.}.  

The SME is a theoretical tool that allows us to find the conditional atomic state for a single measurement record, [\erf{Eq::SingleShotExperiment}].  The state can then be used to calculate conditional moments such as the mean that appears in \erf{Eq::SingleShotExperiment} or the conditional variance that is the focus of QND spin squeezing in Chapter \ref{Ch::SpinSqueezing}.  These conditional moments describe the statistics of \emph{future} measurements.  We present here the SME that results from independent measurements of the position quadrature $\hat{X}_{i}$ in each spatial mode $i$, following the standard prescription given in Refs. \cite{JacSte06, WisMilBook}. Such a multi-mode measurement mathematically generalizes the single-mode measurement in the previous section.  How to physically implement this measurement is not our concern; for our purposes, we ultimately trace over all spatial modes other than that of the probe to get the SME for balanced polarimetry.
	
Prior to measurement, the time evolution operator $\hat{U}(\Delta t)$  describing the interaction of the light and spin waves over a time interval $\Delta t$ can be divided up according to the spatial modes,
	\begin{equation}
		\hat{U}(\Delta t)  = \prod_{i} \hat{U}_{i}(\Delta t). 
	\end{equation}
The quadrature operator that arrives at the detector plane has propagated through then entire ensemble and become entangled with the $z$-local spin waves through the multi-mode Faraday Hamiltonian \erf{Eq::MultiModeFARADAY} as given by \erf{Eq::QuantumXOut}.  In a single spatial mode the unitary interaction is 
	\begin{align} \label{Eq::TimeOrderedExpFaraday}
		\hat{U}_{i}(\Delta t) =\exp  \bigg[  & - i \sqrt{  \frac{ \CstrengthSq }{ 2} }  \int_t^{t+\Delta t} dt' \Big(\mbox{Im} \big\{ \hat{F}_{z}^i  \big\} \hat{X}_{i}(z_D,t') -  \mbox{Re} \big\{ \hat{F}_{i}^z \big\} \hat{P}_{i}(z_D,t') \Big) \bigg]. 
	\end{align}
After this interaction the light and spin waves are entangled so that a polarimetry measurement of each quadrature, $\hat{X}_{i}$, generates quantum backaction on the atomic ensemble.  The evolution of the system conditioned on independent measurements of each mode is determined by the full, multi-mode Kraus operator,
	\begin{equation} \label{Eq::GenKrausOp}
		\hat{K}(\Delta t)= \prod_{i} \hat{K}_{i}(\Delta t).
	\end{equation} 
Here, $\hat{K}_{i}(\Delta t)$ is the Kraus component for outcome $x_{i}$ in the spatial mode $i$ is given by \erf{ApEq::KrausComp0},
	\begin{align} \label{Eq::KrausComp0}
		\hat{K}_{i} (\Delta t) = &   \bra{\hat{X}_i = x_i} \hat{U}_{i}(t, \, t+  \Delta t  )  \ket{0} \nn \\
		= & \exp \bigg[ - \frac{\Delta t }{2} \left( x_i^2 + 2 x_i  \sqrt{  \frac{ \CstrengthSq }{  2 }} \hat{F}^{i}_z  + \frac{ \CstrengthSq }{4} \mbox{Re}\big\{ \hat{F}^{i}_z \big\} \hat{F}^{i}_z  \right) \bigg] .
	\end{align}

As shown in Appendix \ref{Appendix::SMEDerivation}, in the infinitesimal limit, $\Delta t \rightarrow dt$ and $\Delta W_{i} \rightarrow dW_{i}$, we can use the rules of It\={o} calculus to expand the Kraus component in \erf{Eq::KrausComp0} to first order in $dt$,  
	\begin{equation} \label{Eq::KrausComp}
		\hat{K}_i(d t) = \hat{I} - \frac{\CstrengthSq}{2} \hat{F}_z^i  \expt{ \mbox{Re} \{ \hat{F}_z^i \} }_c dt - \frac{\CstrengthSq}{8} \hat{F}_z^{i \dagger} \hat{F}_z^i dt - \sqrt{ \frac{ \CstrengthSq }{4} } \hat{F}_z^i dW_i.
	\end{equation}
Here, we have used the statistical independence of stochastic Wiener processes, $dW_{i} dW_{j} = \delta_{i, j} d t$.  After the measurements are performed, the conditional collective atomic state is updated via the map,
	\begin{align} 
		\hat{\rho}(t+d t)&=\frac{\hat{K}(d t)\hat{\rho}(t)\hat{K}^\dagger(d t)}{\Tr \big[\hat{K}^\dagger (d t)\hat{K}(d t) \hat{\rho}(t) \big] } \label{Eq::KrausUpdate},
	\end{align}
where the full Kraus operator, given in \erf{Eq::GenKrausOp}, is composed of Kraus components for each transverse mode, \erf{Eq::KrausComp}.  Using Eqs. (\ref{Eq::GenKrausOp}) and (\ref{Eq::KrausComp})  with Eq. (\ref{Eq::KrausUpdate}), we derive the conditional atomic state. In differential form, $d\hat{\rho}(t) = \hat{\rho}(t+dt) - \hat{\rho}(t)$, the SME is
	\begin{align}\label{Eq::FullSME}
		d \hat{\rho}&= \sum_{i} \left( \sqrt{ \frac{\CstrengthSq }{4} }  \mathcal{H}_{i}[\hat{\rho} ] \, dW_{i} + \frac{\CstrengthSq}{4} \mathcal{L}_{i}[\hat{\rho}] \, dt \right).
	\end{align}
The terms in this equation describe two distinct effects.  First, the state of the atomic ensemble is conditioned on the measurement outcomes.  This is taken into account by the superoperator $\mathcal{H}_{i}[\hat{\rho}] $, defined
\begin{align}  \label{Eq::HSuperoperator}
		\mathcal{H}_{i}[\hat{\rho}] \equiv \hat{F}^{i}_z \hat{\rho} + \hat{\rho} \hat{F}^{i \dagger}_z - \big\langle\hat{F}^{i}_z + \hat{F}^{i \dagger}_z \big\rangle \hat{\rho}.
	\end{align}
Second, the system undergoes Lindblad-form decoherence via the superoperator 
	\begin{align} \label{Eq::LSuperoperator}
		\mathcal{L}_{i}[\hat{\rho}] \equiv \hat{F}_z^{i} \hat{\rho} \hat{F}_z^{i \dagger} - \frac{1}{2} \hat{F}_z^{i \dagger} \hat{F}_z^{i} \hat{\rho}  - \frac{1}{2} \hat{\rho} \hat{F}_z^{i \dagger} \hat{F}_z^{i}.
	\end{align}
This describes the effect on the atomic ensemble arising from indistinguishable radiation into paraxial modes of the field; Lindblad maps arise for systems interacting with Markovian baths.  

Finally, we return to the physical situation at hand for balanced polarimetry.  Only the fundamental mode is measured, while radiation in higher-order modes is lost.  This is modeled by tracing over (ignoring) the measurement records for modes $i \neq 0$.  Information about the quantum state of the ensemble that is mapped to these unmeasured modes is lost and SME does not preserve state purity.  Including the remaining internal light-shift Hamiltonian, \erf{Eq::ResidualLightShift}, the SME is
	\begin{align} \label{Eq::HomodyneSMEnoDec}
		d \hat{\rho}   =&  - \frac{i}{\hbar} \big[ \hat{H}_{\rm LS}, \hat{\rho} \big]dt +  \sqrt{ \frac{\CstrengthSq}{ 4 } } \mathcal{H}_{0}[\hat{\rho}]  dW + \frac{\CstrengthSq}{4} \sum_{i} \mathcal{L}_{i}[ \hat{\rho}]dt .
	\end{align}  
Finally, we can express the Wiener process as
	\begin{align} \label{Eq::dWi}
		d W = \frac{1}{\sqrt{\dot{N}_L/4}} d\mathcal{M}  - \Big(-\sqrt{ \CstrengthSq } \expt{\mbox{Re} \{ \hat{F}_z^0 \}(t)}_c \Big) dt ,
	\end{align}
where $d\mathcal{M}$ is the infinitesimal measurement increment, [\erf{Eq::dM}], and the conditional mean is calculated from the conditional state at time $t$,
	\begin{align}
		\expt{\mbox{Re} \{ \hat{F}_z^0 \}(t)}_c = \Tr \big[ \mbox{Re} \{ \hat{F}_z^0 \} \hat{\rho}(t)\big].
	\end{align}
The quantity in \erf{Eq::dWi}, known as the \emph{innovations process}, describes deviations of the measurement from the predicted mean, is a measure of the information gained in each infinitesimal time.  For situations in which one does not know the initial state, the innovations process serves to drive the conditional guess towards the ``true" state\footnote{The state that is generating the measurement record.}.

	\subsection{Local decoherence and optical pumping } \label{Sec::Decoherence}
		
The discrete random atomic positions are associated with the density fluctuations that give rise to diffuse scattering into 4$\pi$ steradians \cite{SorSor08}.  We consider light far detuned from any atomic resonance in a highly transparent regime, and thus we can safely neglect the small attenuation of the laser probe associated with this absorption.  The scattering processes, however, cause decoherence of the spin wave due to optical pumping.  This local decoherence breaks the collective symmetry of the problem and adds additional noise, which is detected in the polarimeter and competes with squeezing. 

To treat the atomic decoherence due to diffuse scattering, we divide the dynamics into terms that arise from forward-scattered light, described by the SME in \erf{Eq::HomodyneSMEnoDec}, and terms that arise from diffusely scattered light that lead to optical pumping and decoherence.  This is similar to what was done for a single atom in \erf{Eq::MEbreakdown}, except that we have subjected the paraxial light to measurement.  
The key feature is that the paraxial modes couple to collective spin waves, while the diffuse scattering couples to localized atoms and induces optical pumping according to a decoherent map.  Employing a many-atom map in the form \erf{Eq::MEDec}, we get
	\begin{equation} \label{Eq::GenME}
		\frac{d \hat{\rho}}{dt} \Big|_{\rm dec} = \sum_{n=1}^N \gamma_s(\mathbf{r}_n) \mathcal{D}_n [\hat{\rho}].
	\end{equation}
The map $\mathcal{D}_n$ acts on the $n^{th}$ atom, proportional to the local scattering rate,
\begin{align} \label{Eq::LocalScatRate}
		\gamma_s(\mathbf{r}_n) = I(\mathbf{r}_n) \frac{\sigma_0}{\hbar \omega} \frac{\Gamma^2}{4 \Delta^2} = \gamma_{0}\beta_{0}(\mathbf{r}_n).
	\end{align}
Here, $\gamma_{0}$ is the peak scattering rate defined in \erf{Eq::PeakScat}. We consider here a probe driving an $S_{1/2} \rightarrow P_J$ transition in an alkali atom, with a detuning that is small compared to the ground state hyperfine splitting but large compared to any hyperfine splitting in the excited state.  In this case, the light coherently couples substantially only to atoms in a given ground-electronic hyperfine manifold $f$ and the master equation is restricted to this subspace.  

The master equation for an $x$-polarized probe is given by \erf{Eq::XPolarizedME} with coefficients in \erf{Eq::CsCoefficients} for an $S_{1/2} \rightarrow P_J$ transition in an alkali atom. In the frame co-rotating with the bias magnetic field using the transformations in \erf{Eq::BiasTransformations}, the local decoherence in the master equation due to optical pumping is given by the map for each atom,
	\begin{align} \label{Eq::DiffuseME}
		\mathcal{D}_n [\hat{\rho}]  =  -\frac{2}{9}\hat{\rho} + \frac{g_f^2}{9} \Big[ \hat{f}^{(n)}_z\hat{ \rho} \hat{f}^{(n)}_z + \frac{1}{2} \big( \hat{f}^{(n)}_x \hat{\rho} \hat{f}^{(n)}_x + \hat{f}^{(n)}_y \hat{\rho} \hat{f}^{(n)}_y \big)   \Big].
	\end{align}
The first term on the right-hand side of \erf{Eq::DiffuseME} describes the decaydue to optical pumping, while the second term (in square brackets) represents a feeding that can reduce this decay rate \cite{CohT77}. Note that for $f>1/2$, this master equation is not trace preserving, since atoms can be optically pumped to the other ground hyperfine manifold where they are lost to any further measurement.

Including local decoherence from diffuse scattering, \erf{Eq::DiffuseME}, the full stochastic master equation for homodyne polarimetry measurements of the fundamental, 0-mode is 
	\begin{align} \label{Eq::HomodyneSME}
		d \hat{\rho}   =&  - \frac{i}{\hbar} \big[ \hat{H}_{\rm LS} + \hat{H}_0, \hat{\rho} \big]dt +   \sqrt{ \frac{\CstrengthSq}{ 4 } } \mathcal{H}_{0}[\hat{\rho}]  dW \\
		& + \frac{\CstrengthSq}{4} \sum_{i} \mathcal{L}_{i}[ \hat{\rho}]dt + \sum_{n=1}^N \gamma_s(\mathbf{r}_n) \mathcal{D}_n [\hat{\rho}]  dt. \nn
	\end{align}  
where $dW=dW_{0}.$  The Hamiltonian in the first term consists the remaining light shift on the atoms, \erf{Eq::ResidualLightShift}, as well as any other external Hamiltonian that may be applied to the ensemble.  For example, full coherent control in the ground hyperfine manifolds for each atom can be achieved using applied magnetic fields and RF pulses with properly chosen time-varying phases \cite{ChaJes07}.  This SME is a complete description of the evolution of the collective atomic state, accounting for the three-dimensional nature of the atom-photon modes, decoherence, and measurement backaction.  We see that through its  interaction with the probe, the atomic ensemble undergoes an additional form of \emph{collective} decoherence, \erf{Eq::LSuperoperator}, corresponding to light radiated into paraxial modes $i \neq 0$ that ultimately goes unmeasured.  Thus we have arrived at the same conclusion as in Ref. \cite{DuaZol02}.  That is, decoherence arises through two distinct processes -- first, the inherent mode-mismatch that gives rise to collectively scattered light in spatial modes other than the probe mode and second, the diffuse scattering of photons that acts locally on atoms in the ensemble.

	\subsection{Calculating expectation values of multi-atom observables} \label{Sec::MultiatomExpectations}

The decoherent dynamics generated by diffuse scattering complicate the calculation of expectation values.  This is because \erf{Eq::DiffuseME} is in general not a trace-preserving map, since for spin-$f > 1/2$ it accounts for pumping out of the hyperfine manifold of interest.  We detail here  the effects of diffuse scattering for collective atomic observables that depend on one- and two-point atomic correlations, such as the variance of a collective operator.  Expectation values that depend on higher order correlations follow in a straightforward manner.  

Consider first an inhomogeneous, single-particle collective operator of the form 
	\begin{align}
		\hat{X} = \sum_{n=1}^N c_n \hat{x}^{(n)}.  
	\end{align}
where the weighting coefficient $c_n$ depends on some property of atom $n$, such as its position $\mathbf{r}_n$.  Because $\hat{X}$ is a weighted sum over single atom operators, the equation of motion for its expectation value depends upon the evolution of the single-atom density operator, $\hat{\rho}^{(n)}$. Focusing only on the decoherent dynamics, summing over a single index $n$ in Eq. (\ref{Eq::GenME}) we obtain
	\begin{align}\label{eq::rhoi}
		\frac{d\hat{\rho}^{(n)}}{dt}\Big|_{\rm diff}=\gamma_s(\mbf{r}_n) \mathcal{D}_n[\hat{\rho}^{(n)}],
	\end{align}
from which the evolution of $\expects{\hat{X}}$ is given by 
	\begin{align}
		\frac{d}{dt}\expects{\hat{X}}\Big|_{\rm diff} & =\sum_{n=1}^N c_n \mbox{Tr} \bigg[ \hat{x}^{(n)}\frac{d\hat{\rho}^{(n)}}{dt}\Big|_{\rm diff}\bigg]  \nn \\
		&= \sum_{n=1}^N \gamma_s(\mbf{r}_n) c_n \expect{\mathcal{D}_n [\hat{x}^{(n)}]}. \label{Eq::1stOrderEvol} 
	\end{align}
For inhomogeneous collective operators that depend on pairs of atoms,
	\begin{align}
		\hat{O}  =  \sum_{m \neq n} c_m c_n  \hat{x}^{(m)} \hat{y}^{(n)},
	\end{align}
we require the joint density operator of the $m^{th}$ and $n^{th}$ atoms, $\hat{\rho}^{(m,n)}$, with equation of motion 
\begin{align}\label{Eq::rhoij}
	\frac{d}{dt}\hat{\rho}^{(m,n)}\Big|_{\rm diff} =\gamma_s(\mbf{r}_m)\mathcal{D}_m[\hat{\rho}^{(m,n)}]+ \gamma_s(\mbf{r}_n) \mathcal{D}_n[\hat{\rho}^{(m,n)}].
\end{align}
The evolution of $\expects{\hat{O}}$ due to diffuse scattering is then
\begin{align} \label{Eq::2ObservableEOM}
		\frac{d}{dt} \expects{\hat{O}} \Big|_{\rm diff}  = & \sum_{m \neq n} c_m c_n  \Big\{  \gamma_s(\mbf{r}_m) \expect{ \mathcal{D}_m [\hat{x}^{(m)}] \hat{y}^{(n)}} +  \gamma_s(\mbf{r}_n)\expect{ \hat{x}^{(m)} \mathcal{D}_n[ \hat{y}^{(n)}] } \Big\}. 
	\end{align}	

	This becomes particularly important when calculating the evolution of the variance of a collective spin wave,
	\begin{align}
		\big( \Delta F_z^i \big)^2 = \big\langle \big(\hat{F}_z^i \big)^2 \big\rangle - \big\langle \hat{F}_z^i  \big\rangle^2,
	\end{align}
with respect to some collective atomic state.  The second moment is a combination of one- and two-atom operators,
	\begin{align} \label{Eq::SecondOrderDecay}
		\big(\hat{F}_z^i \big)^2 = \sum_{n=1}^N \big[ \beta_i(\mathbf{r}_n) \big]^2 \hat{f}_z^{(n)} + \sum_{m \neq n} \beta_i(\mathbf{r}_m) \beta_i(\mathbf{r}_n) \hat{f}_z^{(m)} \hat{f}_z^{(n)}.
	\end{align}

	\chapter{Spin squeezing with atomic ensembles} \label{Ch::SpinSqueezing}
%

\section{Introduction }

The three-dimensional quantum light-matter interface developed in the previous chapter was intentionally presented in a general fashion to accommodate a variety of possible physical problems involving atomic ensembles.  In this chapter we apply it to study a particular protocol - collective spin squeezing via QND measurement.  Broadly speaking, spin squeezing refers to the reduction of the variance of a component of an angular momentum observable with respect to a ``standard" variance.  When the angular momentum in question is the collective spin of an ensemble of particles, spin squeezing can be a witness for pairwise entanglement \cite{GuhTot09}.  In this case, a collective spin can be considered squeezed when its variance is reduced below that of spin coherent state, where all of the uncorrelated spins are oriented along the same direction.  From this simple and admittedly limited description, it is clear that spin squeezing can be of both practical and foundational interest.  


Angular momentum, regardless of its specific physical origin, serves as a workhorse for precision quantum metrology in Ramsey spectroscopy, gravimetry, and magnetometry, for example.  These procedures rely on measurement of a phase $\phi$ acquired over some fixed time.  The uncertainty in the outcomes of projective measurements of the spin to determine the acquired phase -- projection noise -- sets the ultimate limit imposed by quantum mechanics on the precision.  The projection-noise limited resolution for Ramsey spectroscopy for mean angular momentum $J_\parallel \equiv |\expects{\hat{\mbf{J}}}|$ and variance orthogonal to the mean, $\Delta J_\perp^2 = \expt{\hat{J}_\perp^2} - \expt{\hat{J}_\perp}^2$ is
	\begin{align}
		\Delta \phi^2 = \frac{\Delta J^2_\perp }{ J_{||}^2 }.
	\end{align}
For the most classical pure state, when the mean value is equal to the total angular momentum oriented along some direction, the phase uncertainty is $\Delta \phi^2 = 1/(2J)$.  When the total angular momentum is the collective spin of a collection of particles, $\hat{\mbf{J}}$ = $\sum_{n=1}^N \hat{j}^{(n)}$, each with total angular momentum $j$, this describes a spin coherent state (SCS), with phase uncertainty
	\begin{align} \label{Eq::SCSUncertainty}
		\Delta \phi_{\rm SCS}^2 = \frac{1}{2Nj}.
	\end{align}
This is known as the standard quantum limit (SQL), as it arises from a fundamental assumption of quantized spins each with inherent quantum uncertainty orthogonal to the mean.  It is not, however, the ultimate limit imposed by quantum mechanics, which  arises from the Heisenberg-Robertson uncertainty relation, 
	\begin{align} \label{Eq::HeisenbergSpin}
		\Delta J_z^2 \Delta J_y^2 \geq \frac{1}{4} \big| \expt{ [\hat{J}_z, \hat{J}_y] } \big|^2 = \frac{1}{4} \big| \expt{\hat{J}_x} \big|^2.
	\end{align}
The mean spin along any direction sets a strict lower bound on the product of the transverse uncertainties.  A SCS, which has maximum mean spin along some direction, satisfies the equality in \erf{Eq::HeisenbergSpin} with equal variances in the transverse components.  Spin squeezing is the reduction of the variance in one transverse component.  Although \erf{Eq::HeisenbergSpin} must always be satisfied, the benefit to metrology comes from the fact that unwanted uncertainty can be siphoned out of the component of interest and fed into another component that does not couple to the measurement.  
	
For phase estimation using Ramsey interferometry, one can quantify the amount of squeezing with the Wineland squeezing parameter \cite{Wineland94}, which compares the phase uncertainty of a given state to that of a SCS [\erf{Eq::SCSUncertainty}], 
	\begin{equation} \label{Eq::StandardParam}
		\zeta \equiv \left(\frac{\Delta \phi}{\Delta \phi_{\text{SCS}}}\right)^2 = 2N j \frac{\Delta J^2_\perp }{ J_{||}^2 }.
	\end{equation}
The Wineland squeezing parameter is just one of a host of spin squeezing measures (see Ref. \cite{MaNor11} for detailed comparisons), but is useful in its operational understanding that for $\zeta < 1$, phase sensitivity improves over that afforded by a SCS.  

Additionally, when the constituent spins are qubits, $j = \smallfrac{1}{2}$, then $\zeta < 1$ also implies many-body entanglement \cite{TotBri09}.  The impact of quantum correlations on collective variance can be seen by representing a collective variance in terms of its atomic components.  For a symmetric collective variance, 
	\begin{align}\label{Eq::1DFundamentalVariance}
		\Delta F_z ^2 =  & \sum_{n=1}^N  \expect{ \big(\Delta \hat{f}_z^{(n)})^2}  + \sum_{m\neq n}  \expect{\Delta \hat{f}_z^{(m)}\Delta \hat{f}_z^{(n)}}.
	\end{align}
For qubits the first term is always $N/4$, and the second term vanishes for a SCS.  Thus, a reduction of the variance below that of a SCS requires negative correlations between the atoms.  This becomes tricker for larger spin ensembles \cite{VigTot14}, but the operational utility of $\zeta$ persists, regardless.  Thus, we will rely on and extend the definition of the spin squeezing parameter in \erf{Eq::StandardParam} throughout this chapter.  

In \srf{Sec::SME} the inhomogeneous collective spin wave measured in the balanced polarimeter was identified,
	\begin{equation} \label{Eq::FundSWQuant}
		\hat{F}_z^{0} \equiv  \sum_i \beta_{0} (\mbf{r}_{i}) \hat{f}_z^{(i)},
	\end{equation}
as were the spin waves that couple to higher-order field modes, $\hat{F}_z^{i}$, defined in \erf{Eq::SpinWave}.  For these inhomogeneous spin waves, we must tie the squeezing parameter directly to the measured quantities. For an initial mean spin polarization along $x$ and a small rotation around $y$, the mean signal by the mean spin wave component addressed by the probe, $\expects{\hat{F}_x^{0}} = \sum_i \beta_{0} (\mbf{r}_i) \expects{\hat{f}^{(i)}_x}$, and the projection noise contribution to the resolution of the measurement will be given by the spin wave variance, $\left( \Delta F^{0}_z \right)^2$. The projection-noise limited resolution of this rotation is therefore $\Delta \phi_{0} = \Delta F_z^{0}/\expects{\hat{F}_x^{0}}$.  

Furthermore, given a SCS initially polarized along $x$, the initial mean spin in the fundamental mode is $\expects{\hat{F}_x^{0}}_{\text{SCS}}= N^{(1)}_\eff f, $ with the effective atom number given by \erf{Eq::N1eff}.  The variance orthogonal to the mean spin is $\big( \Delta F_z^{0} \big)^2_{\text{SCS}}=N^{(2)}_\eff f/2$, and thus the projection noise limited resolution for a SCS preparation is
	\begin{align}
		\big( \Delta \phi^{0}_\text{SCS} \big)^2=\frac{1}{2f} \frac{N_\eff^{(2)}}{  \big(N^{(1)}_\eff \big)^2 }.
	\end{align}
Interestingly, this implies that in the absence of any squeezing, the phase sensitivity of a SCS varies as a function of geometry. Putting this together, the squeezing parameter for the measured spin wave is defined,
	\begin{equation} \label{Eq::SqueezingParam}
		\zeta \equiv \left(\frac{\Delta \phi^{0}}{\Delta \phi^{0}_{\text{SCS}}}\right)^2 = 2f \frac{ \big(N^{(1)}_\eff \big)^2 }{N_\eff^{(2)}} \frac{\big(\Delta F_z^{0}\big)^2}{\expect{\hat{F}_x^{0}}^2}.
	\end{equation}
This parameter quantifies the degree of ``quantum backaction," on a spin coherent state, accounting for the change in projection noise due to measurement as well as the damage done to both the mean spin polarization and variance due to optical pumping.  In the limit of symmetric coupling $N_\eff^{(1)} = N_\eff^{(2)} = N$, and the geometric squeezing parameter, \erf{Eq::SqueezingParam}, reduces to the standard Wineland squeezing parameter in \erf{Eq::StandardParam}.

In a real-world metrological application such as an optically probed atomic magnetometer~\cite{BudRom07, SewMit12}, spin rotations are measured by passing the probe through the atom sample and measuring the resulting Faraday rotation in a polarimeter.  In addition to spin projection noise, the measurement resolution is then subject also to probe shot noise and to ``technical noise," which includes detector electrical noise, atom number fluctuations, initial state preparation uncertainty, etc.  Under those circumstances, optimizing the squeezing parameter as defined in Eq. (\ref{Eq::SqueezingParam}) is distinct from optimizing the magnetometer sensitivity.

	\subsection{Spin squeezing via QND measurement: one-dimensional model} \label{Sec::1DSpinSqueezing}

In this section we present the procedure for generation of spin squeezing using QND measurements with a simplified, single-mode description of the light-matter interface.  For simplicity we consider spin-$\smallfrac{1}{2}$ atoms, for which there is no rank-2 tensor polarizability. Consider an ensemble of $N$ such atoms and a coherent probe with linear polarization along $\mathbf{e}_x$ and photon flux $\dot{N}_L$.  The effective interaction, \erf{Eq::StokesHam}, couples the spin of an atom to the 3-component of the field's Stokes vector -- the Faraday interaction.  A one-dimensional, symmetric, plane-wave model presumes that each atom experiences the same electric field amplitude. The Faraday interaction, discussed in Chapter \ref{Ch::DispersiveInt}, couples the symmetric collective spin, $\hat{F}_z = \sum_{n=1}^N \hat{f}_z^{(n)}$, to the field quadrature, $\hat{P}(z,t) = -i \big(\hat{a}_{y}(z,t) - \hat{a}\dg_y(z,t)\big)/\sqrt{2},$ via the Hamiltonian,
	\begin{align}
		\hat{H}_{\rm int} = - \hbar \sqrt{ \frac{ \kappa}{ 2} } \hat{F}_z \hat{P}(z,t),
	\end{align}
with measurement strength $\kappa = (\sigma_0/A)(4\gamma_0/9)$, corresponding to the rate of scattering into the probe mode per atom\footnote{For spin-$\smallfrac{1}{2}$, $|C^{(1)}|^2 = 4/9.$}.  The collective spin $\hat{F}_z$ is a QND observable because it commutes with the Hamiltonian, $[\hat{F}_z, \hat{H}_{\rm int}] = 0$.  As such, it does not evolve under coherent interaction or its bare Hamiltonian.  

The spin squeezing procedure begins by preparing the collective atomic state in a SCS polarized along the $x$-direction, with $\expt{ \hat{F}_z } = 0$ and $\Delta F_z^2 = N/4$.  The Faraday interaction generates a translation of the orthogonal quadrature $ \hat{X}(z,t) = \big( \hat{a}_{y}(z,t) + \hat{a}\dg_y(z,t) \big) /\sqrt{2 }$ in the Heisenberg picture, as discussed in \srf{Sec::HeisenbergDynamics}.  For each infinitesimal time interval $d t$ such that the interaction is weak ($\kappa d t \ll 1$) the output quadrature is $\hat{X}^{\rm out}(t) = \hat{X}^{\rm in}(t) - \sqrt{ \kappa / 2 } \hat{F}_z$, which follows from \erf{Eq::aoutInputOutput}.  The measurement operator, \erf{Eq::Measurement}, is
	\begin{align} \label{Eq::1DMeasurement}
		\hat{\mathcal{M}} 
		&= \sqrt{\frac{\dot{N}_L}{2} } \int_0^{\Delta t} dt' \hat{X}^{\rm in} (t') - \sqrt{ \frac{\dot{N}_L \CstrengthSq}{4}} \Delta t \hat{F}_z.
	\end{align}
An ensemble of such measurements will have fluctuations around the mean given by \erf{Eq::NoiseBreakdown},  
	\begin{align} 
		\Delta \mathcal{M}^2 
		& = \frac{\dot{N}_L \Delta t}{4} + \frac{\dot{N}_L \kappa}{4} \Delta t^2 \Delta F_z^2. \label{Eq::XoutFluctuations}
	\end{align}
The first term, $\Delta \mathcal{M}_{\rm SN}^2 = \dot{N}_L \Delta t/4$, describes the shot-noise vacuum fluctuations of the input field and the second term, $\Delta \mathcal{M}_{\rm PN}^2 = \big( \dot{N}_L \kappa/4 \big) \Delta t^2 \Delta F_z^2$ describes projection noise mapped onto the quadrature through the Faraday interaction.

For vacuum input and an initial SCS, we can assume Gaussian statistics for the shot noise and for the distribution of projective outcomes in the $z$-basis. The $\hat{X}$-quadrature measurement will initially be Gaussian distributed with zero mean and a variance given by \erf{Eq::XoutFluctuations}.  Using a Schr\"{o}dinger-picture description this can be shown, and the conditional statistics of the next measurement can be found.  The unnormalized Kraus operator for measurement outcome $\hat{X} = x$ is given by \erf{Eq::KrausComp0},
and the probability distribution for measurement outcomes is\footnote{The assumption of Gaussian statistics enters in this calculation when the finite distribution of projective $m$-values is approximated with a continuous Gaussian distribution. }  
	\begin{align}
		\mathbbm{P}\big[ \hat{X}(t) = x \big]  
		& =\exp \left[ \frac{- (x \Delta t)^2 }{2 \left( \frac{\Delta t}{2} +  \frac{\kappa \Delta t^2 }{2} \Delta F_z^2 \right)} \right]  .
	\end{align}
The measurement probabilities have been expressed in terms of $x\Delta t$, since the measurements involve integrating the output quadrature over $\Delta t$, as in \erf{Eq::1DMeasurement}. This result is identical to \erf{Eq::XoutFluctuations}; the variance of measurement outcomes increases through the addition of atomic projection noise.  
	
The (unnormalized) conditional state of the atoms, updated using the Kraus operators, is Gaussian in the $z$-basis,
	\begin{align}
		\hat{\rho}(t+dt) 
			& \approx   \int dm  \exp  \left( \frac{- \big(m- \expt{\hat{F}^{\rm}_z}_{\rm c} \big)^2}{ 2 (\Delta F_z)_c^2 } \right) \op{m}{m}  .
	\end{align}
with conditional mean spin projection along $z$,
	\begin{align}
		\expt{\hat{F}^{\rm}_z}_c = - \sqrt{ \frac{\kappa}{2} }  \frac{ \Delta F^2_z }{\big(\Delta \hat{X}^{\rm out} \big)^2 } x \Delta t, 
	\end{align}
and projection-noise variance,
	\begin{align} \label{Eq::CondVariance}
		\big( \Delta F_{z}\big)_c^2  
		= & \,  \Delta F_{z}^2 \left(  \frac{1}{1 + \xi } \right).
	\end{align}
This reduction in variance -- the conditional spin squeezing -- is determined by the coupling strength $\xi$ in \erf{Eq::CouplingStrength},  
	\begin{align} \label{Eq::Xi}
		\xi =  \frac{\mbox{OD}}{9} \gamma_0 \Delta t,
	\end{align}
where OD is the standard optical density in \erf{Eq::ODdef}. In the well-founded assumption of Gaussian statistics, the squeezing of the variance is not a function of the measurement record, but depends on the coupling strength.  If the input projection-noise fluctuations are not significant compared to the input shot noise, then the conditional spin squeezing will be negligible.  While the direction of the mean spin is stochastically altered depending on the measurement result, its length is not changed (the state remains pure) and the Wineland squeezing parameter decreases in time just as \erf{Eq::CondVariance}; $\zeta(t) = (1 + \xi)^{-1}$.

Equation (\ref{Eq::CondVariance}) describes the conditional variance over time intervals where decoherence can be ignored.  For longer times, the amount of squeezing will ultimately be constrained by incoherent photon scattering, which injects noise into the variance and shortens the mean spin.  In this case, the dynamics of the conditional variance can be found by taking $\Delta t \rightarrow dt$, as shown in Appendix \ref{Appendix::SMEDerivation}, to derive a differential equation \cite{MadMol04, TraDeu10, NorDeu12},
	\begin{align} \label{Eq::VarSpin1/2Symmetric}
		\frac{d}{dt}  \Delta F^2_z   & = - \kappa \big(  \Delta F^2_z \big)^2  - \frac{4\gamma_0}{9}\,  \Delta F^2_z + \frac{\gamma_0 }{9} \, N.
	\end{align}  
Henceforth we drop the subscript $c$ indicating ``conditional", as all time evolution results from measurement. Equation (\ref{Eq::VarSpin1/2Symmetric}) can be solved analytically, with details given in Appendix \ref{Appendix::AnalyticSolution}, 
	\begin{align} \label{Eq::AnalyticSolutionMainText}
	 	\Delta F^2_z(t) = \frac{N}{4} \, \frac{\sqrt{ \mbox{OD}+1} + \tanh \left[ \sqrt{\text{OD}+1}\frac{2}{9} \gamma_0 t \right] }{\sqrt{\mbox{OD}+1} + \big(\frac{ \text{OD} }{2}+1\big)\tanh\left[\sqrt{\text{OD}+1}\frac{2}{9} \gamma_0 t \right]}.
	\end{align}
In the limit of short times, $ \gamma_0 t  \ll 1$, we recover the expression for QND squeezing in the absence of decoherence, \erf{Eq::CondVariance},  $ \Delta F^2_z(t) \approx (N/4)(1+\xi)^{-1}$.  In the opposite limit of long times, $\gamma_0 t \rightarrow \infty$, and large optical density, OD $\gg 1$, we find the expected scaling $ \Delta F^2_z(t\rightarrow \infty) \propto \mbox{OD}^{-1/2}$ \cite{HamPol10}.  

The squeezing parameter in \erf{Eq::StandardParam} is constructed from the magnitude of the mean spin as well as the variance.  The decoherent dynamics result from photon scattering, the equation of motion is $\frac{d}{dt} \expt{ \hat{F}_x} = - \frac{\gamma_0}{3} \expt{ \hat{F}_x},$ which is solved by
	\begin{align}
		\expt{\hat{F}_x(t)} = \frac{N}{2} \exp \left( -\frac{1}{3} \gamma_0 t \right).
	\end{align}
Putting this together with \erf{Eq::AnalyticSolutionMainText}, the analytic solution for the symmetric Wineland squeezing parameter [\erf{Eq::StandardParam}] is
	\begin{align} \label{Eq::AnalyticSolution}
		\zeta(t) = \exp \left( \frac{2}{3} \gamma_0 t \right) \frac{\sqrt{ \mbox{OD}+1} + \tanh \left[ \sqrt{\mbox{OD}+1}\frac{2}{9} \gamma_0 t \right] }{\sqrt{\mbox{OD}+1} + \big(\frac{ \text{OD} }{2}+1\big)\tanh\left[\sqrt{\mbox{OD}+1}\frac{2}{9} \gamma_0 t \right]}.
	\end{align}
There is a slight subtlety that while in a single shot of an experiment the variance is deterministically reduced, in order to make use of the metrological squeezing, one must know the mean which is a function of the stochastic measurement record.  Ignorance of this mean gives no metrological advantage, as it effectively ``undoes" the expected squeezing.  

The analytic solution for the conditional variance, \erf{Eq::AnalyticSolution}, applies for a symmetrically coupled ensemble of spin-$\smallfrac{1}{2}$ atoms.  When the constituent spins are composed of hyperfine spins with $f>\smallfrac{1}{2}$, such as the ground states of $^{133}$Cs, the dynamics are more complicated.  Using a generalization of the Holstein-Primakoff approximation, squeezing of symmetric spin-$f$ ensembles undergoing decoherence was studied in Ref. \cite{NorDeu12}.

\section{Squeezing of spin waves: three-dimensional model}

The intent of the previous section was to give a concrete and relatively simple description of how a light-matter interaction following by projective measurement of the field can give rise to a reduction in the variance of an atomic angular momentum.  The remainder of this chapter focuses on extending this concept to include spatial degrees of freedom that inevitably play a part in a laboratory experiment.  
In Chapter \ref{Ch::3DInterface}, we laid out the mathematical tools necessary to study this problem.   As they are probed by a classical laser probe, a collection of atoms indistinguishably and elastically scatters paraxial electromagnetic radiation in a manner equivalent to a set of linearly polarizable particles.  Thus, a great deal of qualitative and quantitative information can be obtained from classical radiation theory.  Using the semiclassical scattering model from Chapter \ref{Ch::PropagatingFields}, we can characterize the coherent coupling that results from the portion of scattered radiation in the spatial mode of the probe.  However, we are not detecting Faraday rotation per se; rather, when we can resolve the the projection noise fluctuations in the spatial mode of the probe then spin squeezing is generated through measurement backaction.  This spatial dependence is incorporated in the definition of the effective optical density, OD$_{\rm eff}$, defined in \erf{Eq::ODeff}.  Then, with a fully quantum model for the dispersive, three-dimensional light-matter interaction, developed in Chapter \ref{Ch::3DInterface}, we can quantitatively assess the effects of decoherence.  We are ultimately concerned with entanglement between the atoms in the ensemble which manifests through spin squeezing of a collective observable.  The effects of such a measurement on the collective quantum state (in the Schr\"{o}dinger picture) of the atomic ensemble is described by a multi-mode stochastic master equation, developed in \srf{Sec::SME}.  
	

\subsection{The dynamical evolution of squeezing}

The multi-mode Faraday Hamiltonian, [\erf{Eq::MultiModeFARADAY}], serves as the QND interaction which couples the paraxial field to the spin waves.  The signal we seek to measure with balanced polarimetry arises from different spin projections associated with the eigenstates of the \emph{fundamental} spin wave that couples to the spatial mode of the probe, $\hat{F}_z^{0}$.  Whereas in magnetometry these shot-to-shot projection noise variations are the limiting factor, in the context of creating a spin squeezed state, these variations from the mean value represent the ``signal" one seeks to resolve over the laser shot noise. The projection noise in the polarimeter is proportional to the fundamental spin wave variance, \erf{Eq::FundamentalVariance}.  The amount of squeezing, quantified by the geometric squeezing parameter defined in \erf{Eq::SqueezingParam}, requires dynamically evolving this variance as well as the mean spin.  

To determine the squeezing as function of time, we employ the SME in \erf{Eq::HomodyneSME} to track $( \Delta{F}_z^{0})^2$ and $\expects{\hat{F}_x^{0}}$.  For ensembles with large numbers of atoms, we can work in the central-limit approximation where fluctuations in the spin waves are treated as Gaussian random variables~\cite{HamPol10, VasSor12}.  Following  \cite{JacSte06, StocktonThesis}, the SME then couples solely means and covariances. The moments of the fundamental spin wave that characterize the spin squeezing parameter then evolve according to
	\begin{subequations}  \label{Eq::FullSpinWaveEOM}
	\begin{align}
		d\varFz =&  -\kappa  \big[ \left(\Delta F_z^{0} \right)^2 \big]^2 dt + d\varFz\Big|_{\rm dec} , \label{Eq::Var00Evolution}\\
	d \expect{\hat{F}^{0}_x} = & \sqrt{ \frac{\kappa}{4} } \big  \langle\mathcal{H}_{0}\big[\hat{F}^{0}_x\big]\big  \rangle \, dW  + \frac{\kappa}{4} \sum_{i} \big  \langle \mathcal{L}_{i}\big[ \hat{F}^{0}_x \big] \big \rangle dt + d\expect{\hat{F}^{0}_x} \Big|_{\rm dec}, \label{Eq::Mean00Evolution} 
	\end{align}
	\end{subequations}
The dynamical maps that describe measurement backaction, $\mathcal{H}_{0}[\cdot]$, and collective decoherence, $\mathcal{L}_{i}[ \cdot ]$, are defined in Eqs. (\ref{Eq::HSuperoperator}) and (\ref{Eq::LSuperoperator}), respectively.  Because we assume the fundamental mode is measured with unit efficiency, diffuse scattering by local spontaneous emission is the only process contributing to the decoherence of the variance $\varFzText $.  Collective radiation into other transverse modes commutes with $\hat{F}_z^{0}$ and does not contribute to any decay or noise injection into the fundamental variance.  In contrast, the mean spin $\expects{\hat{F}^{0}_x}$ decoheres due to both diffuse scattering and collective scattering into other unmeasured paraxial modes, and further it evolves stochastically from measurement backaction.  However, the contributions to the dynamics from both collective scattering and continuous measurement are small in comparison to diffuse scattering and can be neglected when the radiation pattern of the cloud is well matched to that of the probe.  Additionally, local inhomogeneous coherent local dynamics from the residual light shift in \erf{Eq::ResidualLightShift} can affect the metrologically relevant spin squeezing by dephasing the mean spin.  However, this can be compensated for through additional control techniques\footnote{By employing a two-color probe on the D1 and D2 resonance lines for alkali atoms, one can cancel the residual nonlinear spin dynamics associated with the rank-2 tensor light shift, while retaining the Faraday effect that is used in the QND measurement of the collective spin \cite{EnriquePrivate}.} and thus does not appear in \erf{Eq::Mean00Evolution}.

We consider the moment evolutions, \erf{Eq::FullSpinWaveEOM}, with the initial condition that the ensemble is in a SCS polarized along $x$. The initial mean spin and variance are $\expects{ \hat{F}_{x}^{0} (t_0) }= N_{\rm eff}^{(1)} f$ and $\big( \Delta F_z^{0} (t_0) \big)^2 =  N_{\rm eff}^{(2)}f/2$ respectively.  Along with the cross-sectional area of the probe laser,  $N_{\rm eff}^{(2)}$ specifies the effective optical density, OD$_{\rm eff}$ defined in \erf{Eq::ODeff} (we drop the superscript in this chapter, as we refer here only to one effective optical density).  The  OD$_{\rm eff}$ is the critical geometric parameter for determining how the atomic density distribution influences collective scattering into the probe mode and ultimately leads to spin squeezing. Both of these effective atom numbers are determined solely by the cloud shape and beam geometry, and can be found from the semiclassical model in Sec. \ref{Sec::SME}.  

For times short compared to the photon scattering rate, where decoherence is negligible, the mean spin is essentially constant and the spin variance is affected only by measurement backaction. The solution to \erf{Eq::FullSpinWaveEOM} takes the familiar form
	\begin{align} 
		\big( \Delta F_z^{0} (t) \big)^2 & 
		 = \big( \Delta F_z^{0} (0) \big)^2 \left(  \frac{1}{1+\xi} \right)
	\end{align}
where $\xi$ is the integrated coupling strength in Eq. (\ref{Eq::CouplingStrength}).  In this limit the mean spin is constant, $\expt{\hat{F}_x(t)} = \expt{\hat{F}_x(0)}$,  and the geometric squeezing parameter is
	\begin{align}
		 \zeta = \frac{1}{1+\xi}. \label{Eq::ShortTimeSolution}
	\end{align}
This result is nearly identical to that for symmetric coupling, \erf{Eq::CondVariance}, with the substitution of OD$_{\rm eff}$ (and using $\chi^{(1)}$ for a spin-$f$ atom).  When diffuse scattering can be neglected, the description is effectively single-mode, with geometry serving only to modify the coupling strength.  

For longer times, decoherence due to diffuse photon scattering must be included.  The mean spin will depolarize according to Eq. (\ref{Eq::1stOrderEvol}),
	\begin{equation}\label{Eq::MeanSpinArb}
		\frac{d}{dt} \expect{\hat{F}^{0}_x} = \sum_{n=1}^N \gamma_s(\mbf{r}_n)\beta_{0}(\mbf{r}_i) \big\langle \mathcal{D}_n \big[\hat{f}_x^{(n)}\big] \big\rangle,
	\end{equation}
with local decoherence acting via the map $\mathcal{D}_n[\cdot]$ in \erf{Eq::DiffuseME}.  The variance involves both single atom and pairwise atomic correlations,
	\begin{equation}  \label{Eq::VarianceDecompositionApprox}
		\frac{d}{dt}\varFz =  \sum_{n=1}^N \big[ \beta_{0}(\mbf{r}_n) \big]^2 \frac{d}{dt}\big\langle(\Delta \hat{f}_z^{(n)})^2\big\rangle  +  \sum_{m\neq n} \beta_{0}(\mbf{r}_m) \beta_{0}(\mbf{r}_n)  \frac{d}{dt}\big\langle \Delta \hat{f}_z^{(m)}\Delta \hat{f}_z^{(n)} \big\rangle.  
	\end{equation}
The first term is the spin projection noise of the uncorrelated spins and the second term contains the correlations that generate spin squeezing.   Following Eqs. (\ref{Eq::1stOrderEvol}) and  (\ref{Eq::2ObservableEOM}), these correlations decay due to diffuse scattering according to
	\begin{subequations} \label{Eq::Correlations}
	\begin{align}
& \frac{d}{dt}  \sum_{n=1}^N \big\langle  (\Delta \hat{f}_z^{(n)})^2  \big\rangle  \big|_{\rm dec}  =  \sum_{n=1}^N  \gamma_s(\mbf{r}_n) \Big\{  \big\langle\mathcal{D}_n \big[ \hat{f}_z^{(n)2} \big] \big\rangle  -2 \big\langle \mathcal{D}_n \big[ \hat{f}_z^{(n)} \big] \big\rangle \big\langle \hat{f}_z^{(n)} \big\rangle  \Big\} \label{eq::NoiseTerm2}, \\
		 &\frac{d}{dt}\sum_{m \neq n}  \big\langle \Delta \hat{f}_z^{(m)}\Delta \hat{f}_z^{(n)}   \big\rangle \big|_{\rm dec}  = \nn \\
		 & \quad \quad \sum_{m\neq n} \! \Big\{ \! \gamma_s(\mbf{r}_m) \big\langle \Delta\mathcal{D}_m \big[ \hat{f}_z^{(m)} \big]\Delta \hat{f}_z^{(n)}\big\rangle  \! +\!  \gamma_s(\mbf{r}_n) \big\langle \Delta \hat{f}_z^{(m)}\Delta \mathcal{D}_n \big[\hat{f}_z^{(n)}\big] \big\rangle \! \Big\} \label{eq::correlationDecay2}.
	\end{align}
	\end{subequations}

	\section{Spin-$\frac{1}{2}$ ensembles}

We restrict our attention to ensembles of spin-$\smallfrac{1}{2}$ atoms to focus on spatial effects without the complications that arise for ensembles with larger-spin. Using the fact that the local scattering rate is proportional to the probe intensity, $\gamma_s(\mathbf{r}) = \gamma_{0}\beta_{0}(\mathbf{r})$, the mean spin evolution of \erf{Eq::Mean00Evolution} is 
	\begin{equation} \label{Eq::Mean00GammaBeta}
		\frac{d}{dt} \expect{\hat{F}^{0}_x} = -\frac{\gamma_{0}}{3} \sum_{n=1}^N \big[ \beta_{0}(\mbf{r}_n) \big]^2 \big\langle \hat{f}_x^{(n)} \big\rangle.
	\end{equation}
The local decoherence does respects neither the orthogonality of the transverse paraxial modes nor the collective nature of the coherent interaction \cite{ChaGer08, BarGer10} and we will see that the diffuse scattering acts to couple the fundamental spin wave to higher order spin waves.  
 
Because the transverse modes are orthogonal in a plane at a fixed $z$,  we can  derive a set of coupled equations by decomposing products of the spatial weighting coefficients, \erf{Eq::Beta}, in the basis of mode functions as follows:
	\begin{equation}
		\big[ \beta_{0}(\mbf{r}_\perp, z) \big]^2 =|u_{0}(\mathbf{r}_{\perp}, z)|^4  =  \sum_{i} c^{0}_{i}(z) \beta_{i}(\mathbf{r}_\perp, z) ,
	\end{equation}
with $z$-dependent projection coefficients,
	\begin{equation}
		c^{0}_{i}(z) \equiv \frac{1}{A} \int d^2 \mathbf{r}_\perp \big[ u_{0}(\mathbf{r}_\perp, z) \big]^2 u^*_{0}(\mathbf{r}_\perp, z) u_{i}(\mathbf{r}_\perp, z).
	\end{equation}
Performing the sum over atoms within each coarse-grained slice, it follows that \erf{Eq::Mean00GammaBeta} can be expressed in terms of the $z$-local, coarse-grained collective spin waves just as in \erf{Eq::LocalSpinWave},
	\begin{align}  
		\frac{d}{dt} \big\langle \hat{F}^{0}_{x} \big\rangle 
		= &  - \frac{ \gamma_{0}}{3} \sum_{k} \sum_{i} c^{0}_{i} (z_k) \big\langle \hat{F}_x^{i}(z_k) \big\rangle, \label{Eq::MeanZsliced}
	\end{align}
The total spin wave for a given transverse mode is the sum over the local spin waves, $\hat{F}^{i}_x = \sum_k \hat{F}^{i}_x (z_k)$ [\erf{Eq::OrthogonalSpinWaves}]. The mean spin in the fundamental mode couples, at each longitudinal slice $z_k$, to other $z$-local spin waves $\expect{\hat{F}_x^{i}(z_k)}$.  Thus, in order to find the mean spin in the fundamental mode, we must also track the dynamics of $\expect{\hat{F}_x^{i}(z_k)}$.  In general, the $i$-mode, $z$-local spin wave is also coupled to the other spin waves within its slice $z_k$ and evolves according to
	\begin{align}  \label{Eq::MeanZsliced}
		\frac{d}{dt} \expect{\hat{F}^{i}_{x}(z_k)} =  - \frac{ \gamma_{0}}{3} \sum_{j} c^{i}_{j} (z_k) \expect{\hat{F}_x^{j}(z_k)},
	\end{align}
with projection coefficients $c^{i}_{j} (z_k)$ given in \erf{Eq::ProjCoeff}.  Details of this derivation are found in Appendix \ref{ApSec::SpinHalf} and the form of the projection coefficients for cylindrically symmetric $l=0$ Laguerre-Gauss modes is given in Appendix \ref{Appendix::ProjectionCoefficients}. The initial conditions, \erf{Eq::meanSlice}, account for the matching between the probe mode and cloud geometry.  By projecting onto the spin waves, we obtain a hierarchy of coupled equations within the collective $z$-local spin waves.  

 The effect of diffuse scattering on the evolution of the collective spin variance follows in an analogous manner.  For spin-$\smallfrac{1}{2}$, $\Delta f^2_z =1/4$ for all atoms.  The map for local decoherence, $\Delta \mathcal{D}_n\big[\hat{f}^{(n)}_z\big] = -2 \Delta \hat{f}^{(n)}_z/9$, corresponds to decay of spin-spin correlations with no feeding of coherences.  The evolution of the fundamental spin wave variance, \erf{Eq::Var00Evolution}, simplifies to
	\begin{align} \label{Eq::SpinHalfVariance}
		 \frac{d}{dt}  \varFz  = & -\kappa \Big[ \big( \Delta F_z^{0} \big)^2 \Big]^2   -\frac{2\gamma_{0}}{9}  \sum_{m=1}^N \sum_{n=1}^N \big[ \beta_{0}(\mathbf{r}_m)  +\beta_{0}(\mathbf{r}_n) \big] \beta_{0}(\mbf{r}_m)\beta_{0}(\mbf{r}_n) \expect{\Delta \hat{f}_z^{(m)}\Delta \hat{f}_z^{(n)}} \nn \\
		& + \frac{\gamma_{0}}{9} \sum_{n=1}^N \big[ \beta_{0}(\mbf{r}_n) \big]^3, 
	\end{align}
where again we have used \erf{Eq::LocalScatRate}.  The first term describes squeezing of the variance due to measurement backaction, the second represents decay of correlations due to diffuse scattering, and the third is the noise injected into the variance from spin flips (optical pumping).  Note the similarity to the equation for the variance in the symmetric one-dimensional model [\erf{Eq::VarSpin1/2Symmetric}], which is obtained for $\beta_0(\mathbf{r}) \rightarrow 1$.

Following the same procedure as for the mean spin, the decay terms are projected onto higher order $z$-local spin waves, 
	\begin{equation}  \label{Eq::ProjectedSpin1/2VarianceEOM}
		 \frac{d}{dt} \varFz =  -\kappa \Big[ \big( \Delta F_z^{0} \big)^2 \Big]^2  -\frac{4 \gamma_0}{9} \sum_{i} \sum_{k,k'} c^{0}_{i}(z_k) \big\langle \Delta \hat{F}_z^{0}(z_k) \Delta \hat{F}_z^{i}(z_{k'})\big\rangle  + \frac{ \gamma_{0}}{9}N^{(3)}_\eff. 
	\end{equation}
Here, another effective atom number governing the injection of noise through optical pumping arises,
	\begin{align} \label{Eq::N3}
		N^{(3)}_\eff \equiv \sum_{n=1}^N \big[ \beta_{0}(\mbf{r}_n) \big]^3.
	\end{align}
Equation (\ref{Eq::ProjectedSpin1/2VarianceEOM}) is a covariance description of the dynamics, similar to that commonly employed for spin squeezing \cite{MadMol04,TraDeu10,Tra11}, but which also accounts for local decoherence from first principles.  To solve for the fundamental variance, we must track the evolution of the covariances between coarse-grained slices and between transverse modes, $\langle \Delta \hat{F}_z^{i}(z_k) \Delta \hat{F}_z^{j}(z_{k'})\rangle = \expects{\hat{F}_z^{i}(z_k)\, \hat{F}_z^{j}(z_{k'}) } -\expects{\hat{F}_z^{i}(z_k)}\expects{ \hat{F}_z^{j}(z_{k'}) }$.  Equations of motion for these covariances follow readily from the SME, but the exact form of the equations does not serve to enlighten.  A detailed derivation is given in Appendix \ref{Appendix::SpinWaveEquations}.

\section{Numerical results for spin-$\frac{1}{2}$}

Using our formalism we can calculate the moment dynamics and find the peak achievable squeezing in the presence of decoherence.  We now consider the fundamental effects of geometry and the optimization of experimentally relevant quantities to maximize spin squeezing.  The geometry of the atom-laser system plays two distinct roles in determining the amount of achievable squeezing. First,  OD$_{\rm eff} \propto N_{\rm eff}^{(2)}$, Eq. (\ref{Eq::ODeff}), is a purely geometrical quantity, derivable from the semiclassical model  (see also \cite{MulPol05}).   The OD$_{\rm eff}$  sets the measurement strength, $\xi$, that characterizes the amount of light that is collectively scattered into the spatial mode of the probe.  Second, because of the inhomogeneous intensity profile of the laser mode, the rate of diffuse photon scattering that causes local decoherence and ultimately caps the amount of generated squeezing varies across the cloud.  Further complications arise from the fact that optical pumping both injects noise into the spin wave variance and causes a decay of the mean spin.  

For simulations, we choose the ensemble to be a cylindrically symmetric Gaussian cloud with average density described by \erf{Eq::AtomicDistribution}.  The transverse and longitudinal $1/e^2$ variances are given by $\sigma_\perp^2$ and $\sigma_z^2$, respectively, and $\eta_0$ is the peak density.  The total atom number is found by integrating over the cloud, $N = \int d^3 \mathbf{r}\, \eta(\mathbf{r})$.  To characterize the geometry of the atomic distribution we use the \emph{aspect ratio}, defined as AR $\equiv \sigma_z/\sigma_\perp$.  A longitudinally extended, pencil-shaped cloud commonly employed in cold, dipole-trapped atomic ensemble experiments has an AR $\gg 1$; a pancake-shaped cloud that is much wider than it is long has an AR $\ll 1$.  The cylindrical symmetry of Gaussian clouds motivates the use of Laguerre-Gauss (LG) mode functions, described in Appendix \ref{Appendix::LGModes}.  LG modes are denoted by a radial index $p$ and azimuthal index $l$; $i \rightarrow p,l$ throughout.  The probe is prepared in the fundamental TEM$_{00}$ mode, and since the laser and cloud exhibit no azimuthal dependence, the coupled modes in the descriptions in \erf{Eq::MeanZsliced} and \erf{Eq::ProjectedSpin1/2VarianceEOM} have azimuthal mode index $l=0$.

The quantities of interest, including peak squeezing, are found by solving for the evolution of the collective mean spin and variance, Eqs.  (\ref{Eq::Mean00GammaBeta}) and (\ref{Eq::SpinHalfVariance}), and then calculating the spin squeezing as a function of time.  This requires numerically integrating a set of differential equations for the $z$-local spin waves in all modes, for every longitudinal slice $z_k$.  Diffuse scattering couples spin waves within a slice, and measurement couples spin waves within and between slices.  The initial conditions, found in \srf{ApSec::InitialConditions}, 
	\begin{align}	
		\expt{\hat{F}^{p0}_x(z_k)}_{\rm SCS} &=  \frac{\delta z}{2}  \int d^2 \rperp \eta(\rperp,z_k)\beta_{p0}(\rperp,z_k)
 \\
		\expt{\Delta \hat{F}^{p0}_x(z_k) \Delta \hat{F}^{p0}_z(z_{k'}) }_{\rm SCS} & =  \frac{\delta z}{4}  \int d^2 \rperp \eta(\rperp,z_k)\beta_{p0}(\rperp,z_k) \beta_{p'0}(\mbf{r}_{n_k})
,
	\end{align}
and coupling coefficients, \erf{Eq::ProjCoeff}, both decrease with increasing $p$ and allow a truncation at some mode index $p_{\rm max}$ to get a finite set of equations.  The width of the coarse-grained $z$-slices can be reduced until convergence at a desired level of precision is attained.  In the following numerical analyses, the squeezing parameter \erf{Eq::SqueezingParam} is given in dB, --10 log$_{10} [\zeta^{-1}(t)]$, so that higher values indicate more squeezing.  

	\subsection{Geometric effects of local decoherence for a fixed rate of squeezing}\label{Sec::GeomEffects}

	\begin{figure}[!h]
	\centering
        		\includegraphics[scale=0.75]{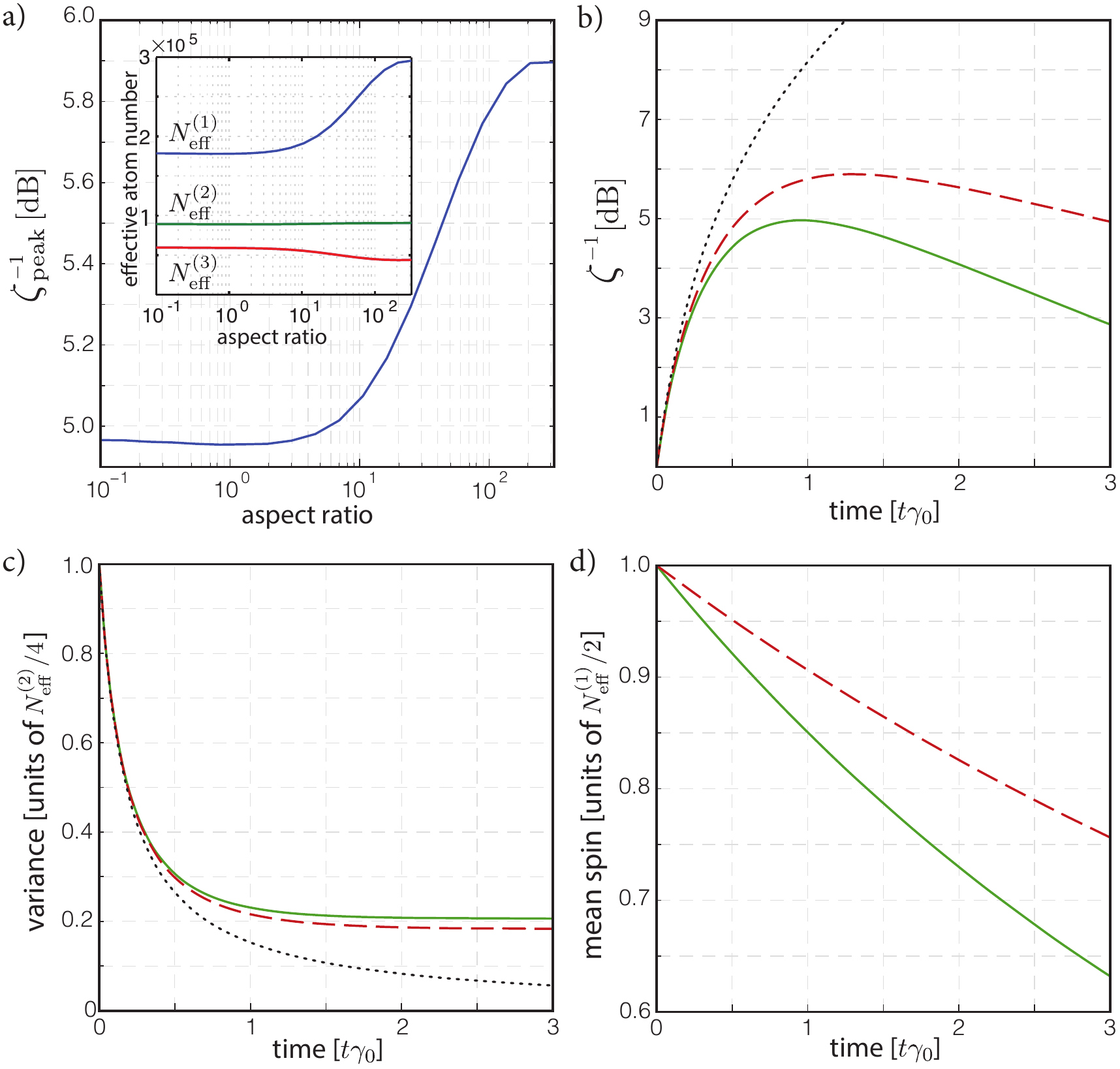}
       		 \caption[Spin squeezing for fixed OD$_{\rm eff}$ with varying geometry]{ Squeezing dynamics for a fixed OD$_\eff = 50$ and different atomic cloud geometries.  The laser probe is a TEM$_{00}$ mode with beam waist $w_0 = 20$ $\mu$m.   a) Peak squeezing denoted as the inverse of the squeezing parameter, $\zeta^{-1}$ in dB, as a function of aspect ratio of the cloud.  The inset shows effective atom numbers as a function of aspect ratio; $N_\eff^{(2)}$ is constant by design.  b) Comparison of squeezing dynamics for clouds with AR = 0.1 (solid green line) and AR $= 316$ (dashed red line).  The behavior in the absence of decoherence, \erf{Eq::ShortTimeSolution} (dotted black line), is plotted at the same OD$_\eff$, showing  agreement for short times. c) Dynamics of the spin wave variance for the two clouds, normalized by dividing each by its initial variance, $N^{(2)}_{\rm eff}/4$.  d) Dynamics of the mean spin for the two clouds, normalized by dividing each by its initial mean spin, $N^{(1)}_{\rm eff}/2$.  For fixed OD$_\eff$, the superior squeezing of the pencil-shaped cloud over the pancake-shaped cloud is attributed to slower decay of the mean spin.}  \label{Fig::SpinHalfCompareSqueezing}
   	\end{figure}

To gain physical insight, in this section we fix the OD$_\eff$ as we vary the geometry in order to isolate the effects of local decoherence as they relate specifically to the squeezing parameter, \erf{Eq::SqueezingParam}. 
This requires varying the peak atomic density, $\eta_0$, as a function of the cloud geometry in order to keep OD$_\eff \propto N_\eff^{(2)}/A$ constant.  Figure \ref{Fig::SpinHalfCompareSqueezing} shows the resulting spin squeezing for different cloud geometries for a fixed beam waist, $w_0 = 20$ $\mu$m.  The effective optical density is held constant, OD$_\eff = 50$, which guarantees identical squeezing in the absence of decoherence for any geometry. Fig. \ref{Fig::SpinHalfCompareSqueezing}(a) shows the peak squeezing as a function of the AR.  An increase in peak squeezing accompanies an increasing aspect ratio, indicating that decoherence is less detrimental to longitudinally extended clouds.  The dynamics of the squeezing parameter are plotted in Fig. \ref{Fig::SpinHalfCompareSqueezing}(b) for the opposing cases of a pancake-shaped cloud with AR = 0.1 and a pencil-shaped cloud with AR = 316.  For comparison, the short-time solution \erf{Eq::ShortTimeSolution} is shown, which describes the squeezing for either cloud in absence of decoherence.  

These results can be understood by separately examining the dynamics of the spin wave variance and the mean spin, as seen in Figs. \ref{Fig::SpinHalfCompareSqueezing}(c-d).  The effects of decoherence lead to different steady state values of the fundamental spin wave variance in Fig. \ref{Fig::SpinHalfCompareSqueezing}(c)  because the noise injection due to optical-pumping-induced spin flips, set by $N_{\rm eff}^{(3)}$, is slightly smaller for the pencil than for the pancake (see subplot in Fig. \ref{Fig::SpinHalfCompareSqueezing}(a)).  More importantly, the decay rate of the mean spin is a strong function of the AR, as seen in  \ref{Fig::SpinHalfCompareSqueezing}(d).  For a fixed OD$_\eff$, under consideration here, different cloud geometries correspond to different  $N_{\rm eff}^{(1)}$, which determines the mean spin of the ensemble addressed by the beam.  The pencil geometry addresses a larger $N_{\rm eff}^{(1)}$ when compared to the pancake geometry, as seen in the subplot of \ref{Fig::SpinHalfCompareSqueezing}(a).  In addition, for the pencil geometry $N_{\rm eff}^{(1)}$  also decays more favorably.  This occurs because for a fixed OD$_\eff$, in the pencil geometry a large fraction of the atoms are spread far from the beam waist where rates of optical pumping are lower.  For the pancake geometry, to achieve the same OD$_\eff$, more of the atoms the we address are concentrated in the high intensity region and more quickly depolarize.

	\subsection{Optimizing geometry for fixed atom number}

	\begin{figure}[!ht]
	\centering
    		 \includegraphics[scale=0.45]{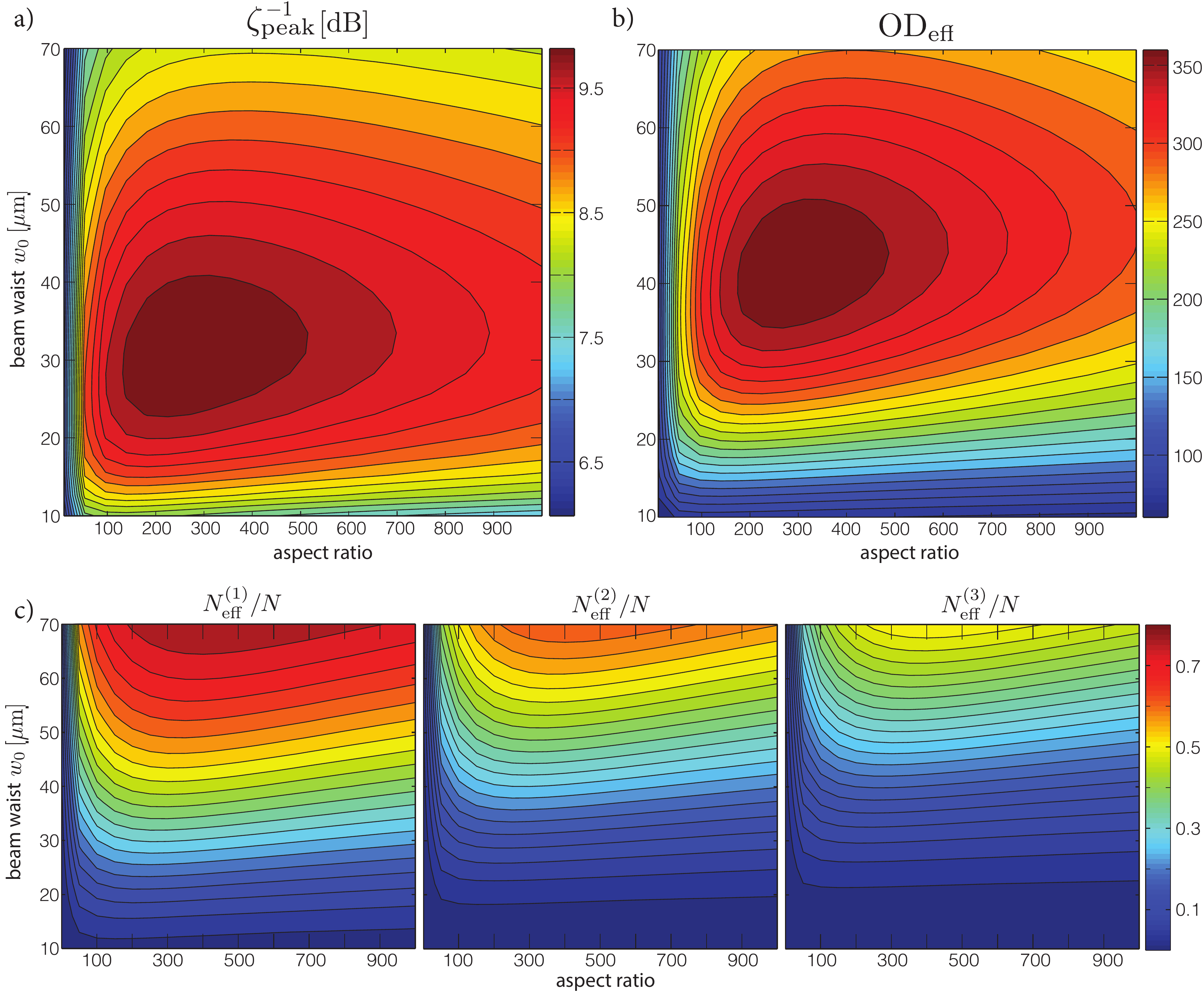}
       		 \caption[Spin squeezing for fixed atom number with varying geometry]{ Squeezing for different cloud geometries with Gaussian atomic density distribution, \erf{Eq::AtomicDistribution}, and a fixed total atom number $N = 9.84 \times 10^6$.  a) Contours of peak squeezing, $\zeta^{-1}$ in dB, as a function of cloud aspect ratio and laser probe beam waist. b) Contours of the coupling strength, OD$_{\rm eff}$.  The difference between the optimal coupling strength and the resulting squeezing depends on the balance between coherent interactions and decoherence, characterized by different effective atom numbers,  $N_{\rm eff}^{(1)}$, $N_{\rm eff}^{(2)}$, and $N_{\rm eff}^{(3)}$, shown in c) on the same scale. }  \label{Fig::SpinHalfScaling}
    	\end{figure}

We gain further insight into the nature of the atom-light interface by keeping the total atom number $N$ fixed and optimizing the cloud dimensions for peak squeezing.  We fix the peak density at $\eta_0 = 5 \times 10^{11}$ cm$^{-3}$, typical of dipole-trapped atoms, and keep the total atom number constant, $N = 9.8 \times 10^6$.  The cloud volume is fixed at $V = 4.19 \times 10^7$ $\mu$m$^3$, chosen to be within typical experimental constraints\footnote{Although arbitrary atomic clouds at this volume may not be physically realizable in a laboratory, this volume was chosen as a reference.  It is the volume for a cigar-shaped cloud with dimensions $\sigma_\perp = 100$ $\mu$m and $\sigma_\perp = 1000$ $\mu$m.}.  In Fig. \ref{Fig::SpinHalfScaling}(a), we plot contours of peak squeezing as a function of aspect ratio and beam waist.  The optimal peak squeezing, $\zeta_{\rm opt}^{-1} = 10.0$ dB, is found for AR $= 256$ at a beam waist of $w^{\rm opt}_0=31$ $\mu$m.  At the optimal geometry, the cloud length extends over several Rayleigh ranges, $\sigma_z/z^{\rm opt}_R = 2.42$, and the transverse width of the cloud is slightly larger than the beam waist, $\sigma_\perp/w^{\rm opt}_0 = 1.09$. 
	
To further understand the region of peak squeezing, in Fig. \ref{Fig::SpinHalfScaling}(b), we plot contours of OD$_{\rm eff}$.  Comparison of Figs. \ref{Fig::SpinHalfScaling}(a-b) shows that the optimal peak squeezing occurs in a parameter region where OD$_{\rm eff}$ is high, as expected.  However, the optimal peak squeezing arises from a balance between high OD$_{\rm eff}$ with low noise injection into the spin wave variance and low decay of the mean spin.   Figure \ref{Fig::SpinHalfScaling}(c) shows the fraction of total atoms contributing to the mean spin, $N_{\rm eff}^{(1)}/N$, to the effective optical density, $N_{\rm eff}^{(2)}/ N$, and to the noise injection $N_{\rm eff}^{(3)}/ N$.  As the cloud becomes too long and narrow, there does not exist a beam waist that can address a sufficiently large number of atoms while keeping a high OD$_{\rm eff}$.  Said another way, when the cloud becomes too long, the diffraction of scattered light is too large to effectively mode match with the probe field, as seen in Fig. \ref{Fig::ModeMatching}(c).  Similarly, too small a waist leaves many atoms outside the Rayleigh range and too large a waist increases the beam area, thus decreasing OD$_{\rm eff}$, both manifestations of poor mode matching of the probe and the scattered field from the atom cloud.

	\subsection{Optimizing the beam waist for a fixed atomic cloud geometry}  

With a better understanding of how cloud geometry influences decoherence, we study the optimization of squeezing in a situation typical of experiments with dipole-trapped cold atoms, where both the trap dimensions and beam waist can be tuned while the peak atomic density $\eta_0$ remains fixed.  In this situation, the total atom number $N$ depends on the trap volume. 

\begin{figure}[!h]
	\centering
    		\includegraphics[scale=0.85]{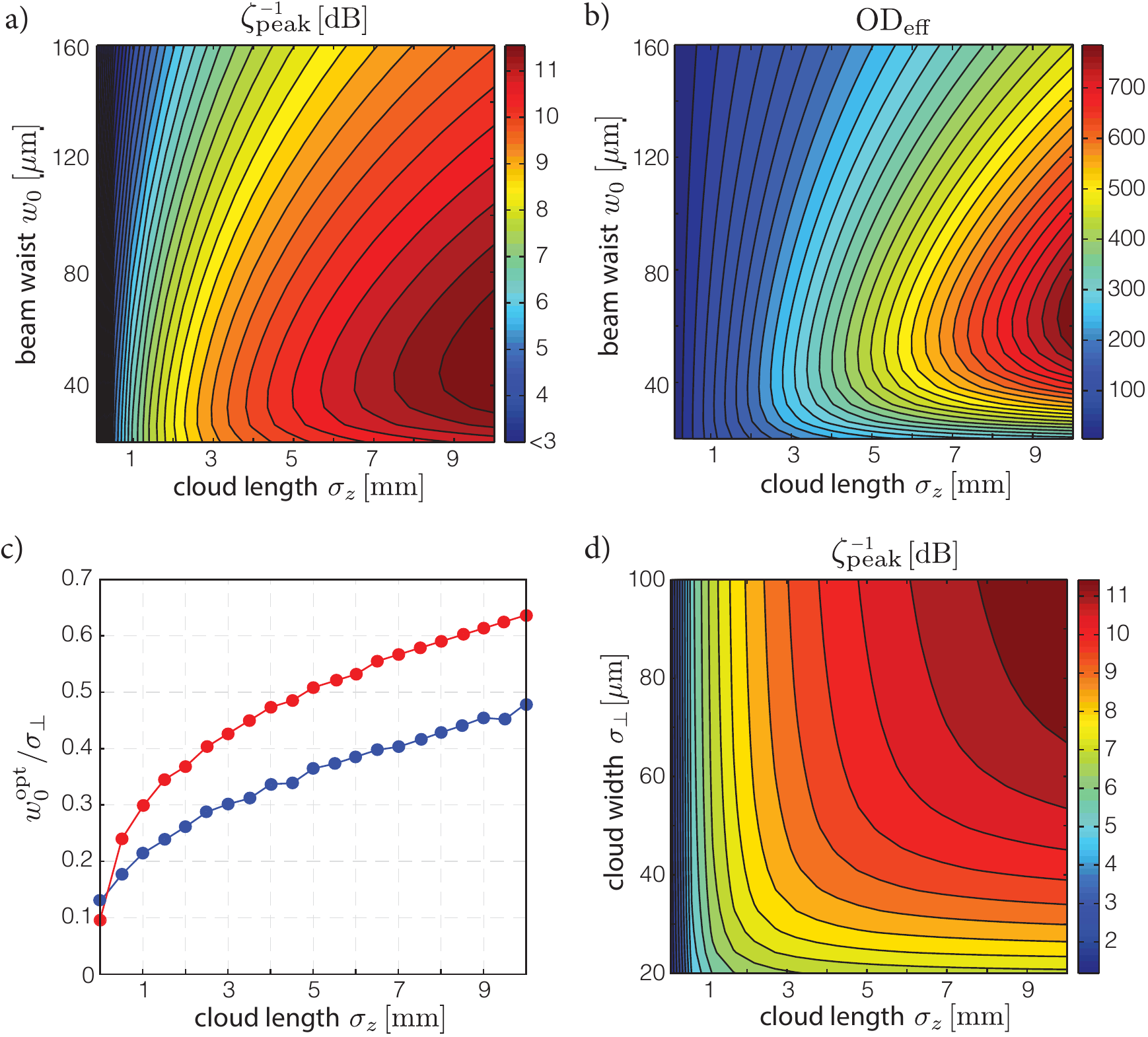}
       		 \caption[Spin squeezing for fixed atomic density with varying geometry]{ Squeezing for a fixed peak density $\eta_0 = 5 \times 10^{11}$ cm$^{-3}$ and variable atom number that fills a dipole trap for cold atoms.  In a), b), and c) the transverse cloud width is fixed at $\sigma_\perp = 100$ $\mu$m and cloud length is taken to be variable.  a) Contours of peak squeezing, $\zeta^{-1}$ in dB.  b) Contours of OD$_{\rm eff}$.  c) Optimal beam waist for maximizing OD$_{\rm eff}$ (upper red dots) and for maximizing peak squeezing (lower blue dots).  For a given atomic geometry, the beam waist that optimizes the OD$_{\rm eff}$ is not the same as that which optimizes peak squeezing.  d) Peak squeezing as a function of cloud size for the optimal beam waist at each point.}  \label{Fig::Fixed_SigmaT}     	
	\end{figure}

For each cloud geometry there exists a beam waist that maximizes OD$_\eff$.  This is seen in Fig. \ref{Fig::Fixed_SigmaT}(b) where contours of OD$_{\rm eff}$ are shown for a cloud with a fixed transverse width of $\sigma_\perp = 100$ $\mu$m as the cloud length $\sigma_z$ and beam waist $w_0$ are varied.  Contours for peak squeezing are shown in Fig. \ref{Fig::Fixed_SigmaT}(a).  Comparison with \ref{Fig::Fixed_SigmaT}(b) demonstrates that for a given cloud geometry, the peak squeezing is achieved with a smaller beam waist than that which optimizes OD$_\eff$. This is seen most clearly in Fig. \ref{Fig::Fixed_SigmaT}(c), where we compare the optimal beam waist for maximizing OD$_\eff$ to the beam waist that maximizes peak squeezing. Optimal squeezing occurs at smaller beam waists where the region of the beam with greatest intensity, the Rayleigh range, is smaller. Because the scattering rate $\gamma_s(\mathbf{r})$ is proportional to the local intensity, atoms outside the Rayleigh range experience a decreased rate of optical pumping.  Although a smaller Rayleigh range implies a decreased OD$_\eff$ and $N_{\rm eff}^{(1)}$ as well, the reduction of the decoherence rate dominates in this regime. This is a direct analogy to  Sec. \ref{Sec::GeomEffects}, in which pencil-shaped clouds with higher mean spins were more robust to decay due to a large number of atoms farther away from the beam waist.  Finally, in Fig. \ref{Fig::Fixed_SigmaT}(d) we plot contours of peak squeezing for different geometries at the optimal beam waist for each point.  Since larger clouds contain more atoms, and in general for properly chosen probe geometry OD$_{\rm eff}$ and peak squeezing increase with more atoms, there is no local maximum in Figs. \ref{Fig::Fixed_SigmaT}(a),  \ref{Fig::Fixed_SigmaT}(b), and  \ref{Fig::Fixed_SigmaT}(d).

	\subsection{Relation to the symmetric one-dimensional model} \label{SubSec::1Dmodel}

	\begin{figure}[!h]
	\centering
    		\includegraphics[scale=0.7]{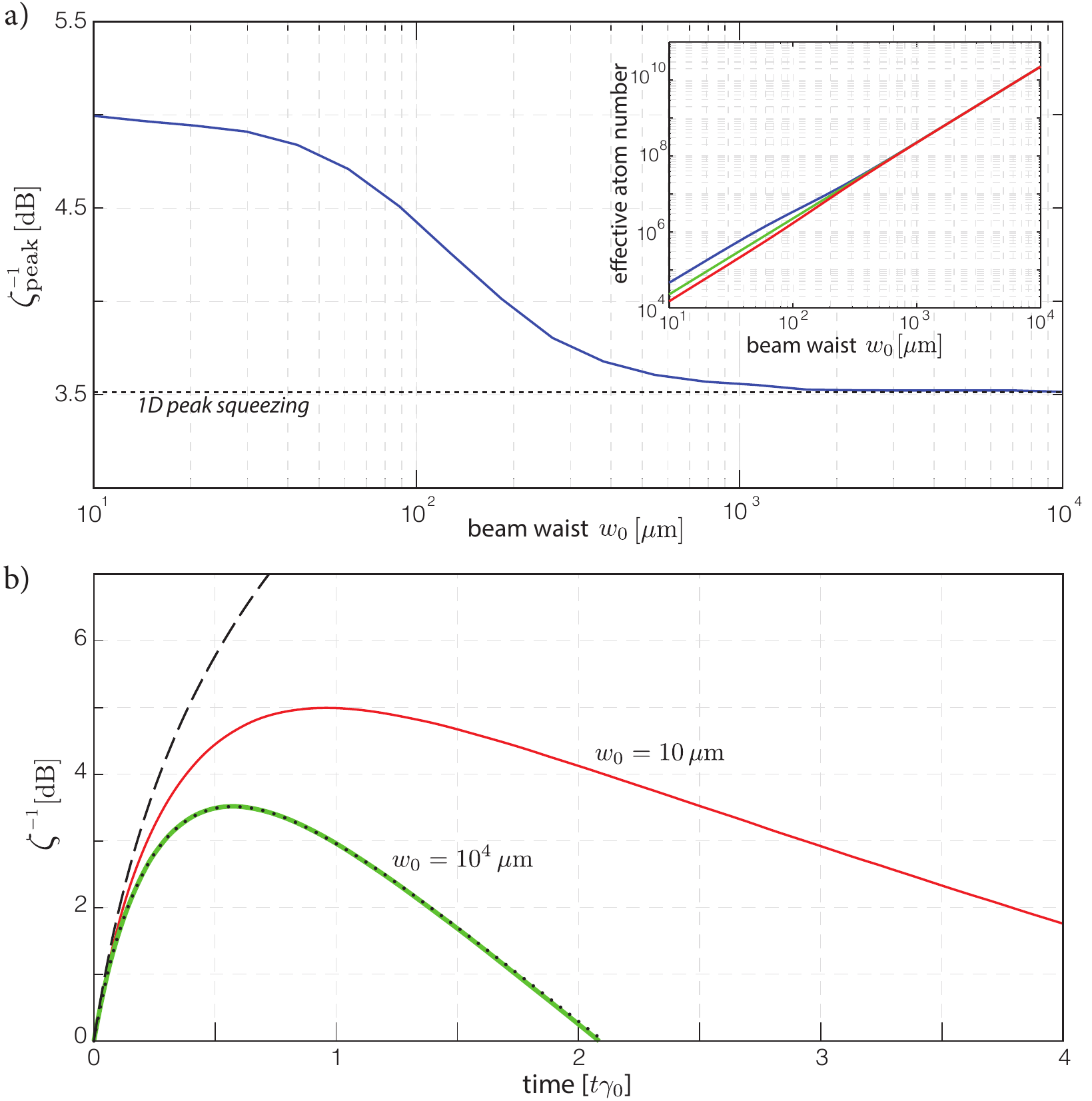} 
       		 \caption[Relation of spin squeezing in the three-dimensional model to a symmetric, plane-wave description]{ Comparison between the symmetric 1D model and the three-dimensional spin wave model.  a) Peak squeezing, $\zeta^{-1}$ in dB, for a spherical cloud with OD$_{\rm eff} = 50$ as the beam waist is increased.  Inset shows the convergence of $N_{\rm eff}^{(1)}$ (upper blue line), $N_{\rm eff}^{(2)}$ (middle green line), and $N_{\rm eff}^{(3)}$ (lower red line) as $w_0$ increases.  b) Comparison of squeezing dynamics for the extremal waists from a): the smallest, $w_0 = 10$ $\mu$m, and the largest, $w_0 = 10^4$ $\mu$m.  For comparison, the symmetric one-dimensional case using \erf{Eq::VarSpin1/2Symmetric} is plotted with decoherence (dotted black line) and without (dashed black line).} \label{Fig::PlaneWaveComp}
	\end{figure}

Spin squeezing by QND measurement is traditionally modeled using a one-dimensional description of the atom-light interface where the ensemble is symmetrically coupled to plane waves with no spatial variations \cite{HamPol10}.  When accounting only for squeezing due to collective scattering and QND measurement, the full three-dimensional system can be effectively described by such a model, with the symmetric OD replaced by OD$_\eff$ [\erf{Eq::ShortTimeSolution}]. When decoherence from local diffuse scattering is included, however, such models become insufficient. In addition, a symmetric description does not account for the difference between the effective atom number contributing to the spin wave variance, $N_{\rm eff}^{(2)}$,  that contributing to the mean spin, $N_{\rm eff}^{(1)}$,  and that contributing to noise injection by spin flips, $N_{\rm eff}^{(3)}$.  

To understand the limit in which we recover the simple symmetric description, recall the symmetric 1D model where an ensemble of spin-1/2 atoms is coupled to a uniform plane wave and scatters collectively into this mode and locally into diffuse modes, studied in \srf{Sec::1DSpinSqueezing}.  In this case a single atom number suffices as every atom contributes equally to the optical density, to the mean spin, and to the injection of noise, $ N_{\rm eff}^{(1)} = N_{\rm eff}^{(2)} = N_{\rm eff}^{(3)}= N$.  The equation of motion for the spin wave variance, \erf{Eq::VarSpin1/2Symmetric}, has an analytic solution given by \erf{Eq::AnalyticSolutionMainText}.  We can compare the symmetric 1D model to a limiting case of the full three-dimensional model developed here.  When the transverse extent of the cloud is much smaller than the beam waist and the longitudinal extent is well within the Rayleigh range, then spatial variations of the field across the cloud are minimal.  That is, the equation of motion for the fundamental variance, Eq. (\ref{Eq::SpinHalfVariance}), becomes exactly Eq. (\ref{Eq::VarSpin1/2Symmetric}) in the limit in which $\beta_{00}(\mathbf{r}_n) \rightarrow 1$ for all atoms.  Although this limiting case replicates the squeezing expected from the symmetric 1D model, it is in fact far from a single-mode description.  As discussed in Sec. \ref{Sec::ClassScatParaxF}, this geometry radiates paraxial light into \emph{many} of transverse modes defined relative to the beam, and the associated spin waves couple together through diffuse scattering, \erf{Eq::ProjectedSpin1/2VarianceEOM}.

We investigate this limit numerically in Fig. \ref{Fig::PlaneWaveComp} for a spherical cloud ($\sigma_\perp, \sigma_z = 100$ $\mu$m) probed by beams of increasing waist $w_0$.  In each case OD$_{\rm eff} = 50$, such that in the absence of decoherence the different geometries would achieve identical squeezing.  In Fig. \ref{Fig::PlaneWaveComp}(a) we see that as the beam waist is increased, the peak squeezing approaches that of the symmetric 1D model.  The inset shows the convergence of the effective atom numbers as the beam waist increases.  Figure \ref{Fig::PlaneWaveComp}(b) shows the dynamics of the squeezing parameter $\zeta^{-1}(t)$ for the spherical cloud at both extremes in Fig. \ref{Fig::PlaneWaveComp}(a).  For comparison, the squeezing parameter for the symmetric 1D model is plotted both with and without decoherence.  The difference between the models is substantial -- the optimal peak squeezing for the symmetric 1D model and full model are $\zeta^{-1}_{\rm peak} = \{ 3.52 \text{ dB}, 4.99 \text{ dB} \}$, respectively.  This difference can be understood in terms of the effective atom numbers.  The advantage for spin squeezing in the three-dimensional model comes from the fact that $N_{\rm eff}^{(1)} \geq  N_{\rm eff}^{(2)} \geq N_{\rm eff}^{(3)}$ due to different dependence on the spatial weightings $\beta_{00}(\mathbf{r})$, while for the symmetric 1D case they are equal.  For the three-dimensional model, not only can the effective number of atoms contributing to the noise injection be smaller than that contributing to the OD$_{\rm eff}$, but the effective number of atoms contributing to the mean spin, and thus the signal, is larger than both.  Inspecting Fig. \ref{Fig::SpinHalfCompareSqueezing}(d) we see an additional advantage for the three-dimensional model -- when geometry is properly chosen, the mean spin decays at a much reduced rate.

\section{Spin-f alkali atom ensembles} \label{Sec::SpinfEnsembles}

The constituent atoms in many spin squeezing experiments are alkali metal atoms whose ground state structure is more complex than spin-$\smallfrac{1}{2}$.  For example, in $^{133}$Cs, the ground electronic subspace is defined by two hyperfine manifolds with total spin angular momentum $f = \{3,4\}$.  Owing to the large ground-state hyperfine splitting (9.2 GHz in $^{133}$Cs), a single hyperfine manifold $f$ is addressed by the coherent interaction with the probe laser.

Though ensembles of higher spin atoms can be squeezed by the same measurement process, spin size affects both the coherent squeezing dynamics and decoherence.  Recall that the strength of the Faraday interaction is quantified by the coupling strength $\xi$, Eq. (\ref{Eq::CouplingStrength}). Because $\xi\propto 1/f^2$, the atom-light coupling decreases with increasing spin size. This decreased coupling strength is partially offset by an increased robustness to the effects of optical pumping. When $f> 1/2$, optical pumping events can be broadly divided into two categories: (i) ``loss" that occurs when an atom is pumped from the $f$ manifold into the other ground hyperfine manifold and (ii) ``spin flips" that leave the atom in the $f$ manifold. Because atoms lost to the other ground manifold are no longer resonant with the probe, loss events decrease the mean spin $\langle\hat{F}_x^{0}\rangle$, though they contribute no excess noise to $(\Delta F_z^{0})^2$. Spin flips are responsible for both a decrease in $\langle\hat{F}_x^{00}\rangle$ and a noise injection into $(\Delta F_z^{0})^2$. For the SCS preparation, the deleterious effects of spin flips are mitigated by ``transfers of coherence" between pairs of magnetic sublevels that reduce the rate of decay of correlations. While the interplay between these effects is complex, the rate of spin flips remains a good indicator of an ensemble's robustness to optical pumping. For an ensemble of spin-$f$ alkalis prepared in a SCS, the spin flip rate is $\gamma_s(\mathbf{r})/(12f)$, thus decreasing for larger spin size.  

Due to these decoherence processes, the dynamics of the squeezing parameter is substantially more complicated for larger spin atoms. Full details of the equations of motion for the mean spin and covariances are given in Appendix \ref{Appendix::SpinWaveEquations}.  For spin-$f$, we obtain the evolution of the mean value of a spin wave in slice $z_k$ by projecting onto the different spin waves in a manner analogous to \erf{Eq::MeanZsliced},
	\begin{align}  \label{Eq::MeanSpins}
		\frac{d}{dt} & \expect{\hat{F}^{i}_{x}(z_k)}  = - \frac{2 \gamma_{0}}{9}  \sum_{j} c^{i}_{j} (z_k) \expect{\hat{F}_x^{j}(z_k)}  + \frac{g_F^2 \gamma_0}{9}  \sum_{j} \sum_{n_k} c^{i}_{j} (z_k) \beta_{j} (\mathbf{r}_{\perp n_k},z_k )\mathcal{C}_{n_k}[\hat{f}_x^{(n_k)}]. 
	\end{align}
 Here, we have defined a local superoperator that arises solely from the ``feeding" terms in the master equation:
	\begin{align}
		\mathcal{C}_n[\hat{X}] \equiv \hat{f}_z^{(n)}\hat{X}\hat{f}_z^{(n)} + \tfrac{1}{2} \big( \hat{f}_x^{(n)}\hat{X}\hat{f}_x^{(n)} + \hat{f}_y^{(n)} \hat{X}\hat{f}_y^{(n)} \big).
	\end{align}
Similarly, we find equations of motion for the fundamental spin wave variance,
	\begin{align} \label{Eq::GenCovarianceEvolution}
		& \frac{d}{dt}  \varFz =- \kappa \big[\left(\Delta F_z^{0} \right)^2\big]^2 \\
		& -\frac{4\gamma_{0}}{9} \sum_{i} \sum_{k',k}c^{0}_{i}(z_k)\big\langle \Delta \hat{F}_{z}^{0}(z_{k'}) \Delta \hat{F}_{z}^{i}(z_k) \big\rangle \nn \\
		&  +  \frac{g_f^2 \gamma_{0}}{9} \sum_{i}\sum_{k',k,n_k} c^{0}_{i}(z_k)\beta_{i}(\mathbf{r}_{n_k})  \Big\{ \big\langle \Delta \hat{F}_{z}^{0}(z_{k'}) \Delta \mathcal{C}_{n_k}[ \hat{f}_{z}^{(n_k)} ] \big\rangle + \big\langle \Delta \mathcal{C}_{n_k}[ \hat{f}_{z}^{(n_k)} ] \Delta \hat{F}_{z}^{0}(z_{k'}) \big\rangle \Big\} \nn \\
		&+\gamma_{0}\sum_{k,n_k} \big[ \beta_{0}(\mathbf{r}_{n_k}) \big]^3 \bigg\{ \frac{2}{9} \expect{(\hat{f}_z^{(n_k)})^2}  + \frac{g_f^2}{9} \left( \big\langle \mathcal{C}_{n_k}[\hat{f}_z^{(n_k)2}]\big\rangle-\big\langle \big\{\hat{f}_z^{(n_k)},\:\mathcal{C}_{n_k}[\hat{f}_z^{(n_k)} ] \big\}_+\big\rangle \right) \bigg\} , \nn
	\end{align}
where $\{ \hat{X}, \hat{Y}\}_+$ denotes the anti-commutator. As for the case of spin-$\smallfrac{1}{2}$, we have an infinite hierarchy of equations that couple spin wave operators in the different $z_k$-slices. In general, the feeding terms in \erf{Eq::GenCovarianceEvolution} couple to covariances outside the set of $z$-local spin waves, $\expects{\Delta\hat{F}_z^{i}(z_k) \Delta\hat{F}_z^{j}(z_{k'})}$. This expands considerably the number of equations that must be solved to reach convergence. Solving these equations, furthermore, requires methods different from those for spin-$\smallfrac{1}{2}$ which are beyond the scope of this dissertation.

\section{Summary} \label{Sec::Conclusion}

%
%
%

We have studied the strength of the atom-photon interface in a traveling wave configuration for spin squeezing via QND measurement.  We developed a description in terms of quantized paraxial modes of the field in order to model the inhomogeneous atom-light coupling across the atomic ensemble, which leads to two distinct effects.  First, the collective coupling describes a generalization of the Faraday interaction that entangles the quantized Stokes vector of the laser field with a spin wave defined by the weighted ensemble of atoms that indistinguishably radiates into the  mode of the probe.  The spin wave that is squeezed is defined by the probe mode we measure in a balanced polarimeter.  Second, diffusely scattered photons lead to optical pumping and decoherence across the ensemble at a rate proportional to the local probe intensity. The delicate balance of these two effects favors certain geometries for spin squeezing.

We numerically investigated the ultimate limits to spin squeezing for spin-$\smallfrac{1}{2}$ based on a stochastic master equation, including the effects of QND measurement backaction and decoherence by photon scattering into unmeasured modes.  Unlike the usual one-dimensional description in which the amount of squeezing is set by a single parameter, the optical density, we find that due to  inhomogeneous coupling, multiple parameters are required. Of particular importance in a metrological context are the mean collective spin and  the projection noise variance, determined by effective atom numbers $N^{(1)}_\eff$ and $N^{(2)}_\eff$ respectively.  Optimal geometries maximize the effective optical density, OD$_\eff$, proportional to $N^{(2)}_\eff$, while minimizing the depolarization of  $N^{(1)}_\eff$  and injected noise into the spin wave by optical pumping.  We found that optimal mode matching occurs for geometries where a large number of atoms are addressed by a beam with a small transverse area, yielding a high OD$_{\rm eff}$, but also where the depolarization rate due to optical pumping is relatively small.  This geometry corresponds to a longitudinally extended, pencil-shaped cloud, with  a probe beam chosen to optimize the tradeoffs between OD$_\eff$ and decoherence.  Such a geometry is far from the regime describing squeezing of a symmetric atomic spin ensemble, as is typically assumed.  One recovers the symmetric description only for ensembles confined with extents much smaller than the beam waist and Rayleigh range, which yield much smaller OD$_\eff$.  
	\chapter{Summary and outlook} \label{Ch::Conclusion}
%

The common core of the results in this dissertation is the interaction between a system and a propagating quantum field.  In this heavily explored realm, we managed to push the boundary of knowledge slightly further in two areas.  The first was an extension to input-output theory for propagating field states of definite photon number, discussed in Chapter \ref{Ch::NPhotonME}.  Such fields are inherently quantum mechanical, and as such possess temporal mode entanglement which drives non-Markovian reduced system dynamics.  I presented a method to model these dynamics using a series of coupled master equations.  

The second was a fully three-dimensional quantum light-matter interface for an ensemble of multi-level atoms, discussed in Chapter \ref{Ch::3DInterface}.  By explicitly including the spatial dependence of the interaction and measurement, the model accounts for dynamics driven both by collective, coherent coupling as well as local, incoherent processes.  In Chapter \ref{Ch::3DInterface} this model was applied to the study of the optimization of measurement-induced spin squeezing for an ensemble of atoms.

\section{Quantum systems interacting with propagating $N$-photon states}

In Chapter \ref{Ch::NPhotonME} we introduced formalism to derive a set of coupled master equations for a quantum system interacting with a continuous-mode Fock state.  The relatively simple derivation reveals that the dynamics of the physical state couples to a set of auxiliary, density matrix-like operators whose dynamics account for the field's evolution during the interaction.  Expectation values of system operators as well as of equal-time field operators can be found simply.  Armed with the basic technique for Fock states, we then generalized to a much broader class of input field states including superpositions and mixtures of general $N$-photon states with arbitrary spectral distribution functions in multiple spatial and/or polarization modes.  

The power of the formalism lies in its direct applicability to general systems of interest in quantum optics such as multi-level atoms, atomic ensembles, and continuous variable systems, i.e. optical or nano-mechanical resonators.  For example, it is possible to reproduce the cavity-mediated, single-photon pulse shaping results of Ref.~\cite{Mil08} by first identifying the coupling operators, $ \hat{H}=0$,  $ \hat{L}= \sqrt{\gamma} \hat{a}$, and $ \hat{S}= \hat{I}_{\rm cavity}$.  Then, our expression for the output photon flux, \erf{Eq::SingleModeFieldME_lambda}, is equivalent to Eq. (22) in Ref.~\cite{Mil08} for a single-photon input.  One can formally describe their dynamics as they interact with novel nonclassical states of light, such as spectrally correlated two-photon states.  Others have shown that this formalism applies in the microwave domain as well \cite{KockumThesis}, and have used it to study cross-Kerr effects in artificial atoms \cite{FanMil13, HoiWil13} and for the design of a QND photon detector in superconducting circuit QED \cite{SatJoh14}.

As arcane formalism can often spark antipathy as much as curiosity, we supplemented the general, purely theoretic results with a variety of pedagogical examples accompanied by numerical simulations.  We began with the study of a two-level atom interacting with Fock states of various photon number in a single mode.  While it is known that an appropriately shaped rising exponential pulse with a single photon is optimal for excitation, such pulses are not easily produced, and thus we focused on Gaussian pulses in \srf{Sec::2LevelExcitation}.  We saw the maximum excitation probability $\mathbbm{P}_e^{\rm max}$ was low for both small ($\Delta_\omega / \Gamma \ll10^{-1}$) and large ($\Delta_\omega / \Gamma \gg10^{2}$) bandwidths. The low $\mathbbm{P}_e^{\rm max}$ for small bandwidths, centered at the atomic resonance, might seem counter intuitive. In the time domain the corresponding wave packet is broad, nevertheless the near-resonant photons all get absorbed, but are immediately reemitted by the vacuum coupling, which leads to a small average $\mathbbm{P}_e$. This intuition is confirmed with simulations in two spatial modes, presented in \srf{Sec::2modeExample}, where wave packets with small bandwidths are nearly perfectly reflected. The reflection is mediated by photon absorption and the consequent reemission, which is directionally unbiased.  However, destructive interference between the incoming wave packet and the transmitted mode results in reflection only; i.e. the atom can act as a perfect reflector. 

A number of interesting applications of our formalism remain to be explored, including the investigation of pulse shaping for few-photon states, temporal mode matching for quantum memories, and mediated photon-photon interactions.  Our formalism is particularly applicable to quantum networks \cite{Kimble08, MoeMauOlm07}. Recently, the theory of cascaded quantum systems \cite{Gar93, Car93} has been formalized to the point where simple rules for composing modular quantum optical systems into a network have been developed \cite{YanKim03a,YanKim03b,GouJam09,GouJam09b,JamGou10}. One needs only the $(S,L,H)$-tuple of each module specified in order to perform network analysis and simplification. As our description of the system, input, and output fields is also in terms of a $(S,L,H)$-tuple, our formalism can be ported to this setting, as has been shown and used in Ref. \cite{SatJoh14}.

The unconditional master equations presented in this dissertation are part of a larger program to fully characterize the interaction of a propagating $N$-photon state with an arbitrary quantum system. There remain many avenues to explore.  

In many cases a light-matter interaction is followed by measurement of the output fields (e.g. the QND spin squeezing in Chapter \ref{Ch::3DInterface}).  As the measurement results are random, the conditional dynamics of the system and remaining unmeasured field are stochastic.  An individual realization of a continuous measurement record generates such stochastic dynamics known as a \emph{quantum trajectory} \cite{CarmichaelBook}.  The unconditional master equations describe the evolution averaged over of a large ensemble of quantum trajectories.  For Fock-state input, the unconditional master equation dynamics require keeping track of the field.  Conditional dynamics are no different, as the temporal mode entanglement in a Fock state requires that the field itself must be nontrivially conditioned on the measurement results.  

For a single photon, a step towards the differential equations for the quantum trajectories was given in Ref. \cite{Zoolander}, where the authors suggested using a cascaded systems approach, summarized in Appendix \ref{Appendix::CascadedApproach}, to determine the conditional evolution. This suggestion has become a standard approach, see e.g. Ref.~\cite{BasAkrMil12}.  An alternative method using coupled stochastic master equations was implemented in Refs. \cite{GJNphoton,GJNCgen}, where the authors derived quantum trajectories\footnote{Technically, the \emph{quantum filters} derived in the mathematics community are different in that they are constructed as optimal estimators for an unknown quantum state, whereas the stochastic master equations describe conditional dynamics for a known state.  However, the resulting equations are identical up to the innovations process, [\erf{Eq::dWi}].} for photon counting and homodyne measurements for single-photon input.  This has been extended to multi-photon fields using yet another technique, that of a non-Markovian embedding \cite{SonXi13}.  Finally, a recent result provides the quantum trajectories for a general class of continuous matrix-product states, which includes time-ordered multi-photon states \cite{GouNur14}.  These derivations proceed in the Heisenberg picture and rely on somewhat arcane mathematical techniques unfamiliar to many physicists.  We have supplemented these results with an accessible \sch{}-picture derivation of the quantum trajectories for photon counting, homodyne, and heterodyne measurements for $N$-photon Fock state input fields.  

Finally, the capstone to this quantum quest is a complete description of the output fields themselves.  A glaring omission in Chaper \ref{Ch::NPhotonME} is the output field correlation functions.  When the input fields are not temporally entangled, field correlation functions can often be directly tied to system correlation functions -- as in the standard problem of resonance fluorescence.  The system reaches a steady state and one can study stationary statistical properties of the output field, such as the fluorescence spectrum.  For $N$-photon states there is, in general, no (nontrivial) steady state\footnote{There is also no steady state for time-dependent Gaussian input fields such as the coherent state wave packet in \srf{SEC::DisplacedFockStates}.}, and further the input fields themselves exhibit temporal correlations.  The first step towards addressing these and other related issues is a complete description of propagating multi-mode quantum field states, including correlated $N$-photon states, squeezed states, etc.  Then, a relatively more difficult task remains: full characterization of $N$-photon scattering for arbitrary quantum systems.  The literature is replete with solved examples for a given system and a specific photon number\footnote{A two-level atom and a single photon \cite{SheFan05, CheWubMor11, Ely12, LiSun13}, a cavity QED system and a single photon \cite{Kos08}, multiple cascaded two-level systems and two photons \cite{Roy13}, a three-level $\Lambda$ or V-system and one and two photons \cite{WitSor10}, two non-identical two-level atoms and two photons \cite{RepFan11}, a cavity QED system and two photons \cite{RepFan12}, an optical cavity and several photons \cite{Aiello00}, a Kerr nonlinear medium and $N$ photons \cite{Koshino08}, a two-level system and $N$ photons \cite{LonBus10}, a three- or four-level system and $N$ photons \cite{ZheBar12}, to name just a few.}  but a general theory is still absent.  A way has been paved in Ref. \cite{FanShe10}, where the direct connection is made between the M\o ller wave operators from scattering theory and input-output theory.  It should be possible to use techniques similar to those in this thesis to calculate the scattering matrix elements in the Fock basis for an arbitrary system and any number of photons.    

For several photons, understanding Fock-state scattering has direct applications in the study of quantum gates for photonic qubits \cite{JohFio12, GeaPed12, VenGae13, ZheHar13}.  The effective mediated interactions require indistinguishable photons at the output.  Our results show that the temporal wave packet of a photon (using the photon flux as proxy) can be significantly modified through interactions, which must be considered if one is to use output wave packets as inputs for the next gate.  With a proper understanding of Fock-state scattering, such temporal effects can be included in the design and implementation of quantum information devices that rely on photonic data channels.

\section{Three-dimensional atom-light interface}

The entangling power of a quantum interface between photons and an ensemble of cold atoms is at the heart of a variety of quantum information processing tasks.  For a spatially-extended atomic cloud, one must consider a full three-dimensional description of the electromagnetic modes and atomic density distribution in order to optimize this entangling power.  Inhomogeneous coupling between atoms and photons is essential to maximize the strength of the quantum interface but comes with substantial complexity in the theoretical description.   In Chapter \ref{Ch::3DInterface} we presented such a description in terms of quantized paraxial modes of the field that describes two distinct effects.  First, the collective coupling gives rise to a generalization of the Faraday interaction that entangles the quantized Stokes vector of the laser field with a collective spin wave defined by the weighted ensemble of atoms that indistinguishably radiates into the spatial mode of the probe.  Balanced polarimetry detects only the light scattered into this mode, with the phase shift, attenuation, Faraday rotation, and birefringence dependent on the fundamental spin wave.  Second, diffusely scattered photons lead to optical pumping and decoherence across the ensemble at a rate proportional to the local probe intensity.  The measurement record can be used to continuously update the collective atomic state, described by a stochastic master equation in \srf{}.  

Using this model, we investigated the ultimate limits to QND spin squeezing for an ensemble of spin-$\smallfrac{1}{2}$ atoms in Chapter \ref{Ch::SpinSqueezing}.  A delicate balance between the competing effects of squeezing through measurement backaction and decoherence by photon scattering favors certain geometries for spin squeezing.  Unlike the usual one-dimensional description in which the amount of squeezing is set by a single parameter, the optical density, we find that due to inhomogeneous coupling, multiple parameters are required. Of particular importance in a metrological context are the mean collective spin and  the projection noise variance, determined by effective atom numbers $N^{(1)}_\eff$ and $N^{(2)}_\eff$, respectively.  Optimal geometries maximize the effective optical density for squeezing, OD$^{(2)}_\eff$, proportional to $N^{(2)}_\eff$, while minimizing depolarization of the mean spin, proportional to $N^{(1)}_\eff$, and injected noise into the spin wave by optical pumping.  We found that optimal mode matching occurs for geometries where a large number of atoms are addressed by a beam with a small transverse area, yielding a high OD$^{(2)}_\eff$, but also where the depolarization rate due to optical pumping is relatively small.  This geometry corresponds to a longitudinally extended, pencil-shaped cloud, with  a probe beam chosen to optimize the tradeoffs between OD$^{(2)}_\eff$ and decoherence.  Such a geometry is far from the regime describing squeezing of a symmetric atomic spin ensemble, as is typically assumed.  One recovers the symmetric description only for ensembles confined with extents much smaller than the beam waist and Rayleigh range, which yield much smaller OD$^{(2)}_\eff$.  

The results could be extended in several ways.  First, recent work has considered ensembles of higher-spin alkali atoms, in which control over the rich internal hyperfine structure can enhance the entangling strength of the atom-light interface \cite{NorDeu12}. Quantifying the gains achievable though such control techniques requires a realistic description of the inhomogeneous interaction between light and atoms. Second, in Chapter \ref{Ch::SpinSqueezing}, we used a stochastic master equation to study the squeezing of the fundamental spin wave from QND measurement.  In the Gaussian approximation, the variance obeys a deterministic equation of motion, and the mean spin is a function of the stochastic measurement record.  In our analysis, we were primarily interested in the the squeezing parameter, which depends only on the length of the mean spin but ignores its orientation.  The relevant mean-spin dynamics arose from diffuse scattering.  In order to make use of the spin squeezing for metrology or to simulate long-time behavior, such as for preparation of Dicke states using feedback control \cite{StoMab04,VanBou11}, the stochastic dynamics play a critical role.  Further, since the decoherence map, \erf{Eq::DiffuseME}, is not symmetric in the angular momentum components, stochastically driven significant $y$- and $z$- components of the mean spin can modify the effects of diffuse photon scattering.  Third, a description of spin squeezing is often accompanied by a study of anti-squeezing, required to maintain Heisenberg uncertainty relations, and as such can be considered \emph{the} core quantum mechanical process.  In the context of multiple coupled spin waves, the possibility may exist to slough off this requirement by pushing the conjugate quantum uncertainty into higher order modes uncoupled to the measurement.  This could allow for simultaneous squeezing in orthogonal spin components, such as in planar quantum squeezing \cite{HeRei11}.

We studied the case of interactions in a highly transparent regime where the optical density is very small at the detuning of the probe.   Collective effects arise solely from the fact that the each of the atoms that scatter photons into the same paraxial mode are indistinguishable.  For much higher densities, and/or lower detunings, multiple scattering effects are non-negligible, and one must go beyond our forward-scattering model in the first Born approximation \cite{ChaKai14}. Under these conditions coherent backscattering \cite{RouKai14}, superradiant scattering \cite{InoKet99, HilMul08}, and stimulated Raman scattering~\cite{RayMos81, SorSor09} become important and can lead to additional collective effects.  At high atomic densities, one must also consider the effects of atomic collisions \cite{KamMul12}. Finally, we considered ultracold ensembles where the atoms are well approximated as fixed point scatterers. Effects of finite temperature can be included by averaging over the positions of the atoms, as in Ref. \cite{DuaZol02}.

While the three-dimensional model developed in this dissertation was specifically tailored to study the problem of spin squeezing by QND measurement, it can be extended to other protocols involving the quantum interface between photons and free-space atomic ensembles.  Mode-matching and spatial effects are important for other spin squeezing protocols including the double-pass counter-twisting interaction \cite{TakTak05,TraDeu10} or the recently proposed planar squeezing protocol \cite{PueMit13}. Understanding spatial effects in order to identify regimes of strong coupling is also essential for quantum memories and repeaters in free-space atomic ensembles.  Finally, a multimode description of the entangling Hamiltonian offers the possibility to exploit spatial modes and their associated spin waves as a resource.   The creation of entanglement between spin waves could lead to novel states with potential application in continuous variable quantum computation and communication \cite{PolzikBook}.

	\appendix	
		\chapter{Laguerre-Gauss modes} \label{Appendix::LGModes}


As discussed in \srf{Sec::ClassicalParaxialFields}, the paraxial field in free space can be decomposed into a set of dimensionless transverse mode functions that satisfy the paraxial wave equation.  When considering geometry where both the atomic cloud and the probe exhibit some degree of cylindrical symmetry, a natural choice are the Laguerre-Gauss modes $\{ u_{pl} (\rperp, z)\}$, where $p$ is the radial mode index and $l$ is the azimuthal index.  In cylindrical coordinates the mode functions are,
	\begin{align} \label{Eq::LGModes}
		u_{pl}(\rperp,z) = & \, \mathcal{N}_{pl} \frac{w_0}{w(z)} \bigg( \frac{\sqrt{2} \rho}{w(z)} \bigg)^{\!|l|} L_p^{|l|}\! \left( \frac{2 \rho^2}{[ w(z) ]^2}\right) \exp \bigg(\!-\frac{ \rho^2}{[w(z)]^2} \bigg) \\
		 & \times  \exp \bigg( \frac{ik_0 \rho^2}{2 R(z)} - i (2p + l + 1)\Phi(z) -il \phi \bigg), \nn
	\end{align}
where $w_0$ is the $1/e^2$ beam waist at the focal plane and $z_R \equiv k_0 w_0^2/2$ is the Rayleigh range.  $\mathcal{N}_{pl} = \sqrt{p!/ (|l| + p)!}$ is the normalization constant, and $L_p^{|l|}(x)$ indicates an associated Laguerre polynomial.  The $z$-dependent beam waist, the radius of curvature of the phase fronts, and the Guoy phase are given respectively by
	\begin{subequations} \label{Eq::GaussianParameters}
	\begin{align}
		w(z) &= w_0 \sqrt{1+\left(\frac{z}{z_R} \right)^2} ,  \\
		R(z) &= z\left[1+\left(\frac{z_R}{z} \right)^2 \right] ,  \\
		\Phi(z) &= \tan^{-1}\left(\frac{z}{z_R} \right). 
	\end{align}
	\end{subequations}
These modes satisfy the properties,
	\begin{subequations} \label{Eq::EssentialProperties}
	\begin{align} 
		\int d^2 \rperp u^*_{pl} (\rperp , z) u_{p'l'} (\rperp , z) &= A \, \delta_{p, p'} \delta_{l, l'}, \\
		\sum_{p,l} u_{pl}(\rperp , z) u^*_{pl}(\rperp' , z)  &=  A \, \delta^{(2)}(\rperp-\rperp'), \\
		\sum_{p,l} u_{pl}(\rperp , z) u^*_{pl}(\rperp' , z')  &=  A \, K(\rperp-\rperp', z-z'), 
	\end{align}
		\end{subequations}
where $K(\rperp-\rperp', z-z')$ is the paraxial propagator.  The Laguerre-Gauss modes are normalized to characteristic transverse area, $A = \pi w_0^2/2$. The TEM$_{00}$ fundamental mode is, 
	\begin{align}
		u_{00}(\rperp,z)=& \frac{w_0}{w(z)} \exp \left[-\frac{ \rho^2}{[w(z)]^2} \right] \exp \left[ \frac{ik_0 \rho^2}{2 R(z)} \right] \exp \Big[-i \Phi(z) \Big].	
	\end{align}

		\chapter[Fock-state It\={o} table]{Fock-state It\={o} table} \label{Appendix::NPhotonItoTable}
%

The It\={o} table gives the rules for products of the quantum noise increments for a given input field state.  Consider the following two combinations of the quantum noise increments that arise in the short-time interaction unitary \erf{},
	\begin{align}
		dB_t dB_t\dg =& \int_{t}^{t+dt} dt' \int_{t}^{t+dt} dt'' \hat{b}(t') \hat{b}\dg (t'') \nn \\
			=& \int_{t}^{t+dt} dt' \int_{t}^{t+dt} dt'' \big( \delta(t' - t'') + \hat{b}\dg(t') \hat{b} (t'') \big) \nn \\
			=& dt + dB_t \dg dB_t
	\end{align}
Under vacuum the normally ordered relation is $\bra{0} dB_t\dg dB_t \ket{0} = 0$, which gives $\bra{0} dB_t dB_t\dg \ket{0} = dt.$  These vacuum relations are used in the derivation of the quantum It\={o}-Langevin equation, [\erf{Eq::dXAppendix}].  For Fock states, the action of the quantum noise increments, [\erf{}], generates two powers of the infinitesimal $dt$, and thus $\bra{m_\xi} dB_t\dg dB_t \ket{n_\xi} = 0$.  One might expect the other term, $dB_t dB_t\dg$, to vanish as well for Fock states of different photon number, but in fact the quantum noise increments in the It\={o}-Langevin equation, \erf{Eq::dXAppendix}, appear in conjunction with system operators.  Unlike Gaussian fields, the Fock states cannot be pulled through the system operators when taking expectations. The results is that the Fock-state It\={o} table is simply a summary of the \emph{order in the infinitesimal} $dt$ of products of quantum noise increments and, as such, is identical to \erf{Eq::ItoTable}.  The consequences of Fock-state expectations are accounted for later, when the master equations are derived.  
It is clear, then, that the It\={o} table for a general $N$-photon state with an arbitrary spectral distribution function (within the quasi-monochromatic approximation) is also identical to the vacuum table. This follows from the occupation number representation, reviewed in Appendix \ref{Appendix::NPhotonStates}, which relies on a decomposition in a basis of orthogonal Fock states, each of which respects its own Fock It\={o} table.
	
In fact, one can always work with the vacuum It\={o} table and consider displacements of the quantum noise increments.  This is an alternate approach to that of using ``non-vacuum" It\={o} tables \cite{GardinerBook, WisMilBook}.  The difference arises in the point of view - for a non-vacuum table, one first displaces the fields and then finds the non-vanishing products of quantum noise increments; in our approach one displaces the increments in the quantum Langevin equation \emph{after} having discarded the vanishing vacuum products.  This approach should also work for thermal and squeezed fields, although it remains to be shown.  

		\chapter{Cascaded single-photon master equation}\label{Appendix::CascadedApproach}
%

Here, we outline a cascaded systems approach for modeling the interaction of a quantum system with a single-photon wave packet.  The idea is to explicitly include an auxiliary system -- in this case an excited two-level atom -- as the ``source" of the photon.  The output of the source is fed, in a one-directional way, into the quantum system.  By  manipulating the decay rate of the source atom, arbitrary single photon wave packets can be generated \cite{GJNphoton, GJNCgen, CarJam12, GouZha14}.  The reduced dynamics of this system can then be related to the Fock-state master equations developed in Chapter \ref{Ch::NPhotonME}.

\section{Model for a single-photon source}

Beginning with an uncorrelated state of the ``source" atom and the multimode vacuum, $\ket{ \psi(0)} = \ket{e} \otimes \ket{ 0  }$, the \sch{} equation for the total state is
	\begin{align} \label{Eq::FullStateSource}
		d \ket{\psi} = \left( \lambda(t) \op{g}{e} \otimes dB\dg_t + \frac{1}{2} |\lambda(t)|^2 \op{e}{e} \otimes \hat{I}_{\rm field} dt \right) \ket{\psi}.
	\end{align}
This has the exact solution \cite{GJNphoton},
	\begin{align}
		\ket{ \psi (t) } = \sqrt{ w(t) } \ket{e} \otimes \ket{0} + \ket{g} \otimes B_{t)}^\dagger(\xi)\ket{  0 },
	\end{align}
where the multimode photon creation operator up to time $t$ is given by
	\begin{align}
		B\dg_{t)}(\xi) \equiv \int_0^t dt' \xi(t') b\dg(t'),
	\end{align}
and $w(t) = \int_t^\infty dt' |\xi(t')|^2 $.  This is related to the time dependent coefficient in the solution, \erf{Eq::FullStateSource}, through the relation
	\begin{align} \label{Eq::QubitDecayRate}
		\lambda(t) = \frac{1}{\sqrt{w(t)}} \xi(t). 
	\end{align}
Thus at time time $t \rightarrow \infty$, the limiting state is that of the source in the ground state and a single photon in the field with temporal envelope $\xi(t)$:
	\begin{align}
		\ket{ \psi(t \rightarrow \infty) } = \ket{g} \otimes \ket{1_\xi}.
	\end{align}
By properly modulating the decay rate of the atom using \erf{Eq::QubitDecayRate}, an arbitrarily shaped photon can be produced.

\section{Cascading the source and system}

We now use the output of the single-photon source as the input to our quantum system of interest using a cascaded quantum systems approach \cite{}.  The system coupling to the input field is described by the $\hat{L}$ operator (we ignore an additional system Hamiltonian and the $\hat{S}$ operator here, as they serve only to complicate the equations).  Within the cascaded systems approach, we identify the following operators on the total system ($\mathcal{H}_{\rm sys} \otimes \mathcal{H}_{\rm source}$),
	\begin{align}
		\hat{H}_{\rm tot} &=\frac{i \hbar}{2 \sqrt{w(t)}} \Big( \xi^*(t) \hat{L} \otimes \op{e}{g} - \xi(t) \hat{L}\dg \otimes \op{g}{e} \Big) \\
		\hat{L}_{\rm tot} & = \hat{L} \otimes \hat{I}_{\rm source} + \frac{\xi(t)}{ \sqrt{ w(t) } } \hat{I}_{\rm sys} \otimes \op{g}{e}
	\end{align}
Now we can write the master equation for the total system as (setting $\hat{H}_{\rm sys} = 0$),
	\begin{align}
		\frac{d}{dt} \hat{\rho}_{\rm tot} =&  - \frac{i}{\hbar} [\hat{H}_{\rm tot}, \hat{\rho}_{\rm tot} ] + \mathcal{L}_{L_{\rm tot}}[ \hat{\rho}_{\rm tot}] \\
			= &  \hat{L} \otimes \hat{I}_{\rm source} \hat{\rho}_{\rm tot} \hat{L}\dg \otimes \hat{I}_{\rm source} -\frac{1}{2}  \hat{L}\dg \hat{L} \otimes \hat{I}_{\rm source} \hat{\rho}_{\rm tot} -\frac{1}{2} \hat{\rho}_{\rm tot}  \hat{L}\dg \hat{L} \otimes \hat{I}_{\rm source} \label{ApEq::FullMasterEquation}\\
			& +  \frac{\xi(t)}{ \sqrt{ w(t) } } \Big(   \hat{I}_{\rm sys} \otimes \op{g}{e} \hat{\rho}_{\rm tot} \hat{L}\dg \otimes \hat{I}_{\rm source} - \hat{L}\dg \otimes \op{g}{e} \hat{\rho}_{\rm tot} \Big) \nn \\
			& +  \frac{\xi^*(t)}{ \sqrt{ w(t) } } \Big(  \hat{L} \otimes \hat{I}_{\rm source} \hat{\rho}_{\rm tot} \hat{I}_{\rm sys} \otimes \op{e}{g} -\hat{\rho}_{\rm tot}  \hat{L} \otimes \op{e}{g}  \Big) \nn \\
			& + \frac{ |\xi(t)|^2 }{  w(t)  } \Big( \hat{I}_{\rm sys} \otimes \op{g}{e} \hat{\rho}_{\rm tot} \hat{I}_{\rm sys} \otimes \op{e}{g}   -\frac{1}{2}  \hat{I}_{\rm sys} \otimes \op{e}{e}  \hat{\rho}_{\rm tot}  -\frac{1}{2}  \hat{\rho}_{\rm tot}  \hat{I}_{\rm sys} \otimes \op{e}{e}  \Big)  \nn
	\end{align}
The key to connecting the Fock-state master equations to the cascaded equation here is to take the partial trace over the source to find the matrix elements in the source subspace,
	\begin{align}
		\hat{\rho}_{ij} = \bra{i} \hat{\rho}_{\rm tot} \ket{j}.
	\end{align}
These matrix elements are operator-valued objects in the system space.   Projecting the full master equation, \erf{ApEq::FullMasterEquation}, into the source subspace this way shows these matrix elements are coupled via the equations,
	\begin{align}
		\frac{d}{dt} \hat{\rho}_{ee} & = \mathcal{L}_L[\hat{\rho}_{ee}]  - \frac{ |\xi(t)|^2 }{w(t)} \hat{\rho}_{ee}, \label{Eq::drho_ee}  \\
		\frac{d}{dt} \hat{\rho}_{ge} & = \mathcal{L}_L[\hat{\rho}_{ge}] + \frac{\xi(t)}{ \sqrt{w(t)}} \big[ \hat{\rho}_{ee}, \hat{L}\dg ] - \frac{1}{2} \frac{|\xi(t)|^2}{w(t)} \hat{\rho}_{ge}, \\
		\frac{d}{dt} \hat{\rho}_{gg} & = \mathcal{L}_L[\hat{\rho}_{gg}] + \frac{\xi(t)}{ \sqrt{w(t)}} \big[ \hat{\rho}_{eg}, \hat{L}\dg \big] + \frac{\xi^*(t)}{ \sqrt{w(t)}} \big[ \hat{L}, \hat{\rho}_{ge} \big]  + \frac{|\xi(t)|^2}{w(t)} \hat{\rho}_{ee} \label{Eq::drho_gg}.  
	\end{align}
The operators $\hat{L}$ and $\hat{L}\dg$ in these equations are those on the system space, as the operators in the source space have already been included in the projection.
	
	\subsection{Connection to the Fock-state master equations}
	
The critical difference between the cascaded approach and the Fock-state master equations is the way in which each keeps track of the photonics degrees of freedom.  That is; the cascaded approach keeps track of the full quantum state of the qubit photon source, and the Fock-state master equations do not.  In order to see the connection, we first note that by tracing over the source in the cascaded equations we get the reduced system state,
	\begin{align}
		\hat{\varrho}_{\rm sys} & = \Tr_{\rm source} \big[ \hat{\rho}_{\rm tot} \big] = \hat{\rho}_{ee} + \hat{\rho}_{gg},
	\end{align}
which is a combination of the diagonal terms of the source.  The physical state's equation of motion is found by combining \erf{Eq::drho_ee} and \erf{Eq::drho_gg},
	\begin{align}
		\frac{d}{dt} \hat{\varrho}_{\rm sys} & = \mathcal{L}_L[\hat{\rho}_{\rm sys}] + \frac{\xi(t)}{ \sqrt{w(t)}} \big[ \hat{\varrho}_{eg}, \hat{L}\dg \big] + \frac{\xi^*(t)}{ \sqrt{w(t)}} \big[ \hat{L}, \hat{\varrho}_{ge} \big].
	\end{align}
This is very similar to the the equation of motion for $\hat{\varrho}_{11}$ in the Fock-state master equations. By making the following connections,
	\begin{align} \label{Eq::EquationRelations}
		\hat{\varrho}_{11} &= \hat{\varrho}_{\rm sys} = \hat{\rho}_{ee} + \hat{\rho}_{gg} \\
		\hat{\varrho}_{10} &= \hat{\rho}_{ge}/\sqrt{w(t)}  \\
		\hat{\varrho}_{00} &= \hat{\rho}_{ee}/w(t),
	\end{align}
and taking time derivatives\footnote{For the ambitious, this calculation is aided by noting that
	\begin{align}
		\frac{d}{dt} w(t)  = \frac{d}{dt} \int_{t}^\infty ds |\xi(s)|^2  = - |\xi(t)|^2.
	\end{align}}, we complete the correspondence by finding the Fock-state master equations
	\begin{align}
		\frac{d}{dt} \hat{\varrho}_{11} & =  \mathcal{L}_L[\hat{\varrho}_{11}] + \xi(t) \big[ \hat{\varrho}_{01}, L\dg \big] + \xi^*(t) \big[ L, \hat{\varrho}_{10} \big]
 \\
		\frac{d}{dt} \hat{\varrho}_{10} & =  \mathcal{L}_L[\hat{\varrho}_{10}] + \xi(t) \big[ \hat{\varrho}_{00}, L\dg ] \\
		\frac{d}{dt} \hat{\varrho}_{00} & =  \mathcal{L}_L[\hat{\varrho}_{00}].
	\end{align}
Note that the relations \erf{Eq::EquationRelations} may be inverted to find 
	\begin{align}
		\hat{\varrho}_{ee} & = w(t) \hat{\varrho}_{00} \\
		\hat{\varrho}_{ge} & = \sqrt{w(t)} \hat{\varrho}_{10} \\
		\hat{\varrho}_{gg} & = \hat{\varrho}_{11} - w(t) \hat{\varrho}_{00}. 
	\end{align}
So in principle, the Fock-state master equations contain full information about the cascaded source density matrix as well.  That is, the full state of the source and system at time $t$ is
	\begin{align}
		\hat{\rho}_{\rm tot}(t) = & \hat{\varrho}_{ee}(t) \op{e}{e} + \hat{\varrho}_{ge}(t) \op{e}{g} + \hat{\varrho}_{eg}(t) \op{g}{e} + \hat{\varrho}_{gg}(t) \op{g}{g} \\
		= & w(t) \hat{\varrho}_{00}(t) \op{e}{e} + \sqrt{w(t)} \hat{\varrho}_{10}(t) \op{e}{g} + \sqrt{w(t)} \hat{\varrho}_{01}(t) \op{g}{e} \\
		& + \big( \hat{\varrho}_{11}(t) - w(t) \hat{\varrho}_{00}(t) \big) \op{g}{g}. \nn
	\end{align}

\section{Initial conditions}		

The curious initial conditions for the Fock-state master equations, \erf{}, have a clear explanation in the cascaded approach.  Consider a source initialized in its excited state, $\hat{\rho}_{\rm tot}(0) = \hat{\rho}_{\rm sys} \otimes \op{e}{e}$.  This means that $\hat{\varrho}_{ee}(0) = \hat{\rho}_{\rm sys}$, and with \erf{Eq::EquationRelations} we see that the initial conditions for the Fock-state master equations are
	\begin{align}
		\hat{\varrho}_{11}(0) & = \hat{\varrho}_{00}(0)  = \hat{\rho}_{\rm sys},\\
		\hat{\varrho}_{10}(0) & = 0 .
	\end{align}	

A final note on this method.  The cascaded systems approach has been used here to study the specific case of a system driven by a single photon field.  Other cases, including superpositions of Fock states and multi-photon states in multiple spatial or temporal modes require different ``source" models.  Much progress has been made recently on this issue; the authors of Ref. \cite{GouZha14} describe source models for general multi-photon states and superpositions of coherent states.  The great advantage of the Fock-state master equations is that a source model is not necessary.

%

	%
		\chapter{Quantum regression theorem}\label{Appendix::QRT}
%

Here I briefly review the quantum regression theory \cite{Lax63, SteckNotes}.  Let us assume an initially unentangled state of the system and field
	\begin{align}
		\hat{\rho}_{\rm tot}(0) = \hat{\rho}_{\rm sys}(0) \otimes \hat{\rho}_{\rm field}(0).
	\end{align}
As time progresses, the total state of the system and field in general becomes entangled through a unitary operator $\hat{U}(t,0)$,
	\begin{align}
		\hat{\rho}_{\rm tot}(t) = \hat{U}(t,0)\hat{\rho}_{\rm tot}(0)\hat{U}\dg(t,0)  = \hat{U}(t,0) \hat{\rho}_{\rm sys}(0) \otimes \hat{\rho}_{\rm field}(0) \hat{U}\dg(t,0)
	\end{align}
and the reduced state of the system	is found by tracing over the field,
	\begin{align}
		\hat{\varrho}_{\rm sys}(t) = \Tr_{\rm field} \big[ \hat{U}(t,0)\hat{\rho}_{\rm tot}(0)\hat{U}\dg(t,0) \big]. 
	\end{align}
The composition property of the unitary evolution operator is
	\begin{align} \label{Eq::UnitaryComposition}
		\hat{U}(t,t'') = \hat{U}(t,t') \hat{U}(t',t'') ,
	\end{align}	
for $t \geq t' \geq t''$.

We are now equipped to calculate two-time correlations between system operators $\hat{A} \otimes I_{\rm field}$ and $\hat{B}\otimes I_{\rm field}$, for times $t$ and $t+\tau$:	
	\begin{align}
	 	\expt{\hat{A}(t)\hat{B}(t+\tau)} & = \Tr_{\rm sys+field}[\hat{A}(t) \hat{B}(t+\tau) \hat{\rho}_{\rm tot}(0)] \\
			& = \Tr_{\rm sys+field}[ \hat{U}\dg(t,0)\hat{A}  \hat{U}(t,0) \hat{U}\dg(t+ \tau,0) \hat{B} \hat{U}(t+ \tau,0) \hat{\rho}_{\rm tot}(0)] 
	 \end{align}
We can now make use of the cyclic property of the trace along with \erf{Eq::UnitaryComposition} to give
	\begin{align}
	 	\expt{\hat{A}(t)\hat{B}(t+\tau)} & = \Tr_{\rm sys+field}[\hat{A}(t) \hat{B}(t+\tau) \hat{\rho}_{\rm tot}(0)] \nn \\
			& = \Tr_{\rm sys+field}[ \hat{A} \hat{U}\dg(t+ \tau,t) \hat{B} \hat{U}(t+ \tau,t) \underbrace{\hat{U}(t,0) \hat{\rho}_{\rm tot}(0) \hat{U}\dg(t,0)}_{\hat{\rho}_{\rm tot}(t)} ] \nn \\
			& = \Tr_{\rm sys+field}[  \hat{B} \hat{U}(t+ \tau,t)  \hat{\rho}_{\rm tot}(t) \hat{A} \hat{U}\dg(t+ \tau,t)  ] \nn \\
			& = \Tr_{\rm sys} \Big[  \hat{B} \, \Tr_{\rm field} \big[ \hat{U}(t+ \tau,t)  \hat{\rho}_{\rm tot}(t) \hat{A} \hat{U}\dg(t+ \tau,t)  \big] \Big]. 
	 \end{align}
In the last step, we have used the fact that $\hat{B} \otimes I_{\rm field}$ is a Schr\"odinger-picture operator on the system and contains no field dependence.  To follow Steck, we can define
	\begin{align}
		\hat{\mathcal{A}}(t+\tau, t) \equiv & \Tr_{\rm field} \big[ \hat{U}(t+ \tau,t)  \hat{\rho}_{\rm tot}(t) \hat{A} \hat{U}\dg(t+ \tau,t)  \big] 
	\end{align}
subject to the boundary condition,
	\begin{align}
		\hat{\mathcal{A}}(t, t) \equiv & \Tr_{\rm field} \big[ \hat{\rho}_{\rm tot}(t) \hat{A} \big] \\
			= & \Tr_{\rm field} \big[ \hat{U}(t,0)  \hat{\rho}_{\rm tot}(0) \hat{U}\dg (t,0) \hat{A} \big] \\
			 = & \hat{\varrho}_{\rm sys}(t) \hat{A}. \label{ApEq::QRTop}
	\end{align}
The two-time correlation function can be written,
	\begin{align}
		\expt{ \hat{A} (t)\hat{B}(t+\tau)} & = \Tr_{\rm sys} \big[ \hat{B} \hat{\mathcal{A}}(t+\tau, t) \big].
	\end{align}
To use this formula for calculations, follows these steps.  First, find $\hat{\varrho}_{\rm sys}(t)$ at time $t$ and use it to find $\hat{\mathcal{A}}(t,t)$ using \erf{ApEq::QRTop}.  Second, we evolve the operator $\hat{\mathcal{A}}(t',t)$ from time $t$ to $t'$.  Finally, at time $t'$ take the trace of $\hat{\mathcal{A}}(t',t)$ with system operator $\hat{B}$.

		\chapter{Fock-state excitation of a two-level atom: formal solution}\label{Appendix::NPhotonExcitation}
The problem of a two-level atom interacting with an $N$-photon Fock state can be treated using a fully unitary method that tracks the joint system-field state similar to the cavity problems considered in Ref. \cite{Bana13}.  This solution was derived by Julio Gea-Banacloche and is presented here with permission \cite{BanaclochePrivate}.   
Here we proceed using the white-noise Hamiltonian in \erf{Eq::WhiteNoiseDipHam} with $\hat{c} = \op{g}{e}$ and $\sqrt{\gamma} = \sqrt{2\pi} \kappa(\omega_0)$. The Markov approximation has already been made and the input field is described by the operators, $\hat{b}(t)$ and  $\hat{b}\dg(t)$.  The joint state at any can time can be written as
	\begin{equation}
		\ket{\psi(t)} = \ket e \otimes \ket{\phi_e(t)} + \ket g \otimes \ket{\phi_g(t)},
		\label{e7}
	\end{equation}
where $\ket{\phi_e(t)}$ and $\ket{\phi_g(t)}$ are photonic wave functions to be determined.  
From the Hamiltonian, the equations of motion are
	\begin{subequations} \label{ApEq::PhotonicEOMs}
		\begin{align}
			d \ket{\phi_e(t)} &= - \sqrt{\gamma} \hat{b}(t) \ket{\phi_g(t)} \label{e8a}, \\
		 	d \ket{\phi_g(t)} &=  \sqrt{\gamma} \hat{b}\dg(t) \ket{\phi_e(t)} .
		\label{e8b}
		\end{align}
	\end{subequations}
Formally integrating for $\ket{\phi_g(t)}$, using the initial conditions $\ket{\psi_e(0)} = 0$ and $\ket{\psi_g(0)} = \ket{N_\xi}$, substituting into the equation for $\ket{\phi_e(t)}$, and normally ordering gives 
	\begin{equation}
		\frac{d}{dt}\ket{\phi_e(t)} = -\frac{\gamma}{2} \ket{\phi_e(t)} - \sqrt{\gamma  N} \xi(t) \ket{N-1_\xi} - \gamma \int_0^t dt^\prime \hat{b}\dg(t') \hat{b}(t)\ket{\phi_e(t^\prime)} . 
\label{e10}
		\end{equation}
This equation can be solved by introducing an integrating factor $e^{- \frac{\gamma}{2} t}$,
	\begin{equation}
		\ket{\phi_e(t)} = -\sqrt{\gamma N} \! \int_0^t \! dt^\prime e^{-\frac{\gamma}{2}(t-t^\prime)}  \xi(t^\prime) \ket{N-1_\xi} - \gamma \! \int_0^t \! dt^\prime \int_0^{t^\prime} \! dt^{\prime\prime} e^{-\frac{\gamma}{2}(t-t^\prime)} \hat{b}\dg(t'') \hat{b}(t')  \ket{\phi_e(t^{\prime\prime})}.
\label{e11}
	\end{equation}
Inserting the expression for $\ket{\phi_e(t)}$ into the integral on the right side, similar to a Born series, we get at the first iteration,
	\begin{align}
		\ket{\phi_e(t)} &  = -\sqrt{\gamma N} \int_0^t e^{-\frac{\gamma}{2}(t-t_1)}  \xi(t_1)\, dt_1 \ket{N-1_\xi} \label{e12} \\
		&\!+\gamma^{3/2} \sqrt{N(N-1)} \! \int_0^t \! dt_1 \! \int_0^{t_1} \!  dt_2 \! \int_0^{t_2} \! dt_3  e^{-\frac{\gamma}{2}(t-t_1)} e^{-\frac{\gamma}{2}(t_2-t_3)} \xi(t_1)  \xi(t_3) \hat{b}^\dagger(t_2) \ket{N-2_\xi}\cr
		&\!+\gamma^2 \! \int_0^t dt_1 \!  \int_0^{t_1} \! dt_2 \!  \int_0^{t_2} \! dt_3 \!  \int_0^{t_3} \! dt_4 \, e^{-\frac{\gamma}{2}(t-t_1)} e^{-\frac{\gamma}{2}(t_2-t_3)}  \hat{b}^\dagger(t_2) \hat{b}(t_1)  \hat{b}^\dagger(t_4) \hat{b}(t_3)\ket{\phi_e(t_4)}. \nn
	\end{align}
Normally ordering the operators in the last line, the term involving $\delta(t_1-t_4)$ vanishes.  This applies at each iteration, and the resulting pattern is that each successive iteration brings two new integrals, a factor $\hat{b}^\dagger(t_n) e^{-\frac{\gamma}{2}(t_n-t_{n+1})}\xi(t_{n+1})$, and acts on a state with one fewer photon.  Because the field starts with a finite number of photons, the series eventually terminates. The analytic solution is then:
	\begin{align}
		\ket{\phi_e(t)} &= -\sqrt{\gamma N} \int_0^t e^{-\frac{\gamma}{2}(t-t_1)}  f(t_1)\, dt_1 \ket{N-1_\xi}\label{e13} \\
		&\!+\gamma^{3/2} \sqrt{N(N-1)} \int_0^t \! dt_1\!  \int_0^{t_1} \! dt_2 \!  \int_0^{t_2} \! dt_3 \, e^{-\gamma(t-t_1)} e^{-\gamma(t_2-t_3)} \xi(t_1)  \xi(t_3) \hat{b}^\dagger(t_2)  \ket{N-2_\xi} \cr
		&\!-\gamma^{5/2}\sqrt{N(N-1)(N-2)}  \int_0^t dt_1 \int_0^{t_1} dt_2 \int_0^{t_2} dt_3 \int_0^{t_3} dt_4 \int_0^{t_4} dt_5 \nn \\
		&\qquad \times e^{-\gamma(t-t_1)} e^{-\gamma(t_2-t_3)} e^{-\gamma(t_4-t_5)} \xi(t_1)  \xi(t_3) \xi(t_5)  \hat{b}^\dagger(t_2) \hat{b}^\dagger(t_4) \ket{N-3_\xi} + \dots \nn
	\end{align}
One can in principle substitute this expression into (\ref{e8b}) and integrate to find $\ket{\psi_g(t)}$, thus giving the full joint state.  In practice, the integrals may prove too difficult to evaluate analytically, except for simple pulses.  From this solution, we can find the excitation probability at any time, $\mathbbm{P}_e(t) = \ip{\psi_e(t)}{\psi_e(t)}$.  

		\chapter{Occupation number representation for $N$-photon states}\label{Appendix::NPhotonStates}
%

\section{$N$-photon states} 

Here we review the occupation number representation of a general $N$-photon state presented in Ref. \cite{Roh07}.  In one dimension and in a single mode, a general quasi-monochromatic $N$-photon state can be written in the time domain, this becomes
	\begin{align} \label{Eq::multiphotonQuantumState}
		\ket{\psi_N} = \frac{1}{\sqrt{\mathcal{N}} }\int dt_1\dots dt_N  \,  {\psi}(t_1,\dots,t_N) b\dg(t_1)\dots b\dg(t_N) \ket{0}\,,
	\end{align}
where the temporal envelope $\psi(t_1,\dots,t_N)$ is the Fourier transform of $\tilde\psi(\omega_1,\dots,\omega_N)$ \cite{BloLou90}. The temporal envelope is in general neither factorable nor symmetric in its indices, $t_i$, indicating that the photons are entangled in the spectral/temporal degrees of freedom.  The normalization factor $\mathcal{N}$ is a function of the permutation symmetry of the temporal envelope,
	\begin{align} \label{ApEq::Normalization}
		\mathcal{N} = \int dt_1 \dots \int dt_N \, \psi^*(t_1,\dots,t_N)  \sum_{\sigma \in \mathcal{S}_N} \psi \big( \sigma[t_1,\dots,t_N] \big),
	\end{align}
where $\sigma$ represents permutations over the indices.  
	
The level of permutation symmetry of $\psi(t_1,\dots,t_N)$ is directly related to the temporal distinguishability of the photons as characterized by the visibility in a generalized Hong-Ou-Mandel experiment \cite{Ou06, Ou08}.  A temporal envelope is fully permutation symmetric if
	\begin{align}
		\psi(t_1,\dots,t_N) = \psi(P[t_1,\dots,t_N]).
	\end{align}
In this case the sum in \erf{ApEq::Normalization} can be done, and such a state has normalization $\mathcal{N} = N! \int dt_1 \dots \int dt_N \, |\psi(t_1,\dots,t_N)|^2 $.  Fock states exhibit full permutation symmetry, and further, the temporal envelope factors, $\psi(t_1,\dots,t_N) = \Pi_{i=1}^N \xi(t_i)$. Thus, $\mathcal{N} = N! $.  For a state that exhibits no permutation symmetry, $\psi(t_1,\dots,t_N)$ is orthogonal to all permutations of its indices with respect to the integration in \erf{ApEq::Normalization}.  

We demonstrate these cases with a simple example.  Consider a 2-photon state,
	\begin{align} \label{ApEq::ExampleTwoPhoton}
		\ket{\psi_2} = \frac{1}{\sqrt{\mathcal{N}}} \int dt_1 \int dt_2 \, \psi(t_1,t_2) \hat{b}\dg(t_1) \hat{b}\dg(t_2) \ket{0},
	\end{align}
with normalization 
	\begin{align}
		\mathcal{N} =  \int dt_1 \int dt_2 \, \psi(t_1,t_2) \big( \psi^*(t_1,t_2) + \psi^*(t_2,t_1) \big).
	\end{align}
For a factorizable envelope, $ \psi(t_1,t_2) = \xi(t_1) \eta(t_2)$, the normalization,
	\begin{align} \label{Eq::NormEx}
		\mathcal{N} 
		=&  \int dt_1 |\xi(t_1)|^2 \int dt_2 \, |\eta(t_2)|^2 + \int dt_1 \xi(t_1)\eta^*(t_1) \int dt_2 \, \eta(t_2)\xi^*(t_2),
	\end{align}
directly shows us that the degree of temporal distinguishability affects the relative size of the integrals in the second term.  If $\xi(t)$ and $\eta(t)$ are orthogonal, meaning that the two photons are perfectly distinguishable, then this second term disappears. In this case one can associate a mode creation operator with both $\xi(t)$ and $\eta(t)$, and the state can be represented as a tensor product of single-photon Fock states $\ket{\psi_2} = \ket{1_\xi} \otimes \ket{1_\eta} = \ket{1,1}$.  If $\xi(t) = \eta(t)$ then the two photons are in the same temporal wave packet and the second term in \erf{Eq::NormEx} is identical to the first, giving a factor of 2.  In this case the state is a two-photon Fock state, $\ket{\psi_N} = \ket{2_\xi}$.  Partial overlap of $\xi(t)$ and $\eta(t)$ indicates partial distinguishability of the photons, and some work is required to write the state in a basis of Fock states over temporal modes.  That is the subject of the following subsection.

\section{Occupation number representation of an $N$-photon state}

The general $N$-photon state can be expressed in a set of Fock states defined on a basis of orthogonal temporal modes.  Associated with each mode is a mode creation operator, repeated application of which to vacuum produces Fock states.  We first identify a set of complex-valued, orthonormal basis functions that satisfy $\int dt \, \xi_i^*(t) \xi_j(t) = \delta_{i,j}$.  Expanded in this basis, the temporal envelope is
	\begin{align} \label{Eq::BasisWavepackets}
		\psi(t_1,\dots,t_N) = \sum_{i_1} \cdots \sum_{i_N}  \lambda'_{i_1,\dots,i_N} \xi_{i_1}(t_1)...\xi_{i_N}(t_N).
	\end{align}
Each subscript runs over the labels for the basis functions, i.e. $i_k \in \{ \alpha,\dots \zeta\}$\footnote{Note that there is no relationship between the number of basis functions and the number of photons -- even a single photon can be expanded in a countably infinite set of basis functions.}. The expansion coefficients are given by the projection of the temporal envelope onto the basis functions,
	\begin{align} \label{Eq::MPExpansionC}
		\lambda'_{\alpha,...\zeta} = \int  dt_1\dots dt_N  \,  \xi_{\alpha}^*(t_1)\dots\xi_{\zeta}^*(t_N) {\psi}
			(t_1,\dots,t_N).
	\end{align}
Defining a creation operator for a single photon in basis mode $\xi_{\alpha}(t)$ as $ B\dg (\xi_{\alpha}) =\int  dt \, \xi_{\alpha}(t) b\dg(t)$, and using Eq. (\ref{Eq::multiphotonQuantumState}-\ref{Eq::MPExpansionC}), we write the $N$-photon state as
	\begin{align}
		\ket{\psi_N} = \frac{1}{ \sqrt{\mathcal{N} }} \sum_{i_1,\dots,i_N} \lambda'_{i_1,\dots,i_N} B\dg(\xi_{i_1}) \dots B\dg(\xi_{i_N}) \ket{0}. 
	\end{align}
Acting these operators on vacuum yields an expression for the $N$-photon state in terms of basis Fock states, \erf{Eq::ContModeFockState}, in the basis functions,
	\begin{align} \label{Eq::FirstOcRep}
		\ket{\psi_N} = \frac{1}{\sqrt{\mathcal{N}}} \sum_{i_1,\dots,i_N} \lambda'_{i_1,\dots,i_N} \sqrt{n_1! n_2! \dots} \ket{ {n_1}_{\xi_1}} \ket{{n_2}_	{\xi_2} } ...
	\end{align}
	Counting the number of subscripts of $\lambda'$ gives the total photon number $N$, which can be distributed among the basis Fock states in \erf{Eq::FirstOcRep}.  The number of photons $n_\alpha$ in a particular basis function $\xi_\alpha(t)$ is found by counting the number of indices of $\lambda'$ that are equal to $\alpha$.  For example, since they have 3 indices, the coefficients $\{ \lambda_{i_1,i_2,i_3}' \}$ all describe a 3-photon state.  The coefficient $\lambda_{1,1,4}'$ refers to the state $ \ket{ {2}_{\xi_1}} \ket{ {1}_{\xi_4}} $, in which the first and second photons are in $\xi_1(t)$ and the third in $\xi_4(t)$.  Due to the indistinguishability of photons, $\lambda_{1,4,1}'$ and $\lambda_{4,1,1}'$ are also coefficients for the state $ \ket{ {2}_{\xi_1}} \ket{ {1}_{\xi_4}} $, although they need not have the same value.  In general, $\lambda_{\alpha,\dots,\zeta}'$ is not invariant under permutation of its indices. The degree to which index permutations are is a function of the symmetry in the temporal envelope $\psi(t_1,...,t_N)$ \cite{Ou06, Ou08}.

We define a new set of coefficients,
	\begin{align} \label{Eq::MPAmplitudes}
		c_{n_1, n_2, ...} = \sqrt{ \frac{ n_1!n_2!\dots }{\mathcal{N} } } \sum_{\sigma \in \mathcal{S}_N} \lambda'_{\sigma( i_1,\dots,i_N ) },
	\end{align}
that sum over all permutations $\sigma$ (in the symmetric group $\mathcal{S}_N$) of the indices of coefficients of the type in \erf{Eq::MPExpansionC}.  The subscripts $n_i$ are the number of photons in mode $i$, the number of subscripts is the number of temporal modes, and the normalization factor has been absorbed.  With these coefficients the $N$-photon state in \erf{Eq::multiphotonQuantumState} can be written,
	\begin{align} \label{AEq::NPhotonState}
		\ket{\psi_N} = \sum_{n_1,n_2,...} c_{n_1, n_2,...} \ket{ {n_1}, {n_2}, \dots }
	\end{align}
The state itself has been written such that the order of its elements specifies the temporal modes in the basis, for compactness.  It is clear that these algebraic acrobatics have culminated in a set of expansion coefficients that are precisely probability amplitudes,
	\begin{align}
		\sum_{n_1,n_2,...} |c_{n_1, n_2, ...}|^2 = 1,
	\end{align}
and \erf{AEq::NPhotonState} is the \emph{occupation number representation}\footnote{We make a departure here from Refs. \cite{Roh07, TeamAwesome}, wherein the final probability amplitude coefficients were defined similar to \erf{Eq::FirstOcRep}.  Those coefficients have the advantage that for a finite number of photons they have a finite number of indices, even when the number of temporal modes is infinite.  However, for the Fock-state formalism the coefficients here [\erf{Eq::MPAmplitudes}] are more convenient, since the subscripts of each generalized density matrix specify the number of photons in each temporal mode, even those in vacuum.} of a general $N$-photon state.  This representation of the state will be most useful in our multimode Fock-state formalism.

	\subsection{Two-photon example}
	
To illustrate the formalism of the occupation number representation, consider the two-photon state in \erf{ApEq::ExampleTwoPhoton}.  For the sake of example, assume that the state can be represented in a basis consisting of two temporal modes, $\{ \xi_1(t), \xi_2(t) \}$.  Projecting the temporal function $\psi(t_1, t_2)$ onto those modes, we can represent the state as
	\begin{align}
		\ket{\psi_2} = & \frac{1}{ \sqrt{\mathcal{N} }} \sum_{i_1=1}^2 \sum_{i_2=1}^2 \lambda'_{i_1,i_2} B\dg(\xi_{i_1}) B\dg(\xi_{i_2}) \ket{0} \\ 
		= & \frac{1}{ \sqrt{\mathcal{N} }} \Big\{ \lambda'_{1,1} \big[ B\dg(\xi_{1}) \big]^2 + \big( \lambda'_{1,2} +  \lambda'_{2,1}\big) B\dg(\xi_{1}) B\dg(\xi_{2}) + \lambda'_{2,2} \big[ B\dg(\xi_{2}) \big]^2 \Big\} \ket{0} \\ 
		= & \frac{1}{ \sqrt{\mathcal{N} }} \Big\{  \sqrt{2} \lambda'_{1,1}  \ket{2_{\xi_1}} \ket{0_{\xi_2}}  + \big( \lambda'_{1,2} +  \lambda'_{2,1}\big) \ket{1_{\xi_1}} \ket{1_{\xi_2}} + \sqrt{2}  \lambda'_{2,2} \ket{0_{\xi_1}} \ket{2_{\xi_2}}  \Big\} \ket{0}  \\
			=& c_{2,0} \ket{2,0} + c_{1,1} \ket{1,1} + c_{0,2} \ket{0,2} . \label{ApEq::TwoPhotonOcNum} 
	\end{align}
The square root factors in the third line come from the definitions of the two-photon Fock states.  The state in the final line is the occupation mode representation using the coefficients in \erf{Eq::MPAmplitudes}.

		\chapter[Corrections to the master equation]{Higher-order corrections to the master equation} \label{Appendix::MECorrections}
%

In this appendix we derive the terms in the master equation, \erf{Eq::StandarME}, to order $1/\Delta^2$.  Even in this limit, a fully quantum treatment of the master equation reveals that decoherence acts not only on the atom but also on the paraxial field  \cite{VasSor12, Tra11}.  The aim of this appendix is to show that when one of the polarization components of the paraxial field is displaced to a coherent state, the standard master equation is recovered.  

To describe entanglement generated between the atom and the paraxial field, we consider here a fully quantum field, \erf{Eq::SimpleField}.  For compactness, we designate the single photon amplitude as $E_0 \equiv \sqrt{2 \pi \hbar \omega_0 / Ac}$ and suppress the position, time, and mode labels on the field operators, $\hat{a}_{i, \lambda}(z,t) \rightarrow \hat{a}_\lambda$, leaving only the polarization index $\lambda$.  
First, we expand the coefficient in the atomic polarizability tensor, \erf{Eq::PTensorGen}, into real and imaginary parts to order $1/\Delta^2$,
	\begin{align}
		\frac{1}{\Delta_{f'} + i \Gamma/2} & = \frac{\Delta_{f'}}{\Delta^2_{f'} + \Gamma^2/4} + \frac{-i\Gamma/2}{\Delta^2_{f'} + \Gamma^2/4}	 \\
		& \approx \frac{1}{\Delta} \left( 1 - \frac{ \delta_{f'}}{\Delta}\right) - i \frac{1}{\Delta} \left( \frac{\Gamma}{2 \Delta} \right).\label{Eq::DetExpansion}
	\end{align}
The approximation applies for detunings larger than the spontaneous emission rate, $\Delta \gg \Gamma$, 
and we have decomposed the detuning as $\Delta_{f'} = \Delta + \delta_{f'}$.

	\subsubsection{Real part of the interaction: coherent terms}

The coherent evolution is given by the real part -- the first term in Eq. (\ref{Eq::DetExpansion}) -- of the effective interaction, \erf{Eq::LightShiftHam}.  Written using the irreducible tensor decomposition, \erf{Eq::IrreducibleDecomp}, the interaction is
	\begin{align}
		\! \hat{H}_{\rm coh}  
			\!=\!  & |E_0|^2  \alpha_0(\Delta) \! \sum_{f'} \! \Big( 1 \! - \! \frac{ \delta_{f'}}{\Delta} \Big) \bigg\{ C^{(0)}_{j' f f'} \hat{ I } \big( \hat{a}_x^\dagger \hat{a}_x \!+\!  \hat{a}_y^\dagger \hat{a}_y \big) \! + i C^{(1)}_{j' f f'} \hat{f}_z \! \left( \hat{a}_x^\dagger \hat{a}_y \!-\! \hat{a}_y^\dagger \hat{a}_x \right)  \\
			   & + C^{(2)}_{j' f f'} \bigg[ \Big( \hat{f}_x^2 - \frac{\hat{\mathbf{f}}^2}{3}   \Big) \hat{a}_x^\dagger \hat{a}_x +  \Big( \hat{f}_y^2 - \frac{\hat{\mathbf{f}}^2}{3} \Big) \hat{a}_y^\dagger \hat{a}_y+ \frac{ \hat{f}_x \hat{f}_y + \hat{f}_y \hat{f}_x}{2} \big(\hat{a}_x^\dagger \hat{a}_y + \hat{a}_y^\dagger \hat{a}_x \big) \bigg] \bigg\} .\nn
	\end{align} 
We now displace the $x$-polarized field to the coherent state $\ket{\alpha_x}$ with photon flux $\dot{N}_L$.  For a sufficiently powerful probe, the terms proportional to $\hat{a}_y^\dagger \hat{a}_y$ can be dropped, and the coherent Hamiltonian becomes
	\begin{align} \label{Eq::HcohXfield}
		\bra{\alpha_x} \hat{H}_{\rm coh}  \ket{\alpha_x}  
			=& |E_0|^2 \alpha_0(\Delta)\sum_{f'}\Big( 1 - \frac{ \delta_{f'}}{\Delta} \Big) \bigg\{ \dot{N}_L C^{(0)}_{j' f f'} \hat{ I}+ i \sqrt{\dot{N}_L} C^{(1)}_{j' f f'}  \hat{f}_z \left( \hat{a}_y - \hat{a}_y^\dagger  \right)  \nn \\
			  & \quad  + C^{(2)}_{j' f f'} \bigg[ \dot{N}_L \Big( \hat{f}_x^2 - \frac{\hat{\mathbf{f}}^2}{3} \Big) + \sqrt{\dot{N}_L} \frac{\hat{f}_x \hat{f}_y + \hat{f}_y \hat{f}_x}{2} \left( \hat{a}_y + \hat{a}_y^\dagger \right) \bigg] \bigg\}  . 
	\end{align}

	\subsubsection{Imaginary part of the interaction: loss terms}
	
The anti-Hermitian part of the Hamiltonian, Eq. (\ref{Eq::DetExpansion}), to order $1/\Delta^2$ is
	\begin{align}
		\hat{H}_{\rm loss} 
		& =  - i \frac{\Gamma}{2 \Delta} |E_0|^2 \alpha_0(\Delta) \left[ C^{(0)}_{j' f}\hat{ I } \big( \hat{a}_x^\dagger \hat{a}_x + \hat{a}_y^\dagger \hat{a}_y \big) + i C^{(1)}_{j' f} \hat{f}_z \big( \hat{a}_x^\dagger \hat{a}_y - \hat{a}_y^\dagger \hat{a}_x \big) \right] . 
	\end{align}
Putting the $x$-polarized field in a coherent state gives
	\begin{align} \label{ApEq::CoherentLoss}
		\bra{\alpha_x}\hat{H}_{\rm loss}\ket{\alpha_x} = - i \frac{\Gamma}{2 \Delta} |E_0|^2 \alpha_0(\Delta) \Big[ \dot{N}_L  C^{(0)}_{j' f} \hat{ I }  + i \sqrt{ \dot{N}_L } C^{(1)}_{j' f} \hat{f}_z ( \hat{a}_y - \hat{a}_y^\dagger  \big) \Big].
	\end{align}

		\subsubsection{Jump operators}
		
In addition to loss driven by the non-Hermitian Hamiltonian, we must also account for incoherent feeding.  The Cartesian jump operators, $\lambda \in \{x,y,z\}$, are
	\begin{align}
		\hat{W}_\lambda &=  \sum_{f'} \frac{ \bra{j'}| d | \ket{j}/\hbar }{ \Delta_{f'} + i\Gamma/2 } \left( \mathbf{e}_\lambda \cdot \hat{\mathbf{D}}_{f f'} \right) \left(\hat{ \mathbf{D}}_{f'f}^\dagger \cdot ( \hat{a}_x \mathbf{e}_x + \hat{a}_y  \mathbf{e}_y    ) \right). 
	\end{align}
In the feeding terms, the jump operators appear in conjugate pairs. To order $1/\Delta^2$,
	\begin{align}
		 \sum_{f'} \frac{1}{| \Delta_{f'} + i\Gamma/2 |^2} 
			\approx \frac{1}{\Delta^2},
	\end{align} 
and thus this gives feeding terms,
	\begin{align}
		\Gamma \sum_\lambda \hat{W}_\lambda \hat{\rho} \hat{W}_\lambda^\dagger 
		   =  \frac{\Gamma}{\Delta} |E_0|^2  \alpha_0(\Delta) \Big\{ \big( C^{(0)}_{j' f} \hat{ I } \hat{a}_x + i C^{(1)}_{j' f} \hat{f}_z  \hat{a}_y \big) & \hat{\rho} \big( C^{(0)}_{j' f} \hat{ I } \hat{a}_x^\dagger - i C^{(1)}_{j' f} \hat{f}_z  \hat{a}_y^\dagger \big)   \\
		  + \big( \! - \! i C^{(1)}_{j' f} \hat{f}_z  \hat{a}_x + C^{(0)}_{j' f} \hat{ I } \hat{a}_y \big) & \hat{\rho} \big( i C^{(1)}_{j' f} \hat{f}_z  \hat{a}^\dagger_x + C^{(0)}_{j' f}\hat{ I } \hat{a}^\dagger_y \big) \nonumber \\
		  + \big( i C^{(1)}_{j' f} \hat{f}_y  \hat{a}_x - i C^{(1)}_{j' f} \hat{f}_x  \hat{a}_y \big) & \hat{\rho} \big( \! - \! i C^{(1)}_{j' f} \hat{f}_y  \hat{a}^\dagger_x + i C^{(1)}_{j' f} \hat{f}_x  \hat{a}^\dagger_y \big) \Big\} \nn.
	\end{align}	
With the $x$-polarized field in a coherent state these terms become,
	\begin{align}
		\Gamma \sum_\lambda \bra{\alpha_x}\hat{W}_\lambda \hat{\rho} \hat{W}_\lambda^\dagger \ket{\alpha_x}
		   =  \frac{\Gamma}{\Delta} |E_0|^2 & \alpha_0(\Delta) \bigg\{ \dot{N}_L \Big[ |C^{(0)}_{j' f}|^2  \hat{\rho} + |C^{(1)}_{j' f}|^2 \big(  \hat{f}_z \hat{\rho} \hat{f}_z + \hat{f}_y \hat{\rho} \hat{f}_y \big) \Big] \nn \\
		   &+ i \sqrt{\dot{N}_L}C^{(0)}_{j' f} C^{(1)}_{j' f} \big( \hat{f}_z  \hat{a}_y \hat{\rho} - \hat{\rho} \hat{f}_z  \hat{a}_y\dg + \hat{a}_y  \hat{\rho} \hat{f}_z -  \hat{f}_z \hat{\rho}  \hat{a}_y\dg\big) \nn \\
		   & - \sqrt{\dot{N}_L} | C^{(1)}_{j' f} |^2 \big( \hat{a}_y \hat{f}_x \hat{\rho}  \hat{f}_y +  \hat{f}_y  \hat{\rho} \hat{f}_x \hat{a}_y\dg \big)  \bigg\}. \label{ApEq::JumpCoherent}
	\end{align}

	\section{Comparing terms for realistic parameters}
		
We now take the expressions for the coherent, loss, and feeding terms -- Eqs. (\ref{Eq::HcohXfield}),  (\ref{ApEq::CoherentLoss}), and (\ref{ApEq::JumpCoherent}) -- and compare the orders of magnitude of the terms for realistic experimental parameters\footnote{It in interesting to note that, to order $1/\Delta^2$, the rank-2 tensor terms only appear in the coherent dynamics, due to the extra power of $1/\Delta$ in the other terms.}.  The typical order of the parameters for $^{133}$Cs are
	\begin{align}
		\Delta & \approx 2500 \, \mbox{MHz}, \quad \delta_{F'}  \approx 250 \, \mbox{MHz}, \quad \Gamma/2  \approx 2.5 \, \mbox{MHz}, 
	\end{align}
noting that $\Delta$ can vary in experiments.  Now consider a short coherent pulse containing $N_L = 10^6$ photons \cite{SewMit12b}.  The relative coefficients that show up in the master equation are
	\begin{align}
		\frac{ \delta_{F'} }{\Delta} N_L & \rightarrow 10^5 ,\\
		\frac{ \delta_{F'} }{\Delta} \sqrt{N_L} & \rightarrow 10^2 ,\\
		\frac{ \Gamma }{2 \Delta} N_L &  \rightarrow 10^3 ,\\
		\frac{ \Gamma }{2 \Delta} \sqrt{N_L} &  \rightarrow 1.
	\end{align}
This comparably negligible coefficient in the final line is the prefactor in the loss Hamiltonian and feeding terms that involves paraxial field operators.  The resulting coherent and loss parts of the effective Hamiltonian are
		\begin{align} \label{Eq::Coh}
		\hat{H}_{\rm coh} & =  \sqrt{\dot{N}_L} |E_0|^2 \alpha_0(\Delta) \Big( C^{(1)}_{j' f} - \sum_{f'} \frac{ \delta_{f'}}{\Delta} C^{(1)}_{j' f f'} \Big) i \hat{f}_z ( \hat{a}_y - \hat{a}_y^\dagger )  \\
			& - |E_0|^2 \alpha_0(\Delta)\sum_{f'} \frac{ \delta_{f'}}{\Delta} C^{(2)}_{j' f f'}  \Big( \dot{N}_L \left( \hat{f}_x^2 - \hat{\mathbf{f}}^2/3  \right) + \frac{ \sqrt{\dot{N}_L}  }{2}\left(\hat{f}_x \hat{f}_y + \hat{f}_y \hat{f}_x\right) \left( \hat{a}_y + \hat{a}_y^\dagger \right) \Big) , \nn \\
		\hat{H}_{\rm loss} & = - i \dot{N}_L \frac{\Gamma}{2 \Delta} |E_0|^2 \alpha_0(\Delta) C^{(0)}_{j' f} \hat{ I },
	\end{align}
and the feeding terms are
	\begin{align}
		\Gamma \sum_i \hat{W}_\lambda \hat{\rho} \hat{W}_\lambda^\dagger &  =  N_L \frac{\Gamma}{\Delta} |E_0|^2 \alpha_0(\Delta) \left( |C^{(0)}_{j' f}|^2 \hat{\rho} + |C^{(1)}_{j' f}|^2 \big(  \hat{f}_z \hat{\rho} \hat{f}_z +  \hat{f}_y \hat{\rho} \hat{f}_y \big) \right).
	\end{align}	
With this approximation, all of the decoherence that acts directly on the paraxial field has dropped out, leaving loss and feeding terms that act solely on the atom.  This recovers the master equation in \erf{Eq::XPolarizedME}.  The relative importance of neglected terms changes as the probe is moved closer to resonance.  Specifically, when the detuning approaches the hyperfine splitting, then higher-order terms in the decoherence as well as tensor effects in the coherent dynamics will become important.

		 \chapter[Continuous polarimetry stochastic master equation]{Derivation of the stochastic master equation for continuous polarimetry measurements} 
%


To project onto $\hat{X}$-eigenstates in each mode $i$, we first use the Zassenhaus formula,
	\begin{align}
		e^{\hat{A}+\hat{B}} = e^{\hat{A}} e^{\hat{B}} e^{-\half [\hat{A},\hat{B}]},
	\end{align}
to rewrite the time evolution operator in mode $i$. With the associations, 
	\begin{align}
		\hat{A} &= -i \sqrt{ \frac{ \CstrengthSq }{  2 }} \int_{t_0}^t dt' \, \mbox{Im} \big\{ \hat{F}_{z}^i  \big\} \hat{X}_{i}(z_D,t'), \\
		\hat{B} &= i \sqrt{ \frac{ \CstrengthSq }{  2 }} \int_{t_0}^t dt' \, \mbox{Re} \big\{ \hat{F}_{z}^i  \big\} \hat{P}_{i}(z_D,t'),
	\end{align}
which have a commutation relation,
	\begin{align}
		[ \hat{A},\hat{B}] = i \frac{\Cstrength (t-t_0)}{2}  \mbox{Re} \big\{ \hat{F}_{z}^i \big\}  \mbox{Im} \big\{ \hat{F}_{z}^i \big\}, 
	\end{align}
the time evolution operator between the atomic ensemble and the propagating field modes at the detector is
	\begin{align} \label{ApEq::EvolutionOperator}
		\hat{U}_{i}(t_0,t) = & \exp  \bigg[ -i \frac{\Cstrength (t-t_0)}{4}  \mbox{Re} \big\{ \hat{F}_{z}^i \big\}  \mbox{Im} \big\{ \hat{F}_{z}^i \big\} \bigg] \\
		&\times \overleftarrow{\mathcal{T}} \Bigg\{   \exp  \bigg[ -i \sqrt{ \frac{ \CstrengthSq }{  2 }} \int_{t_0}^t dt' \, \mbox{Im} \big\{ \hat{F}_{z}^i  \big\} \hat{X}_{i}(z_D,t') \bigg] \Bigg\}   \nn \\
		& \times \overleftarrow{\mathcal{T}} \Bigg\{   \exp  \bigg[ i \sqrt{ \frac{ \CstrengthSq }{  2 }} \int_{t_0}^t dt' \, \mbox{Re} \big\{ \hat{F}_{z}^i  \big\} \hat{P}_{i}(z_D,t') \bigg] \Bigg\} 
		\nn .
	\end{align}
	
The Kraus component for continuous measurements of the $\hat{X}$ quadrature in mode $i$ is obtained by evolving via \erf{ApEq::EvolutionOperator} for a small time interval $\Delta t$ and projecting onto an $\hat{X}_i$-eigenstate over that interval,	
	\begin{align} \label{Eq::GaussianKraus}
		\bra{\hat{X}_i = x_i} \hat{U}_{i}(t, \, & t+  \Delta t  )  \ket{0} =  \exp  \bigg[ -i \frac{\Cstrength \Delta t}{4}  \mbox{Re} \big\{ \hat{F}_{z}^i \big\}  \mbox{Im} \big\{ \hat{F}_{z}^i \big\} \bigg] \\
		& \times \exp \bigg[- i \sqrt{ \frac{ \CstrengthSq }{  2 }} \Delta t \, \mbox{Im} \big\{ \hat{F}_{z}^i  \big\} x_i \bigg]  \nn \\
		& \times \bra{\hat{X}_i = x_i} \overleftarrow{\mathcal{T}} \Bigg\{   \exp  \bigg[ i \sqrt{ \frac{ \CstrengthSq }{  2 }}\int_{t}^{t+\Delta t} dt' \, \mbox{Re} \big\{ \hat{F}_{z}^i  \big\} \hat{P}_{i}(z,t') \bigg] \Bigg\} \ket{0} \nn .
	\end{align}
The final term is a translation of the $\hat{X}$-eigenstate and can be written
	\begin{align}
		\big\langle \hat{X}_{i} = x_{i} +    \sqrt{\smallfrac{\CstrengthSq}{2 } } \mbox{Re} \big\{ \hat{F}_z^i  \big\} | 0 \big\rangle 
		&= \exp \bigg[ -\frac{\Delta t}{2} \Big( x_i +   \sqrt{\smallfrac{\CstrengthSq}{2 } } \mbox{Re} \big\{ \hat{F}_z^i \big\} \Big)^2 \bigg]. 
		\end{align}
Putting this together with \erf{Eq::GaussianKraus}, we get
	\begin{align} \label{ApEq::KrausComp0}
		\hat{K}_{i} (\Delta t)  = & \exp \bigg[ - \frac{\Delta t }{2} \left( x_i^2 + 2 x_i  \sqrt{ \frac{ \CstrengthSq }{  2 }} \hat{F}^{i}_z  + \frac{\kappa}{2} \mbox{Re}\big\{ \hat{F}^{i}_z \big\} \hat{F}^{i}_z  \right) \bigg] .
	\end{align}	
		
From the (unnormalized) POVM elements, $ \hat{E}_i(\Delta t) = \hat{K}\dg_i( \Delta t) \hat{K}_i(\Delta t)$, we can calculate the probability density for the measurement outcomes \cite{JacSte06}\footnote{There is a subtlety in calculating the outcome probabilities using \erf{ApEq::OutcomeProbabilities}. },
	\begin{align} \label{ApEq::OutcomeProbabilities}
		\mathbbm{P}(x_i) & \propto \Tr \left[ \hat{E}_i(\Delta t) \right] \nn \\
			& = \exp \bigg[ - \Delta t \Big( x_i +\sqrt{ \smallfrac{\CstrengthSq}{2} } \big\langle \mbox{Re} \big\{ \hat{F}_z^i \big\} \big\rangle_c \Big)^2 \bigg]. 
	\end{align}
Because we defined the quantized field operators (and quadratures) in terms of photon flux, we see that the measurement outcome $x_i$ is a Gaussian-distributed random variable with conditional mean, $-\sqrt{\CstrengthSq/2} \expt{ \mbox{Re} \{ \hat{F}_z^i \} }_c$, and variance $1/(2 \Delta t)$. We can thus describe the measurement outcomes as a stochastic process driven by white noise,
	\begin{align} \label{Eq::MeasurementResults}
		x_i =	- \sqrt{ \frac{\CstrengthSq}{2} } \expt{ \mbox{Re} \{ \hat{F}_z^i \} }_c+ \frac{ \Delta W_i }{ \sqrt{ 2} \Delta t}.
	\end{align}
Plugging \erf{Eq::MeasurementResults} into \erf{Eq::KrausComp0} and expanding the exponential to highest nonvanishing order in the infinitesimal limit $\Delta t \rightarrow dt$, $\Delta W_i \rightarrow dW_i$ gives the Kraus operator we seek for continuous measurements,
	\begin{align} \label{Eq::KrausAppendix}
		\hat{K}(d t) = \hat{I} - \frac{\kappa}{2} \hat{F}_z^i  \expt{ \mbox{Re} \{ \hat{F}_z^i \} }_cdt - \frac{\kappa}{8} \hat{F}_z^{i \dagger} \hat{F}_z^i dt - \sqrt{ \frac{\kappa}{4} } \hat{F}_z^i dW_i,
	\end{align}
where conditional mean at time $t$.

The integrated classical measurement record\footnote{Here we use the standard notation $y(t)$ for quadrature measurements.  It is related to the physical measurement operator, \erf{Eq::Measurement}, which has units of integrated photon flux (total photon number), by a scaling factor $\sqrt{\dot{N}_L/4}$.} in mode $i$ is,
	\begin{align} \label{Eq::MeasurementAppendix}
		y_i(t) = - \sqrt{ \CstrengthSq } \int_{t_0}^t dt' \expt{\mbox{Re} \{ \hat{F}_z^i \}(t')}_c + d W_i.
	\end{align}
where $\expt{\mbox{Re} \{ \hat{F}_z^i \}(t')}$ is the conditional mean, determined by the previous stochastic measurement record.  The infinitesimal measurement increments are
	\begin{align}
		dy_i(t) = - \sqrt{ \CstrengthSq } \expt{\mbox{Re} \{ \hat{F}_z^i \}(t)}_c dt + d W_i,
	\end{align}
as seen in the increments for polarimetry of the fundamental mode, \erf{Eq::dM}.  Inverting this relationship, we can write the Wiener process in terms of the measurement results and the expected mean value - the innovations process,
	\begin{align}
		 d W_i =  dy_i(t) - \Big( -\sqrt{\CstrengthSq}  \expt{\mbox{Re} \{ \hat{F}_z^i \}(t)}_c dt \Big).
	\end{align}  	
Note that \erf{Eq::MeasurementAppendix} in the fundamental mode is equivalent to \erf{Eq::dM}.

	\label{Appendix::SMEDerivation}
	
		\chapter{Analytic solution for the symmetrically coupled variance} \label{Appendix::AnalyticSolution}
%

The equation of motion for the collective variance in a one-dimensional description, $\Delta F_z^2$, can be derived in various ways.  Here we provide a derivation directly from the SME in \erf{Eq::HomodyneSME}, in order to prepare for the more general multimode calculation.  Within a single-mode assumption, the SME for an ensemble of spin-$\smallfrac{1}{2}$ atoms is
	\begin{align} \label{Eq::HomodyneSMESingleMode}
		d \hat{\rho}   =&   \sqrt{\frac{\kappa}{4}} \mathcal{H}[\hat{\rho}]  dW  +   \frac{\kappa}{4} \mathcal{L}_{F_z}[ \hat{\rho}]dt + \gamma_0 \sum_{n=1}^N \mathcal{D}_n [\hat{\rho}] dt,
	\end{align} 
with measurement strength per atom, $\kappa = (\sigma_0/A)(4\gamma_0/9)$.

From \erf{Eq::HomodyneSMESingleMode} we can write down the stochastic equations of motion for the first and second moments.  To include the decay we decompose the second moment according to \erf{Eq::SecondOrderDecay}.  The effects of local decoherence are given found using \erf{Eq::1stOrderEvol} and \erf{Eq::2ObservableEOM}, along with the fact that $\mathcal{D}[\hat{f}_z] = -2\hat{f}_z/9$ and $\mathcal{D}[\hat{f}_z^2] = 0$ for spin-$\smallfrac{1}{2}$.  The equations of motion are,
	\begin{align}
		d \expt{\hat{F}_z} =&  \sqrt{\kappa} \big(  \expt{\hat{F}^2_z} -  \expt{\hat{F}_z}^2 \big)dW - \frac{2 \gamma_0}{9}  \expt{\hat{F}_z} dt \label{Eq::FirstMomentSym} \\
		d \expt{\hat{F}^2_z} = & \sqrt{\kappa} \big(  \expt{\hat{F}^3_z} -  \expt{\hat{F}_z^2}\expt{\hat{F}_z} \big)dW - \frac{4 \gamma_0}{9}  \expt{\hat{F}_z} dt - \frac{ \gamma_0}{9} N  dt. \label{Eq::SecondMomentSym} 
	\end{align}
For Gaussian statistics the third moment can be written in terms of the first and second, $\expt{\hat{F}^3_z} = 3 \expt{\hat{F}^2_z} \expt{\hat{F}_z} - 2\expt{\hat{F}_z}^3$.  Using the rules of It\={o} calculus, the equation of motion for the variance is expressed in terms \erf{Eq::FirstMomentSym} and \erf{Eq::SecondMomentSym} as
	\begin{align} \label{ApEq::ItoVariance}
		d \Delta F_z^2 = & d \expt{\hat{F}^2_z} - 2 \expt{\hat{F}_z} d \expt{\hat{F}_z} - d \expt{\hat{F}_z} d \expt{\hat{F}_z}.
	\end{align}
The stochastic terms cancel, and the equation of motion for the variance can be expressed as an ordinary differential equation, 
	\begin{align} \label{ApEq::VarianceODE}
		\frac{d}{dt} \Delta F_z^2= & -\kappa \big( \Delta F_z^2 \big)^2 - \frac{4 \gamma_0}{9} \Delta F_z^2  - \frac{ \gamma_0}{9} N. 
	\end{align}
	
We wish to express \erf{ApEq::VarianceODE} in terms of the standard optical density in \erf{Eq::ODdef}.  We first introduce a new variable by dividing by the initial variance for a SCS,
	\begin{align}
		\zeta \equiv \frac{\Delta F_z^2}{N/4}.
	\end{align}
and a dimensionless time $\tau = \gamma_0 t$.  Then, we can write \erf{ApEq::VarianceODE} as
	\begin{align} \label{Eq::ZetaODE}
		\frac{d}{d \tau} \zeta = - \frac{\mbox{OD}}{9} \zeta^2 - \frac{4}{9} \zeta + \frac{4}{9},
	\end{align}
with initial condition $\zeta(0) = 1$.

	\subsection{Solving the differential equation} 

	The general form of the differential equation is
	\begin{align}
		\frac{d x}{d t} = -A x^2 - B x + C.
	\end{align}
This equation is separable,
	\begin{align} \label{Eq::VarianceODE}
		\int_{x_0}^{x(t)} \frac{dx'}{-Ax^{'2} - Bx' + C} = \int_0^t dt'.
	\end{align}
with initial condition $x_0$.  Focusing on the left side of the equation, we complete the square and integrate
	\begin{align}
		&\int_{x_0}^{x(t)}  \frac{dx'}{-A \big( x'+ \frac{B}{2A} \big)^2 +\left( C + \frac{B^2}{4 A}\right) } \nn \\
		& = \frac{2}{\sqrt{ B^2 + 4 A C}} \left[ \tanh^{-1} \left( \frac{B + 2Ax(t)}{\sqrt{B^2 + 4 AC}}  \right) - \tanh^{-1} \left( \frac{B + 2Ax_0}{\sqrt{B^2 + 4 AC}}  \right)   \right].
	\end{align}
Combining this with the solution to the right side of \erf{Eq::VarianceODE}, we get
	\begin{equation}
		x(t) = - \frac{B}{2A} + \frac{\sqrt{B^2 + 4 AC}}{2A} \tanh \left[  \sqrt{B^2 + 4 AC}  \frac{t}{2} + \tanh^{-1} \left( \frac{B + 2Ax_0}{\sqrt{B^2 + 4 AC}}  \right)  \right].
	\end{equation}
Using the relation,
	\begin{align}
		\tanh (A+B) = \frac{\tanh A + \tanh B}{1 + \tanh A \tanh B },
	\end{align}
the solution can be rewritten in a more pleasing form:
	\begin{align}
		x(t) & = - \frac{B}{2A} +  \frac{\sqrt{B^2 + 4 AC}}{2A} \left( \frac{ \frac{B + 2Ax_0}{\sqrt{B^2 + 4 AC}} + \tanh \left( \sqrt{B^2 + 4 AC}  \frac{t}{2} \right) }{1 + \frac{B + 2Ax_0}{\sqrt{B^2 + 4 AC}} \tanh \left( \sqrt{B^2 + 4 AC}  \frac{t}{2} \right)} \right) \nonumber \\
			&  = \frac{ \sqrt{B^2 + 4 AC}x_0 - (Bx_0-2C) \tanh \left( \sqrt{B^2 + 4 AC}  \frac{t}{2} \right) }{\sqrt{B^2 + 4 AC} + (2Ax_0 + B) \tanh \left( \sqrt{B^2 + 4 AC}  \frac{t}{2} \right) } .
	\end{align}

From \erf{Eq::ZetaODE} we identify the coefficients as
	\begin{equation}
		A = \frac{\mbox{OD}}{9}, \quad 
		B = \frac{4}{9}, \quad 
		C =  \frac{4}{9}. 
	\end{equation}
This gives a solution
	\begin{align}
	 \Delta F^2_z(t) = \frac{N}{4} \, \frac{\sqrt{ \mbox{OD}+1} + \tanh \left[ \sqrt{\text{OD}+1}\frac{2}{9} \gamma_0 t \right] }{\sqrt{\mbox{OD}+1} + \big(\frac{ \text{OD} }{2}+1\big)\tanh\left[\sqrt{\text{OD}+1}\frac{2}{9}\gamma_0 t \right]}. 
	\end{align}

		\chapter{Derivation of the spin wave equations of motion} \label{Appendix::SpinWaveEquations}	
%

While the squeezing parameter, \erf{Eq::SqueezingParam}, depends solely upon the mean and variance of the fundamental spin wave defined by the spatial mode of the laser probe, the diffuse scattering by individual atoms is not collective in nature and acts to couple the different spin waves to one another.  In order to model the dynamical evolution of the squeezing, including decoherence, we must track the evolution of  a hierarchy of differential equations coupling the means and covariances of spin waves in all spatial modes. This appendix provides a detailed derivation of these equations and the numerical methods used in their solution for the case of an ensemble of spin-$\smallfrac{1}{2}$ atoms.

	\subsection{Mean spin}
	
The mean spin wave measured in the polarimeter, $\expt{\hat{F}_x^0}$ is coupled to the $z$-local mean spins through diffuse scattering.  In general, we need the equation of motion for each of the $z$-local mean spins at each coarse-grained longitudinal slice $z_k$.  The equation of motion for a single-atom operator is given in \erf{Eq::1stOrderEvol} and gives
	\begin{align}
		d\expt{\hat{F}_x^i(z_k)} 
		= & \gamma_0 \sum_{n_k=1}^{N_k} \beta_0(\mathbf{r}_{n_k}) \beta_i(\mathbf{r}_{n_k}) \bigg(  -\frac{2}{9} \expt{ \hat{f}_x^{(n_k)} } + \frac{g_f^2}{9} \expt{ \mathcal{C}_{n_k}[ \hat{f}_x^{(n_k)} ] } \bigg) \label{EqAp::MeanSpin}
	\end{align}
where in the second line we have used \erf{Eq::LocalScatRate}.  The sum is only over those atoms $n_k$ within the slice $z_k$.  We have defined a local superoperator that arises solely from the ``feeding" terms in the master equation,
	\begin{align}
		\mathcal{C}_n[\hat{\rho}] = \hat{f}_z^{(n)} \hat{\rho} \hat{f}_z^{(n)} + \frac{1}{2} \left( \hat{f}_x^{(n)} \hat{\rho} \hat{f}_x^{(n)} + \hat{f}_y^{(n)} \hat{\rho} \hat{f}_y^{(n)}\right). \label{Eq::Csuperoperator}
	\end{align}
Technically, since the decoherence map $\mathcal{D}_n[\hat{\rho}]$ (as well as $\mathcal{C}_n[\hat{\rho}]$) is defined by its action on states in the Schr\"{o}dinger picture, in the Heisenberg picture it should act on operators as $\mathcal{D}\dg_n[\hat{x}]$.  However, since it is comprised entirely of Hermitian operators, this makes no difference.  

Summing over the $i=0$ solutions at each slice gives the mean of the fundamental spin wave,
	\begin{align}\label{Eq::FundSpinwave}
		\expect{\hat{F}_x^{0}(t)}=\sum_{k} \expect{\hat{F}_x^{0}(z_k,t)},
	\end{align} 
which is the mean spin in the definition of the squeezing parameter \erf{Eq::SqueezingParam}.

	\subsection{Covariances}

	To solve for the variance of the fundamental spin wave, we follow a similar procedure. As shown in \erf{Eq::SpinHalfVariance}, the fundamental variance couples through diffuse scattering to covariances between spin waves in slices $z_k$ and $z_{k'}$:
	\begin{align} \label{Eq::GenCovariance}
		\expect{\Delta\hat{F}_z^{i}(z_k)\Delta\hat{F}_z^{j}(z_{k'})} =\expect{\hat{F}_z^{i}(z_k)\hat{F}_z^{j}(z_{k'})}-\expect{\hat{F}_z^{i}(z_k)}\expect{\hat{F}_z^{j}(z_{k'})} .
	\end{align}
From the SME in \erf{Eq::HomodyneSME}, we find the equations of motion for these covariances.  Unlike the mean spin, the effects of continuous measurement must be included along with diffuse scattering. However, decoherence from collective scattering, described by the map $\mathcal{L}_{i}$ in \erf{Eq::LSuperoperator}, does not affect these covariances since the $\hat{F}_z^{i}$ commute with one another.  From the SME in \erf{Eq::HomodyneSME} and the rule of It\={o} calculus that differentials must be taken to second order \cite{JacSte06}, i.e. $d(XY) = (dX) Y + X (dY) + (dX)(dY)$, we find that the covariances in \erf{Eq::GenCovariance} evolve according to:
	\begin{align} \label{Eq::GenCovarianceODE}
		d\expect{\Delta\hat{F}_z^{i}(z_k)\Delta\hat{F}_z^{j}(z_{k'})} = & d \big[ \expect{\hat{F}_z^{i}(z_k)\hat{F}_z^{j}(z_{k'})} \Big] -  \Big[ d\expect{\hat{F}_z^{i}(z_k)} \Big] \expect{\hat{F}_z^{j}(z_{k'})} \\
		&- \expect{\hat{F}_z^{i}(z_k)} \Big[ d\expect{\hat{F}_z^{j}(z_{k'})} \Big] - \Big[d\expect{\hat{F}_z^{i}(z_k)}\Big]\Big[d\expect{\hat{F}_z^{j}(z_{k'})}\Big]  .\nn
	\end{align}

	\subsubsection{Coherent dynamics from measurement}	
	
First, we examine the dynamics due to continuous measurement. The contributions from the spin waves in each coarse-grained slice $z_k$ to the light measured in the polarimeter and indistinguishable, and such measurement serves to generate correlations between them.  This portion of the covariance dynamics is given by  
	\begin{align} 
		 d\expect{\Delta\hat{F}_z^{i}(z_k) & \Delta\hat{F}_z^{j}(z_{k'})} \Big|_{\text{meas}} =  \sqrt{ \frac{\kappa}{4} } \bigg\{  \big\langle \mathcal{H}_{0}[\hat{F}_z^{i}(z_k)\hat{F}_z^{j}(z_{k'})]\big\rangle  - \big\langle\mathcal{H}_{0}[\hat{F}_z^{i}(z_k)]\big\rangle \big\langle\hat{F}_z^{j}(z_{k'})\big\rangle \nn \\
		 &- \big\langle\hat{F}_z^{i}(z_k)\big\rangle \big\langle\mathcal{H}_{0}[\hat{F}_z^{j}(z_{k'})]\big\rangle
 \bigg\} dW -\frac{\kappa}{4}\big\langle\mathcal{H}_{0}[\hat{F}_z^{i}(z_k)]\big\rangle\big\langle\mathcal{H}_{0}[\hat{F}_z^{j}(z_{k'})]\big\rangle dt. \label{Eq::CovHomodyne}
	\end{align} 
The map $\mathcal{H}_{0}$, \erf{Eq::HSuperoperator}, couples the first- and second-order moments of the spin waves to higher-order moments, just as in Appendix \ref{Appendix::AnalyticSolution} for the one-dimensional case.  For the initial SCS along $x$ and during its subsequent evolution, the spin waves $\hat{F}_z^{i}$ are Gaussian distributed, both over the entire cloud and within each coarse-grained slice $z_k$. Thus, third-order moments of commuting observables can be expressed in terms of first- and second-order moments with the relation, $\expects{XYZ} = \expects{XY}\expects{Z} + \expects{XZ}\expects{Y} + \expects{YZ}\expects{X} - 2\expects{X}\expects{Y}\expects{Z}$ \cite{JacSte06, Habib04}.
In this regime, all stochastic terms in \erf{Eq::CovHomodyne} cancel, leaving the deterministic equation: 
	\begin{align}
		\frac{d}{dt}\expect{\Delta\hat{F}_z^{i}(z_k)  \Delta\hat{F}_z^{j} (z_{k'})}\Big|_{\text{meas}} &= -\kappa\big\langle\Delta\hat{F}_z^{i}(z_k)\Delta\hat{F}_z^{0}\big\rangle\big\langle\Delta\hat{F}_z^{j}(z_{k'})\Delta\hat{F}_z^{00}\big\rangle \nn \\
		&=-\kappa\sum_{k'',k'''}\big\langle\Delta\hat{F}_z^{i}(z_k)\Delta\hat{F}_z^{0}(z_{k''})\big\rangle\big\langle\Delta\hat{F}_z^{j}(z_{k'})\Delta\hat{F}_z^{0}(z_{k'''})\big\rangle. \label{Eq::CovarianceBackaction}
	\end{align}
These dynamics, which arise from continuous polarimetry measurements, serve to generate the correlations at rate $\kappa$ that produce spin squeezing.  In the single-mode approximation, taking $i,j=0$ and summing over all $k$ and $k'$, we recover the measurement term in \erf{ApEq::VarianceODE}.

	\subsubsection{Decoherent dynamics from diffuse scattering}

	The correlations that develop from measurements according to \erf{Eq::CovarianceBackaction} are degraded by diffuse photon scattering.  In the general equation of motion for the covariances, \erf{Eq::GenCovarianceODE}, there are first- and second-order terms.  The dynamics due to diffuse scattering must be treated carefully using the results in Section \ref{Sec::MultiatomExpectations}. 
	
The first moment's evolution due to diffuse scattering is just as in \erf{EqAp::MeanSpin},
	\begin{align}
		\frac{d}{dt} \expt{ \hat{F}^i_z(z_k) } \Big|_{\rm dec} & = \gamma_0 \sum_{n_k=1}^{N_k} \beta_0(\mathbf{r}_{n_k}) \beta_i(\mathbf{r}_{n_k}) \bigg(  -\frac{2}{9} \expt{ \hat{f}_z^{(n_k)} } + \frac{g_f^2}{9} \expt{ \mathcal{C}_{n_k}[ \hat{f}_z^{(n_k)} ] } \bigg). 
	\end{align}
The second moment's evolution is complicated by the fact that it consists of both one- and two-atom terms
	\begin{align} \label{ApEq::Covariances}
		\expt{ \hat{F}^i_z(z_k)\hat{F}^j_z(z_{k'}) } =  & \sum_{n_k=1}^{N_k} \beta_i(\mathbf{r}_{n_k}) \beta_j(\mathbf{r}_{n_{k'}}) \expt{ \hat{f}_z^{(n_{k})2} } \delta_{k,{k'}}\\
		&+ \sum_{n_k \neq n_{k'}} \beta_i(\mathbf{r}_{n_k}) \beta_j(\mathbf{r}_{n_{k'}}) \expt{ \hat{f}_z^{(n_k)} \hat{f}_z^{(n_{k'})} } \nn,
	\end{align}
where the $\delta_{k,k'}$ in the first line indicates that the single-atom terms only appear for covariances within a single coarse grained slice.  By isolating the single-atom terms, the sums in the second line are strictly over pairs of atoms; $n_k \neq n_{k'}$.  The second moment's evolution follows from \erf{Eq::1stOrderEvol} and \erf{Eq::2ObservableEOM},
	\begin{align} \label{ApEq::2ndMoment}
		 \frac{d}{dt} & \expt{ \hat{F}^i_z(z_k)\hat{F}^j_z(z_{k'}) } \Big|_{\rm dec} \\
		 = &\delta_{k,k'}  \gamma_0 \sum_{n_k=1}^{N_k} \beta_0(\mathbf{r}_{n_k}) \beta_i(\mathbf{r}_{n_k}) \beta_j(\mathbf{r}_{n_{k}}) \\
		 &\times \bigg\{ \frac{2}{9} \expt{ \hat{f}_z^{(n_k)2} } + \frac{g_f^2}{9} \Big( \expt{ \mathcal{C}_{n_k}[ \hat{f}_z^{(n_k)2} ] } - \expt{\{ \hat{f}_z^{(n_k)}, \mathcal{C}_{n_k}[\hat{f}_z^{(n_k)}] \}_+}\Big) \bigg\} 
\nn \\
& - \frac{2 \gamma_0}{9} \sum_{n_k=1}^{N_k} \sum_{n_{k'}=1}^{N_{k'}}   \beta_i(\mathbf{r}_{n_k}) \beta_j(\mathbf{r}_{n_{k'}}) \bigg\{ \beta_0(\mathbf{r}_{n_k}) \expt{  \hat{f}_z^{(n_k)} \hat{f}_z^{(n_{k'})} } + \beta_0(\mathbf{r}_{n_{k'}}) \expt{ \hat{f}_z^{(n_k)}  \hat{f}_z^{(n_{k'})} } \bigg\}. \nn \\
		& + \frac{g_f^2 \gamma_0}{9} \sum_{n_k=1}^{N_k} \sum_{n_{k'}=1}^{N_{k'}}   \beta_i(\mathbf{r}_{n_k}) \beta_j(\mathbf{r}_{n_{k'}}) \bigg\{ \beta_0(\mathbf{r}_{n_k}) \expt{ \mathcal{C}_{n_k} [\hat{f}_z^{(n_k)}] \hat{f}_z^{(n_{k'})} } + \beta_0(\mathbf{r}_{n_{k'}}) \expt{ \hat{f}_z^{(n_k)} \mathcal{C}_{n_{k'}} [\hat{f}_z^{(n_{k'})}] } \bigg\}. \nn
	\end{align}
By adding and subtracting the $n_k = n_{k'}$ terms, the sums in the two final lines of \erf{ApEq::2ndMoment} are now free to run over all atom indices.  This is the origin of the positive sign on the first term in the second line and the extra anti-commutator term in the second line.  

Combining the dynamics from measurement and diffuse scattering, we have equations of motion for the covariances, 
	\begin{align}
		\frac{d}{dt} & \expt{ \Delta \hat{F}^i_z(z_k) \Delta \hat{F}^j_z(z_{k'}) } \nn \\
		= & -\kappa\sum_{k'',k'''}\big\langle\Delta\hat{F}_z^{i}(z_k)\Delta\hat{F}_z^{0}(z_{k''})\big\rangle\big\langle\Delta\hat{F}_z^{j}(z_{k'})\Delta\hat{F}_z^{0}(z_{k'''})\big\rangle \label{ApEx::GeneralCovariance} \\
		& - \frac{2 \gamma_0}{9} \sum_{n_k=1}^{N_k}  \beta_0(\mathbf{r}_{n_k}) \beta_i(\mathbf{r}_{n_k}) \expt{ \Delta \hat{f}_z^{(n_k)} \Delta \hat{F}_z^j(z_{k'}) }  \nn \\
		& - \frac{2 \gamma_0}{9} \sum_{n_{k'}=1}^{N_{k'}}   \beta_0(\mathbf{r}_{n_{k'}}) \beta_j(\mathbf{r}_{n_{k'}}) \expt{ \Delta \hat{F}_z^i (z_k)  \Delta \hat{f}_z^{(n_{k'})} } \nn \\
		& + \frac{g_f^2 \gamma_0}{9} \sum_{n_k=1}^{N_k}  \beta_0(\mathbf{r}_{n_k})   \beta_i(\mathbf{r}_{n_k}) \expt{ \Delta\mathcal{C}_{n_k} [\hat{f}_z^{(n_k)}] \Delta\hat{F}_z^j(z_k') } \nn \\
		& + \frac{g_f^2 \gamma_0}{9} \sum_{n_{k'}=1}^{N_{k'}} \beta_0(\mathbf{r}_{n_{k'}}) \beta_j(\mathbf{r}_{n_{k'}}) \expt{ \Delta\hat{F}_z^i(z_k) \Delta\mathcal{C}_{n_{k'}} [\hat{f}_z^{(n_{k'})}] }. \nn \\
		& + \delta_{k,k'} \frac{g_f^2 \gamma_0}{9} \sum_{n_k=1}^{N_k} \beta_0(\mathbf{r}_{n_k}) \beta_i(\mathbf{r}_{n_k}) \beta_j(\mathbf{r}_{n_{k}}) \Big( \expt{ \mathcal{C}_{n_k}[ \hat{f}_z^{(n_k)2} ] } - \expt{\{ \hat{f}_z^{(n_k)}, \mathcal{C}_{n_k}[\hat{f}_z^{(n_k)}] \}_+} \Big) \nn \\
		& + \delta_{k,k'}  \frac{2 \gamma_0}{9} \sum_{n_k=1}^{N_k} \beta_0(\mathbf{r}_{n_k}) \beta_i(\mathbf{r}_{n_k}) \beta_j(\mathbf{r}_{n_{k}}) \expt{ \hat{f}_z^{(n_k)2} } \nn
	\end{align}

The covariance of the spin wave measured in the polarimeter is found by summing the $i=0$ solutions over all coarse-grained slices,
	\begin{align}\label{Eq::FundSpinwaveVariance}
		\big( \Delta F^0_z (t) \big)^2 =\sum_{k,k'} \expt{ \Delta \hat{F}^0_z(z_k) \Delta \hat{F}^0_z(z_{k'}) }(t). 
	\end{align} 
This fundamental spin wave variance determines the projection noise in the measurements and is variance in the definition of the squeezing parameter, \erf{Eq::SqueezingParam}.

	\subsection{Initial conditions} \label{ApSec::InitialConditions}
	
Solving the resulting system of coupled differential equations for the means, \erf{EqAp::MeanSpin}, and covariances, \erf{ApEx::GeneralCovariance}, requires the initial conditions.  In applications, the state of the ensemble is prepared by optically pumping the atoms into a SCS.  
For an initial SCS oriented along $x$, each atom has a spin projection $\expect{\hat{f}_x}_{\rm SCS}=f$.  The initial state of the $z$-local mean spin in mode $i$ at coarse-grained longitudinal slice $z_k$ is
	\begin{align}\label{Eq::meanSlice}
		\expect{\hat{F}_x^{i}(z_k)}_{\rm SCS} & = \sum_{n_k=1}^{N_k} \beta_{i}(\mbf{r}_{n_k})\expect{\hat{f}_x^{(n_k)}}_{\rm SCS} \nn \\
		 & = f \sum_{n_k=1}^{N_k} \beta_{i}(\mbf{r}_{n_k}).
	\end{align} 
For a average atomic density, $\eta(\mathbf{r})$, the sum becomes an integral, 
	\begin{align}
		\expect{\hat{F}_x^{i}(z_k)}_{\rm SCS} \rightarrow f \times \delta z  \int d^2 \rperp \eta(\rperp,z_k)\beta_{i}(\rperp,z_k). \label{Eq::meanSliceCont}
	\end{align} 
where $\delta z$ is the width of the coarse-grained longitudinal slice.  The $z$-slices are chosen such that both $\eta(\mathbf{r})$ and $\beta(\mathbf{r})$ have little longitudinal variation within a slice.  Note that \erf{Eq::meanSliceCont} is proportional to the effective atom number $N_{\rm eff}^{(1)}$ in the slice $z_k$.
		
For an initial SCS oriented along $x$, each atom's variance for $z$-projective measurements is $\expect{ \Delta \hat{f}_z^2}_{\rm SCS}=f/2$, and there are no correlations between atoms.  Thus, there are no initial correlations between $z$-local spin waves in different $z_k$-slices.  The initial condition is then,
	\begin{align}\label{ApEq::InitialCovariances}
		\expect{ \Delta \hat{F}_z^{i}(z_k) \Delta \hat{F}_z^{j}(z_{k'}) }_{\rm SCS} &= \delta_{k,k'}\sum_{n_k=1}^{N_k} \beta_{i}(\mbf{r}_{n_k}) \beta_{j}(\mbf{r}_{n_k}) \expect{ \Delta \hat{f}_z^{(n_k)2}}_{\rm SCS}  \\
		&= \frac{f}{2}  \sum_{n_k=1}^{N_k} \beta_{i}(\mbf{r}_{n_k}) \beta_{j}(\mbf{r}_{n_k}),
	\end{align} 
For a average atomic density, $\eta(\mathbf{r})$, the sum becomes an integral within the slice $z_k$, 
	\begin{align}
		\expect{ \Delta \hat{F}_z^{i}(z_k) \Delta \hat{F}_z^{j}(z_{k'}) }_{\rm SCS} \rightarrow \frac{f}{2} \times \delta z  \int d^2 \rperp \eta(\rperp,z_k)\beta_{i}(\rperp,z_k) \beta_{j}(\mbf{r}_{n_k}). \label{ApEq::InitialCovariancesCont}
	\end{align} 
Note that this is proportional to the effective atom number $N_{\rm eff}^{(2)}$ within the slice $z_k$.
	
The initial means, \erf{Eq::meanSlice}, and covariances, \erf{ApEq::InitialCovariances}, give an initial value for the squeezing parameter, \erf{Eq::SqueezingParam}, $\zeta(0) = 1$.

\section{Spin-$\frac{1}{2}$ ensembles} \label{ApSec::SpinHalf}

	When the ensemble is composed of spin-$\smallfrac{1}{2}$ particles, significant simplifications allow the equations of motion for the means and covariances to be expressed entirely in terms of the $z$-local, coarse-grained spin waves.  The terms that drop out allow the remaining mean spin and covariance equations of motion to be expressed entirely in terms of the $z$-local spin waves, thus forming a closed set of equations which can be solved numerically for a given mean atomic distribution, $\eta(\mathbf{r})$, which sets the initial conditions, \erf{Eq::meanSliceCont} and \erf{ApEq::InitialCovariancesCont}.  For arbitrary spin-$f$, such a closed set is not accessible through these projections, and one must turn to other approximate methods to solve the equations \cite{NorDeu12}.  Numerical solutions for spin squeezing in ensembles of spin $f>\smallfrac{1}{2}$ atoms are not treated in this thesis.

	\subsection{Mean spin}

For spin-1/2 the local feeding term simplifies to $\mathcal{C}_n[\hat{f}_x^{(n)}] =-\hat{f}_x^{(n)}/4$, and the equation of motion in \erf{EqAp::MeanSpin} becomes
	\begin{align}\label{Eq::SimplifiedMeanFx}
		\frac{d}{dt}\expect{\hat{F}_x^{i}(z_k)}=-\frac{\gamma_0}{3} \sum_{n_k=1}^{N_k} \beta_{0}(\mbf{r}_{n_k}) \beta_{i}(\mbf{r}_{n_k})\expect{\hat{f}_x^{(n_k)}}.
	\end{align} 
By decomposing the spatial weighting $\beta_{0}(\mbf{r}) \beta_{i}(\mbf{r})$ in terms of orthogonal mode functions, the right hand side of \erf{Eq::SimplifiedMeanFx} can be expressed in terms of $z$-local spin waves. In terms of the mode functions,
	\begin{align}
		 \beta_{0}(\mbf{r}_\perp,z)  \beta_{i}(\mbf{r}_\perp,z)& =|u_{0}(\mbf{r}_\perp,z)|^2u_{i}^*(\mbf{r}_\perp,z)u_{ 0}(\mbf{r}_\perp,z) 
		= \sum_{j} c^{i}_{j}(z) \beta_{j}(\mbf{r}_\perp,z), \label{Eq::ProjCoeff_proto}
\end{align}
where we have made use of orthogonality and completeness to define projection coefficients from mode $i$ to mode $j$ at longitudinal plane $z$,
	\begin{align} \label{Eq::ProjCoeff}
		c^{i}_{j}(z)  \equiv  \frac{1}{A} \int d^2 \mathbf{r}_\perp \left[ u_{0}(\mathbf{r}_\perp, z)\right]^2 u^*_{i}(\mathbf{r}_\perp, z) u_{j}(\mathbf{r}_\perp, z).
	\end{align}
The explicit form for the projection coefficients when cylindrically symmetric $l=0$ Laguerre-Gauss modes are considered is given in Appendix \ref{Appendix::ProjectionCoefficients}.  Using this projection in \erf{Eq::SimplifiedMeanFx}, we obtain an infinite hierarchy of differential equations that couple mean spin waves in a given slice to one another,
	\begin{align}
		\frac{d}{dt}\expect{\hat{F}_x^{i}(z_k)} = & -\frac{\gamma_0}{3}  \sum_{j} c^{i}_{j}(z_k) \sum_{n_k=1}^{N_k} \beta_j(\mathbf{r}_{n_k}) \expect{\hat{f}_x^{(n_k)}} \\
			= & -\frac{\gamma_0}{3} \sum_{j} c^{i}_{j}(z_k) \expect{\hat{F}_x^{j}(z_k)}. \label{Eq::ZkSliceFx}
	\end{align} 
With the projection, the evolution has now been expressed entirely in terms of the $z$-local, collective mean spins.  

An approximate solution to \erf{Eq::ZkSliceFx} is found for each slice by choosing $\delta z$ and truncating the number of spin waves at some index $i_{\rm max}$ determined by the initial conditions, the projection coefficients, and the required precision.

	\subsection{Covariances}
	
	For spin-$\smallfrac{1}{2}$, the relations $\mathcal{C}_n[\hat{f}_z^{(n)}] = 0$ and $\mathcal{C}_n[\hat{f}_z^{(n)2}]  = \frac{1}{8} \hat{I}$ simplify the covariance equation of motion, \erf{ApEx::GeneralCovariance}, to 
	\begin{align} \label{ApEq::CovarianceBeforeProjection}
		\frac{d}{dt} & \expt{ \Delta \hat{F}^i_z(z_k) \Delta \hat{F}^j_z(z_{k'}) } \nn \\
		= & -\kappa\sum_{k'',k'''}\big\langle\Delta\hat{F}_z^{i}(z_k)\Delta\hat{F}_z^{0}(z_{k''})\big\rangle\big\langle\Delta\hat{F}_z^{j}(z_{k'})\Delta\hat{F}_z^{0}(z_{k'''})\big\rangle \\
		& - \frac{2 \gamma_0}{9} \sum_{n_k=1}^{N_k}  \beta_0(\mathbf{r}_{n_k}) \beta_i(\mathbf{r}_{n_k}) \expt{ \Delta \hat{f}_z^{(n_k)} \Delta \hat{F}_z^j(z_{k'}) }  - \frac{2 \gamma_0}{9} \sum_{n_{k'}=1}^{N_{k'}}   \beta_0(\mathbf{r}_{n_{k'}}) \beta_j(\mathbf{r}_{n_{k'}}) \expt{ \Delta \hat{F}_z^i (z_k)  \Delta \hat{f}_z^{(n_{k'})} } \nn \\
		& + \delta_{k,k'} \frac{ \gamma_0}{9} \sum_{n_k=1}^{N_k} \beta_0(\mathbf{r}_{n_k}) \beta_i(\mathbf{r}_{n_k}) \beta_j(\mathbf{r}_{n_{k}}).  \nn
	\end{align}
By performing projections onto the mode functions at each longitudinal plane, these equations of motion can be expressed entirely in terms of the corase-grained, $z$-local spin waves,
	\begin{align}
		\frac{d}{dt} & \expt{ \Delta \hat{F}^i_z(z_k) \Delta \hat{F}^j_z(z_{k'}) } \nn \\
		= & -\kappa\sum_{k'',k'''}\big\langle\Delta\hat{F}_z^{i}(z_k)\Delta\hat{F}_z^{0}(z_{k''})\big\rangle\big\langle\Delta\hat{F}_z^{j}(z_{k'})\Delta\hat{F}_z^{0}(z_{k'''})\big\rangle \label{ApEq::CovarianceSpinHalf} \\
		& - \frac{2 \gamma_0}{9} \sum_{l} c^i_l(z_k)  \expt{ \Delta \hat{F}_z^{l}(z_k) \Delta \hat{F}_z^j(z_{k'}) } - \frac{2 \gamma_0}{9} \sum_{l} c^j_l(z_{k'})  \expt{ \Delta \hat{F}_z^{i}(z_k) \Delta \hat{F}_z^l(z_{k'}) }  \nn \\
		&+ \frac{ \gamma_0}{9} N_{i,j }(z_k).  \nn
	\end{align}
with projection coefficients defined in \erf{Eq::ProjCoeff}.  The sum over atoms in the final term has been reexpressed as
	\begin{align} \label{ApEq::N3sum}
		N_{i,j}(z_k) &= \sum_{n_k=1}^{N_k} \beta_0(\mathbf{r}_{n_k}) \beta_i(\mathbf{r}_{n_k}) \beta_j(\mathbf{r}_{n_{k}}). 
	\end{align}  
For the fundamental mode, when $i,j = 0$, this is exactly the $N^{(3)}_\eff$ atom number, defined in \erf{Eq::N3}, within the slice $z_k$.  For a continuous average atomic density distribution $\eta(\mathbf{r})$, \erf{ApEq::N3sum} can be expressed as an integral within the longitudinal slice $z_k$,
	\begin{align}
		N_{i,j}(z_k)  \rightarrow \delta z \int d^2 \mathbf{r}_\perp \eta(\mathbf{r}_\perp, z_k) \beta_0(\mathbf{r}_\perp, z_k)\beta_i(\mathbf{r}_\perp, z_k) \beta_j(\mathbf{r}_\perp, z_k).
	\end{align}  
The fundamental spin wave variance measured in the polarimeter and factoring into the squeezing parameter is found by summing over $k,k'$ in \erf{ApEq::CovarianceSpinHalf} with $i,j = 0$,
\begin{align}
		\frac{d}{dt}  \big( \Delta \hat{F}^0_z \big)^2 = & -\kappa \Big[  \big( \Delta \hat{F}^0_z \big)^2 \Big]^2  - \frac{4 \gamma_0}{9} \sum_{i} \sum_{k,k'}c^0_i(z_k) \expt{ \Delta \hat{F}_z^{i}(z_k) \Delta \hat{F}_z^0(z_{k'}) }  + \frac{ \gamma_0}{9} N^{(3)}_{\rm eff}.  \nn
	\end{align}

An approximate solution to \erf{ApEq::CovarianceSpinHalf} is found by choosing $\delta z$ and truncating the number of spin waves at indices $i_{\rm max}, j_{\rm max}$ determined by the initial conditions, the projection coefficients, and the required precision.

\section{Local decoherence on arbitrary matrix elements}

We can understand the effects of the local decoherence map by writing the map on matrix elements in the $z$-basis.  The local map from diffuse scattering couples the matrix elements of a single atom's internal state according to
	\begin{align}
		\mathcal{D}_n \big[\op{m}{m'}^{(n)} \big] 
		 = &\Big( - \frac{2}{9} + m \, m' \frac{g_f^2}{9} \Big) \op{m}{m'}^{(n)} \\
		&+ \frac{g_f^2}{36}\Big( c^+_{m,m'} \op{m+1}{m'+1}^{(n)} +c^-_{m,m'}  \op{m-1}{m'-1}^{(n)}  \Big) \nn,
	\end{align}
with coupling coefficients that come from the action of  raising and lowering operators on $z$-eigenstates,
	\begin{align}
		c^+_{m,m'} &= \sqrt{(f+m)(f+m+1)(f+m')(f+m'+1)}, \\
		c^-_{m,m'} &= \sqrt{(f+m)(f-m+1)(f+m')(f-m'+1)}.
	\end{align}

The effect of local decoherence on collective operators depends on their specific form.  In some cases it can be useful to define coarse-grained, $z$-local collective matrix elements, which couple together through the coherent and incoherent dynamics.  Since they involve sums over atoms within each coarse-grained longitudinal slice, there will still generally be a significant reduction in the number of equations required to track the evolution as compared to following every atom individually.  For instance, consider the $z$-component of the fundamental spin wave.  It can be written in terms of $z$-local spin waves, $ \hat{F}_z^0 = \sum_k\hat{F}_z^0(z_k)$, which in turn can be written in terms of collective matrix elements, 
	\begin{align} \label{Eq::MatrixElementDecomposition}
		\hat{F}_z^0(z_k) & = \sum_{n_k=1}^{N_k} \beta_0(\mathbf{r}_{n_k}) \hat{f}_z^{(n_k)} = \sum_m m \op{m}{m}_0(z_k).
	\end{align}
In the final equality, we have defined the $z$-local, diagonal collective matrix elements in the fundamental spatial mode as\footnote{The awkwardness of the notation is balanced by the fact that it is sealed within a deep appendix that nearly no one will ever see.}
	\begin{align} \label{Eq::CollectiveMatrixElement}
		\op{m}{m}_0(z_k) \equiv \sum_{n_k=1}^{N_k} \beta_0(\mathbf{r}_{n_k}) \op{m}{m}^{(n_k)}.
	\end{align}
Since the collective operator $\hat{F}_z^0(z_k)$ commutes with the coherent multi-mode Faraday interaction, the diagonal collective matrix elements evolve only from diffuse photon scattering.  One finds a set of coupled differential equations for the $z$-local collective matrix elements in mode $i$ in the $k^{th}$ longitudinal coarse-grained slice,
	\begin{align}
		\frac{d}{dt} \op{m}{m}_i(z_k) =& \gamma_0  \sum_{n_k=1}^{N_k}  \beta_0(\mathbf{r}_{n_k}) \beta_i(\mathbf{r}_{n_k}) \mathcal{D}_n \big[\op{m}{m}^{(n_k)} \big] \\
		 =& \sum_j c^i_j(z_k)  \bigg\{  \Big( - \frac{2}{9} +  \frac{g_f^2}{9} m^2 \Big)\op{m}{m}_j(z_k) \\
		&+ \frac{g_f^2}{36} c^+_{m,m} \op{m+1}{m+1}_j(z_k) + \frac{g_f^2}{36} c^-_{m,m} \op{m-1}{m-1}_j(z_k) \bigg\} \nn.
	\end{align}
with the projection coefficients defined in \erf{Eq::ProjCoeff}.

Specifying a particular state corresponds to specifying initial conditions.  For example, to find the mean spin along $z$ for an initial $z$-polarized SCS of spin-$f$ atoms, one solves the above equations with initial conditions, $\op{m=f}{m=f}^{(n)} = f$ and all others vanishing.  This gives for the initial $z$-local collective matrix elements,
	\begin{align}
		\expt{ \op{m=f}{m=f}_i(z_k)}_{\rm SCS} = f \sum_{n_k=1}^{N_k} \beta_i(\mathbf{r}_{n_k}) = f \delta z \int d^2 \mathbf{r}_\perp \beta_i(\mathbf{r}_\perp, z_k),
	\end{align}
with all other initialized to zero.  The solutions are then recombined using \erf{Eq::MatrixElementDecomposition}.  
 
Clearly, this approach could be used to find the evolution of other collective operators that are not diagonal in the $z$-basis, although there is no guarantee that the total complexity will be reduced.

		\chapter[Projection coefficients for $l=0$ Laguerre-Gauss modes]{Projection coefficients for cylindrically symmetric Laguerre-Gauss modes} \label{Appendix::ProjectionCoefficients}	
%

When cylindrical symmetry is preserved, we can work with a smaller set of Laguerre-Gauss mode functions, defined in Appendix \ref{Appendix::LGModes}, for which $l = 0$:
	\begin{align}
		u_{p0} ( \mathbf{r}_\perp , z) = \frac{w_0}{w(z)}L_p^{0} \! \left( \frac{2 \rho^2}{[w(z)]^2} \right) \exp \bigg( - \frac{ \rho^2 }{[w(z)]^2 } \bigg) \exp \bigg(  i \frac{k_0 \rho^2}{2 R(z)} -i(2p + 1) \Phi(z) \bigg). 
	\end{align}
The beam waist $w(z)$, radius of curvature $R(z)$, and Guoy phase $\Phi(z)$ are given in \erf{Eq::GaussianParameters}.
The decay terms that appear, for example, in the equations of motion for the spin-$\smallfrac{1}{2}$ mean spin \erf{Eq::SimplifiedMeanFx} and covariance \erf{ApEq::CovarianceBeforeProjection} can be projected onto the basis functions:
	\begin{align}
		\beta_{00}(\mathbf{r}_\perp, z) \beta_{p0}(\mathbf{r}_\perp, z) & = u_{00}(\mathbf{r}_\perp , z) u^*_{00}(\mathbf{r}_\perp, z) u_{00}(\mathbf{r}_\perp , z) u^*_{p0}(\mathbf{r}_\perp, z) \nonumber \\
		& = \Big\{ u_{00}(\mathbf{r}_\perp , z) u^*_{00}(\mathbf{r}_\perp, z) u^*_{p0}(\mathbf{r}_\perp, z) \Big\} u_{00}(\mathbf{r}_\perp , z) \nonumber \\
		& = \bigg\{ \sum_{p'} c^p_{p'} (z) u^*_{p'0}(\mathbf{r}_\perp, z) \bigg\}  u_{00}(\mathbf{r}_\perp , z) \nonumber \\
		& = \sum_{p'} c^p_{p'} (z) \beta_{p'0}(\mathbf{r}_\perp, z).
	\end{align}
From the second to the third line, we project onto basis function $u^*_{p'}(\mathbf{r}_\perp, z)$ by multiplying the term in braces by the complex conjugate,  $u_{p'}(\mathbf{r}_\perp, z)$, and integrating (with the mode area $A$ included to maintain normalization):	
	\begin{align}
		c^p_{p'} (z) & \equiv \frac{1}{A} \int d^2 \mathbf{r}_\perp |u_{00}(\mathbf{r}_\perp, z_k )|^2 u^*_{p0}(\mathbf{r}_\perp, z_k ) u_{p'0} (\mathbf{r}_\perp, z_k ) \nonumber \\
		& = \frac{2 \pi}{A} \left( \frac{w_0}{w(z)} \right)^{\!4}  e^{  -2i(p' - p) \Phi(z) } \int_0^\infty \! d\rho \, \rho \,  L_p^{0} \! \left( \frac{2 \rho^2}{[w(z)]^2 } \right) L_{p'}^{0} \! \left( \frac{2 \rho^2}{[w(z)]^2 } \right) e^{ - \frac{4 \rho^2 }{[w(z)]^2 } }. \nn
	\end{align}
Substitution of the dimensionless variable $x = \sqrt{2} \rho / w(z)$ produces the form,
	\begin{align}		
		c^p_{p'} (z) 
		 & =\frac{ e^{-2i(p'-p) \tan^{-1} \left( z/z_R \right) } }{1 + \left( z / z_R \right)^2} \int_0^\infty dx \,  2 x  L_{p}^0 (x^2) L^0_{p'}(x^2) e^{- 2x^2 } . \label{ApEq::Coefs}
	\end{align}	
We see that the coefficients have two parts: an integral that describes the overlap between the modes and a multiplicative factor that captures the $z$-dependence.  Following is a table of values for the integral in \erf{ApEq::Coefs} for $p,p' \in \{0,4\}$\footnote{It seems highly likely that there is an analytic solution to the integral in \erf{ApEq::Coefs}, but I could not find it and neither could \emph{Mathematica} 9.}:
\begin{align}
\begin{array}{c || ccccc}
 &p'\!=\!0 & p'\!=\!1 & p'\!=\!2 & p'\!=\!3 & p'\!=\!4 \\
\hline
\hline
 p\!=\!0 & \frac{1}{2} & \frac{1}{4} & \frac{1}{8} & \frac{1}{16} & \frac{1}{32} \\
 p\!=\!1 &\frac{1}{4} & \frac{1}{4} & \frac{3}{16} & \frac{1}{8} & \frac{5}{64} \\
 p\!=\!2 &\frac{1}{8} & \frac{3}{16} & \frac{3}{16} & \frac{5}{32} & \frac{15}{128} \\
 p\!=\!3 &\frac{1}{16} & \frac{1}{8} & \frac{5}{32} & \frac{5}{32} & \frac{35}{256} \\
 p\!=\!4 &\frac{1}{32} & \frac{5}{64} & \frac{15}{128} & \frac{35}{256} & \frac{35}{256} \\
\end{array}
\end{align}
As both $p$ and $p'$ become larger, the coupling between them becomes smaller.  It is this relationship, along with the fact that the initial conditions also decrease for larger $p$ and $p'$, that allows a truncation of the hierarchy of differential equations.  

The multiplicative factor in \erf{ApEq::Coefs} contains a portion that arises from the intensity diffraction of the modes away from the beam waist at $z = 0$ and a portion that comes from the relative Guoy phases of the modes at the $z$-plane.  The real and imaginary parts of these factors are shown in \frf{Fig::ProjCoefs} for $|p-p'| \in \{0,1,2,3,4\}.$  At the focal plane ($z=0$) the curves all coincide, since both the intensity is maximum and the Guoy phase is zero for all modes.  Away front the focal plane, we see decay of the coefficients due to the intensity falloff and oscillations that come from the Guoy phase mismatch.

	\begin{figure}[!t]
	\centering
    		\includegraphics[scale=0.95]{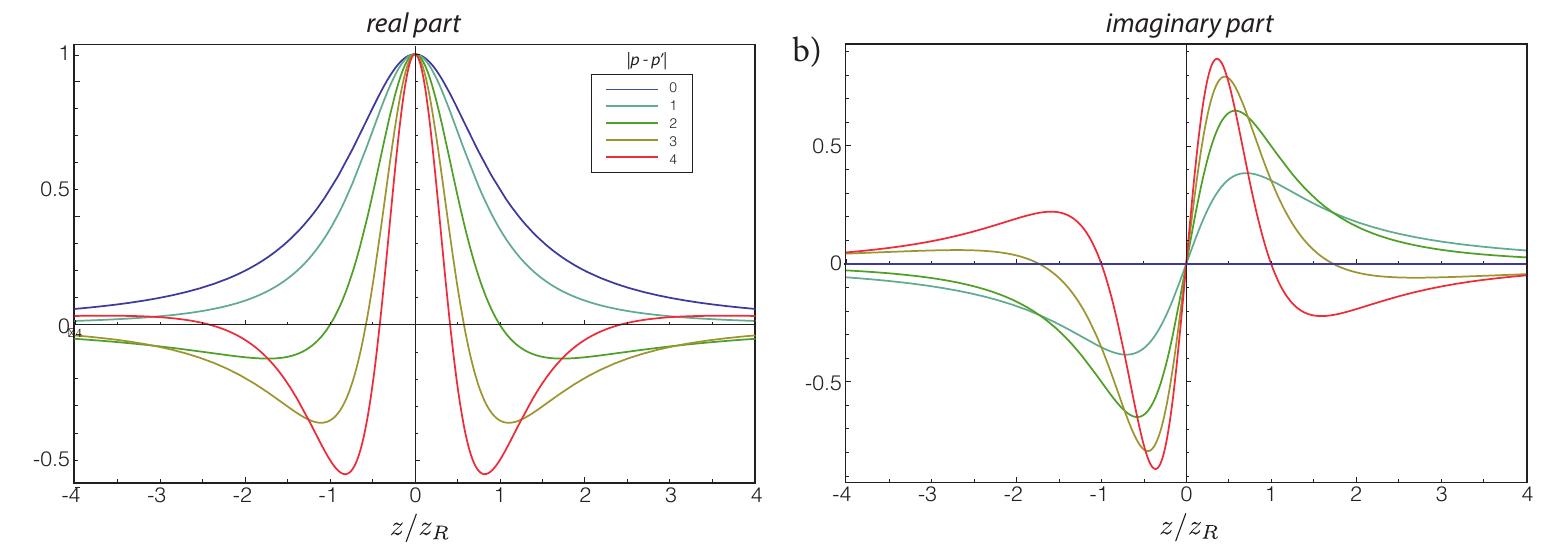}
       		 \caption[Multiplicative factor in the projection coefficients]{Real and imaginary parts of the factor that multiplies the integral in the expression for the projection coefficients, \erf{ApEq::Coefs}.  The longitudinal distance from the focal plane is plotted in Rayleigh ranges}  \label{Fig::ProjCoefs}     	
	\end{figure}

	 \addcontentsline{toc}{chapter}{{\bf Bibliography}}
	 \bibliographystyle{abbrv-alpha-letters-links}
	 \bibliography{BQBdissertationBib}
	 \twocolumn
	 \addcontentsline{toc}{chapter}{{\bf Index}}

\end{document}